\documentclass[twocolumn,superscriptaddress,amsmath,amssymb,floatfix,aps]{revtex4}
\usepackage{amsmath}    

\usepackage{graphicx}   
\usepackage{verbatim}   
\usepackage{color}      
\usepackage{subfigure}  
\usepackage{hyperref}   

\usepackage{amsmath}    
\usepackage{graphicx}   
\usepackage{verbatim}   
\usepackage{color}      
\usepackage{subfigure}  
\usepackage{hyphenat}
\usepackage[normalem]{ulem}
\usepackage{array}
\usepackage{booktabs}

\usepackage{titlesec}	
\usepackage{lettrine}	
\titleformat{\section}	
{\normalfont\scshape}{\thesection}{1em}{}	
	
\usepackage{color}      
\usepackage{soul}	
\newcommand{\dropcapfont}{\fontfamily{lmss}\bfseries\fontsize{26pt}{28pt}\selectfont}
\newcommand{\dropcap}[1]{\lettrine[lines=2,lraise=0.05,findent=0.1em, nindent=0em]{{\dropcapfont{#1}}}{}}

\begin{document}
\bibliographystyle{unsrt} 

\title{Robust replication initiation from coupled homeostatic mechanisms
}
\author{Mareike Berger} \affiliation{Biochemical Networks Group, Department of Living Matter, AMOLF, 1098 XG Amsterdam, The Netherlands}
 \author{Pieter
  Rein ten Wolde} \affiliation{Biochemical Networks Group, Department of Living Matter, AMOLF, 1098 XG Amsterdam, The Netherlands}

\begin{abstract}
	The bacterium \textit{Escherichia coli} initiates replication once per cell cycle at a precise volume per origin and adds an on average constant volume between successive initiation events, independent of the initiation size. Yet, a molecular model that can explain these observations has been lacking. Experiments indicate that \textit{E. coli} controls replication initiation via titration and activation of the initiator protein DnaA. Here, we study by mathematical modelling how these two mechanisms interact to generate robust replication-initiation cycles. We first show that a mechanism solely based on titration generates stable replication cycles at low growth rates, but inevitably causes premature reinitiation events at higher growth rates. In this regime, the DnaA activation switch becomes essential for stable replication initiation. Conversely, while the activation switch alone yields robust rhythms at high growth rates, titration can strongly enhance the stability of the switch at low growth rates. Our analysis thus predicts that both mechanisms together drive robust replication cycles at all growth rates. In addition, it reveals how an origin-density sensor yields adder correlations.
\end{abstract}

\maketitle
\begin{figure*}
	\centering
	\includegraphics[width=11.4cm,height=11.4cm]{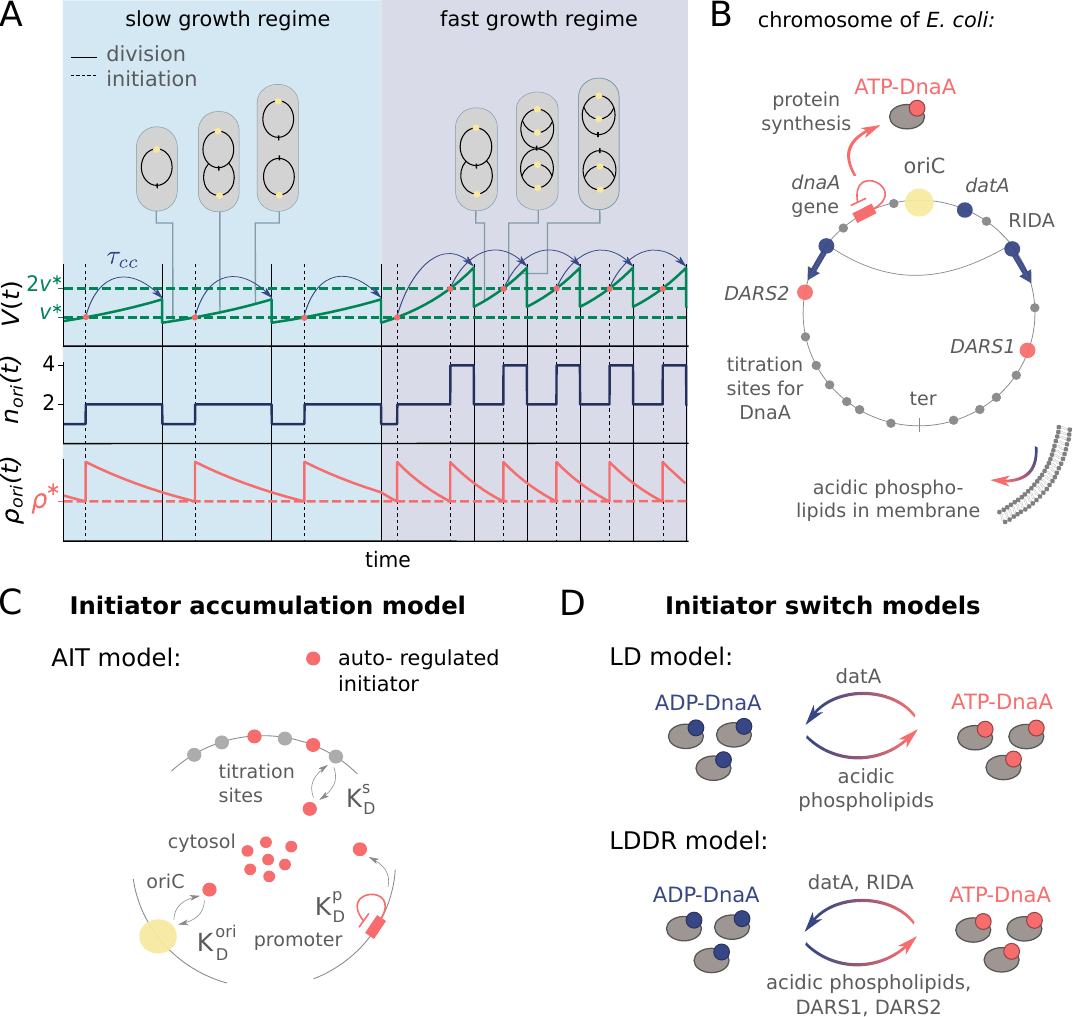}
	\caption{
		\textbf{We present two distinct models to elucidate the molecular mechanism by which \textit{E. coli} initiates replication at an on average constant volume per origin.}
		(A) The volume $V(t)$, the number of origins $n_{\rm ori}(t)$ and the origin density $\rho_{\rm ori}(t) = n_{\rm ori}(t)/ V(t)$ as a function of time. Initiating replication at a constant origin density $\rho^\ast$ (dashed red line) and division a constant time $\tau_{\rm cc}$ later (blue arrows) ensures that the cell initiates replication once per division cycle and that it maintains cell size homeostasis at slow (light blue regime) and fast (dark blue regime) growth rates. (B) Schematic representation of an \textit{E. coli} chromosome: Replication starts at the origin (oriC, yellow circle) and proceeds via two replication forks to the terminus (ter, grey bar). Replication is initiated by the ATP-bound form of the initiator protein DnaA. DnaA is activated via the acidic phospholipids in the cell membrane and via the two chromosomal sites \textit{DARS1} and \textit{DARS2}, and deactivated via the chromosomal site \textit{datA} and via regulatory inactivation of DnaA (RIDA), a process coupled to active DNA replication. DnaA also has a high affinity for titration sites (grey circles) located on the DNA. (C) Scheme of the AIT model: In \textit{E. coli}, the initiator DnaA (red circles) is negatively autoregulated with the dissociation constant $K_{\rm D}^{\rm p}$, and can bind both to the oriC and the titration sites with dissociation constants $K_{\rm D}^{\rm ori}$ and $K_{\rm D}^{\rm s}$, respectively. (D) Scheme of the initiator switch models: In the LD model, ATP-DnaA is mainly activated via the acidic phospholipids and deactivated via the site \textit{datA}. In the LDDR model, replication forks overlap and RIDA is the main deactivator in combination with the activators \textit{DARS1} and \textit{DARS2}.}
	\label{fig:fig_1}
\end{figure*}

\dropcap{T}{o} maintain stable cell cycles over many generations, living cells must coordinate DNA replication with cell growth and cell division. 
Intriguingly, in nutrient-rich environments, the model organism \textit{Escherichia coli} can even divide faster than the time it takes to replicate its entire chromosome \cite{Maaloe1966, Helmstetter1968507, Wallden2016, Si2017}. This apparent paradox was resolved by the model of Cooper and Helmstetter in which new rounds of replication are initiated before the previous round has finished \cite{Cooper1968} (Fig. \ref{fig:fig_1} A). Donachie then predicted that replication is initiated at a constant volume per origin $v^\ast$ \cite{Donachie1968}. 
Initiating replication at a constant origin density ensures that DNA replication is initiated once per cell cycle per origin, which is a necessary condition for maintaining stable cell cycles at all growth rates (Fig. \ref{fig:fig_1} A).
Recent experiments at the population level showed that the average initiation volume per origin $v^\ast$ varies within a $\sim 50 \%$ range over a tenfold change in the growth rate \cite{Zheng2020}. Moreover, single-cell measurements revealed that the initiation volume is one of the most tightly controlled cell-cycle parameters, varying by about $10\%$ for any measured growth rate \cite{Wallden2016,Si2019}. Yet, how the initiation volume is controlled so precisely, and what molecular mechanism gives rise to robust cell cycles over many generations remains despite extensive studies poorly understood \cite{Dewachter2018, Skarstad2013, Willis2017, Katayama2017, Riber2016}.

To obtain insight into the mechanisms that control DNA replication and cell division, fluctuations in cell size have been studied \cite{Campos2014, Taheri-Araghi2015}. These experiments revealed that cells obey an adder principle, which states that cells add an on average constant volume independent of the birth volume during each cell cycle. It has been proposed that cell division control is tightly coupled to the control over replication initiation \cite{Wallden2016, Amir2014, Ho2015}, via a sizer on replication initiation and a timer for cell division. Yet, recent experiments revealed the existence of two adders, one on cell division and the other on replication initiation, and that these two processes are more loosely coupled than hitherto believed \cite{Si2019, Witz2019, LeTreut2021, Witz:2020fb, Micali2018, Micali2018_2, Adiciptaningrum2015}. While these phenomenological observations are vital because they constrain any model on the molecular mechanism for initiation and cell division control, no such molecular model has yet been presented that is consistent with the experimental data. 

So far, two distinct classes of models for replication initiation control have been proposed. In the here called initiator accumulation models \cite{Barber2017, Ho2015, Amir2014, Sompayrac1973, Atlung1991, Basan2015, Hansen2018},
an initiator protein accumulates during the cell cycle proportional to the cell volume, and replication is initiated when a threshold amount per origin has accumulated. As a fixed amount of initiators per origin needs to be accumulated per replication cycle, models of this class are often seen as a mechanistic implementation of an adder \cite{Ho2015,Taheri-Araghi2015, Amir2014, Basan2015}.
Many variations of this idea with different degrees of detail have been proposed \cite{Amir2014, Sompayrac1973, Atlung1991, Basan2015}. Hansen et al. \cite{Atlung1991, Hansen2018} identified the initiator protein as the protein DnaA, which can be titrated away from the origin by DnaA boxes, high-affinity binding sites on the chromosome \cite{Katayama2017, Schaper1995}. This constant number of titration sites per chromosome sets the critical threshold number of initiator proteins required for initiating replication.

In this manuscript, we consider a mechanistic implementation of the initiator accumulation model (Fig. \ref{fig:fig_1} C). 
In \textit{E. coli}, the initiator protein DnaA is negatively autoregulated and can be bound to titration sites on the chromosome. Following Hansen et al. \cite{Atlung1991, Hansen2018}, we therefore consider a model in which the initiator is autoregulated, the Autoregulated Initiator-Titration (AIT) model. While the AIT model indeed gives rise to stable cell cycles at low growth rates, it exhibits reinitiation events at high growth rates. We thus argue that the initiator titration model is not sufficient to explain the experimental data on replication initiation in \textit{E. coli}.

The second class of models is based on a switch of the initiator protein DnaA between an active and an inactive form (Fig. \ref{fig:fig_1} D) \cite{Wallden2015, Dewachter2018, Katayama2017, Kasho2013, Kasho2014, Kurokawa1999}. In \textit{E. coli}, the initiator protein DnaA forms a tight complex with ATP or ADP, but only ATP–DnaA can initiate replication by forming a complex with the chromosomal replication origin (oriC) \cite{Katayama2010, Nishida:2002dp, Speck2001}. While the total DnaA concentration is approximately constant at different growth rates \cite{Hansen1991, Zheng2020}, the cellular level of ATP–DnaA oscillates over the course of the cell cycle, with a peak at the time of replication initiation \cite{Kurokawa1999, Fujimitsu2009, Katayama2001}.
It has been suggested that the oscillations in the fraction of ATP-DnaA in the cell are the key to understanding how replication is regulated in \textit{E. coli}, but a quantitative description that is consistent with experiments is currently lacking \cite{Grant2011, Katayama2017, Riber2016, Grimwalde2011, Kasho2014, Donachie2003, Grant2011}. Intriguingly, the level of ATP-DnaA is strictly regulated by multiple systems in the cell. DnaA is activated via acidic phospholipids in the cell membrane \cite{Sekimizu1988} and via two chromosomal regions called DnaA-Reactivation Sequence 1 (\textit{DARS1}) and \textit{DARS2} \cite{Fujimitsu2009, Kasho2014}, and deactivated via the chromosomal site \textit{datA} in a process called \textit{datA}-dependent DnaA-ATP Hydrolysis (DDAH) \cite{Kasho2013} and via a mechanism coupled to active DNA replication, called Regulatory Inactivation of DnaA (RIDA) \cite{Katayama2010, Kurokawa1999, Kato2001} (Fig. \ref{fig:fig_1} B). 
Deleting or modifying any of these systems can lead to untimely initiation, asynchrony of initiation, and changes in the initiation volume \cite{Ogawa2002, Camara2005, Riber2006, Riber2016, Kasho2013, Xia1995,Saxena2013}. 

To dissect how these multiple mechanisms give rise to a stable cell cycle, we first study the Lipid-\textit{DatA} (LD) model, which consists of only the acidic lipids and \textit{datA} (Fig. \ref{fig:fig_1} D). This model reveals that the interplay between a constant rate of activation and a rate of deactivation that depends on the origin density gives rise to stable cell cycles. Yet, at higher growth rates these two reactions alone, based on the experimentally estimated rates of activation and deactivation, respectively, are not sufficient to generate large amplitude oscillations in the fraction of ATP-DnaA. Simulations of our Lipid-\textit{DatA}-\textit{DARS1/2}-RIDA (LDDR) model show that in this regime, activation via \textit{DARS2} and deactivation via RIDA become essential.

Importantly, in our mean-field switch models, DNA replication is initiated at a threshold origin density and mechanistically they should arguably be qualified as a sizer. Yet, we show that a stochastic version of the switch model naturally gives rise to the experimentally observed adder correlations in the initiation volume \cite{Si2019, Witz2019}. Fluctuations in the components that control the DnaA activation switch (lipids, HdA, Fis, IHF) are transmitted from mother to daughter cells and this generates mother-daughter correlations in the initiation volume that can explain the observed adder correlations \cite{Si2019}. 

Finally, while the AIT model inevitably fails at higher growth rates, the LDDR model is less robust at low growth rates. Yet, combining titration with the activation switch yields robust DnaA oscillations over the full range of growth rates. We thus argue that \textit{E. coli} has evolved an elaborate set of mechanisms that act synergistically to create robust replication-initiation cycles at all growth rates.

\section*{Models and Results}
\noindent
\textbf{A titration-based mechanism is not sufficient to ensure stable cell cycles at high growth rates.}
Figure \ref{fig:fig_1} C shows the key ingredients of the AIT model. It consists of a negatively autoregulated initiator protein $p$, such that the change in copy number $N_{\rm p}$ is given by
\begin{equation}
\frac{dN_{\rm p}}{dt}= \frac{\tilde{\phi}_{\rm p}^0 \, \lambda \, V}{1+ \left( \frac{[p]}{K_{\rm D}^{\rm p}}\right)^{n}}
\label{eq:number_proteins_changes_volume}
\end{equation}
following the growing cell model of gene expression of Lin et al. \cite{Lin2018} \textcolor{blue}{(SI section S1)} with gene allocation density $\tilde{\phi}_{\rm p}^0$, dissociation constant of the promoter $K_{\rm D}^{\rm p}$, Hill coefficient $n$ and concentration of the initiator protein $[p]= N^{\rm f}/V$ in the cytoplasm.
The model also includes a number $N_{\rm s}$ of high-affinity titration sites that are distributed randomly on the chromosome \cite{Hansen2018, Roth1998}. The volume $V(t)$ of the cell grows exponentially, $V(t)= V_{\rm b} \, e^{\lambda \, t}$, where the growth rate $\lambda = \rm{ln}(2)/\tau_{\rm d}$, with cell-doubling time $\tau_{\rm d}$, is a model parameter. A new round of replication is initiated when the free initiator concentration $[p]$ reaches the dissociation constant for binding to the origin, $K_{\rm D}^{\rm ori}$. Based on the general growth law, the cell divides a constant cycling time $\tau_{\rm cc}$ after initiation of replication \cite{Si2017, Wallden2016}. This choice is convenient, as it directly couples cell division to replication, thus eliminating the need for implementing an
additional mechanism for cell division, yet does not affect our results, as we discuss later.
\begin{figure*}
	\centering
	\includegraphics[width=\linewidth]{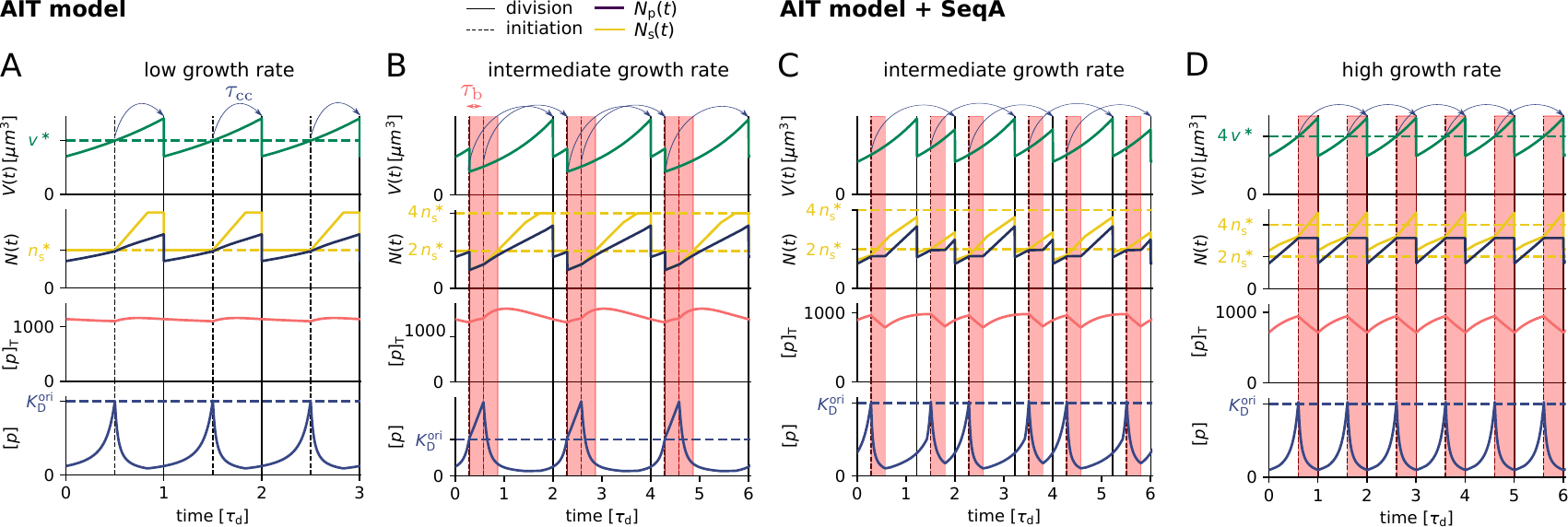}
	\caption{\textbf{While the AIT model ensures stable cell cycles at low growth rates (A), it gives rise to premature reinitiation events at high growth rates (B). Adding SeqA, which transiently blocks DnaA synthesis after replication initiation, prevents reinitiation events at high (D) but not at intermediate growth rates (C).} (A, B, C, D) The volume $V(t)$, the number of initiator proteins $N_{\rm p}(t)$ and titration sites $N_{\rm s}(t)$, the total concentration of initiator proteins $[p]_{\rm T}(t)$, and the concentration of initiator proteins in the cytoplasm $[p](t)$ as a function of time (in units of the doubling time of the cell $\tau_{\rm d}$) for $\tau_{\rm d}=2$~h (A), $\tau_{\rm d}=35$~min (B, C) and $\tau_{\rm d}=25$~min (D), respectively. (A) When the number of initiator proteins per origin $n_{\rm p}(t)$ exceeds the number of titration sites per origin $n_{\rm s}$ (yellow dashed line), the free concentration $[p](t)$ rapidly rises to reach the threshold concentration $K_{\rm D}^{\rm ori}$ (blue dashed line), initiating a new round of replication.  Due to the homogeneous distribution of titration sites on the chromosome of \textit{E. coli} and the constant DNA constant replication rate, the number of titration sites then increases linearly in time. At low growth rates, new titration sites are synthesized faster than new initiator proteins and the free concentration $[p](t)$ rapidly drops after initiation. After a fixed cycling time $\tau_{\rm cc}$ (blue arrows) the cell divides. The initiation volume per origin $v^\ast$ (green dashed line) at low growth rates is constant in time.
		(B) When the doubling time is however smaller than the time to replicate the entire chromosome, $\tau_{\rm d} <T_{\rm C}$, new proteins are synthesized faster than new titration sites are formed.
		After a short period $\tau_{\rm b}=10$~min (shaded red area) during which initiation at oriC is blocked via the protein SeqA, replication is reinitiated prematurely, dramatically raising the variation in the initiation volume (see Fig. \ref{fig:fig_5} C, green line). (C, D) Blocking also transiently DnaA synthesis via SeqA during $\tau_{\rm b}=10$~min (shaded red area) can prevent reinitiation at high (D), but not at intermediate growth rates (C).
		\textcolor{blue}{(See Table S1 for all parameters.)}
	}
	\label{fig:fig_2}
\end{figure*}

The AIT model generates stable cell cycles at low growth rates (Fig. \ref{fig:fig_2} A and \textcolor{blue}{Fig. S3}).
Because the dissociation constant of the initiator protein for the titration sites $K_{\rm D}^{\rm s}$ is smaller than that for the origin $K_{\rm D}^{\rm ori} > K_{\rm D}^{\rm s}$, the cytoplasmic initiator concentration $[p]$ \textcolor{blue}{(SI section S2B)} remains below the critical initiation threshold $K_{\rm D}^{\rm ori}$ as long as there are still unoccupied titration sites (Fig. \ref{fig:fig_2} A, lowest panel). Yet, when the total number of proteins $N_{\rm p}$ exceeds the total number of titration sites $N_{\rm s}$, the free concentration $[p]$ rapidly rises. 
When the free initiator concentration $[p]$ reaches the threshold $K_{\rm D}^{\rm ori}$, a new round of replication is initiated. New titration sites are now being synthesized faster than new proteins are being produced and therefore the free initiator concentration $[p]$ drops rapidly far below $K_{\rm D}^{\rm ori}$ (Fig. \ref{fig:fig_2} A, lowest graph). The cell then divides a constant time $\tau_{\rm cc}$ after replication initiation, during which the volume, the number of initiator proteins, and the number of titration sites are halved. In fact, in this mean-field description cell division does not change the concentrations of the components and it therefore does not affect the replication cycle. Importantly, this mechanism ensures stable cell cycles also in the presence of \textit{dnaA} expression noise and gives rise to the experimentally observed adder correlations in the initiation volume \textcolor{blue}{(Fig. S4)}.

At higher growth rates, the titration mechanism, however, breaks down. Because the titration sites are homogeneously distributed over the chromosome \cite{Roth1998, Hansen2018}, the rate at which new titration sites are formed after replication initiation is given by the DNA duplication rate, which is, to a good approximation, independent of the growth rate \cite{Si2017}. In contrast, the protein synthesis rate increases with the growth rate $\lambda$, see Eq. \ref{eq:number_proteins_changes_volume}. As a result, when the system enters the regime of overlapping replication forks, where the cell division time $\tau_{\rm d}$ is shorter than the time $T_{\rm C}$ to replicate the DNA  \textcolor{blue}{(SI section S2B4)}, the mechanism will fail to sequester the initiator after replication initiation, leading to premature reinitiation. Even when the system contains the protein SeqA, which protects the cell against immediate reinitiation events for `an eclipse period' of about 10 minutes \cite{Campbell:1990it, Lu:1994ee, Waldminghaus:2009em}, reinitiation happens as soon as this period is over (Fig. \ref{fig:fig_2} B). Also varying the number of titration sites and their affinity can not prevent premature reinitiation at high growth rates \textcolor{blue}{(Fig. S3)}; only placing the titration sites near the origin would \textcolor{blue}{(Fig. S3)}, but this is not consistent with experiments \cite{Roth1998, Hansen2018}. These observations show that the \textit{E. coli} replication cycle is not regulated via titration only.

Interestingly, experiments indicate that after replication initiation SeqA not only blocks the origin, preventing immediate reinitiation, but also transiently lowers the DnaA synthesis rate \cite{Campbell:1990it,Lu:1994ee,Waldminghaus:2009em}. The combination of periodic suppression of DnaA synthesis with DnaA titration enables robust DnaA rhythms at sufficiently high growth rates ($\lambda > 1.5$~h$^{-1}$) (Fig. \ref{fig:fig_2} D). But at lower growth rates, corresponding to longer doubling times, the effect of SeqA becomes weaker because of the fixed duration of the eclipse period. As a result, at intermediate growth rates ($1 > \lambda > 1.5$~h$^{-1}$) this combination cannot prevent premature reinitiation events (Fig. \ref{fig:fig_2} C). In this regime, another mechanism is needed.\\
\begin{figure}
	\centering
	\includegraphics[width=\linewidth]{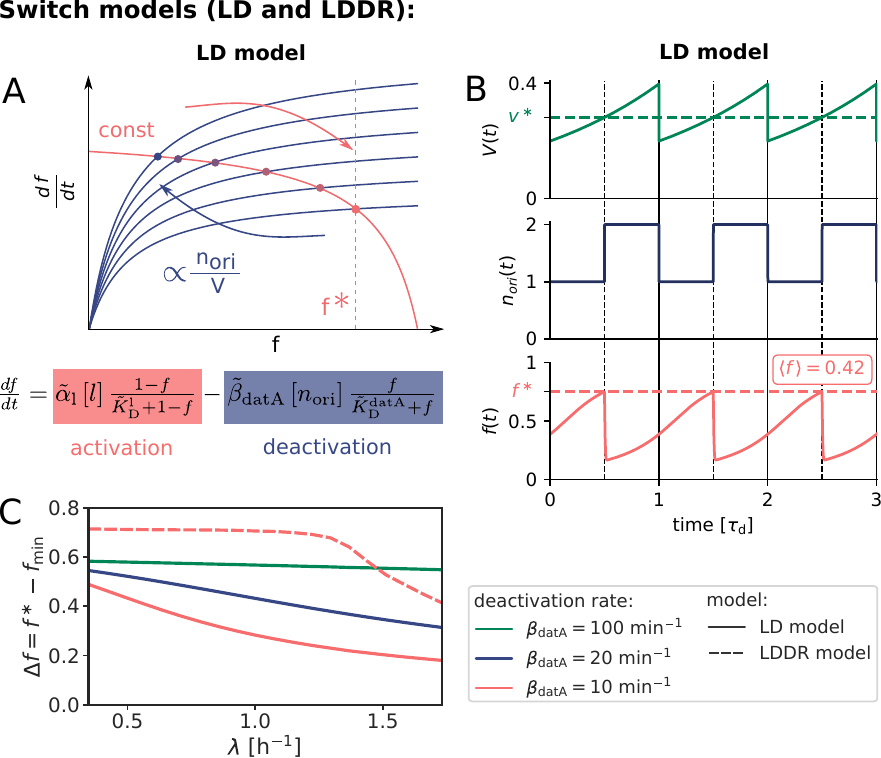}
	\caption{\textbf{An ultra-sensitive switch between ATP-DnaA and ADP-DnaA gives rise to stable cell cycles.} (A) LD model: The constant activation rate (red curve) and the origin density-dependent deactivation rate (blue curve) as a function of the active fraction of the initiator protein $f$ at different moments of the cell cycle. The steady-state active fractions are given by the intersection of the activation and deactivation rates (colorful dots) and when $f$ equals the critical initiator fraction $f^\ast$, replication is initiated. A doubling of the number of origins leads to a decrease of the active fraction $f$. (B) LD model: The volume of the cell $V(t)$, the number of origins $n_{\rm ori}(t)$ and the fraction of ATP-DnaA $f(t)$ from equation \ref{eq:switch_simple_fraction} as a function of time (in units of the doubling time $\tau_{\rm d}=2$~h). The average active fraction over one cell cycle $\langle f \rangle$ is indicated in red in the third panel. Replication is initiated at a critical initiator fraction $f^\ast$ (red dashed line) and the system gives rise to a constant initiation volume per origin $v^\ast$ over time (green dashed line). (C) The amplitude $\Delta f$ of the oscillations in the active fraction $f$ as a function of the growth rate for different magnitudes of the (de)activation rates ($\alpha_{\rm l} = 4.6 \times \beta_{\rm datA}$). The amplitude of the oscillations $\Delta f$ becomes small for biologically realistic values of the (de)activation rates in the LD model (red solid curve), but not in the LDDR model (red dashed line). \textcolor{blue}{(See Table S2 for all parameters and Fig. S8 for time traces of LDDR model.)}}
	\label{fig:fig_3}
\end{figure}
\\
\\
\noindent
\textbf{An ultra-sensitive switch between ATP- and ADP-DnaA  gives rise to an origin-density sensor.}
In the second class of models, not the total number of DnaA is the key variable that controls replication initiation, but the concentration or fraction of DnaA that is bound to ATP \cite{Wallden2015, Donachie2003}. While DnaA has a high affinity for both ATP and ADP, only ATP-DnaA can initiate replication at the origin \cite{Katayama2010, Nishida:2002dp, Speck2001}. The switch between these two states is controlled by several mechanisms, which, we will argue, play distinct roles in different growth-rate regimes. 

We first focus on the regime of slow growth in which the replication forks are non-overlapping. RIDA, a mechanism promoting ATP hydrolysis in a replication-coupled manner, becomes active upon
replication initiation, but, since there are no overlapping forks, is inactive {\em before} replication initiation \cite{Katayama2010}. The chromosomal locus \textit{datA} can hydrolyze ATP-DnaA via DDAH and is crucial for repressing untimely initiation events (Fig. \ref{fig:fig_1} B) \cite{Kasho2013}. The two chromosomal DNA regions \textit{DARS1} and \textit{DARS2} can regenerate ATP-DnaA from ADP-DnaA \cite{Katayama2010, Kasho2014, Riber2016}. The activating site \textit{DARS2} is reported to be only active at high growth rates and the activity of \textit{DARS1} was reported to be ten times weaker than \textit{DARS2} \textit{in vitro} \cite{Kasho2014}. In addition to \textit{DARS1/2}, both in vitro \cite{Sekimizu1988,Crooke1992,Castuma.1993} and in vivo \cite{Xia1995,Fingland2012} experiments indicate that acidic phospholipids can rejuvenate DnaA by promoting the exchange of ADP for ATP.
Moreover, as we show in \textcolor{blue}{section S3C3}, for a switch-based system, activation by \textit{DARS1/2} is not sufficient, while lipid-mediated activation of DnaA is vital to generate stable cell cycles. In summary, our modelling in combination with experiments indicates that at slow growth, the dominant DnaA cycle of the switch setting the initiation volume consists of activation by the phospholipids and deactivation via DDAH. This cycle forms the basis of the Lipid-\textit{DatA} (LD) model \textcolor{blue}{(SI section S3B)}. 

Since the growing cell model \cite{Lin2018} predicts that the total DnaA concentration is nearly constant in time while experiments show that it is nearly independent of the growth rate \cite{Zheng2020}, we make the simplifying assumption that the total DnaA concentration is strictly constant as a function of time and the growth rate. This allows us to focus on the fraction $f=[D_{\rm ATP}]/[D]_{\rm T}$ of DnaA that is bound to ATP \cite{Speck1999}. 
Exploiting that DnaA is predominantly bound to either ATP or ADP \cite{Katayama2010}, the change of the active fraction $f$ in the LD model is given by
\begin{align}
\frac{df}{dt}= & \frac{d[D]_{\rm ATP}}{dt} \, \frac{1}{[D]_{\rm T}} 
\\
= & \tilde{\alpha}_{\rm l} \,[l] \, \frac{1-f}{\tilde{K}_{\rm D}^{\rm l} + 1 - f} - \tilde{\beta}_{\rm datA} \, [n_{\rm ori}] \, \frac{f}{\tilde{K}_{\rm D}^{\rm datA}+f} + \lambda \, (1-f)
\label{eq:switch_simple_fraction}
\end{align}
with the constant, re-normalized activation and deactivation rates $\tilde{\alpha}_{\rm l}= \alpha_{\rm l}/ [D]_{\rm T}$ and $\tilde{\beta}_{\rm datA}= \beta_{\rm datA}/ [D]_{\rm T}$ and the Michaelis-Menten constants $\tilde{K}_{\rm D}^{\rm l}= K_{\rm D}^{\rm l}/ [D]_{\rm T}$ and $\tilde{K}_{\rm D}^{\rm datA}= K_{\rm D}^{\rm datA}/ [D]_{\rm T}$. Note that because \textit{datA} is located close to the origin, we have used here that their concentrations are equal. We further assume that the concentration of the acidic phospholipids $[l]$ is constant. The last term describes the effect of protein synthesis \textcolor{blue}{(Fig. S5)}. Since ATP is tenfold more abundant than ADP, new DnaA will predominantly bind ATP \cite{Katayama2010}. This term is however small at low growth rates ($\lambda \ll \tilde{\alpha}_{\rm l}, \tilde{\beta}_{\rm datA}$). 

Our switch model gives rise to stable cell cycles. The crux of the model is that while the activation rate is independent of the volume of the cell, the deactivation rate decreases with the volume because it is proportional to the density of oriC (Fig. \ref{fig:fig_3} A). The ATP-DnaA fraction $f(t)$ therefore increases with increasing volume $V(t)$ as the origin density decreases (Fig. \ref{fig:fig_3} B). When the critical initiator fraction $f^\ast= [D]_{\rm ATP}^\ast/[D]_{\rm T}$ is reached, replication is initiated. As soon as the origin and thus the site \textit{datA} have been replicated, the maximum of the deactivation rate doubles and the active initiator fraction $f$ decreases strongly, preventing reinitiation. As the cell continues to grow, the active initiator fraction rises again. This simple mechanism directly senses the origin density and ensures stable cell cycles (Fig. \ref{fig:fig_3} B).

At high (de)activation rates, the amplitude of the oscillations $\Delta f = f^\ast-f_{\rm min}$ is very large (Fig. \ref{fig:fig_3} C). At smaller and more biologically realistic rates ($\beta_{\rm datA} \approx 10~{\rm min}^{-1} $) \cite{Kasho2013} \textcolor{blue}{(see section S3A)}, 
the amplitude of the oscillations becomes very small especially at high growth rates (Fig. \ref{fig:fig_3} C); this continues to hold, even when the activation-deactivation system is deeper in the zero-order regime \textcolor{blue}{(Fig. S6)}. Such small amplitudes do not agree with the experiments \cite{Kurokawa1999} and are likely to be harmful, as even small fluctuations in the active fraction could result in untimely initiation of replication. 
\\
\\
\noindent
\textbf{LDDR model with all known activators and deactivators allows for larger amplitude oscillations even at high growth rates.}
Because at biologically realistic (de)activation rates the LD model fails to generate large amplitude oscillations in the active DnaA fraction at high growth rates, the question arises how the cell cycle is regulated in this regime. Interestingly, in the fast growth regime $\lambda > \rm{ln(2)}/ T_{\rm C} \approx 1.04$/h, where the doubling time $\tau_{\rm d}$ is shorter than the time to replicate the entire chromosome $T_{\rm C}$, replication is still proceeding when a new round of replication is initiated. This means that at the moment of replication initiation, the deactivation mechanism RIDA, which is associated with active replication forks, is active \cite{Katayama1998}. Importantly, since RIDA is a potent deactivator \cite{Camara2005}, its activity must be balanced by another activation mechanism to maintain a roughly constant initiation volume independent of the growth rate \cite{Zheng2020, Si2017, Elf2007}. We argue that this is the principal role of \textit{DARS2}. 

We therefore included the effects of RIDA and DARS1/2 in our full Lipid-\textit{DatA}-\textit{DARS1/2}-RIDA (LDDR) model
\textcolor{blue}{(SI section S3C)}. The RIDA deactivation rate is proportional to the total number of active replisomes. The activation rates of \textit{DARS1} and \textit{DARS2} are proportional to the copy numbers of their loci, which are located in the middle of the chromosome and are replicated at constant times after replication initiation \textcolor{blue}{(see Fig. S7)}. The LDDR model also takes into account the temporal regulation of the activities of DDAH and \textit{DARS2} via the
Integrating Host Factor (IHF) \cite{Katayama2017, Kasho2013, Kasho2014, Riber2016} \textcolor{blue}{(see Fig. S7)}.

The LDDR model gives rise to stable cell cycles at all growth
rates. Moreover, in contrast to the LD model, the LDDR model gives rise to large amplitude oscillations at all
growth rates, even for realistic parameter values (Fig. \ref{fig:fig_3} C) \textcolor{blue}{(see Fig. S8 for time traces)}. This is because after a new round of replication, the RIDA deactivation rate is raised immediately while the activation rates of \textit{DARS1/2} are increased only later, after the loci have been duplicated. This differential temporal dependence of the activation and deactivation rates is key to establishing large-amplitude oscillations at all growth rates.
\\
\\
\noindent
\textbf{A stochastic model can recover the experimentally observed adder correlations in the initiation volume per origin.}
In the titration-based system, a new round of replication is initiated when the number of DnaA proteins that have been accumulated since the last initiation event equals roughly the number of titration sites, irrespective of the previous initiation volume; moreover, DnaA proteins are accumulated proportionally to the volume of the cell. These two elements together naturally give rise to adder correlations \textcolor{blue}{(see section S2B6 and Fig. S4)}.
Yet, our switch model is a sizer at the mean-field level: replication is initiated when the origin density reaches a critical threshold. Do the experimentally observed adder correlations \cite{Si2019, Witz2019} rule out our switch model?

\begin{figure}
	\centering
	\includegraphics[width=\linewidth]{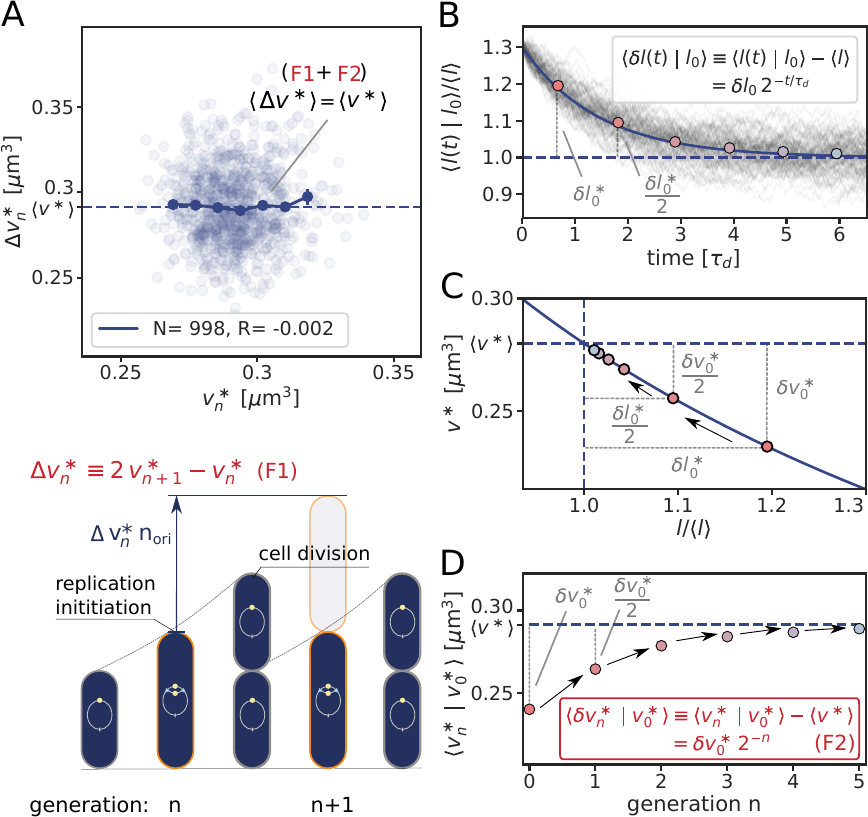}
	\caption{\textbf{Fluctuations in the switch components can give rise to the experimentally observed adder correlations in the initiation volume per origin $v^\ast$, illustrated using the LD model with lipid concentration fluctuations (Eq. \ref{eq:lipid_concentration}).} (A) The added volume per origin between successive initiation events, $\Delta v^\ast_{\rm n}= 2 \, v^\ast_{\rm n+1}-v^\ast_{\rm n}$, is independent of the initiation volume $v^\ast_{\rm n}$ per origin and on average equal to the average initiation volume, $\langle\Delta v^\ast \rangle=\langle v^\ast\rangle$, as expected for an initiation volume adder. (B) Lipid-concentration fluctuations $l(t)\equiv [l](t)$ regress to the mean on a timescale $\tau_{\rm d}=\ln(2)/\lambda$ set by the growth rate $\lambda$, such that an initial perturbation $l_0-\langle l\rangle$ is halved every subsequent cell cycle. (C) The initiation volume depends on the lipid concentration \textcolor{blue}{(Eq. S35 and Fig. S10)}. (D) The initiation volume relaxes on the same timescale $\tau_{\rm d}$ as the lipid concentration, such that a perturbation $v_0^\ast-\langle v^\ast \rangle$ is halved every cell cycle, giving rise to adder correlations. In (A) the dark blue line shows the mean of the binned data and the error bars represent the standard error of the mean (SEM) per bin. The number of data points $N$ and the Pearson correlation coefficient $R$ are indicated. The model includes an eclipse period of about 10 minutes following replication initiation to prevent immediate reinitiation. \textcolor{blue}{(See Table S2 for all parameters.)}}
	\label{fig:fig_4}
\end{figure}

To address this question, we systematically studied the effect of fluctuations in the individual components of our switch model. Consider fluctuations in the lipid concentration, modelled as
\begin{equation}
\frac{d[l]}{dt} = \alpha - \lambda \, [l] + \xi(t),
\label{eq:lipid_concentration}
\end{equation}
where $\alpha$ is the production rate, the second term describes the effect of dilution set by the growth rate $\lambda$ and $\xi(t)$ models the noise resulting from protein production and partitioning upon cell division \textcolor{blue}{(SI section S3D)}. 
Fig. \ref{fig:fig_4} illustrates our findings using the LD model, but \textcolor{blue}{Fig. S11} shows that the principal result also holds for the full LDDR model: the added initiation volume between consecutive initiation events $\Delta v^\ast_{\rm n}= 2 \, v^\ast_{\rm n+1}-v^\ast_{\rm n}$ is indeed independent of the volume at initiation $v^\ast_{\rm n}$, in agreement with experiments \cite{Witz2019, Si2019}.

\begin{figure}
	\centering
	\includegraphics[width=\linewidth]{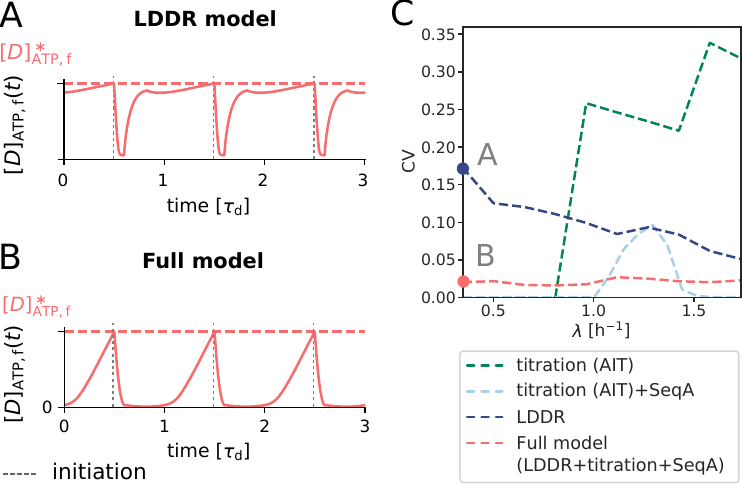}
	\caption{\textbf{Combining the DnaA activation switch with titration and SeqA generates robust replication-initiation cycles over a wide range of growth rates.} (A, B) The concentration of free ATP-DnaA $[D]_{\rm ATP, f}(t)$ as a function of time (in units of the doubling time $\tau_{\rm d}$) for $\lambda = 0.35$~h$^{-1}$ as indicated in panel C. The dashed red line is the critical free ATP-DnaA concentration $[D]_{\rm ATP, f}^\ast$ at which replication is initiated. While in the LDDR model the free ATP-DnaA fraction is high during a large fraction of the cell cycle (A, \textcolor{blue}{see also section S3C2}), combining it with titration sites and SeqA gives rise to a much sharper increase of the free ATP-DnaA concentration at low growth rates (B). (C) The coefficient of variation ${\rm CV}= \sigma /\mu$ with the standard deviation $\sigma$ and the average initiation volume $\mu=\langle v^\ast \rangle$ as a function of the growth rate for different models in the presence of noise in the lipid concentration. Even in the absence of biochemical noise in DnaA synthesis, the titration model gives rise to a very high CV at high growth rates, due to premature reinitiation (Fig. \ref{fig:fig_2} B). Adding SeqA to the titration model can reduce the CV at high, but not at intermediate growth rates (Fig. \ref{fig:fig_2} C). The large coefficient of variation in the LDDR model at low growth rates is reduced significantly by the titration sites. Conversely, the LDDR model prevents the reinitiation events that inevitably occur at intermediate growth rates in the AIT+SeqA model. Combining DnaA activation with titration thus enhances the robustness of replication initiation at all growth rates, also in the presence of noise in DnaA synthesis \textcolor{blue}{(Fig. S13)}. All models include an eclipse period of about 10 minutes following replication initiation to prevent immediate reinitiation \cite{Campbell:1990it,Lu:1994ee,Waldminghaus:2009em}.  \textcolor{blue}{(See Table S2 for all parameters.)}}
	\label{fig:fig_5}
\end{figure}

The concentrations of cellular components will fluctuate
inevitably, and unless the components are degraded actively or produced with negative feedback control, the fluctuations will persist over several generations, regressing to the mean on a timescale set by the growth rate (Fig. \ref{fig:fig_4} B). The components that control the threshold of the DnaA activation switch are no exception to this rule. Moreover, their concentration fluctuations will give rise to fluctuations in the initiation volume $v^\ast$ (Fig. \ref{fig:fig_4} C) that, to a good approximation, relax on the same timescale because (de)activation is fast compared to the growth rate and the mapping between these components and the initiation volume is roughly linear. If this timescale is set by the growth rate, then deviations of $v^\ast$ from its mean are on average halved every cell cycle (Fig. \ref{fig:fig_4} D), and this gives rise to adder correlations (SI section S3D) \cite{Si2019}. Fluctuations in switch components that relax with the growth rate, be they lipids or proteins that modulate the activity of \textit{datA}, RIDA, or \textit{DARS1/2}  like IHF and Hda \cite{Katayama2017, Kasho2013, Kasho2014, Riber2016}, thus give rise to adder correlations \textcolor{blue}{(Fig. S12)}.  
\\
\\
\noindent
\textbf{Coupling titration with DnaA activation enhances robustness.} All our systems are stable in the presence of biochemical noise. The concentrations do not diverge, also not in the titration-based system at high growth rates (Fig. \ref{fig:fig_2}). Yet, the precision of replication initiation differs markedly between the respective models, see Fig. \ref{fig:fig_5}. The protein synthesis and the titration-site formation rate scale differently with the growth rate, which means that a titration-based mechanism inevitably breaks down at sufficiently high growth rates, causing premature reinitiation events and a dramatic rise of the coefficient of variation (CV) in the initiation volume; even in the absence of any biochemical noise, the CV becomes larger than that reported experimentally \cite{Wallden2016, Si2019} (Fig. \ref{fig:fig_5} C). The transient suppression of DnaA synthesis by SeqA after replication initiation can prevent these premature reinitiation events, but only at high growth rates: at intermediate growth rates, the CV of a system based on only titration and SeqA still rises strongly. This indicates that the activation switch is essential  (Fig. \ref{fig:fig_5} C). But could it be sufficient?  Our modelling predicts it could because the LDDR model can generate robust oscillations at all growth rates. Yet, our modelling also predicts that titration helps the switch by shaping the oscillations in the {\em free concentration} of ATP-bound DnaA (Fig. \ref{fig:fig_5} A and B), such that the precision of replication initiation in the presence of noise is significantly enhanced (Fig. \ref{fig:fig_5} C). In section S4A2 we show that a concentration cycle, as generated by titration and SeqA, can generically enhance an activation cycle, as driven by the switch, by increasing the steepness of the oscillations; this tames the propagation of fluctuations in the free concentration of active DnaA to the initiation volume \textcolor{blue}{(Fig. S14)}. Combining the switch with titration can thus protect the system against fluctuations in the switch components.

\section*{Discussion}
\noindent
While the two mechanisms of titration and protein activation have so far been mostly studied independently \cite{Hansen1991, Hansen2018, Si2019, Kurokawa1999, Kasho2013, Kasho2014, Fujimitsu2009}, our manuscript indicates that the robustness arises from the coupling of the two. Interestingly, recent experiments, which show that replication is neither controlled by titration only nor by a DnaA activation switch only, support this prediction from our model \cite{Knoppel2021}. Moreover, the idea that coupling an oscillation in the concentration with an oscillation in the fraction gives rise to more robust rhythms than either oscillation alone, is very generic. Our results are thus expected to apply to any cell-cycle control system that combines titration with protein activation or modification. This finding is of particular interest given the recent observation that also higher organisms employ not only protein modification but also titration for cell-cycle control \cite{DArio2021, Rhind2021}. In fact, the evidence is accumulating that also oscillatory systems, most notably circadian clocks in cyanobacteria and higher organisms, derive their robustness to changes in the growth rate by intertwining a protein modification cycle with a protein concentration cycle \cite{Zwicker.2010,Teng:2013cf,Larrondo:2015cb}.


The mechanisms of titration and activation belong to distinct classes of replication initiation control. The titration-based AIT model is an example of an initiator accumulation model, in which an initiator protein needs to accumulate to a threshold number to initiate replication \cite{Sompayrac1973, Si2017, Hansen1991, Si2019, Hansen2018}. 
In contrast, the DnaA activation switch is an example of a push-pull network in which the regulator switches between an inactive and an active state. Conceptually, this switch model is different from the accumulation model because replication is triggered at a critical concentration or fraction and not at a critical number of accumulated initiator proteins. In the switch model, the concentration of ATP-DnaA is set by the balance between DnaA activation and deactivation. Because the (de)activation rates depend on the origin density, the critical initiator concentration maps onto a critical origin density for replication initiation. This switch system is thus a bonafide origin-density sensor.

In recent years, single-cell tracking data have revealed that not only {\it E. coli} but also other evolutionary divergent organisms like {\it Bacillus subtilis} \cite{Taheri-Araghi2015}, {\it Caulobacter crescentus} \cite{Campos2014}, the archaeon {\it Halobacterium salinarum} \cite{Eun2018}, and even budding yeast \cite{Soifer2016}, obey a division adder principle. Our study gives a new perspective on the question whether a cell cycle is controlled via a sizer or adder. While the titration mechanism naturally qualifies as an adder, our switch model should be characterised as a sizer at the mean-field level: the mechanism is based on sensing the origin density. Yet, the inevitable fluctuations in the components that control the density threshold for replication give rise to adder correlations. This idea is general and likely applies to other organisms that obey the adder principle: adder behavior may result from size sensing. Our prediction could be tested by measuring the critical active DnaA concentration for replication initiation and how its fluctuations relax. Since ATP binding induces a conformational switch of DnaA \cite{Erzberger:2006jd}, developing a FRET-based ATP-DnaA sensor may be feasible.

While our models are built on a wealth of data, they all make the simplifying assumption that the cell divides a constant time $\tau_{\rm cc}$ after replication initiation, independent of the growth rate. Experiments indicate, however, that this is an oversimplification \cite{Michelsen:2003ku, Adiciptaningrum2015, Wallden2016, Micali2018,Micali2018_2, Si2019, Witz2019} and that cell division is more loosely coupled to replication initiation \cite{Si2019, Witz2019}. Importantly, our results on replication initiation control are robust to the assumption of a constant $\tau_{\rm cc}$, because on average cell division does not change the densities of the components. Indeed, while this assumption will affect the correlations between the cell volume at birth and the initiation volume, it does not change the correlations between the initiation volume and the volume added until the next initiation event \textcolor{blue}{(Fig. S19)}.

Our model is supported by many experimental observations. Of particular interest are mutants in which the (de)activation mechanisms are modified or even deleted, because these allow us to test the prediction that replication initiation is controlled by the activation switch \textcolor{blue}{(SI section S4B1)}. Naturally, our model can reproduce the observations on which it is built: deleting {\it datA} \cite{Kasho2013} and deactivating RIDA \cite{Katayama2010,Kurokawa1999, Kato2001, Kasho2013} raises the active fraction of DnaA, while deleting {\it DARS1/2} \cite{Fujimitsu2009, Kasho2014} reduces it \textcolor{blue}{(Fig. S16)}. Our model then predicts that impeding activation increases the average volume per origin, while weakening deactivation has the opposite effects. Many experiments support these predictions: 
deleting {\it DARS1} and/or {\it DARS2} increases the initiation volume per origin \cite{Frimodt-Moller2015}, while deleting {\it datA} decreases it \cite{Frimodt-Moller2015}. Our model cannot only reproduce these observations, but also the effect of combinations of deletions of these chromosomal loci on the initiation volume \textcolor{blue}{(Fig. S16)}. Moreover, it can describe how the initiation volume per origin changes when {\it datA} or {\it DARS2} is translocated towards the terminus \cite{Kitagawa1998, Frimodt-Moller2016, Inoue2016} \textcolor{blue}{(Fig. S16)}. In addition, our model can reproduce the observation that increasing the number of titration sites via multicopy plasmids increases the initiation volume per origin \cite{Christensen1999} \textcolor{blue}{(section S4B2)}, while increasing the DnaA concentration reduces it \cite{Si2017, Knoppel2021, Hill2012, Atlung1993} (section S4B3, Fig. S17). Taken together, these experiments support the idea that replication initiation is controlled by both titration and DnaA activation.

Intriguingly, the relative position of
\textit{DARS2} with respect to the origin and the terminus is conserved in various genomes of different sizes and strains \cite{Frimodt-Moller2015}, suggesting it plays an important role. Our modelling provides the following rationale:
In the high growth-rate regime of overlapping replication forks, \textit{DARS2} not only serves to balance the strong deactivation by RIDA to yield a roughly constant initiation volume, but also needs to generate oscillations in concert with RIDA. Because the activities of both \textit{DARS2} and RIDA are proportional to the origin density,
\textit{DARS2} can only play this dual role if its position meets two constraints: On the one hand, the activity of \textit{DARS2} should rise as late as possible in order to push the active initiator fraction down right after initiation. On the other hand, to achieve a nearly constant initiation volume independent of the growth rate, the activity of \textit{DARS2} must be high to counteract RIDA before the next initiation event; indeed, moving {\it DARS2} towards the terminus increases the initiation volume \cite{Frimodt-Moller2016, Inoue2016} \textcolor{blue}{(Fig. S16I)}. The shortest period until replication is set by the highest doubling time of \textit{E. coli}, $\tau_{\rm d} \approx 18$~min. The position of \textit{DARS2} in the middle of the chromosome ($\tau_{\rm d2} \approx 16$~min) therefore naturally results from our model.

Arguably the most enigmatic element of our model is the role of
the lipids in rejuvenating DnaA. In vitro experiments have shown
that acidic phospholipids in the cell membrane promote
dissociation of nucleotides from DnaA very effectively
\cite{Sekimizu1988}, and can restore replication activity of DnaA
bound to ADP \cite{Crooke1992, Castuma.1993}. Depleting acidic
phospholipids \textit{in vivo} can lead to growth arrest
\cite{Xia1995} and inhibit initiation at oriC
\cite{Fingland2012}. These experiments support the idea that lipids
can reactivate DnaA by promoting the exchange of bound ADP for
ATP. On the other hand, it has been observed that the lethal effect
of a {\it pgsA} null mutation, which causes a complete lack of the
major acidic phospholipids, is alleviated by mutations that change
the membrane structure \cite{Shiba2004}. More recently, it has been
reported that while downregulating {\it pgsA} reduced the growth
rate, the initiation volume was not significantly altered
\cite{Camsund2020}. We have therefore also studied models in which
lipid-mediated DnaA is absent \textcolor{blue}{(SI section S5A)}. Our modelling
predicts that lipid-mediated DnaA activation is essential for the
switch \textcolor{blue}{(Fig. S20A-D)}. The capacity of the switch to act as an origin-density sensor hinges on the idea that the activation and deactivation rates scale differently with the origin density. Without the lipids, only protein synthesis remains as an activation mechanism that does not scale with the origin density (Eq.~\ref{eq:switch_simple_fraction}). Consequently, to obtain a stable switch-based system, the rates of all other (de)activation mechanisms must be comparable to or smaller than the growth rate. This
dramatically lowers the amplitude of the oscillations. The full model, which combines the switch with titration and SeqA, is, however, surprisingly resilient to the removal of lipids, although the latter does compromise the precision of replication initiation \textcolor{blue}{(Fig. S20E-G)}. It has also been suggested that DnaA rejuvenation is contingent on oriC \cite{Crooke1992} (SI section S5B). However, a lipid-mediated DnaA activation rate that scales with the origin density effectively reduces the datA-mediated deactivation rate; this yields a switch that behaves similarly to that of the lipid-devoid system, because protein synthesis is again the only DnaA activation mechanism that is independent of the origin density. In summary, lipids enhance replication initiation, but only if their effect is independent of the origin density.

Perhaps the most non-trivial prediction of our model is that the relaxation timescale of the switch components governs whether the switch generates adder or sizer correlations in the inter-initiation volume. The experiments of Si et al. provide strong support for this prediction: by expressing DnaA in an oscillatory fashion, the adder is turned into a sizer \cite{Si2019}, precisely as our model predicts \textcolor{blue}{(Fig. S18)}. 


Our modelling predicts that negative autoregulation does not play a direct role in replication initiation. This is supported by recent experiments, which show that the average initiation volume and precision of replication initiation are only weakly affected in strains with constitutive {\it dnaA}  expression\cite{Knoppel2021}. Following Hansen et al. \cite{Hansen1991}, we believe that negative autoregulation only plays an indirect role, by setting the growth-rate dependence of the DnaA concentration. Experiments have revealed that the total DnaA concentration varies with the growth rate, anticorrelating with the initiation volume \cite{Zheng2020}. However, the variation of both the total DnaA concentration and the initiation volume is rather weak, i.e. about 50\% over a tenfold change of the growth rate \cite{Zheng2020}. It seems likely that negative autoregulation is crucial for constraining the growth-rate dependence of the total DnaA concentration  \cite{Klumpp2009, Scott2010} and hence the initiation volume  \cite{Wallden2016, Si2017}. How negative autoregulation with a differential sensitivity of the DnaA promoter to DnaA-ATP and DnaA-ADP \cite{Speck1999, Sclavi2013} and titration conspire to shape the growth-rate dependence of the DnaA concentration and the initiation volume, we leave for future work.

Another open question remains why \textit{E. coli} has evolved two different switch systems, Lipid-DatA (LD) and DARS1/2-RIDA (DR). In principle, a switch based on activating lipids and deactivating \textit{datA} would be sufficient to control replication initiation at all growth rates. Yet, to ensure high amplitude oscillations in the active DnaA fraction at high growth rates, the (de)activation rates would have to be higher than observed (Fig. \ref{fig:fig_3}~C). This would require higher turnover rates of ATP, which may not be achievable when the growth rate is low. Our model thus suggests that \textit{E. coli} has evolved a slow system to control the initiation volume at low growth rates, the lipids-\textit{datA} system, and then switches on a faster, more energy-consuming system at higher growth rates, based on RIDA and \textit{DARS2}.

Finally, our model predicts that in the regime of non-overlapping replication forks it should be possible to move the system from a switch-dominated regime to a titration-based one by increasing the number of titration sites or decreasing the basal synthesis rate of DnaA. Our model predicts that the dependence of the initiation volume on the number of titration sites or basal synthesis rate exhibits a marked, characteristic crossover when the system transitions between these two regimes \textcolor{blue}{(Fig. S15)}. This is a strong prediction that could be tested experimentally.

We thank Lorenzo Olivi, Sander Tans, Suckjoon Jun, Erik van Nimwegen and Johan Elf for a careful reading of the manuscript. This work is part of the Dutch Research Council (NWO) and was performed at the research institute AMOLF.

{\bf Code Availability} The code is publicly available at the Github repository MareikeBerger/Cellcycle via https://github.com/MareikeBerger/Cellcycle or https://zenodo.org/record/5913722.

{\bf Data Availability} The datasets generated during and analysed during the current study are available at Zenodo via https://zenodo.org/record/5911070.

\bibliography{Switch_Resub}

\begin{thebibliography}{100}

\bibitem{Maaloe1966}
O.~Maal{\o}e and N.~O. Kjeldgaard.
\newblock {\em Control of macromolecular synthesis : a study of DNA, RNA, and
  protein synthesis in bacteria}.
\newblock New York (N.Y.) : Benjamin, 1966.

\bibitem{Helmstetter1968507}
Charles~E. Helmstetter and Stephen Cooper.
\newblock {DNA synthesis during the division cycle of rapidly growing
  Escherichia coli Br}.
\newblock {\em Journal of Molecular Biology}, 31(3):507--518, 1968.

\bibitem{Wallden2016}
Mats Wallden, David Fange, Ebba~Gregorsson Lundius, {\"{O}}zden Baltekin, and
  Johan Elf.
\newblock {The Synchronization of Replication and Division Cycles in Individual
  E. coli Cells}.
\newblock {\em Cell}, 166(3):729--739, 2016.

\bibitem{Si2017}
Fangwei Si, Dongyang Li, Sarah~E. Cox, John~T. Sauls, Omid Azizi, Cindy Sou,
  Amy~B. Schwartz, Michael~J. Erickstad, Yonggun Jun, Xintian Li, and Suckjoon
  Jun.
\newblock {Invariance of Initiation Mass and Predictability of Cell Size in
  Escherichia coli}.
\newblock {\em Current Biology}, 27(9):1278--1287, 2017.

\bibitem{Cooper1968}
Stephen Cooper and Charles~E. Helmstetter.
\newblock {Chromosome replication and the division cycle of Escherichia
  coliBr}.
\newblock {\em Journal of Molecular Biology}, 31(3):519--540, feb 1968.

\bibitem{Donachie1968}
W~D Donachie.
\newblock {Relationship between Cell Size and Time of Initiation of DNA
  Replication}.
\newblock {\em Nature}, 219:1077--1079, sep 1968.

\bibitem{Zheng2020}
Hai Zheng, Yang Bai, Meiling Jiang, Taku~A. Tokuyasu, Xiongliang Huang, Fajun
  Zhong, Yuqian Wu, Xiongfei Fu, Nancy Kleckner, Terence Hwa, and Chenli Liu.
\newblock {General quantitative relations linking cell growth and the cell
  cycle in Escherichia coli}.
\newblock {\em Nature Microbiology}, 5(8):995--1001, 2020.

\bibitem{Si2019}
Fangwei Si, Guillaume {Le Treut}, John~T. Sauls, Stephen Vadia, Petra~Anne
  Levin, and Suckjoon Jun.
\newblock {Mechanistic Origin of Cell-Size Control and Homeostasis in
  Bacteria}.
\newblock {\em Current Biology}, 29(11):1760--1770.e7, 2019.

\bibitem{Dewachter2018}
Liselot Dewachter, Natalie Verstraeten, Maarten Fauvart, and Jan Michiels.
\newblock {An integrative view of cell cycle control in Escherichia coli}.
\newblock {\em FEMS Microbiology Reviews}, 42(2):116--136, 2018.

\bibitem{Skarstad2013}
Kirsten Skarstad and Tsutomu Katayama.
\newblock {Regulating DNA replication in bacteria.}
\newblock {\em Cold Spring Harbor perspectives in biology}, 5(4):a012922, apr
  2013.

\bibitem{Willis2017}
Lisa Willis and Kerwyn~Casey Huang.
\newblock {Sizing up the bacterial cell cycle}.
\newblock {\em Nature Reviews Microbiology}, 15(10):606--620, 2017.

\bibitem{Katayama2017}
Tsutomu Katayama, Kazutoshi Kasho, and Hironori Kawakami.
\newblock {The DnaA cycle in Escherichia coli: Activation, function and
  inactivation of the initiator protein}.
\newblock {\em Frontiers in Microbiology}, 8(DEC):1--15, 2017.

\bibitem{Riber2016}
Leise Riber, Jakob Frimodt-M{\o}ller, Godefroid Charbon, and Anders
  L{\o}bner-Olesen.
\newblock {Multiple DNA Binding Proteins Contribute to Timing of Chromosome
  Replication in E. coli}.
\newblock {\em Frontiers in Molecular Biosciences}, 3(June):1--9, 2016.

\bibitem{Campos2014}
Manuel Campos, Ivan~V. Surovtsev, Setsu Kato, Ahmad Paintdakhi, Bruno Beltran,
  Sarah~E. Ebmeier, and Christine Jacobs-Wagner.
\newblock {A constant size extension drives bacterial cell size homeostasis}.
\newblock {\em Cell}, 159(6):1433--1446, 2014.

\bibitem{Taheri-Araghi2015}
Sattar Taheri-Araghi, Serena Bradde, John~T. Sauls, Norbert~S. Hill, Petra~Anne
  Levin, Johan Paulsson, Massimo Vergassola, and Suckjoon Jun.
\newblock {Cell-size control and homeostasis in bacteria}.
\newblock {\em Current Biology}, 25(3):385--391, 2015.

\bibitem{Amir2014}
Ariel Amir.
\newblock {Cell size regulation in bacteria}.
\newblock {\em Physical Review Letters}, 112(20):1--5, 2014.

\bibitem{Ho2015}
Po-Yi Ho and Ariel Amir.
\newblock {Simultaneous Regulation of Cell Size and Chromosome Replication in
  Bacteria}.
\newblock {\em Frontiers in microbiology}, 6:662, 07 2015.

\bibitem{Witz2019}
Guillaume Witz, Erik van Nimwegen, and Thomas Julou.
\newblock {Initiation of chromosome replication controls both division and
  replication cycles in \textit{E. coli} through a double-adder mechanism}.
\newblock {\em eLife}, 8:e48063, nov 2019.

\bibitem{LeTreut2021}
Guillaume Le~Treut, Fangwei Si, Dongyang Li, and Suckjoon Jun.
\newblock Quantitative examination of five stochastic cell-cycle and cell-size
  control models for escherichia coli and bacillus subtilis.
\newblock {\em Frontiers in Microbiology}, 12:3278, 2021.

\bibitem{Witz:2020fb}
Guillaume Witz, Thomas Julou, and Erik van Nimwegen.
\newblock {Response to comment on {\textquoteleft}Initiation of chromosome
  replication controls both division and replication cycles in E. coli through
  a double-adder mechanism{\textquoteright}}, August 2020.

\bibitem{Micali2018}
Gabriele Micali, Jacopo Grilli, Jacopo Marchi, Matteo Osella, and Marco
  {Cosentino Lagomarsino}.
\newblock {Dissecting the Control Mechanisms for DNA Replication and Cell
  Division in E. coli}.
\newblock {\em Cell Reports}, 25(3):761--771.e4, 2018.

\bibitem{Micali2018_2}
Gabriele Micali, Jacopo Grilli, Matteo Osella, and Marco~Cosentino Lagomarsino.
\newblock {Concurrent processes set E. coli cell division}.
\newblock {\em Science Advances}, 4(11):1--8, 2018.

\bibitem{Adiciptaningrum2015}
Aileen Adiciptaningrum, Matteo Osella, M.~Charl Moolman, Marco
  Cosentino~Lagomarsino, and Sander~J. Tans.
\newblock {Stochasticity and homeostasis in the E. coli replication and
  division cycle}.
\newblock {\em Scientific Reports}, 5(1):18261, Dec 2015.

\bibitem{Barber2017}
Felix Barber, Po~Yi Ho, Andrew~W. Murray, and Ariel Amir.
\newblock {Details matter: Noise and model structure set the relationship
  between cell size and cell cycle timing}.
\newblock {\em Frontiers in Cell and Developmental Biology}, 5(NOV):1--16,
  2017.

\bibitem{Sompayrac1973}
L.~Sompayrac and O.~Maaloe.
\newblock {Autorepressor Model for Control of DNA Replication}.
\newblock {\em Nature New Biology}, 241(January):133--135, 1973.

\bibitem{Atlung1991}
F.~G. Hansen, B.~B. Christensen, and T.~Atlung.
\newblock {The initiator titration model: computer simulation of chromosome and
  minichromosome control}.
\newblock {\em Research in Microbiology}, 142(2-3):161--167, 1991.

\bibitem{Basan2015}
Markus Basan, Manlu Zhu, Xiongfeng Dai, Mya Warren, Daniel S{\'{e}}vin,
  Yi‐Ping Wang, and Terence Hwa.
\newblock {Inflating bacterial cells by increased protein synthesis}.
\newblock {\em Molecular Systems Biology}, 11(10):836, 2015.

\bibitem{Hansen2018}
Flemming~G. Hansen and Tove Atlung.
\newblock {The DnaA tale}.
\newblock {\em Frontiers in Microbiology}, 9(FEB):1--19, 2018.

\bibitem{Schaper1995}
Sigrid Schaper and Walter Messer.
\newblock {Interaction of the Initiator Protein DnaA of Escherichia coli with
  Its DNA Target}.
\newblock {\em Journal of Biological Chemistry}, 270(29):17622--17626, 1995.

\bibitem{Wallden2015}
Mats Wallden, David Fange, {\"{O}}zden Baltekin, and Johan Elf.
\newblock {Fluctuations in growth rates determine the generation time and size
  distributions of E. coli cells}, 2015.

\bibitem{Kasho2013}
Kazutoshi Kasho and Tsutomu Katayama.
\newblock {DnaA binding locus datA promotes DnaA-ATP hydrolysis to enable cell
  cycle-coordinated replication initiation}.
\newblock {\em Proceedings of the National Academy of Sciences},
  110(3):936--941, 2013.

\bibitem{Kasho2014}
Kazutoshi Kasho, Kazuyuki Fujimitsu, Toshihiro Matoba, Taku Oshima, and Tsutomu
  Katayama.
\newblock {Timely binding of IHF and Fis to DARS2 regulates ATP-DnaA production
  and replication initiation}.
\newblock {\em Nucleic acids research}, 42(21):13134--13149, Dec 2014.

\bibitem{Kurokawa1999}
Kenji Kurokawa, Satoshi Nishida, Akiko Emoto, Kazuhisa Sekimizu, and Tsutomu
  Katayama.
\newblock {Replication cycle-coordinated change of the adenine nucleotide-bound
  forms of DnaA protein in Escherichia coli}.
\newblock {\em The EMBO Journal}, 18(23):6642--6652, 1999.

\bibitem{Katayama2010}
Tsutomu Katayama, Shogo Ozaki, Kenji Keyamura, and Kazuyuki Fujimitsu.
\newblock {Regulation of the replication cycle: Conserved and diverse
  regulatory systems for DnaA and oriC}.
\newblock {\em Nature Reviews Microbiology}, 8(3):163--170, 2010.

\bibitem{Nishida:2002dp}
Satoshi Nishida, Kazuyuki Fujimitsu, Kazuhisa Sekimizu, Tadahiro Ohmura,
  Tadashi Ueda, and Tsutomu Katayama.
\newblock {A Nucleotide Switch in the Escherichia coli DnaA Protein Initiates
  Chromosomal Replication}.
\newblock {\em The Journal of biological chemistry}, 277(17):14986--14995,
  April 2002.

\bibitem{Speck2001}
Christian Speck and Walter Messer.
\newblock {Mechanism of origin unwinding: Sequential binding of DnaA to double-
  and single-stranded DNA}.
\newblock {\em EMBO Journal}, 20(6):1469--1476, 2001.

\bibitem{Hansen1991}
F.~G. Hansen, T.~Atlung, R.~E. Braun, A.~Wright, P.~Hughes, and M.~Kohiyama.
\newblock {Initiator (DnaA) protein concentration as a function of growth rate
  in Escherichia coli and Salmonella typhimurium}.
\newblock {\em Journal of Bacteriology}, 173(16):5194--5199, 1991.

\bibitem{Fujimitsu2009}
Kazuyuki Fujimitsu, Takayuki Senriuchi, and Tsutomu Katayama.
\newblock {Specific genomic sequences of E. coli promote replicational
  initiation by directly reactivating ADP-DnaA}.
\newblock {\em Genes and Development}, 23(10):1221--1233, 2009.

\bibitem{Katayama2001}
Tsutomu Katayama, Kazuyuki Fujimitsu, and Tohru Ogawa.
\newblock {Multiple pathways regulating DnaA function in Escherichia coli:
  Distinct roles for DnaA titration by the datA locus and the regulatory
  inactivation of DnaA}.
\newblock {\em Biochimie}, 83(1):13--17, 2001.

\bibitem{Grant2011}
Matthew Grant, Chiara Saggioro, Ulisse Ferrari, Bruno Bassetti, Bianca Sclavi,
  and Marco~Cosentino Lagomarsino.
\newblock {DnaA and the timing of chromosome replication in Escherichia coli as
  a function of growth rate}.
\newblock {\em BMC Systems Biology}, 5(1):201, 2011.

\bibitem{Grimwalde2011}
Alan~C. Leonard and Julia~E. Grimwade.
\newblock {Regulation of DnaA Assembly and Activity: Taking Directions from the
  Genome}.
\newblock {\em Annual Review of Microbiology}, 65(1):19--35, 2011.
\newblock PMID: 21639790.

\bibitem{Donachie2003}
William~D Donachie and Garry~W Blakely.
\newblock {Coupling the initiation of chromosome replication to cell size in
  Escherichia coli}, 2003.

\bibitem{Sekimizu1988}
K.~Sekimizu and A.~Kornberg.
\newblock {Cardiolipin activation of dnaA protein, the initiation protein of
  replication in Escherichia coli}.
\newblock {\em Journal of Biological Chemistry}, 263(15):7131--7135, 1988.

\bibitem{Kato2001}
Jun~Ichi Kato and Tsutomu Katayama.
\newblock {Hda, a novel DnaA-related protein, regulates the replication cycle
  in Escherichia coli}.
\newblock {\em EMBO Journal}, 20(15):4253--4262, 2001.

\bibitem{Ogawa2002}
Tohru Ogawa, Yoshitaka Yamada, Takao Kuroda, Tetsuya Kishi, and Shigeki Moriya.
\newblock {The datA locus predominantly contributes to the initiator titration
  mechanism in the control of replication initiation in Escherichia coli}.
\newblock {\em Molecular Microbiology}, 44(5):1367--1375, 2002.

\bibitem{Camara2005}
Johanna~E. Camara, Adam~M. Breier, Therese Brendler, Stuart Austin, Nicholas~R.
  Cozzarelli, and Elliott Crooke.
\newblock {Hda inactivation of DnaA is the predominant mechanism preventing
  hyperinitiation of Escherichia coli DNA replication}.
\newblock {\em EMBO Reports}, 6(8):736--741, 2005.

\bibitem{Riber2006}
Leise Riber, Jan~A. Olsson, Rasmus~B. Jensen, Ole Skovgaard, Santanu Dasgupta,
  Martin~G. Marinus, and Anders L{\o}bner-Olesen.
\newblock {Hda-mediated inactivation of the DnaA protein and dnaA gene
  autoregulation act in concert to ensure homeostatic maintenance of the
  Escherichia coli chromosome}.
\newblock {\em Genes and Development}, 20(15):2121--2134, 2006.

\bibitem{Xia1995}
Weiming Xia and William Dowhan.
\newblock {In vivo evidence for the involvement of anionic phospholipids in
  initiation of DNA replication in Escherichia coli}.
\newblock {\em Proceedings of the National Academy of Sciences of the United
  States of America}, 92(3):783--787, 1995.

\bibitem{Saxena2013}
Rahul Saxena, Nicholas Fingland, Digvijay Patil, Anjali~K. Sharma, and Elliott
  Crooke.
\newblock {Crosstalk between DnaA protein, the initiator of Escherichia coli
  chromosomal replication, and acidic phospholipids present in bacterial
  membranes}.
\newblock {\em International Journal of Molecular Sciences}, 14(4):8517--8537,
  2013.

\bibitem{Lin2018}
Jie Lin and Ariel Amir.
\newblock {Homeostasis of protein and mRNA concentrations in growing cells}.
\newblock {\em Nature Communications}, 9(1), 2018.

\bibitem{Roth1998}
Angelika Roth and Walter Messer.
\newblock {High-affinity binding sites for the initiator protein DnaA on the
  chromosome of Escherichia coli}.
\newblock {\em Molecular Microbiology}, 28(2):395--401, 1998.

\bibitem{Campbell:1990it}
J~L Campbell and N~Kleckner.
\newblock {E. coli oriC and the dnaA gene promoter are sequestered from dam
  methyltransferase following the passage of the chromosomal replication fork.}
\newblock {\em Cell}, 62(5):967--979, September 1990.

\bibitem{Lu:1994ee}
Min Lu, Joseph~L. Campbell, Erik Boye, and Nancy Kleckner.
\newblock {SeqA: A negative modulator of replication initiation in E. coli}.
\newblock {\em Cell}, 77(3):413--426, 1994.

\bibitem{Waldminghaus:2009em}
Torsten Waldminghaus and Kirsten Skarstad.
\newblock {The Escherichia coli SeqA protein}.
\newblock {\em Plasmid}, 61(3):141--150, 2009.

\bibitem{Crooke1992}
E.~Crooke, C.~E. Castuma, and A.~Kornberg.
\newblock {The chromosome origin of Escherichia coli stabilizes DnaA protein
  during rejuvenation by phospholipids}.
\newblock {\em Journal of Biological Chemistry}, 267(24):16779--16782, 1992.

\bibitem{Castuma.1993}
C~E Castuma, E~Crooke, and A~Kornberg.
\newblock {Fluid membranes with acidic domains activate DnaA, the initiator
  protein of replication in Escherichia coli}.
\newblock {\em Journal of Biological Chemistry}, 268(33):24665 -- 24668, 01
  1993.

\bibitem{Fingland2012}
Nicholas Fingland, Ingvild Fl{\aa}tten, Christopher~D. Downey, Solveig
  Fossum-Raunehaug, Kirsten Skarstad, and Elliott Crooke.
\newblock {Depletion of acidic phospholipids influences chromosomal replication
  in Escherichia coli}.
\newblock {\em MicrobiologyOpen}, 1(4):450--466, 2012.

\bibitem{Speck1999}
Christian Speck, Christoph Weigel, and Walter Messer.
\newblock {ATP- and ADP-DnaA protein, a molecular switch in gene regulation}.
\newblock {\em EMBO Journal}, 18(21):6169--6176, 1999.

\bibitem{Katayama1998}
Tsutomu Katayama, Toshio Kubota, Kenji Kurokawa, Elliott Crooke, and Kazuhisa
  Sekimizu.
\newblock {The initiator function of DnaA protein is negatively regulated by
  the sliding clamp of the E. coli Chromosomal replicase}.
\newblock {\em Cell}, 94(1):61--71, 1998.

\bibitem{Elf2007}
Johan Elf, Gene-Wei Li, and X.~Sunney Xie.
\newblock Probing transcription factor dynamics at the single-molecule level in
  a living cell.
\newblock {\em Science}, 316(5828):1191--1194, 2007.

\bibitem{Knoppel2021}
Anna Kn{\"{o}}ppel, Oscar Brostr{\"{o}}m, Konrad Gras, David Fange, and Johan
  Elf.
\newblock {The spatial organization of replication is determined by cell size
  independently of chromosome copy number}, 2021.

\bibitem{DArio2021}
Marco D'Ario, Rafael Tavares, Katharina Schiessl, B{\'{e}}n{\'{e}}dicte
  Desvoyes, Crisanto Gutierrez, Martin Howard, and Robert Sablowski.
\newblock {Cell size controlled in plants using DNA content as an internal
  scale}.
\newblock {\em Science}, 372(6547):1176--1181, 2021.

\bibitem{Rhind2021}
Nicholas Rhind.
\newblock {Cell-size control}.
\newblock {\em Current Biology}, 31(21):R1414--R1420, 2021.

\bibitem{Zwicker.2010}
D~Zwicker, D~K Lubensky, and P~R~ten Wolde.
\newblock {Robust circadian clocks from coupled protein-modification and
  transcription–translation cycles}.
\newblock {\em Proceedings of the National Academy of Sciences}, 107(52):22540
  -- 22545, 2010.

\bibitem{Teng:2013cf}
S~W Teng, S~Mukherji, J~R Moffitt, S~de~Buyl, and E~K O'Shea.
\newblock {Robust Circadian Oscillations in Growing Cyanobacteria Require
  Transcriptional Feedback}.
\newblock {\em Science}, 340(6133):737--740, May 2013.

\bibitem{Larrondo:2015cb}
L~F Larrondo, C~Olivares-Yanez, C~L Baker, J~J Loros, and J~C Dunlap.
\newblock {Decoupling circadian clock protein turnover from circadian period
  determination}.
\newblock {\em Science}, 347(6221):1257277--1257277, January 2015.

\bibitem{Eun2018}
Ye-Jin Eun, Po-Yi Ho, Minjeong Kim, Salvatore LaRussa, Lydia Robert, Lars~D.
  Renner, Amy Schmid, Ethan Garner, and Ariel Amir.
\newblock {Archaeal cells share common size control with bacteria despite
  noisier growth and division}.
\newblock {\em Nature Microbiology}, 3(2):148--154, Feb 2018.

\bibitem{Soifer2016}
Ilya Soifer, Lydia Robert, and Ariel Amir.
\newblock {Single-cell analysis of growth in budding yeast and bacteria reveals
  a common size regulation strategy}.
\newblock {\em Current Biology}, 26(3):356--361, 2016.

\bibitem{Erzberger:2006jd}
Jan~P Erzberger, Melissa~L Mott, and James~M Berger.
\newblock {Structural basis for ATP-dependent DnaA assembly and
  replication-origin remodeling.}
\newblock {\em Nature Structural {\&} Molecular Biology}, 13(8):676--683,
  August 2006.

\bibitem{Michelsen:2003ku}
Ole Michelsen, M~Joost Teixeira~de Mattos, Peter~Ruhdal Jensen, and Flemming~G
  Hansen.
\newblock {Precise determinations of C and D periods by flow cytometry in
  Escherichia coli K-12 and B/r.}
\newblock {\em Microbiology (Reading, England)}, 149(Pt 4):1001--1010, April
  2003.

\bibitem{Frimodt-Moller2015}
Jakob Frimodt-M{\o}ller, Godefroid Charbon, Karen~A. Krogfelt, and Anders
  L{\o}bner-Olesen.
\newblock {Control regions for chromosome replication are conserved with
  respect to sequence and location among Escherichia coli strains}.
\newblock {\em Frontiers in Microbiology}, 6(SEP):1--15, 2015.

\bibitem{Kitagawa1998}
Risa Kitagawa, Toru Ozaki, Shigeki Moriya, and Tohru Ogawa.
\newblock {Negative control of replication initiation by a novel chromosomal
  locus exhibiting exceptional affinity for Escherichia coli DnaA protein}.
\newblock {\em Genes and Development}, 12(19):3032--3043, 1998.

\bibitem{Frimodt-Moller2016}
Jakob Frimodt-M{\o}ller, Godefroid Charbon, Karen~A. Krogfelt, and Anders
  L{\o}bner-Olesen.
\newblock {DNA Replication Control Is Linked to Genomic Positioning of Control
  Regions in Escherichia coli}.
\newblock {\em PLoS Genetics}, 12(9):1--27, 2016.

\bibitem{Inoue2016}
Yukie Inoue, Hiroyuki Tanaka, Kazutoshi Kasho, Kazuyuki Fujimitsu, Taku Oshima,
  and Tsutomu Katayama.
\newblock {Chromosomal location of the DnaA-reactivating sequence DARS2 is
  important to regulate timely initiation of DNA replication in Escherichia
  coli}.
\newblock {\em Genes to Cells}, 21(9):1015--1023, 2016.

\bibitem{Christensen1999}
Bjarke~Bak Christensen, Tove Atlung, and Flemming~G. Hansen.
\newblock {DnaA boxes are important elements in setting the initiation mass of
  Escherichia coli}.
\newblock {\em Journal of Bacteriology}, 181(9):2683--2688, 1999.

\bibitem{Hill2012}
Norbert~S. Hill, Ryosuke Kadoya, Dhruba~K. Chattoraj, and Petra~Anne Levin.
\newblock {Cell size and the initiation of DNA replication in bacteria}.
\newblock {\em PLoS Genetics}, 8(3):14--16, 2012.

\bibitem{Atlung1993}
T.~Atlung and F.~G. Hansen.
\newblock {Three distinct chromosome replication states are induced by
  increasing concentrations of DnaA protein in Escherichia coli}.
\newblock {\em Journal of Bacteriology}, 175(20):6537--6545, 1993.

\bibitem{Shiba2004}
Yasuhiro Shiba, Yasuko Yokoyama, Yoshiko Aono, Takashi Kiuchi, Jin Kusaka,
  Kouji Matsumoto, and Hiroshi Hara.
\newblock {Activation of the Rcs signal transduction system is responsible for
  the thermosensitive growth defect of an Escherichia coli mutant lacking
  phosphatidylglycerol and cardiolipin}.
\newblock {\em Journal of Bacteriology}, 186(19):6526--6535, 2004.

\bibitem{Camsund2020}
Daniel Camsund, Michael~J. Lawson, Jimmy Larsson, Daniel Jones, Spartak Zikrin,
  David Fange, and Johan Elf.
\newblock {Time-resolved imaging-based CRISPRi screening}.
\newblock {\em Nature Methods}, 17(1):86--92, 2020.

\bibitem{Klumpp2009}
Stefan Klumpp, Zhongge Zhang, and Terence Hwa.
\newblock {Growth Rate-Dependent Global Effects on Gene Expression in
  Bacteria}.
\newblock {\em Cell}, 139(7):1366--1375, 2009.

\bibitem{Scott2010}
Matthew Scott, Carl~W Gunderson, Eduard~M Mateescu, Zhongge Zhang, and Terence
  Hwa.
\newblock {Interdependence of cell growth and gene expression: origins and
  consequences.}
\newblock {\em Science (New York, N.Y.)}, 330(6007):1099--102, nov 2010.

\bibitem{Sclavi2013}
Chiara Saggioro, Anne Olliver, and Bianca Sclavi.
\newblock {Temperature-dependence of the DnaA–DNA interaction and its effect
  on the autoregulation of dnaA expression}.
\newblock {\em Biochemical Journal}, 449(2):333--341, 12 2012.

\bibitem{Panlilio2021}
Mia Panlilio, Jacopo Grilli, Giorgio Tallarico, Ilaria Iuliani, Bianca Sclavi,
  Pietro Cicuta, and Marco~Cosentino Lagomarsino.
\newblock {Threshold accumulation of a constitutive protein explains E. coli
  cell-division behavior in nutrient upshifts}.
\newblock {\em Proceedings of the National Academy of Sciences of the United
  States of America}, 118(18), 2021.

\bibitem{Brenner.2015}
Naama Brenner, Erez Braun, Anna Yoney, Lee Susman, James Rotella, and Hanna
  Salman.
\newblock {Single-cell protein dynamics reproduce universal fluctuations in
  cell populations}.
\newblock {\em The European Physical Journal E}, 38(9):102, 2015.

\bibitem{Kempe.2015}
Hermannus Kempe, Anne Schwabe, Frédéric Crémazy, Pernette~J. Verschure, and
  Frank~J. Bruggeman.
\newblock {The volumes and transcript counts of single cells reveal
  concentration homeostasis and capture biological noise}.
\newblock {\em Molecular Biology of the Cell}, 26(4):797--804, 2015.

\bibitem{Padovan-Merhar.2015}
Olivia Padovan-Merhar, Gautham P. Nair, Andrew G. Biaesch, Andreas Mayer,
  Steven Scarfone, Shawn W. Foley, Angela R. Wu, L. Stirling Churchman,
  Abhyudai Singh, and Arjun Raj.
\newblock {Single Mammalian Cells Compensate for Differences in Cellular Volume
  and DNA Copy Number through Independent Global Transcriptional Mechanisms}.
\newblock {\em Molecular Cell}, 58(2):339--352, 2015.

\bibitem{Ietswaart.2017}
Robert Ietswaart, Stefanie Rosa, Zhe Wu, Caroline Dean, and Martin Howard.
\newblock {Cell-Size-Dependent Transcription of FLC and Its Antisense Long
  Non-coding RNA COOLAIR Explain Cell-to-Cell Expression Variation}.
\newblock {\em Cell Systems}, 4(6):622--635.e9, 2017.

\bibitem{Zheng:2017hm}
Xiao-yu Zheng and Erin~K O’Shea.
\newblock {Cyanobacteria Maintain Constant Protein Concentration despite Genome
  Copy-Number Variation}.
\newblock {\em CellReports}, 19(3):497 -- 504, 04 2017.

\bibitem{Paulsson2005}
Johan Paulsson.
\newblock {Models of stochastic gene expression}.
\newblock {\em Physics of Life Reviews}, 2(2):157--175, 2005.

\bibitem{Thattai2001}
M.~Thattai and A.~{Van Oudenaarden}.
\newblock {Intrinsic noise in gene regulatory networks}.
\newblock {\em Proceedings of the National Academy of Sciences of the United
  States of America}, 98(15):8614--8619, 2001.

\bibitem{Friedman2006}
Nir Friedman, Long Cai, and X.~Sunney Xie.
\newblock {Linking stochastic dynamics to population distribution: An
  analytical framework of gene expression}.
\newblock {\em Physical Review Letters}, 97(16):1--4, 2006.

\bibitem{Shahrezaei2008}
Vahid Shahrezaei and Peter~S. Swain.
\newblock {Analytical distributions for stochastic gene expression}.
\newblock {\em Proceedings of the National Academy of Sciences of the United
  States of America}, 105(45):17256--17261, 2008.

\bibitem{Milo2013}
Ron Milo.
\newblock {What is the total number of protein molecules per cell volume? A
  call to rethink some published values}.
\newblock {\em BioEssays}, 35(12):1050--1055, 2013.

\bibitem{Paijmans2014}
Joris Paijmans and Pieter~Rein {Ten Wolde}.
\newblock {Lower bound on the precision of transcriptional regulation and why
  facilitated diffusion can reduce noise in gene expression}.
\newblock {\em Physical Review E - Statistical, Nonlinear, and Soft Matter
  Physics}, 90(3):1--14, 2014.

\bibitem{Schenk2017}
Katrin Schenk, Ana~B. Hervás, Thomas~C. Rösch, Marc Eisemann, Bernhard~A.
  Schmitt, Stephan Dahlke, Luise Kleine-Borgmann, Seán~M. Murray, and Peter~L.
  Graumann.
\newblock Rapid turnover of dnaa at replication origin regions contributes to
  initiation control of dna replication.
\newblock {\em PLOS Genetics}, 13(2):1--32, 02 2017.

\bibitem{Kitagawa1996}
Risa Kitagawa, Hironobu Mitsuki, Tuneko Okazaki, and Tohru Ogawa.
\newblock {A novel DnaA protein-binding site at 94.7 min on the Escherichia
  coll chromosome}.
\newblock {\em Molecular Microbiology}, 19(5):1137--1147, 1996.

\bibitem{Blaesing2000}
Franca Blaesing, Christoph Weigel, Michaela Welzeck, and Walter Messer.
\newblock {Analysis of the DNA-binding domain of Escherichia coli DnaA
  protein}.
\newblock {\em Molecular Microbiology}, 36(3):557--569, 2000.

\bibitem{Kawakamii2005}
Hironori Kawakamii, Kenji Keyamura, and Tsutomu Katayama.
\newblock {Formation of an ATP-DnaA-specific initiation complex requires DnaA
  arginine 285, a conserved motif in the AAA+ protein family}.
\newblock {\em Journal of Biological Chemistry}, 280(29):27420--27430, 2005.

\bibitem{Flatten2015}
Ingvild Fl{\aa}tten, Solveig Fossum-Raunehaug, Riikka Taipale, Silje Martinsen,
  and Kirsten Skarstad.
\newblock {The DnaA Protein Is Not the Limiting Factor for Initiation of
  Replication in Escherichia coli}.
\newblock {\em PLoS Genetics}, 11(6):1--22, 2015.

\bibitem{Nakamura2010}
Kenta Nakamura and Tsutomu Katayama.
\newblock {Novel essential residues of Hda for interaction with DnaA in the
  regulatory inactivation of DnaA: Unique roles for Hda AAA<sup>+</sup> Box VI
  and VII motifs}.
\newblock {\em Molecular Microbiology}, 76(2):302--317, 2010.

\bibitem{Moolman2014}
M.~Charl Moolman, Sriram~T.iruvadi Krishnan, Jacob~W.J. Kerssemakers, Aafke
  van~den Berg, Pawel Tulinski, Martin Depken, Rodrigo Reyes-Lamothe, David~J.
  Sherratt, and Nynke~H. Dekker.
\newblock {Slow unloading leads to DNA-bound $\beta$2-sliding clamp
  accumulation in live Escherichia coli cells}.
\newblock {\em Nature communications}, 5:5820, 2014.

\bibitem{Yung1988}
B.~Y.M. Yung and A.~Kornberg.
\newblock {Membrane attachment activates dnaA protein, the initiation protein
  of chromosome replication in Escherichia coli}.
\newblock {\em Proceedings of the National Academy of Sciences of the United
  States of America}, 85(19):7202--7205, 1988.

\bibitem{Zheng2001}
Weidong Zheng, Zhenya Li, Kirsten Skarstad, and Elliott Crooke.
\newblock {Mutations in DnaA protein suppress the growth arrest of acidic
  phospholipid-deficient Escherichia coli cells}.
\newblock {\em EMBO Journal}, 20(5):1164--1172, 2001.

\bibitem{Shiba2012}
Yasuhiro Shiba, Hiroyoshi Miyagawa, Hideki Nagahama, Kenji Matsumoto, Daitetsu
  Kondo, Satoshi Matsuoka, Kouji Matsumoto, and Hiroshi Hara.
\newblock {Exploring the relationship between lipoprotein mislocalization and
  activation of the Rcs signal transduction system in Escherichia coli}.
\newblock {\em Microbiology}, 158(5):1238--1248, 2012.

\bibitem{Mallik.2006}
Prabhat Mallik, Brian~J Paul, Steven~T Rutherford, Richard~L Gourse, and Robert
  Osuna.
\newblock {DksA Is Required for Growth Phase-Dependent Regulation, Growth
  Rate-Dependent Control, and Stringent Control of fis Expression in
  Escherichia coli}.
\newblock {\em Journal of Bacteriology}, 188(16):5775--5782, 2006.

\bibitem{Flatten2013}
Ingvild Fl{\aa}tten and Kirsten Skarstad.
\newblock {The Fis protein has a stimulating role in initiation of replication
  in Escherichia coli in vivo}.
\newblock {\em PLoS ONE}, 8(12):1--9, 2013.

\bibitem{Keyamura2009}
Kenji Keyamura, Yoshito Abe, Masahiro Higashi, Tadashi Ueda, and Tsutomu
  Katayama.
\newblock {DiaA dynamics are coupled with changes in initial origin complexes
  leading to helicase loading}.
\newblock {\em Journal of Biological Chemistry}, 284(37):25038--25050, 2009.

\bibitem{Elowitz.2002}
Michael~B Elowitz, Arnold~J Levine, Eric~D Siggia, and Peter~S Swain.
\newblock {Stochastic gene expression in a single cell.}
\newblock {\em Science}, 297(5584):1183 -- 1186, 08 2002.

\bibitem{Govern:2014ef}
Christopher~C Govern and Pieter~Rein ten Wolde.
\newblock {Optimal resource allocation in cellular sensing systems}.
\newblock {\em Proceedings of the National Academy of Sciences of the United
  States of America}, 111(49):17486--17491, December 2014.

\bibitem{Nozaki2009}
Shingo Nozaki, Yoshitaka Yamada, and Tohru Ogawa.
\newblock {Initiator titration complex formed at datA with the aid of IHF
  regulates replication timing in Escherichia coli}.
\newblock {\em Genes to Cells}, 14(3):329--341, 2009.

\bibitem{Fujimitsu2008}
Kazuyuki Fujimitsu, Masayuki Su'etsugu, Yoko Yamaguchi, Kensaku Mazda, Nisi Fu,
  Hironori Kawakami, and Tsutomu Katayama.
\newblock {Modes of overinitiation, dnaA gene expression, and inhibition of
  cell division in a novel cold-sensitive hda mutant of Escherichia coli}.
\newblock {\em Journal of Bacteriology}, 190(15):5368--5381, 2008.

\bibitem{Heacock1989}
P.~N. Heacock and W.~Dowhan.
\newblock {Alteration of the phospholipid composition of Escherichia coli
  through genetic manipulation}.
\newblock {\em Journal of Biological Chemistry}, 264(25):14972--14977, 1989.

\bibitem{Aranovich2006}
Alexander Aranovich, Garik~Y. Gdalevsky, Rivka Cohen-Luria, Itzhak Fishov, and
  Abraham~H. Parola.
\newblock {Membrane-catalyzed nucleotide exchange on DnaA: Effect of surface
  molecular crowding}.
\newblock {\em Journal of Biological Chemistry}, 281(18):12526--12534, 2006.

\bibitem{Garner1998}
Jennifer Garner, Peter Durrer, Jennifer Kitchen, Josef Brunner, and Elliott
  Crooke.
\newblock {Membrane-mediated release of nucleotide from an initiator of
  chromosomal replication, Escherichia coli DnaA, occurs with insertion of a
  distinct region of the protein into the lipid bilayer}.
\newblock {\em Journal of Biological Chemistry}, 273(9):5167--5173, 1998.

\bibitem{Nishiwaki-Ohkawa2014}
Taeko Nishiwaki-Ohkawa, Yohko Kitayama, Erika Ochiai, and Takao Kondo.
\newblock {Exchange of ADP with ATP in the CII ATPase domain promotes
  autophosphorylation of cyanobacterial clock protein KaiC.}
\newblock {\em Proceedings of the National Academy of Sciences of the United
  States of America}, 111(12):4455 -- 4460, 03 2014.

\bibitem{Paijmans2017}
Joris Paijmans, David~K Lubensky, and Pieter~Rein ten Wolde.
\newblock {A thermodynamically consistent model of the post-translational Kai
  circadian clock}.
\newblock {\em PLoS Computational Biology}, 13(3):e1005415, March 2017.

\end{thebibliography}


\widetext
\pagebreak
\begin{center}
\textbf{\large Supplemental Material: \\ 
    Robust replication initiation from coupled homeostatic mechanisms}
\end{center}

\setcounter{equation}{0}
\setcounter{figure}{0}
\setcounter{table}{0}
\setcounter{page}{1}
\setcounter{section}{0}
\makeatletter

\renewcommand{\thefigure}{S\arabic{figure}}
\renewcommand{\theequation}{S\arabic{equation}}
\renewcommand{\thetable}{S\arabic{table}}
\renewcommand{\thesection}{S\arabic{section}}
\titleformat{\section}
{\normalfont\normalsize\bfseries\filcenter}{\thesection.}{1em}{}

{\bf Overview.} Two classes of mechanistic models for the regulation of replication initiation in \textit{E. coli} have been proposed in the literature: Initiator accumulation models \cite{Barber2017, Ho2015, Amir2014, Sompayrac1973, Atlung1991, Basan2015} and initiator switch models \cite{Grant2011, Katayama2017, Riber2016, Grimwalde2011, Kasho2014, Donachie2003, Grant2011}. We propose mechanistic models out of each class and test whether they are consistent with experiments. Then we combine a titration with a switch model and show that it can increase the robustness of the system in the presence of noise. This Supporting Information is structured into four parts: In the first part, we present the gene expression model we are using throughout this work (section \ref{sec:ribo_limiting}).
In the second part, we present a model from the {\bf initiator accumulation class} (section \ref{sec:accumulation_models}) that is based on the accumulation of an initiator protein up to a threshold number, which is set by the fixed
number of titration sites per chromosome. First, we show that in order to maintain stable cell cycles with the initiator accumulation model, the initiator production rate must be proportional to the volume of the cell (section \ref{sec:init_accum_stability_criteria}).
Then we demonstrate that while the Autoregulated Initiator Titration (AIT) model would ensure stable cell cycles at all growth rates if all titration sites were located at the origin, it exhibits over-initiation events in the overlapping replication-fork regime at high growth rates because, as experiments show, the sites are distributed randomly over the chromosome (section \ref{sec:AIT_model}).
In the third part of this Supporting Information, we present {\bf two initiator switch models} based on a switch between an active and an inactive form of the initiator protein DnaA (section \ref{sec:switch_models}): The Lipid-\textit{DatA} (LD) model is based on an origin density-dependent ultra-sensitivity switch of DnaA (section \ref{sec:LD_model}). The Lipid-\textit{DatA}-\textit{DARS1/2}-RIDA (LDDR) model includes all known activators and deactivators in \textit{E. coli} and generates high amplitude oscillations at realistic activation and deactivation rates (section \ref{sec:LDDR_model}). 
In section \ref{sec:switch_noise} we elucidate the origin of adder and sizer correlations using the LD model, and we also show that the same correlations are observed in the full LDDR model. In section \ref{sec:model_validation_and_prediction} we {\bf validate our model and present testable predictions}. 
We first combine titration with an activation switch and show how titration sharpens the oscillations of the activation switch, increasing the precision of replication initiation (section \ref{sec:switch_titration_combined}). While a titration-based mechanism initiates replication precisely only at low growth rates and the activation switch does so only at higher growth rates, the combined titration-switch model initiates replication accurately at all growth rates. We then discuss the role of SeqA. We show that suppression of {\it dnaA} expression by SeqA can rescue the titration-based mechanism at high growth rates, but not at intermediate growth rates: in this regime, the switch is essential. In section \ref{sec:model_validation} we then validate our theoretical model by comparing key predictions to experimental observations and we make several novel experimentally testable predictions (section \ref{sec:model_predictions}). In this section, we also
show that our results are robust to the precise type of coupling of the replication cycle to the cell division cycle (section \ref{sec:division_separate}). In the last section, we study two variants of our models where the lipid activation is either oriC-dependent or is removed entirely (section \ref{sec:role_of_lipids}).

\section{Growing cell model of gene expression}
\label{sec:ribo_limiting}

In this section, we present the gene expression model,  which underlies all our models.
In the recently developed growing-cell model by Lin et  al. \cite{Lin2018}, transcription is limited by the availability of RNAPs while  translation is limited by the ribosomes. In this model, the mRNA and  protein copy numbers are proportional to the cell volume, as recent  experiments indicate  \cite{Panlilio2021, Lin2018, Brenner.2015, Kempe.2015, Padovan-Merhar.2015, Ietswaart.2017,Zheng:2017hm}. Concomitantly,  the protein synthesis rate is, as observed very recently  \cite{Panlilio2021}, proportional to the volume, which is a crucial requirement for the stability of the initiator accumulation model (see section \ref{sec:init_accum_stability_criteria}).
We start this section by deriving the basal protein  synthesis rate in the growing-cell model (section \ref{sec:ribo_limiting_model}). In section \ref{sec:ribo_limiting_temporal_variation}, we  show how the synthesis rate of a constitutively expressed protein is proportional to the volume, such that its concentration increases exponentially in time over the course of the cell cycle. 
In section \ref{sec:ribo_limiting_neg_auto} we then describe how gene regulation can be included in the growing cell model.

\subsection{Basal gene expression}
\label{sec:ribo_limiting_model}
In the gene expression model of Lin et al. \cite{Lin2018}, the genes and the mRNAs compete for the limiting pool of RNAPs and ribosomes, respectively \cite{Lin2018}. Therefore, the transcription rate of a gene $i$ is directly proportional to the total number RNAPs $n$ times the fraction of RNAPs $\phi_{\rm i}$ that are transcribing gene $i$. To quantify the gene allocation fraction $\phi_{\rm i}$, Lin et al. define an effective gene copy number $g_{\rm i}$ that accounts for its copy number and the binding strength of its promoter \cite{Lin2018}. The gene allocation fraction of gene $i$ is then given by the effective gene copy number $g_{\rm i}$ divided by the sum over all effective gene copy numbers in the cell $\phi_{\rm i} = g_{\rm i}/\sum_{\rm j} \, g_{\rm j}$. As the number of ribosomes is assumed to limit translation, the protein synthesis rate of gene $i$ is proportional to the number of ribosomes $N_{\rm R}$ times the fraction of ribosomes translating the mRNA of gene $i$. Assuming that the affinity of ribosomes binding to mRNA is equal for all types of mRNA $m_{\rm i}$, the ribosome allocation fraction $f_{\rm i}$ of gene $i$ is given by the number of mRNAs $m_{\rm i}$ of gene $i$ divided by the total amount of mRNAs, thus $f_{\rm i}= m_{\rm i} / \sum_{\rm j} m_{\rm j}$. The growing cell model then gives rise to the following set of equations for the change in the number of mRNAs $m_{\rm i}$ and the number of proteins $p_{\rm i}$ of gene $i$:
\begin{align}
\frac{dm_{\rm i}}{dt}=& k_{\rm m} \, \phi_{\rm i} \, n - \frac{m_{\rm i}}{\tau_m}
\label{eq:ribo_lim_mRNA_change}
\\
\frac{dp_{\rm i}}{dt}=& k_{\rm R} \, f_{\rm i} \, f_{\rm a} \, N_{\rm R}
\label{eq:ribo_lim_protein_change}
\end{align}
where $k_{\rm m}$ is the transcription rate of a single RNAP, $\tau_{\rm m}$ is the degradation time of the mRNA (taken to be equal and constant for all mRNAs), $k_{\rm R}$ is the translation rate of a ribosome, $f_{\rm a}$ is the fraction of actively translating ribosomes and $N_{\rm R}$ is the number of ribosomes.
Due to the fast production and degradation rate of the mRNA compared to the growth rate of the cell, we can approximate the mRNA number to be at a steady state such that
\begin{equation}
\langle m_{\rm i} \rangle = k_{\rm m} \, \phi_{\rm i} \langle n \rangle \, \tau_{\rm m}
\label{eq:ribo_lim_mRNA_steady_state}
\end{equation}
Plugging equation \ref{eq:ribo_lim_mRNA_steady_state} into equation \ref{eq:ribo_lim_protein_change} and using that $\sum_{\rm j} \phi_{\rm j}=1$ gives the following general expression for the change in the number of proteins:
\begin{equation}
\frac{dp_{\rm i}}{dt}= k_{\rm R} \, \phi_{\rm i} \, f_{\rm a} \, N_{\rm R}
\label{eq:ribo_lim_protein_change_shorter}
\end{equation}
The protein production rate of any gene $i$ is therefore proportional to the number of ribosomes $N_{\rm R}$ times the gene allocation fraction $\phi_{\rm i}$ of gene $i$. The gene allocation fraction $\phi_{\rm i}$ is a measure of the relative affinity and amount of gene $i$ with respect to all other genes in the cell. In the simplified scenario of an instantaneous replication of the entire DNA after replication initiation, replication of the DNA does not affect the gene allocation fraction. If the gene $i$ is not regulated, the affinity of gene $i$ is constant in time. If at a given growth rate the total affinity of all genes remains approximately constant in time, the gene allocation fraction $\phi_{\rm i}$ is constant in time too.

\subsection{Constitutively expressed proteins}
\label{sec:ribo_limiting_temporal_variation}
In this section, we will first demonstrate that in the growing cell model, the protein production rate is directly proportional to the volume of the cell, which, as we will see in section \ref{sec:AIT_model}, ensures the stability of the AIT model.
The total number of proteins $N$ in the cell is given by the sum over all proteins $p_{\rm j}$
\begin{equation}
N = \sum_{\rm j} p_{\rm j}
\label{eq:total_number_sum}
\end{equation}
and the fraction of proteins that are ribosomes is 
\begin{equation}
\Phi_{\rm R} = \frac{N_{\rm R}}{N}.
\label{eq:ribosome_sector_definition}
\end{equation}
From equations \ref{eq:ribo_lim_protein_change_shorter}, \ref{eq:total_number_sum} and \ref{eq:ribosome_sector_definition}, and using that $\sum_{\rm j} \Phi_{\rm j}=1$,  we find that the change in the total number of proteins in time is
\begin{equation}
\frac{dN}{dt}= \sum_{\rm j} \frac{dp_{\rm j}}{dt} = k_{\rm R} \, f_{\rm a} \, N_{\rm R} = k_{\rm R} \, f_{\rm a} \, \Phi_{\rm R} \, N
\end{equation}
while, defining the total number density $\rho\equiv N/V$, the change in the volume is
\begin{equation}
\frac{dV}{dt}= \frac{1}{\rho} \, \frac{dN}{dt} =  k_{\rm R} \, f_{\rm a} \, \Phi_{\rm R} \, V
\label{eq:ribo_lim_volume_change}
\end{equation}
Hence, the cell grows exponentially with a growth rate 
\begin{equation}
\lambda = \frac{1}{N}\frac{dN}{dt} = \frac{1}{V} \frac{dV}{dt} = k_{\rm R} \, f_{\rm a} \, \Phi_{\rm R} 
\label{eq:growth_rate_ribo_derived}
\end{equation}
Using equation \ref{eq:growth_rate_ribo_derived} we can then derive the change in the number of a protein of gene $i$:
\begin{equation}
\frac{dp_{\rm i}}{dt}= \phi_{\rm i} \, k_{\rm R} \, f_{\rm a} \, N_{\rm R} = \phi_{\rm i} \, k_{\rm R} \, f_{\rm a} \, \Phi_{\rm R} \, N = \phi_{\rm i} \, \lambda \, N = \phi_{\rm i} \, \lambda \, \rho \, V
\label{eq:ribo_lim_protein_propto_volume}
\end{equation}

Therefore, while in the standard model of gene expression the copy number of a constitutively expressed protein $i$ increases bi-linearly in time, in the growing cell model it increases exponentially over the course of the cell cycle.
The change in the protein concentration of gene $i$ is then given by
\begin{equation}
\frac{d[p_{\rm i}]}{dt}= \frac{dp_{\rm i}}{dt} \, \frac{1}{V} - p_{\rm i} \, \frac{1}{V^2} \, \frac{dV}{dt}= \phi_{\rm i} \, \lambda \, \rho - \lambda \, [p_{\rm i}]
\end{equation}
At steady state, we find that the growth rate drops out and the steady state protein concentration is given by:
\begin{equation}
[p_{\rm i}]^\ast = \phi_{\rm i} \, \rho
\label{eq:ribo_steady_state_protein_conc}
\end{equation}

\begin{figure}
	\centering
	\includegraphics[width =0.6\textwidth]{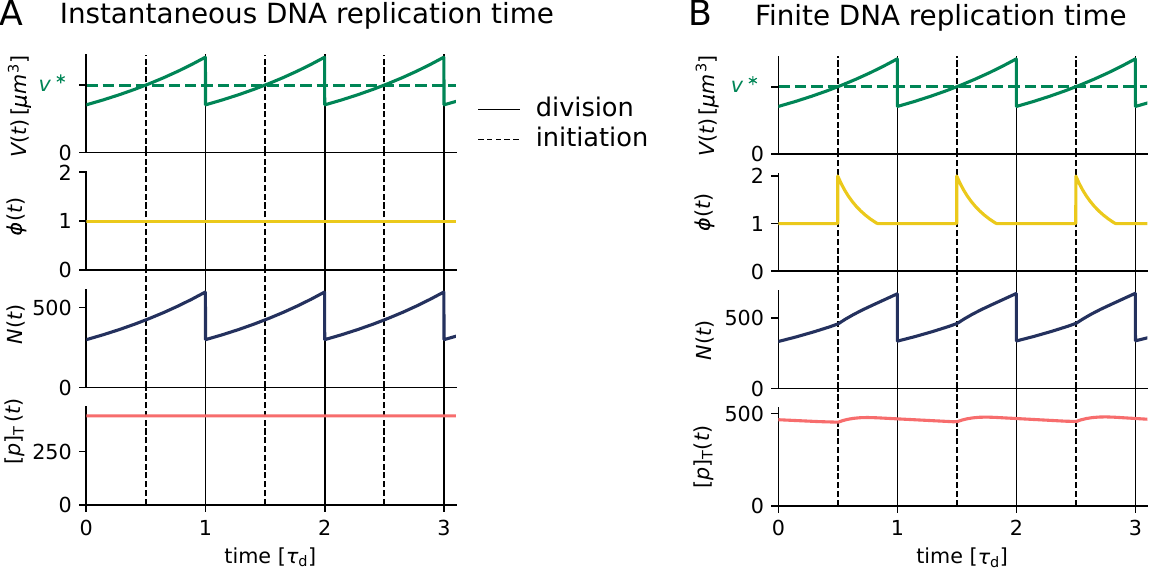}
	\caption{\textbf{The concentration of differently regulated proteins in the growing cell model of gene expression} (A, B) The volume
		$V(t)$, the gene allocation fraction $\phi(t)$, the number of proteins $N(t)$ and the total
		concentration $[p]_{\rm T}$ of a constitutively expressed protein within the growing cell model. The volume and the protein number are evolved according to equations \ref{eq:ribo_lim_volume_change} and \ref{eq:ribo_lim_protein_propto_volume}, respectively. (A) While the protein number increases exponentially in time, the total concentration remains perfectly constant. (B) The change of the number and concentration of a constitutively expressed protein when the gene allocation fraction changes in time due to a finite time to replicate the entire chromosome. The gene is assumed to be located at the origin which causes a doubling of the allocation fraction at the moment of replication initiation. When the entire chromosome has been replicated, the gene allocation fraction is again constant. As a consequence, the concentration of a constitutively expressed gene exhibits weak oscillations due to the changes in the gene allocation fraction. 
		The parameters in all simulations  are $v^\ast=1\, \mu$m$^3$, $\tau_{\rm cc}= 1$~h, $T_{\rm C}= 2/3$~h, $\rho=10^6\, \mu$m$^{-1}$, $\tau_{\rm d}=2$~h and $\phi_{\rm i}= 2 \times 10^{-3}$. 
	}
	\label{fig:S2_growing_cell_model}
\end{figure}

In order to investigate how the protein number and concentration of an unregulated protein changes over the course of the cell cycle, we evolve the volume of a cell according to to $dV/dt=\lambda V$ (see \ref{eq:ribo_lim_volume_change} and \ref{eq:growth_rate_ribo_derived}) and the protein number according to equation \ref{eq:ribo_lim_protein_propto_volume}. Replication is initiated at a fixed volume per origin $v^\ast$ and the cell divides a fixed time $\tau_{\rm cc}$ after replication initiation. The exponential increase in the number of proteins over the course of the cell cycle can be seen in Figure \ref{fig:S2_growing_cell_model}~A. In the scenario where the entire chromosome is replicated instantaneously and the gene is not regulated, the gene allocation fraction $\phi_{\rm i}$ remains constant (Fig. \ref{fig:S2_growing_cell_model}~A, yellow line). While the number of a protein $p$ increases proportional to the volume of the cell (Fig. \ref{fig:S2_growing_cell_model}~A, blue line), the concentration remains perfectly constant in time (Fig. \ref{fig:S2_growing_cell_model}~A, red line).

In reality the chromosome is not replicated instantly. This means that when the part that houses gene $i$ is replicated, the gene allocation fraction $\phi_{\rm i}$ rises transiently, as illustrated in the second panel of Figure \ref{fig:S2_growing_cell_model}~B. The transiently higher gene allocation fraction results in a temporal increase of the production rate (Figure \ref{fig:S2_growing_cell_model}~B, third panel), which gives rise to weak oscillations in the protein concentration over the course of the cell cycle.

\subsection{Negatively autoregulated proteins}
\label{sec:ribo_limiting_neg_auto}
Regulation of gene $i$ can be included by modifying the gene affinity $g_{\rm i}$. If gene $i$ is for example negatively autoregulated, the gene affinity becomes
\begin{equation}
g_{\rm i}= g_{\rm i}^0 \, \frac{1}{1+\left( \frac{[p_{\rm i}]}{K_{\rm D}^{\rm p}}\right)^n}
\end{equation}
where $g_{\rm i}^0$ is the basal gene affinity if the promoter is not repressed at all, $[p_{\rm i}]$ the free initiator concentration, $K_{\rm D}^{\rm p}$ is the dissociation constant of the promoter and $n$ is the Hill coefficient. The protein production rate then becomes dependent on the protein concentration via the modified gene allocation fraction $\phi_{\rm i}$:
\begin{align}
\frac{dp_{\rm i}}{dt}&= \phi_{\rm i} \, \lambda \, \rho \, V = \frac{g_{\rm i}}{\sum_{\rm j} g_{\rm j}} \, \lambda \, \rho \, V \\
&=  \phi_{\rm i}^0\,\frac{1}{1+\left( \frac{[p_{\rm i}]}{K_{\rm D}^{\rm p}}\right)^n} \, \lambda \, \rho  \, V
\label{eq:ribo_lim_protein_change_neg_auto}
\end{align}
where we defined the {\em basal} gene allocation fraction $\phi_{\rm i}^0 \equiv g_{\rm i}^0 / \sum_{\rm j} g_{\rm j}$. By defining the gene allocation density as $\tilde{\phi}_{\rm i}^0 =  \phi_{\rm i}^0 \, \rho$, we obtain Eq. 1 of the main text for the production rate of a negatively autoregulated protein $p$ (with $i=p$):
\begin{equation}
\frac{dN_{\rm p}}{dt}= \frac{\tilde{\phi}_{\rm p}^0 \, \lambda \, V}{1+ \left( \frac{[p]}{K_{\rm D}^{\rm p}}\right)^{n}}
\label{eq:number_proteins_changes_volume_SI}
\end{equation}

\section{Initiator accumulation model}
\label{sec:accumulation_models}
In the initiator accumulation model, an initiator protein accumulates over the course of the cell cycle and replication is initiated when a threshold amount per origin is attained. We first show that a volume-dependent production rate is required to ensure stable replication cycles (section \ref{sec:init_accum_stability_criteria}). 
We then present the Autoregulated Initiator Titration (AIT) model and investigate under what conditions the AIT model can ensure stable cell cycles (section \ref{sec:AIT_model}). In the AIT model, a fixed number of titration sites per chromosome sets the critical number of initiators $n_{\rm p}^\ast$ that need to be accumulated in order to initiate replication (section \ref{sec:titration_sites}). We first show that the model ensures stable cell cycles at all growth rates when all titration sites are located at the origin (section \ref{sec:AIT_ribo_limiting}). When the titration sites are however homogeneously distributed on the chromosome, which is a good approximation for the experimentally reported random distribution \cite{Roth1998, Hansen2018}, reinitiation events occur at high growth rates (section \ref{sec:AIT_homogeneous_titration}). Finally, we derive an analytical expression for the initiation volume in the AIT model and investigate under what conditions the initiation volume becomes independent of the growth rate of the cell (section \ref{sec:titration_growth_rate_dependence}). All parameters used in the AIT in the main part of the paper and in the SI are discussed in section \ref{sec:AIT_parameters} and can be found in Table \ref{tab:RIT_AIT_parameters}.

\subsection{Stability of the initiator accumulation model}
\label{sec:init_accum_stability_criteria}
In this section, we demonstrate that a volume-dependent protein production rate is essential to obtain stable cell cycles with the initiator accumulation model. The bacterium \textit{E. coli} must initiate replication once per division cycle in order to be able to distribute two copies of the chromosome in the two daughter cells. In good nutrient conditions, \textit{E. coli} grows exponentially with a growth rate $\lambda$ such that the volume is given by
\begin{equation}
V(t) = V_{\rm b} \, e^{\lambda t}
\label{eq:accum_stability_volume}
\end{equation}
The growth rate $\lambda$ can fluctuate due to noise, but on average cells double their entire volume after the cell-doubling time $\langle \tau_{\rm d} \rangle = \ln(2) / \langle \lambda \rangle$. As in \textit{E. coli} replication is initiated synchronously at all origins also in the overlapping fork regime at high growth rates, we can define the inter-initiation time $\tau_{\rm ii}$ as the time between two consecutive initiation events. Any molecular mechanism for replication initiation must ensure that the average inter-initiation time $\langle \tau_{\rm ii} \rangle$ equals the average cell-doubling time $\langle \tau_{\rm d} \rangle$. If that is not the case, the average origin density, $\langle \rho \rangle =\langle n_{\rm ori} \rangle /\langle V \rangle$, does not remain constant over the course of several generations. 

In the initiator accumulation models, an initiator protein is accumulated up to a fixed threshold per origin at which replication is initiated. In the AIT model in section \ref{sec:AIT_model} we will show that a constant number of high-affinity binding sites for the initiator on the chromosome can ensure such a constant number threshold per origin. Given that this threshold per origin is fixed, the time from one initiation event to the next is determined by how fast the initiator proteins are synthesized. In contrast to the recently proposed growing cell model presented in section \ref{sec:ribo_limiting}, in an arguably more traditional model of gene expression, the protein production rate of a constitutively expressed gene is given by a constant basal $\alpha$ rate times the gene copy number $g$ \cite{Klumpp2009, Paulsson2005, Thattai2001, Friedman2006, Shahrezaei2008}:
\begin{equation}
\frac{dN}{dt}= \alpha \, g
\label{eq:accum_stability_constit_expr}
\end{equation}
Assuming again that the gene is located at the origin, the number of genes $g$ equals the number of origins $n_{\rm ori}$. Thus, a constant number of initiators per origin $\Delta n = \Delta N/ n_{\rm ori}$ is accumulated in a time interval $\Delta t$:
\begin{equation}
\Delta n =  \alpha \, \Delta t
\end{equation}
As in the initiator accumulation model replication is initiated after a constant amount of proteins per origin $\Delta n^\ast$ has been accumulated, we find that the inter-initiation time $\tau_{\rm ii}$ in this model is given by
\begin{equation}
\tau_{\rm ii}= \frac{\Delta n^\ast}{\alpha}
\label{eq:accum_instability_t_ii}
\end{equation}
As the number of initiators that need to be accumulated per origin $\Delta n^\ast$ is constant and the basal rate does not explicitly depend on the volume in the traditional model of gene expression, the inter-initiation time thus is constant. If the basal production rate is not set such that the average replication period exactly equals the doubling time of the cell, $\tau_{\rm ii}=\tau_{\rm d}$, this system gives rise to an instability in the chromosome density.

\begin{figure}
	\centering
	\includegraphics[width =0.6\textwidth]{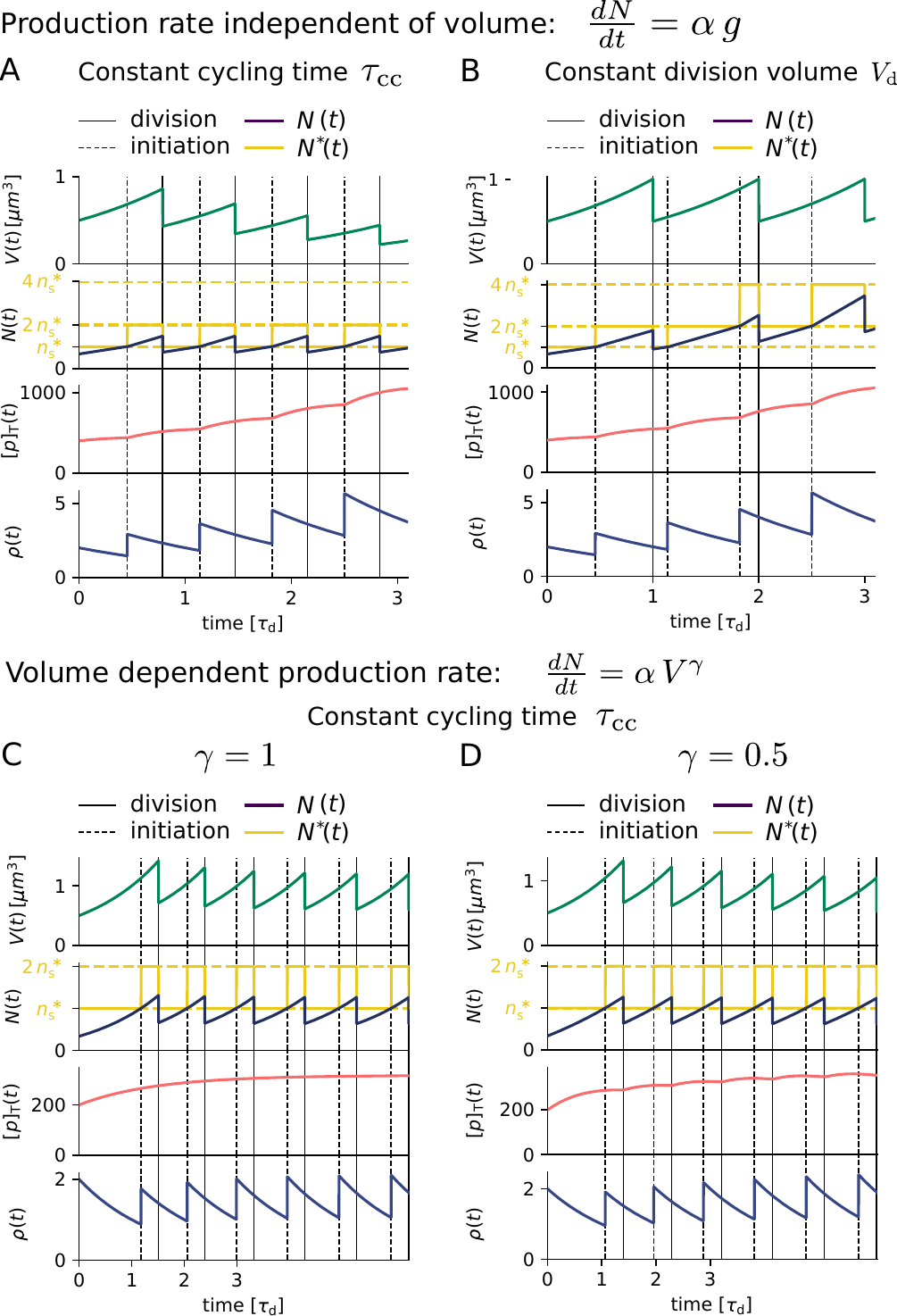}
	\caption{\textbf{For the initiator accumulation model to become stable, the production rate needs to depend on the growth rate of the cell} (A, B, C, D) The volume
		$V(t)$ (according to
		equation \ref{eq:accum_stability_volume}), the number of proteins $N(t)$ together with the critical threshold $N^\ast= n^\ast \, n_{\rm ori}$, the total
		concentration $[p]_{\rm T}= N(t) / V(t)$, and the origin density $\rho(t)= n_{\rm ori}(t) / V(t)$ as a function of time. (A, B) The protein is produced at a constant rate times the number of genes (according to equation \ref{eq:accum_stability_constit_expr}). This gives rise to an unstable chromosome density independent of the division mechanism. (A) Cell division is triggered a constant cycling time $\tau_{\rm cc}$ after replication initiation. The time between consecutive replication events $\tau_{\rm ii}$ is given by equation \ref{eq:accum_instability_t_ii} and is shorter than the doubling time of the cell. Thus, the origin density increases in time. (B) Cell division is triggered at a fixed division volume $V_{\rm d}= 1$ $\mu m^3$ and is thus independent of the replication cycle. Again, the replication period $\tau_{\rm ii}$ is shorter than the doubling time of the cell and the origin density increases in time. (C, D) Now, the initiator protein is produced proportional to the volume of the cell according to equation \ref{eq:prod_prop_to_volume_gamma} with an exponent $\gamma$. Cell division is triggered a fixed time $\tau_{\rm cc}$ after replication initiation. For any positive exponent that is larger $\gamma>0$, the gene density stabilizes after an initial perturbation. (C) For an exponent of $\gamma = 1$, the gene density relaxes to a constant average density after an initial perturbation. The total initiator concentration becomes perfectly constant in time. (D) For an exponent of $\gamma = 0.5$ the relaxation time increases and the total concentration oscillates weakly over the course of the cell cycle. The system relaxes to a stable gene density and initiator concentration. The parameters of all simulations are $\tau_{\rm cc}= 1$~h, $\alpha= 110$~h$^{-1}$, $\tau_{\rm d}=2$~h, $n^\ast= 300$.}
	\label{fig:S3_accum_stability}
\end{figure}

We verify this prediction by performing simulations. The cell volume and the number of initiators are evolved according to equations \ref{eq:accum_stability_volume} and \ref{eq:accum_stability_constit_expr} and replication is initiated when the number of initiators per origin $n(t)= N(t) / n_{\rm ori}(t) $ equals the critical number per origin $n^\ast$. At initiation, the number of origins doubles and the number of initiators per origin in generation $i$ right after initiation thus becomes $n_{\rm i}=n^\ast/2$. The number of initiators per origin that needs to be accumulated until the next initiation event is therefore $\Delta n_{\rm i} = n^\ast-n_{\rm i}= n^\ast/2$. 
Following the Cooper-Helmstetter model \cite{Cooper1968}, the cell divides a constant cycling time $\tau_{\rm cc}$ after replication initiation. In Figure \ref{fig:S3_accum_stability}~A, the replication period $\tau_{\rm ii}$ is chosen to be shorter than the doubling time $\tau_{\rm d}$ of the cell. As every replication initiation event triggers a cell division event, the division period $\tau_{\rm div}$ equals the replication period $\tau_{\rm div}= \tau_{\rm ii} < \tau_{\rm d}$. As the replication period and thus the division period is smaller than the doubling time of the cell, the volume of the cell decreases over several generations while the gene density increases. We emphasise that even when $\tau_{\rm ii}$ is chosen to be equal to $\tau_{\rm d}$, any noise, even that coming from the finite machine-precision, will cause the gene density to eventually become unstable.  To show that this instability does not depend on the choice of the division control, we also study another model in which cell division is triggered at a fixed division volume $V_{\rm d}$ instead of a fixed time $\tau_{\rm cc}$ after replication initiation. Because in this model the division cycle is independent of the replication cycle and division is triggered at a fixed division volume $V_{\rm d}$, the division cycle naturally remains stable (Fig. \ref{fig:S3_accum_stability}~B). The replication cycle is however not coupled to this division cycle, because the synthesis rate of the accumulator and the replication threshold are constant, i.e. do not depend on the volume. Replication is therefore initiated at a period that is again shorter than the doubling time of the cell $\tau_{\rm ii} < \tau_{\rm d}$. Also in this scenario, the gene density increases over the course of several generations. 

The initiator accumulation model becomes stable by introducing a volume-dependent production rate, which couples the replication cycle to the cell division cycle. Taking the production rate to be
\begin{equation}
\frac{dN}{dt}= \alpha \, V^\gamma
\label{eq:prod_prop_to_volume_gamma}
\end{equation}
where $\gamma$ is an exponent quantifying the strength of the volume dependence of the production rate. For $\gamma = 0$ the production rate becomes independent of the volume. We show that for the exponents $\gamma = 1$ (Fig. \ref{fig:S3_accum_stability}~C) and  $\gamma = 0.5$ (Fig. \ref{fig:S3_accum_stability}~D) the system recovers from an initial perturbation and becomes stable. The relaxation time increases with decreasing volume dependence. 

We have demonstrated that the initiator accumulation model requires a volume dependent production rate. In the traditional model of gene expression, the production rate of an unregulated protein is proportional to the gene copy number times a constant production rate \cite{Klumpp2009, Paulsson2005, Thattai2001, Friedman2006, Shahrezaei2008} and thus cannot fulfill this requirement (it corresponds to $\gamma=0$). In the previous section, we showed that in the growing cell model, which we use throughout this work, the production rate is directly proportional to the volume of the cell, thus corresponding to the scenario $\gamma=1$.  

\begin{table}
	\centering
	\caption{Parameters used in the AIT model}
	\begin{tabular}{lrrrr}
		Parameter & name & value & Motivation \\
		\midrule
		$\phi_{\rm 0}$ & gene allocation fraction &  $10^{-3}$ & set to match initiation volume \\
		& & & reported in \cite{Si2017} \\
		$K_{\rm D}^{\rm p}$ [$\mu \rm{m}^{-3}$] & dissociation constant initiator promoter & 200 & \cite{Speck1999} \\
		$n$ & Hill coefficient initiator & 5 & \cite{Speck1999} \\
		$n_{\rm s}$ & number of titration sites per chromosome & 300 & \cite{Roth1998, Hansen2018} \\
		$K_{\rm D}^{\rm ori}$ [$\mu \rm{m}^{-3}$] & dissociation constant origin & 20 & \cite{Schaper1995} \\
		$K_{\rm D}^{\rm s}$ [$\mu \rm{m}^{-3}$] & dissociation constant titration sites & 1 & \cite{Schaper1995} \\
		$\rho$ [$\mu \rm{m}^{-3}$] & number density & $10^6$ & \cite{Milo2013} \\
		$D_{\rm D}$ & noise strength DnaA & 100 & set to match CV from \cite{Wallden2016} \\
		$T_{\rm C}$ [h] & C-period & 2/3 & \cite{Cooper1968} \\
		$T_{\rm D}$ [h] & D-period & 1/3 & \cite{Cooper1968} \\
		$\lambda$ [h$^{-1}$] & growth rate & 0.35-1.73 & \cite{Si2017, Wallden2016} \\
		\bottomrule
	\end{tabular}
	\\
	{* One molecule per cubic micrometer corresponds to approximately one nM ($1~\mu \rm{m}^{-3}= 1.67$~nM).}
	\label{tab:RIT_AIT_parameters}
\end{table}

\subsection{The AIT model}
\label{sec:AIT_model}
In this section, we present the AIT model that is consistent with the experimental data on the cell-cycle network of \textit{E. coli}. In the AIT model, the initiator protein is DnaA, which is negatively autoregulated and binds to high-affinity titration sites on the DNA. Here we first discuss the parameters used in the AIT model (section \ref{sec:AIT_parameters}). Then we show how a fixed number of titration sites per chromosome can set the critical number of initiators required for replication initiation (section \ref{sec:titration_sites}). Next, in section \ref{sec:AIT_ribo_limiting}, we show that the AIT model ensures stable cell cycles at all growth rates when the titration sites are located closely to the origin. Then we show that the experimentally reported random titration site distribution on the chromosome can give rise to premature reinitiation events at high growth rates (section \ref{sec:AIT_homogeneous_titration}). 
In section \ref{sec:titration_growth_rate_dependence}, we derive an analytical expression for the initiation volume in the AIT model and discuss its growth rate dependence. Finally, we show that gene expression noise in the production rate of the initiator protein DnaA naturally gives rise to the experimentally observed initiation adder (section \ref{sec:AIT_adder}). 
All parameters used in the AIT model in the main part of the paper and in the SI can be found in Table \ref{tab:RIT_AIT_parameters}.

\subsubsection{Biological parameters of the AIT model}
\label{sec:AIT_parameters}
In this section, we discuss the experimentally found parameters and compare them to the ones used in the simulations of the AIT model. The parameters of the AIT model used both in the main figures and in the Supplementary Information can be found in Table \ref{tab:RIT_AIT_parameters}.

The protein DnaA in \textit{E. coli} is generally referred to as the initiator protein, as its ATP-bound form is required to bind to the origin for initiating replication \cite{Katayama2017}. Both forms of the protein DnaA, ATP-DnaA and ADP-DnaA, have strong affinity for an asymmetric 9 bp consensus sequence on the DNA, the DnaA box \cite{Katayama2017}. In the replication origin region of \textit{E. coli} several DnaA boxes are present, including R1-R4 and M. \cite{Roth1998}. In total, 308 DnaA boxes of the stringent definition (5'- TT $\rm ^A/_T$ TNCACA) have been found on the \textit{E. coli} genome \cite{Roth1998}. The dissociation constant of DnaA binding to the DnaA boxes on the DNA lies in the range of $K_{\rm D}^{\rm s} = 1-50$~nM, depending on the flanking sequences \cite{Schaper1995}. While for some DnaA boxes, the binding was non-specific $K_{\rm D}^{\rm s} \geq  200$~nM, the highest affinity was found for the DnaA boxes R1 and R4 in the origin with $K_{\rm D}^{\rm s} = 1$~nM. In \textit{E. coli}, the approximately three hundred 9-mer DnaA boxes are randomly distributed on the \textit{E. coli} chromosome \cite{Hansen2018, Roth1998}. The \textit{dnaA} gene is regulated by two promoters, \textit{dnaAp1} and \textit{dnaAp2}, with a DnaA box located between them. \textit{dnaAp2} is the stronger promoter and contributes 60–80 \% of the dnaA transcripts \cite{Speck1999}. Both ATP-DnaA and ADP-DnaA bind cooperatively to these two promoters, but the repression via ATP-DnaA is more efficient \cite{Speck1999}. As there are five binding sites for DnaA in the promoter region \cite{Speck1999}, we choose a Hill coefficient of $n=5$ in the simulations.

In the AIT model we used $n_{\rm s}= 300$ titration sites per chromosome with a dissociation constant of $K_{\rm D}^{\rm s}=1$~nM (Table \ref{tab:RIT_AIT_parameters}). We approximate the experimentally reported random distribution of titration sites on the chromosome \cite{Hansen2018, Roth1998} by a homogeneous distribution.
At a concentration of ATP-DnaA of approximately $[D]_{\rm ATP}= 100$~nM, the expression of DnaA was reduced by 50 \% \cite{Speck1999}. Therefore, we used in the AIT model for the promoter a dissociation constant of $K_{\rm D}^{\rm p}=100$~nM. The dissociation constant of DnaA for the origin was chosen to be $K_{\rm D}^{\rm ori}=20$~nM, reflecting the combination of high and intermediate affinity of the titration sites required to be filled by ATP-DnaA in order to initiate replication. Using the experimentally reported topology of the biochemical network in combination with the growing cell model of gene expression, we obtain stable cell cycles with the AIT at low growth rates, but not at high growth rates as explained in the main text of the paper and in section \ref{sec:AIT_homogeneous_titration}.

\subsubsection{The titration sites}
\label{sec:titration_sites}
In this section, we present how the titration sites set a fixed replication threshold, such that a fixed number of initiator proteins needs to be accumulated per number of origin between consecutive replication initiation events. We discuss why the quasi-equilibrium assumption is appropriate and calculate the concentration of free initiator proteins as a function of the total initiator protein concentration $[p]_{\rm T}$ and of the total titration site concentration $[s]_{\rm T}$ in the cell.
\\
\\
\textbf{Binding and unbinding rates of DnaA binding to the titration sites are fast.} 
In the main text, we assumed that the binding and unbinding of the initiator proteins to the titration sites is well described by a quasi-steady-state. Here we show that the binding and
unbinding dynamics are relatively fast compared to the doubling time of the cell, such that this assumption is well justified. It seems reasonable to assume that DnaA finds its target sites in a way that is similar to that of other transcription factors, such as the lac repressor whose binding dynamics has been well characterized \cite{Elf2007}. These transcription factors move by facilitated diffusion, i.e. combining 3D with 1D diffusion along the DNA. Elf et al. \cite{Elf2007} have measured that the effective diffusion constant of transcription factors in \textit{E. coli} is of the order of $D_{\rm eff}=0.4 \, \mu$m$^2$/s. Assuming the binding rate is diffusion-limited, the binding rate is given by $k_{\rm on}=4 \pi \sigma D_{\rm eff}$. For an estimated cross section in the order of $\sigma \approx 10^{-2} \mu $m \cite{Paijmans2014}, the binding rate therefore becomes $k_{\rm on} \approx 0.05 \, \mu$m$^3$/s. The time for a transcription factor to bind to its target site is given by one over the concentration of the transcription factor $[c]$ times the binding rate: $\tau_{\rm on}=([c]\times k_{\rm on})^{-1}$. With a typical volume of an \textit{E. coli} cell of $V=1$ $\mu$m$^3$, the search time of one transcription factor for finding its target site on the DNA should then be $\tau_{\rm on}= k_{\rm on}^{-1} \times V = 20$~s. This estimate compares well to the measured value of $\tau_{\rm on}= 65-360$~s by Elf et al. \cite{Elf2007}. The dissociation constant of DnaA binding to the DnaA boxes on the DNA is in the range of $K_{\rm D}^{\rm s} = 1-50$~nM \cite{Schaper1995}. Using $K_{\rm D}^{\rm s}= k_{\rm off} / k_{\rm on}$ allows us to estimate $k_{\rm off} = K_{\rm D}^{\rm s} \times k_{\rm on} \approx 0.015-0.8$~s$^{-1}$. With an average concentration of the initiator protein DnaA in \textit{E. coli} of $[D]_{\rm T} \approx$~400~$\mu \rm{m}^{-3}$ \cite{Hansen1991}, the correlation time for binding and unbinding then becomes $\tau = 1 / (k_{\rm on} \, [D]_{\rm T} + k_{\rm off})\approx 0.16$~s. This is much faster than the timescale at which the volume changes, set by the growth rate. Recent FRAP experiments combined with single molecule tracking experiments show that DnaA rapidly moves between chromosomal binding sites and has a residence time of less than a second \cite{Schenk2017}. Thus, the quasi-equilibrium approximation of the initiator binding to the titration sites we make is well justified.
\\
\\
\textbf{Concentration of free initiator proteins in the
	quasi-equilibrium assumption.} As binding and unbinding dynamics of
the initiator protein to the titration sites are relatively fast, we
can assume for simplicity a quasi-equilibrium state of the
concentration of free initiator proteins
$[p] = K_{\rm D}^{\rm s}\,[sp] / [s]$ with the dissociation constant
$K_{\rm D}^{\rm s}$. At every given total titration site concentration
$[s]_{\rm T}= [s]+ [sp]$ and total initiator protein concentration
$[p]_{\rm T}= [p] + [ps]$, the average free initiator protein
concentration $[p]$ is given by the quadratic equation
\begin{align}
[p]([s]_{\rm T}, [p]_{\rm T})=&[p]_{\rm T}- \frac{K_{\rm D}^{\rm s} + [s]_{\rm T} + [p]_{\rm T}}{2} 
\nonumber \\
&+ \frac{\sqrt{(K_{\rm D}^{\rm s} + [s]_{\rm T} + [p]_{\rm T})^2-4\,[s]_{\rm T} \, [p]_{\rm T}}}{2}
\label{eq:free_proteins}
\end{align}
We use this expression in the main text to calculate at every given total titration site concentration and total initiator concentration in a cell the concentration of initiators freely diffusing in the cytoplasm. As can be seen in Figure 2A of the main text (and in Fig. \ref{fig:SI_1C3_RIT_homogeneous}~A and C), as long as there are more titration sites than proteins in the cell, the free DnaA concentration remains low. When the total number of DnaA proteins exceeds the total number of titration sites, the free concentration quickly rises and replication is initiated when the critical free initiator concentration $K_{\rm D}^{\rm ori}$ is attained. The fixed number of titration sites per chromosome therefore sets the critical number of initiators that need to be accumulated in order to reach the critical free initiator concentration $K_{\rm D}^{\rm ori}$ in the cytoplasm.

\subsubsection{The AIT model ensures stable cell cycles at all growth rates when all titration sites are located closely to the origin}
\label{sec:AIT_ribo_limiting}

The three key variables of the AIT model are the volume of the cell $V(t)$, the total number of DnaA proteins $N_{\rm p}(t)$ and the total number of titration sites $N_{\rm s}(t)$ in the cell. In the following we derive expressions for these three quantities and show that the AIT model gives rise to stable replication cycles at all growth rates when all titration sites are located at the origin.
From the growing cell gene expression model we derived the following volume-dependent expression for the change in the number of a negatively autoregulated protein $p$ (see section \ref{sec:ribo_limiting_neg_auto}):
\begin{equation}
\frac{dN_{\rm p}}{dt}= \frac{\phi_{\rm p}^0 \, \lambda \, \rho}{1+\left( \frac{[p]}{K_{\rm D}^{\rm p}}\right)^{\rm n}} \, V
\end{equation}
with the gene allocation fraction $\phi_{\rm p}^0$, the growth rate $\lambda$, the number density $\rho= N/V$, the free initiator concentration $[p]$, the dissociation constant of the promoter $K_{\rm D}^{\rm p}$ and the Hill coefficient $n$. As the DnaA gene is located very closely to the origin, we assume that at the moment of replication initiation the gene number doubles instantaneously. We summarize the terms that do not depend on the cell volume or the growth rate in the gene allocation density
\begin{equation}
\tilde{\phi}_{\rm p}^0 = \phi_{\rm p}^0 \, \rho
\label{eq:alphap}
\end{equation}
and obtain Eq. 1 of the main text 
\begin{equation}
\frac{dN_{\rm p}}{dt}= \frac{\tilde{\phi}_{\rm p}^0 \, \lambda \, V}{1+ \left( \frac{[p]}{K_{\rm D}^{\rm p}}\right)^{n}}.
\label{eq:AIT_production_rate_growing_cell}
\end{equation}

As explained in the main text, in the AIT model we explicitly model the exponentially growing cell with the growth rate $\lambda$:
\begin{equation}
\frac{dV}{dt}=\lambda \, V
\label{eq:volume_growth}
\end{equation}

Replication is initiated when the amount of initiators exceeds the number of titration sites per chromosome (see section \ref{sec:titration_sites}). Here, for simplicity, we assume that all titration sites are located at the origin and therefore the total number of titration sites is doubled instantaneously after replication initiation. In the next section, we present the more realistic scenario that the titration sites are distributed homogeneously along the chromosome. Based on the experimental observation that the cell divides an approximately constant time $\tau_{\rm cc}$ after replication initiation, we assume here that $\tau_{\rm cc}$ is constant (see section \ref{sec:division_separate} for scenario where $\tau_{\rm cc}$ is not constant). At cell division, not only the volume, but also the total number of initiators and of titration sites is divided by two. 

Evolving the number of initiator proteins and the volume according to equations \ref{eq:AIT_production_rate_growing_cell} and \ref{eq:volume_growth}, respectively, we find that the total DnaA concentration remains approximately constant in time (Fig. \ref{fig:SI_1C3_RIT_homogeneous}~A and B). The weak oscillations in the total concentration arise from the effect of a finite replication time of the chromosome as explained in section \ref{sec:ribo_limiting_temporal_variation}. When the total number of DnaA proteins exceeds the total number of titration sites on the chromosome, the free DnaA concentration rises and at the critical initiation concentration $K_{\rm D}^{\rm ori}$, replication is initiated. As here all titration sites are located at the origin, the number of titration sites doubles and the free concentration drops immediately after replication initiation both at high and at low growth rates (Fig. \ref{fig:SI_1C3_RIT_homogeneous}~A and B). Only when again enough initiator proteins have been accumulated is a new round of replication initiated. The AIT model therefore gives rise to stable cell cycles at all growth rates (Fig. \ref{fig:SI_1C3_RIT_homogeneous}~A and B), when all titration sites are located on the origin.

An open question remains what role negative autoregulation plays in the AIT model. In order to attain the critical initiation concentration $K_{\rm D}^{\rm ori}$ at the origin, the dissociation constant of the promoter of DnaA $K_{\rm D}^{\rm p}$ must be higher than $K_{\rm D}^{\rm ori}$. At the same, the mechanism of titration requires that the affinity of the titration sites is higher than that of the origin: $K_{\rm D}^{\rm s} < K_{\rm D}^{\rm ori}$. Combining these two requirements yields: $K_{\rm D}^{\rm s} < K_{\rm D}^{\rm ori} < K_{\rm D}^{\rm p}$. The free protein concentration $[p]$ thus remains (far) below the promoter dissociation constant $K_{\rm D}^{\rm p}$, which means the latter is repressed only weakly and proteins are produced approximately at the maximal rate. Therefore, equation \ref{eq:AIT_production_rate_growing_cell} can be approximated by
\begin{equation}
\frac{dN_{\rm p}}{dt}\approx \tilde{\phi}_{\rm p}^0 \, \lambda \, V
\end{equation}
The stability of the AIT model arises from the volume dependence in the initiator production rate in equation \ref{eq:AIT_production_rate_growing_cell} as explained in section \ref{sec:init_accum_stability_criteria}.

\subsubsection{The homogeneous titration site distribution causes reinitiation events at high growth rates}
\label{sec:AIT_homogeneous_titration}
In the previous section, we assumed out of simplicity that all titration sites are located at the origin. Yet, experiments indicate that the titration sites are distributed approximately homogeneously on the chromosome \cite{Hansen2018, Roth1998}. Here, we investigate how a homogeneous titration site distribution on the chromosome affects the stability of the cell cycles. 
\begin{figure}
	\centering
	\includegraphics[width =\textwidth]{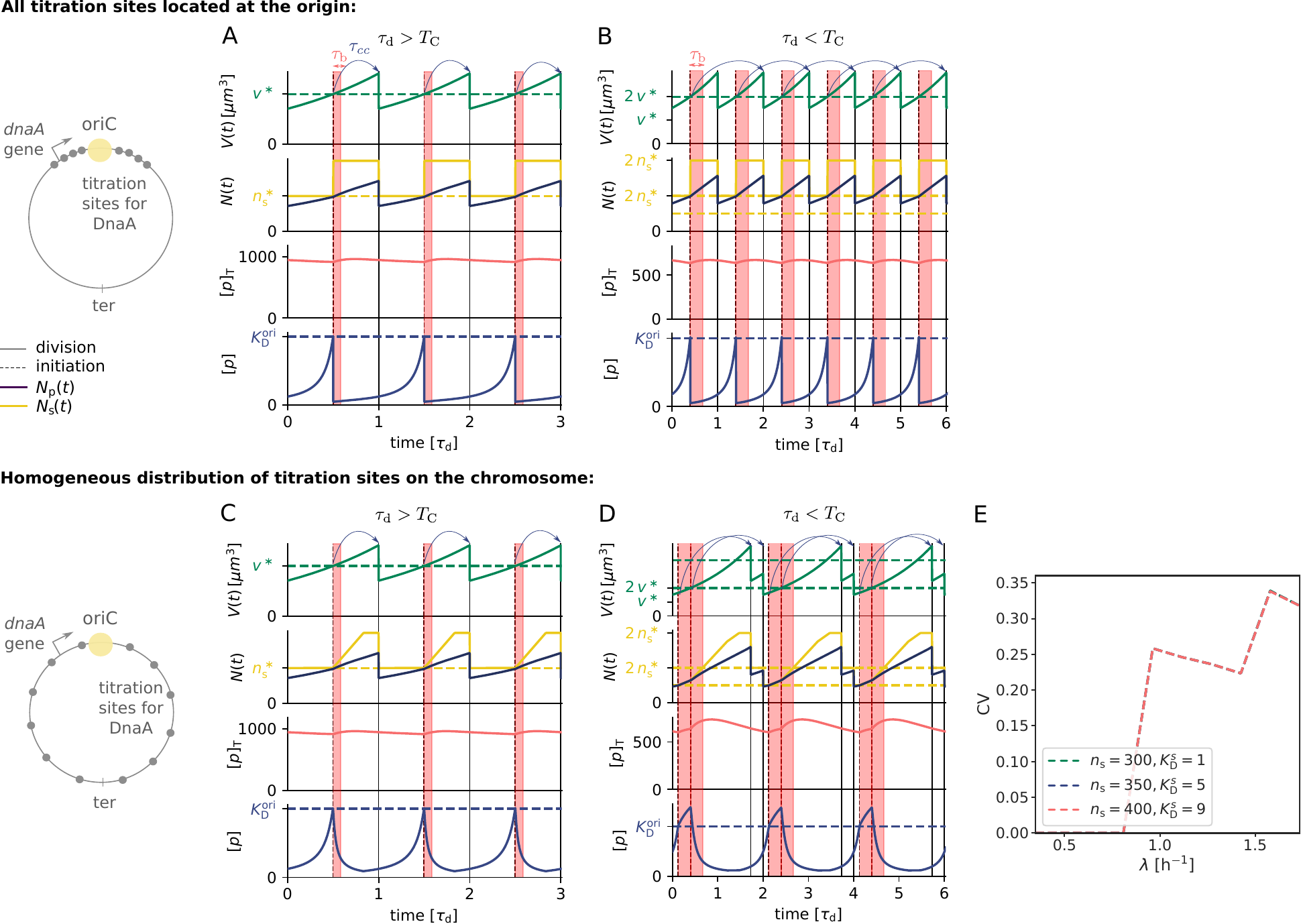}
	\caption{\textbf{A homogeneous titration site distribution on the chromosome in the AIT model causes reinitiation events at high growth rates} (A, B, C, D): The volume $V(t)$, the number of initiator proteins $N_{\rm p}(t)$ (black line) and titration sites $N_{\rm s}(t)$ (yellow line), the total concentration of initiator proteins $[p]_{\rm T}(t)$ together with the dissociation constant of the regulator $K_{\rm D}^{\rm r}$ (dotted red line), and the concentration of initiator proteins in the cytoplasm $[p](t)$ as a function of time (in units of the doubling time of the cell $\tau_{\rm d}$) for $\tau_{\rm d}=2$~h (A, C) and $\tau_{\rm d}=35$~min (B, D), respectively. When the number of initiator proteins per origin $n_{\rm p}(t)$ exceeds the number of titration sites per origin $n_{\rm s}$ (yellow dashed line), the free concentration $[p](t)$ rapidly rises to reach the threshold concentration $K_{\rm D}^{\rm ori}$ (blue dashed line) for initiating a new round of replication. The blue arrows indicate that the cell divides a constant cycling time $\tau_{\rm cc}$ after replication initiation. During the blocked period $\tau_{\rm b}$(red shaded area), no new round of replication can be initiated. (A, B) If all titration sites are located at the origin, the free initiator concentration $[p](t)$ decreases immediately after replication is initiated, independent of whether the doubling time of the cell $\tau_{\rm d}$ is smaller (A) or larger (B) than the time $T_{\rm C}$ to replicate the entire chromosome. (C) When the titration sites are distributed homogeneously along the chromosome, the free initiator concentration decreases during the entire replication time $T_{\rm C}$ at low growth rates. As the time to produce new titration sites is still faster than the time to synthesize new initiator proteins, we obtain regular stable cell cycles in this regime. (D) When the doubling time is however smaller than the time to replicate the entire chromosome, $\tau_{\rm d} <T_{\rm C}$, newly replicated titration sites are being filled faster with new proteins than they are replicated. After a short blocked period $\tau_{\rm b}$, replication is reinitiated. As a result, each long (sub)cycle is followed by a very short one, together forming the cell cycle. Moreover, replication is not initiated at a constant volume per origin anymore, but oscillates over time. 
		The appearance of premature reinitiation events suggest that replication initiation in \textit{E. coli} can not fully be explained by a titration-based mechanism. (E) The coefficient of variation ${\rm CV}= \sigma /\mu$ with the standard deviation $\sigma$ and the average initiation volume $\mu=\langle v^\ast \rangle$ as a function of the growth rate for the AIT model with homogeneous titration site distribution. Due to the rapid reinitiation events shown in (D), the coefficient of variation increases strongly in the overlapping fork regime at high growth rates. Varying the total number of titration sites in concert with the dissociation constant of the titration sites $K_{\rm D}^{\rm s}$ such that the initiation volume remains constant and equal to the experimentally observed initiation volume \cite{Si2017} (by solving equation \ref{eq:free_proteins} for $v^\ast$) cannot prevent these reinitiation events. This demonstrates that the failure of the titration model at high growth rates is independent of the precise parameter choice; it is thus a robust result.
	}
	\label{fig:SI_1C3_RIT_homogeneous}
\end{figure}

When the titration sites are distributed homogeneously along the chromosome, the number of titration sites $N_{\rm s}(t)$ is not directly proportional to the number of origins anymore but increases linearly from the moment of initiation of replication $t_{\rm i}$ until the end of replication at $t_{\rm i} + T_{\rm C}$:
\begin{equation}
N_{\rm s}(t) = \begin{cases}
N_{\rm 0} & \text{for } t < t_{\rm i} \\
N_{\rm 0} +N_{\rm 0}\, \frac{t-t_{\rm i}}{T_{\rm C}}\, & \text{for } t_{\rm i} \leq t < t_{\rm i} + T_{\rm C} \\
2 \, N_{\rm 0} & \text{for } t \ge t_{\rm i} + T_{\rm C}
\end{cases}
\label{eq:titration_sites_changes_hom}
\end{equation}
with the C-period $T_{\rm C} \approx 40$ min being the time to replicate the entire chromosome and $N_{\rm 0} = n_{\rm s} \, n_{\rm ori}$ is the total number of titration sites before replication initiation, given by the number $n_{\rm s}$ of titration sites per chromosomes times the number $n_{\rm ori}$ of origins before replication initiation. In the main part of the paper we used the experimental observation that the cell divides an approximately constant cycling time $\tau_{\rm cc}$ after replication has been initiated \cite{Si2017}. This cycling time can be split into two times $\tau_{\rm cc} = T_{\rm C} + T_{\rm D}$, the C-period and the D-period: During the C-period, the DNA is being replicated and during the D-period the chromosomes are being separated and the cell divides \cite{Cooper1968, Wallden2016, Si2017}. 
The total number of binding sites before initiation $N_0$ will only be doubled, when the entire chromosome has been replicated, thus after the end of the C-period $T_{\rm C}$. 

In the low growth regime, the time to replicate the entire chromosome
$T_{\rm C}$ is shorter than the time to double the volume of the cell
$\tau_{\rm d}$. The time it takes to double the number of titration sites upon replication initiation is therefore shorter than the time to double the number of initiation proteins. This results in a gradual decrease of the free initiator concentration upon replication initiation (Fig. \ref{fig:SI_1C3_RIT_homogeneous} C, lowest panel). 

In favorable growth conditions, the doubling time of
\textit{E. coli} can however be shorter than the time it takes to replicate the entire chromosome $T_{\rm C}$. As a result, the rate at which new titration sites are formed upon the first replication initiation event (marked by the dashed vertical lines) is therefore lower than the rate at which initiator proteins are produced (Fig. \ref{fig:SI_1C3_RIT_homogeneous} D); the number of titration sites (yellow line) rises slower than the number of initiators (black line). This means that after the first replication initiation event, the free initiator concentration continues to {\em rise} (lower row). To prevent immediate reinitiation, we introduce a refractory or `eclipse' period of $\tau_{\rm b} \approx 10 $~min after replication initiation during which replication initiation is blocked (red shaded area), mimicking the effect of SeqA
\cite{Campbell:1990it,Lu:1994ee,Waldminghaus:2009em, Katayama2017}. When this eclipse period is over, a new round of replication is initiated, which triples the rate at which new titration sites are formed. Now the rate of titration-site formation is higher than the rate at which new initiator proteins are produced, causing the concentration of free
initiator to go down. At some point, the first round of
replication is finished, causing a small decrease in the rate at
which new titration sites are formed and some time later also the next round is finished, causing the number of titration sites to become constant. Then the time $\tau_{\rm cc}$ after the first initiation event is reached and the cell divides. After this division event it grows briefly and then it divides again, a time $\tau_{\rm cc}$ after the second initiation event in the previous cycle. A given cell cycle thus consists of a long and a short cycle, such that the average division time (time from birth to death) equals the doubling time $\tau_{\rm d} = \ln(2)/\lambda$.
These unnatural time traces of the volume, namely the oscillation between a short and a long (sub) cycle, have not been observed experimentally and can be prevented by decoupling cell division from replication initiation as described in section \ref{sec:division_separate}. The reinitiation events, which are caused by the excess of initiators after the first initiation event are however not affected by the choice of how the replication and the division cycle are coupled. Because in the long (sub) cycle two initiation events are triggered in rapid succession, the initiation volume {\em per origin} flip-flops between a high and a low initiation volume per origin. This causes a dramatic rise of the Coefficient of Variation (CV) in the initiation volume at higher growth rates  (see Fig. \ref{fig:SI_1C3_RIT_homogeneous} E). Importantly, the CV  becomes much larger than that observed experimentally, {\em even though the system is deterministic and no biochemical noise is present; adding noise would only make the CV even higher.}

We emphasize that the breakdown of the titration mechanism arises from the different scaling of two timescales with the growth rate: The rate at which the initiator DnaA is synthesized scales with the growth rate, see Eqs. \ref{eq:alphap} and \ref{eq:AIT_production_rate_growing_cell}. In contrast, the titration-site formation rate is nearly independent of the growth rate: when the titration sites are homogeneously distributed, as experiments show \cite{Roth1998}, then the titration-site formation rate per origin is set by the DNA duplication rate, which indeed varies only little with the growth rate \cite{Si2017}. The protein synthesis rate thus increases faster with the growth rate than the titration-site formation rate, which means that at sufficiently high growth rates the mechanism fails to sequester DnaA proteins after replication initiation; to a good approximation, this breakdown happens when the system enters the overlapping replication-fork regime with $\tau_{\rm d}\lesssim T_{\rm C}$, because the rate at which titration sites are formed is given by $n_{\rm s}/T_{\rm C}$, while the rate at which proteins are produced right after replication initiation is given by $dN_{\rm p} / dt = \phi_{\rm p}^0 \, \lambda \, \rho \, V = \lambda \, N_{\rm p} = \ln(2) \, N_{\rm p} / \tau_{\rm d} \simeq \ln(2) \, n_{\rm s} / \tau_{\rm d}$, where we have used that the fraction of initiator equals the gene (ribosome) allocation fraction $\phi_{\rm p}^0$ (assuming all proteins are made with the same rate) and right after replication initiation $N_{\rm p} \simeq n_{\rm s}$. Since this prediction follows from the scaling of two timescales, it is robust, i.e. insensitive to the details of the model.

Indeed, this prediction is insensitive to how the other key parameters in the AIT model are varied: the number of titration sites per origin $n_{\rm s}$ and their affinity $K_{\rm D}^{\rm s}$. Fig. \ref{fig:SI_1C3_RIT_homogeneous} E shows that exactly the same rise in the CV of the initiation volume is observed for different values of $n_{\rm s}$ and $K_{\rm D}^{\rm s}$, which are varied together to keep the average initiation volume constant and within the range observed experimentally \cite{Si2017}. The fact that the curves nearly fully overlap is because a new replication round is initiated as soon as the eclipse period is over.

Naturally, if the affinity of the titration sites located at the origin is higher than the affinity of titration sites at the rest of the chromosome, we can recover the behavior of the inhomogeneous titration site distribution. Interestingly, it had been proposed that the site \textit{datA} which is located close to the origin has a very high affinity and can titrate large numbers of proteins, of up to 60-370 \cite{Kitagawa1996, Blaesing2000}. These numbers had been inferred indirectly, from experiments that analyzed the de-repression of dnaA or mioC transcription upon introduction of plasmids containing \textit{datA} sequences \cite{Kasho2013}. It remained however unclear by which mechanism \textit{datA} would be able to absorb so many DnaA molecules. The discovery that the site \textit{datA} can deactivate the initiator protein ATP-DnaA by promoting ATP hydrolysis provides a more likely explanation for this indirect observation \cite{Kasho2013}. In the original initiator titration paper by Hansen et al. \cite{Hansen1991}, a bias of titrating DnaA boxes towards the oriC region was assumed. Roth and Messer \cite{Roth1998} find however that while boxes of the R1 type indeed show such a bias, the high-affinity DnaA boxes show a distribution on the chromosome as random as possible.

\subsubsection{Growth-rate dependence of the cell cycle in the AIT model}
\label{sec:titration_growth_rate_dependence}
In this section, we discuss how the key cell cycle parameters---the initiation time and volume, and the volume at birth and division---vary with the growth rate $\lambda$ in the AIT model. The initiation time is given by $t^\ast = \tau_{\rm d} - \tau_{\rm cc}$, where $\tau_{\rm d} = \ln(2) / \lambda$ is the cell division time and $\tau_{\rm cc}$ is the constant time between initiation and division. The volume at birth, $V_{\rm b}$, and the initiation volume $V^\ast$ are related via $V^\ast = V_{\rm b} e^{\lambda t^\ast}$, and the volume at division $V_{\rm d}$ is simply twice the birth volume $V_{\rm b}$. The central quantity is thus the total initiation volume $V^\ast$, or its value per origin $v^\ast = V^\ast / n_{\rm ori}$, with  $n_{\rm ori}^\ast$ the number of origins at initiation: from this and the initiation time, $V_{\rm b}$ and $V_{\rm d}$ follow.

To obtain the initiation volume, we exploit that at the moment of replication initiation the free initiator concentration $[p]$ equals the dissociation constant for binding the origin: $[p]=K_{\rm D}^{\rm ori}$. For a given total intiator concentration $[p]_{\rm T}$, we can then combine $[p]=K_{\rm D}^{\rm ori}$ with Eq. \ref{eq:free_proteins} to obtain the total titration site concentration $[s]_{\rm T}$. The latter is given by $[s]_{\rm T} = n_{\rm s} / v^\ast$, where $n_{\rm s}$ is the known number of titration sites per origin. Hence, for a given $[p]_{\rm T}$ we can obtain the initiation volume $v^\ast$ from Eq. \ref{eq:free_proteins}.

To understand how the initiation volume $v^\ast$ depends on the total initiator concentration $[p]_{\rm T}$ it is illuminating to consider the limit in which the binding of the initiator proteins to the titration sites is very strong. We connect the critical number of initiators per origin $n^\ast$ to the initiation volume per origin $v^\ast$ via the total concentration of initiators at the moment of initiation:
\begin{equation}
[p]_{\rm T}^\ast = \frac{N_{\rm p}^\ast}{V^\ast} = \frac{N_{\rm p}^\ast/n_{\rm ori}^\ast}{V^\ast/{n_{\rm ori}^\ast}} = \frac{n^\ast}{v^\ast}
\end{equation}
where $N_{\rm p}^\ast$ is the total number of initiators at initiation. Hence,
\begin{equation}
v^\ast = \frac{n^\ast}{[p]_{\rm T}^\ast}
\label{eq:v_init_titration_ait_model}
\end{equation}
In the limit of very tight binding, the critical number of initiators per origin $n^\ast$ is set by the fixed number of titration sites per origin, $n^\ast \approx n_{\rm s}$, and thus is constant when a new round of replication is initiated in the non-overlapping replication fork regime. Now we see that if the total concentration is maintained approximately constant in time, the initiation volume is also constant and the replication cycle becomes stable. Furthermore, the total concentration could be maintained approximately constant in time for a given growth rate, but vary as a function of the growth rate $\lambda$. Then, the growth-rate dependence of the total concentration directly translates into a growth-rate dependence of the initiation volume:
\begin{equation}
v^\ast (\lambda) = \frac{n_{\rm s}}{[p]_{\rm T}(\lambda)}
\label{eq:growth_rate_dependence_v_init_titration}
\end{equation}
Experiments indicate that both the initiation volume and the total DnaA concentration vary by about 50\% over a tenfold change in the growth rate, yet in an opposite, anti-correlated fashion \cite{Zheng2020}. This is consistent with Eq. \ref{eq:growth_rate_dependence_v_init_titration}. Yet, how the total concentration varies by only 50\% over this range in growth rates remains unclear. As we have seen in section \ref{sec:AIT_ribo_limiting}, in the AIT model titration interferes with negative autoregulation such that the initiator is effectively constitutively expressed. Experiments indicate that while the concentration of negatively autoregulated proteins is relatively independent of the growth rate, the concentration of constitutively expressed proteins typically decreases linearly with the growth rate \cite{Klumpp2009, Scott2010}; Eq. \ref{eq:growth_rate_dependence_v_init_titration} would then predict that the initiation volume increases with the growth rate. How the total DnaA concentration and hence the initiation volume are kept within a 50\% range over the tenfold change in the growth rate thus remains an open question. 

\subsubsection{Adder correlations in the AIT model}
\label{sec:AIT_adder}
In sections \ref{sec:init_accum_stability_criteria} and \ref{sec:AIT_homogeneous_titration}, we showed that a volume-dependent production rate of DnaA in combination with homogeneously distributed titration sites yields stable cell cycles at low growth rates. In this section, we confirm that the AIT model also ensures stable cell cycles at low growth rates in the presence of gene expression noise in DnaA and we show that the AIT model gives rise to the experimentally observed adder correlations in the initiation volume in this growth-rate regime.

We again model the change in the number of DnaA proteins according to equation \ref{eq:AIT_production_rate_growing_cell} plus a noise term $\xi_{\rm D}(t)$ accounting for noise in gene expression
\begin{equation}
\frac{dN_{\rm D}^{\rm T}}{dt}= \frac{\phi^0\,\lambda \, \rho }{1+\left( \frac{[D]_{\rm T}}{K_{\rm D}^{\rm p}}\right)^n} \, V + \xi_{\rm D} (t).
\label{eq:AIT_noise_total_conc}
\end{equation}
The noise is modelled as Gaussian white noise, $\langle \xi(t) \xi(t^\prime) \rangle = 2 D_{\rm l} \delta (t-t^\prime)$, with the noise strength $D_{\rm l}$ chosen to match the measured variance in the initiation volume of $CV=0.1$ \cite{Wallden2016} (see Table \ref{tab:LD_LDDR_parameters}). 

\begin{figure}
	\centering
	\includegraphics[width =0.4\textwidth]{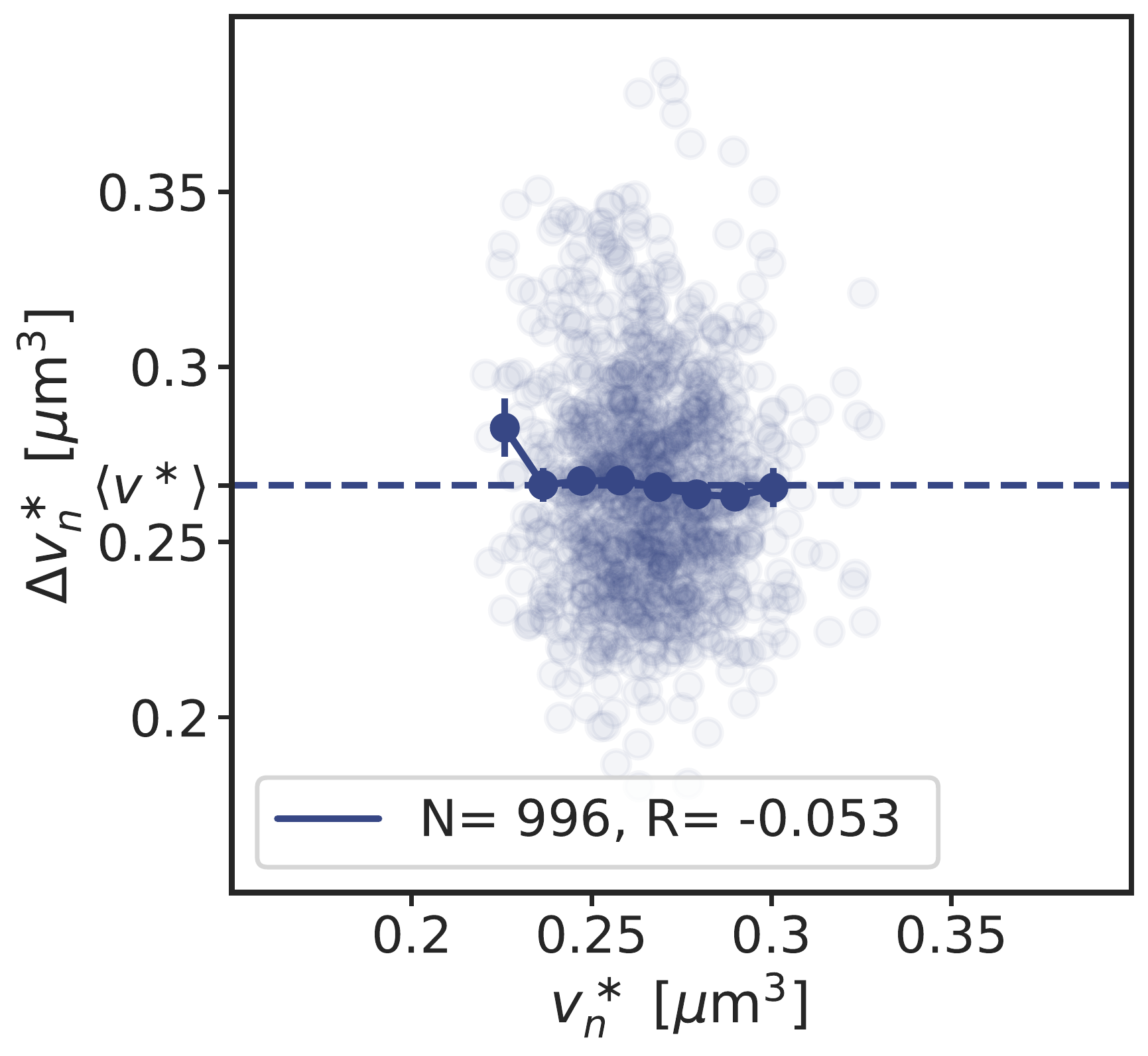}
	\caption{\textbf{Gene expression noise in the DnaA concentration gives rise to adder correlations in the initiation volume in the AIT model} The added volume per origin between successive initiation events, $\Delta v^\ast_{\rm n}= 2 \, v^\ast_{\rm n+1}-v^\ast_{\rm n}$, is independent of the initiation volume $v^\ast_{\rm n}$ per origin and on average equal to the average initiation volume, $\langle\Delta v^\ast \rangle=\langle v^\ast\rangle$, as expected for an initiation volume adder. The doubling time is $\tau_{\rm d}=2$~h and the number of data points $N$ and the Pearson correlation coefficient $R$ are indicated. }
	\label{fig:S4_AIT_adder}
\end{figure}

Figure \ref{fig:S4_AIT_adder} shows that the AIT model maintains a stable initiation volume in the presence of gene expression noise. The fluctuations in the total concentration result in fluctuations in the initiation volume. The volume dependence of the production rate in equation \ref{eq:AIT_noise_total_conc} ensures however that the total concentration regresses back to a stable fixed point as derived in section \ref{sec:ribo_limiting_temporal_variation}. Therefore, also the initiation volume regresses back to the initiation volume derived in section \ref{sec:titration_growth_rate_dependence}. Plotting the volume that is added between two consecutive initiation events, $\Delta v^\ast_{\rm n}= 2 \, v^\ast_{\rm n+1}-v^\ast_{\rm n}$, as a function of the initiation volume $v^\ast_{\rm n}$, reveals that the AIT model exhibits adder correlations in the initiation volume. This observation can be understood intuitively. The key feature of an adder is that the added volume is on average constant, independent of the last initiation volume \cite{Campos2014, Taheri-Araghi2015, Si2019}. As explained in section \ref{sec:init_accum_stability_criteria}, in the initiator accumulation models replication is initiated when a critical number of initiators per origin have been accumulated. In the AIT model, this critical number is set by the fixed number of titration sites per chromosome. Replication is thus initiated when the number of proteins that have been produced since the last initiation event equals the number of titration sites, irrespective of the magnitude of the last initiation volume. As the number of initiators is accumulated proportionally to the volume of the cell, a fixed amount of accumulated proteins maps directly to a fixed volume that needs to be accumulated (Eq. \ref{eq:AIT_noise_total_conc}). The added volume since the last initiation event is thus, on average, always the same, irrespective of the last initiation volume---the hallmark of an adder.

\section{Initiator switch models}
\label{sec:switch_models}
The initiator protein DnaA binds strongly to the nucleotides ATP and ADP, but only the ATP-bound form of DnaA can form the initiation complex at the origin \cite{Katayama2017}. The switch between ATP-DnaA and ADP-DnaA is tightly regulated via several activators and deactivators in \textit{E. coli} \cite{Katayama2017} and the ATP-DnaA fraction increases before initiation of replication and decreases rapidly afterwards \cite{Kurokawa1999, Katayama2001}. Mutations or deletions of one or several activators and deactivators strongly affect the initiation volume per origin and can even lead to non-viable cells \cite{Ogawa2002, Camara2005, Riber2006, Riber2016, Kasho2013, Xia1995, Saxena2013}. Based on this experimental evidence, we present and analyse two models in which replication initiation is regulated via a switch of the active form of DnaA. First, we give an overview of the experimental data known about the ATP/ADP-switch of DnaA so far (section \ref{sec:switch_parameters}).  We first present the LD model, a minimal model consisting only of the activating acidic-phospholipids in the cell membrane and the deactivator \textit{datA} located on the chromosome (section \ref{sec:LD_model}). We argue that these are the main activator and deactivator at low growth rates. At high growth rates, several additional mechanisms are known to play an important role in regulating replication initiation in \textit{E. coli}. We include all so far known activators and deactivators in the LDDR model and show that we obtain stable cell cycles with high amplitude oscillations in the ATP-DnaA fraction (section \ref{sec:LDDR_model}). 
Then we show that the characteristic adder correlations can be obtained in the LD and LDDR model by including noise in the switch components (section \ref{sec:switch_noise}).
All parameters used in the LD and LDDR model in the main part of the paper and in the SI can be found in Table \ref{tab:LD_LDDR_parameters}. 

\subsection{Experimental findings and parameters of the switch models}
\label{sec:switch_parameters}
Here we give an overview of the experimental findings on the regulation of replication initiation in \textit{E. coli} and discuss realistic parameter ranges of the parameters of the switch models. As the protein DnaA binds very strongly to both ATP and ADP with a dissociation constant of $K_{\rm D}=10-100$~nM \cite{Kawakamii2005, Katayama2017}, we assume that DnaA is always bound to either ATP or ADP. In the following we refer to the total DnaA concentration as $[D]_{\rm T}$, to the ATP-DnaA as $[D]_{\rm ATP}$ and to the ADP-DnaA as $[D]_{\rm ADP}$. The total DnaA concentration $[D]_{\rm T}= [D]_{\rm ATP} + [D]_{\rm ADP}$ varies within a 50\% range over a tenfold change in the growth rate \cite{Zheng2020}. Hansen et al. reported a typical number of 330 molecules in an \textit{E. coli} cell with the doubling rate $1/\tau_{\rm d}= 0.58$~h$^{-1}$ \cite{Hansen1991}. Combining this number with the estimated volume at this doubling rate using the data from Si et al. \cite{Si2017} (for the same \textit{E. coli} strain K-12) of $V(\tau_{\rm d}^{-1}=0.58$ h$^{-1}) \approx 0.7 \, \mu$m$^{3}$, we obtain an estimated concentration of DnaA of $[D]_{\rm T}\approx 471 \, \mu$m$^{-3}$. We use throughout this work a total DnaA concentration of $[D]_{\rm T}=400 \, \mu$m$^{-3}$ (See Table \ref{tab:LD_LDDR_parameters}). As discussed in the previous section, the protein DnaA can be bound to the DnaA boxes on the DNA and at the origin or be diffusing in the cytoplasm. In a first step, we neglect the effect of the titration sites on the chromosome and assume that all DnaA proteins are present in the cytosol (entire section \ref{sec:switch_models}). In the full model in section \ref{sec:switch_titration_combined}, we relax this constraint and investigate the effect of a negatively autoregulated DnaA protein that can also bind to titration sites.

In \textit{E. coli}, replication is initiated once per cell cycle at the origin region by the binding of ATP-DnaA to two high-affinity DnaA boxes (R1 and R4) and several low-affinity DnaA boxes together with two other proteins, the DnaA-initiator-associating protein DiaA and the integration host factor (IHF) \cite{Katayama2017}. While ADP-DnaA can bind to the DnaA boxes on the origin, it does not form the cooperative complex required for the initiation of replication. The fraction of ATP-DnaA is maintained at a low level during most of the cell cycle and increases to approximately 80\% at the moment of replication initiation \cite{Kurokawa1999, Katayama2001}. An interesting and strongly debated question is whether replication is initiated at a critical amount, concentration or fraction of ATP-DnaA in the cell \cite{Donachie2003, Flatten2015, Hansen2018}. In our LD and LDDR model, replication is initiated when the ATP-DnaA concentration in the cell attains a critical concentration $[D]_{\rm ATP}^\ast$. We exploit that the total concentration of DnaA is maintained approximately constant and take $[D]_{\rm T}=$~constant in our model such that a critical initiation concentration $[D]_{\rm ATP}^\ast$ corresponds to a critical fraction $f^\ast=[D]_{\rm ATP}^\ast/ [D]_{\rm T}$.
In section \ref{sec:switch_mapping_total_conc} we also analyse the implications of the difference between initiating replication at a critical fraction versus a critical concentration of active DnaA. In section \ref{sec:switch_noise_total} we investigate the effect of fluctuations in the total concentration of DnaA.
\\
\\
So far, several activators and deactivators of DnaA have been
identified and characterized in great detail. Here, we briefly
summarize these experimental results, starting with the
deactivators. \\
\\
\textbf{Deactivation mechanisms: datA and RIDA} Regulatory Inactivation of DnaA (RIDA) was the first deactivation mechanism of DnaA that could be identified \cite{Katayama1998}: The DNA polymerase clamp on newly synthesized DNA forms a complex with ADP and the Hda protein. The resultant ADP-Hda-clamp-DNA complex interacts
with ATP-DnaA molecules catalytically and stimulates ATP hydrolysis
yielding ADP-DnaA. This system is predominant in the inactivation of
DnaA after replication initiation as it strongly represses
over-initiation of replication \cite{Katayama2017}. Importantly, at
low growth rates ($\tau_{\rm d}> T_{\rm C}$) the replication forks are
not overlapping and RIDA is inactive at the moment of replication
initiation. RIDA can hydrolyze at least 0.9 molecules of ATP-DnaA per DNA-clamp-Hda complex per minute \textit{in vitro} \cite{Nakamura2010, Kasho2013}. Single-cell experiments have shown that the number of DNA-bound sliding clamps increases during the cell cycle, peaking at more than 8 per replication fork \cite{Moolman2014}. We therefore use a deactivation rate of $\beta_{\rm RIDA}= 8$~min$^{-1}$ in the LDDR model (See Table \ref{tab:LD_LDDR_parameters}). This RIDA deactivation rate $\beta_{\rm RIDA}$ is only non-zero during active
replication for $T_{\rm C}$ after initiation and is taken to be constant during this time period. Besides RIDA, a chromosomal site named \textit{datA} can hydrolyze ATP-DnaA via a process called
\textit{datA}-dependent DnaA-ATP Hydrolysis (DDAH)
\cite{Kasho2013}. DDAH is temporally regulated over the course of the
cell cycle via the protein IHF. The binding of IHF to \textit{datA}
increases within about 5-10 minutes, peaks at about 15 minutes, and
decreases again about 20-30 minutes after initiation of replication
\cite{Katayama2017, Kasho2013}. In the LD model, we neglect this
temporal variation in the deactivation rate and take it to be constant
for simplicity. In the LDDR model, we have two activity states of
DDAH, a high deactivation rate $\beta_{\rm datA}^+$ from the moment of
replication initiation onward ($\tau_{\rm datA}^+=0$~h) until 0.2~h
after replication initiation ($\tau_{\rm datA}^-=0.2$~h) and a low
deactivation rate $\beta_{\rm datA}^-$ during the rest of the cell
cycle (See Table \ref{tab:LD_LDDR_parameters} and Fig. 4A/B of the main text). \textit{In vitro}, the hydrolysis rate of ATP-DnaA can be at least 1.6 molecules per minute per \textit{datA}, and \textit{in vivo} the deactivation strength of DDAH is about $20-30\%$ of that of RIDA \cite{Kasho2013}. Therefore, we use a deactivation rate of $\beta_{\rm datA}= 10$~min$^{-1}$ in the LD and $\beta_{\rm datA}^+= 10$~min$^{-1}$ in the LDDR model. From the
experimental findings on the temporal variation in the IHF binding to \textit{datA} \cite{Kasho2013}, we estimate that the activity of DDAH goes down by a factor of two or three in the low activity state and we use a low deactivation rate of $\beta_{\rm datA}^-=5$~min$^{-1}$ (See Table \ref{tab:LD_LDDR_parameters}).\\
\\
Concerning activation, at least three mechanisms for the production of
ATP-DnaA have been characterized: de novo DnaA synthesis; nucleotide
dissociation from ADP-DnaA by acidic phospholipids in the cell
membrane; and a mechanism involving specific chromosomal DNA sequences
termed \textit{DARS} sites \cite{Katayama2017}.\\
\\
\textbf{Activation mechanism: acidic phospholipids} \textit{In vitro}, acidic lipids such as cardiolipin (CL) and phosphotidylglycerol (PG) can enhance the release of ADP and ATP from DnaA, but Phosphatidylethanolamine (PE), comprising nearly 80\% of the phospholipids, is inert \cite{Sekimizu1988, Crooke1992}. DnaA can bind ATP and ADP in the absence of phospholipids \cite{Sekimizu1988}. \textit{In vitro}, CL and PG can restore replication activity of DnaA bound to ADP \cite{Sekimizu1988, Saxena2013, Katayama2017,Yung1988}. \textit{In vivo}, reducing the concentration of acidic lipids leads to growth arrest \cite{Saxena2013, Zheng2001, Fingland2012} and inhibited replication initiation \cite{Fingland2012}; in section \ref{sec:switch_validation_mutations}, we discuss in more detail the effect of mutations in the acidic lipid synthesis. Based on these experimental observations, we envision the following lipid-mediated activation scenario: DnaA-ADP generated by \textit{datA} and RIDA binds the lipids, causing the ADP to dissociate \cite{Sekimizu1988, Crooke1992}. After DnaA then dissociates from the lipids, it rapidly binds ATP in the cytoplasm \cite{Sekimizu1988}, and subsequently oriC \cite{Katayama2010, Nishida:2002dp, Speck2001}, initiating replication. This scenario gives rise to the model of the main text, in which the lipid-mediated activation rate is independent of the origin density.  We emphasize, however, that while the experiments clearly demonstrate that acidic phospholipids can enhance the release of ADP \cite{Sekimizu1988, Crooke1992}, it remains unclear how important the lipids are for DnaA re-activation \textit{in vivo} \cite{Shiba2004, Shiba2012,Camsund2020}. Moreover, it is not understood how and where DnaA would be re-activated after it has released ADP. In particular, there is evidence that the rejuvenation of DnaA is contingent on oriC \cite{Crooke1992}. In section \ref{sec:role_of_lipids}, we discuss these experiments in more detail; here we also analyze a scenario in which lipid-mediated DnaA activation depends on the origin density, and one in which DnaA activation is entirely independent of the lipids. The activation rates of the different types of acidic phospholipids have so far not been characterized experimentally. We combine the experimentally characterized deactivation rates of \textit{datA} and RIDA with the activation rates of \textit{DARS1/2} and the experimentally observed initiation volume per origin $v^\ast$ to infer reasonable activation rates of the lipids of $\alpha_{\rm l} \, [l] = 46$~min$^{-1} \, \mu$m$^{-3}$ in the LD model and $\alpha_{\rm l} \, [l] = 12.5$~min$^{-1} \, \mu$m$^{-3}$ in the LDDR model.  \\
\\
\textbf{Activation mechanism: DARS1/2} Experiments have found two activation sites located on the
chromosome of \textit{E. coli}: \textit{DARS1} and
\textit{DARS2}. ADP-DnaA can form oligomers at \textit{DARS1} and
\textit{DARS2}, resulting in the dissociation of ADP and the release
of nucleotide-free apo-DnaA, which then binds ATP. \textit{DARS2} requires the binding of the proteins Fis and IHF. The binding of IHF to \textit{DARS2} is cell-cycle regulated: It increases after 10 minutes, peaks after 20 minutes and decreases again 30-40 minutes after initiation of replication \cite{Kasho2014, Riber2016}. We model this observation via step functions that switch from a low to a high activity state at $\tau_{\rm d2}^+= 0.2$~h and back to a low activity state at $\tau_{\rm d2}^-= 2/3 \, \rm{h} = T_{\rm C}$ (See Fig. 4A/B of the main text). As we could not find an experimental value for the activation rate of \textit{DARS2}, but deleting \textit{DARS2} had a similarly strong effect as deleting RIDA \cite{Riber2016}, we used a high activation rate of $\alpha_{\rm d2}^+=33$~min$^{-1}$ and an arbitrarily low activation rate of $\alpha_{\rm d2}^-=0.83$~min$^{-1}$ in the LDDR model (See Table \ref{tab:LD_LDDR_parameters}). Concerning the binding of Fis to \textit{DARS2}, there is no experimental evidence that it is cell-cycle regulated. We therefore do not model Fis explicitly and assume its effect is contained in the values of $\alpha_{\rm d2}^+$ and $\alpha_{\rm d2}^-$, respectively. Experiments do indicate that the activity of Fis increases with the growth rate \cite{Mallik.2006,Flatten2013, Kasho2014}, but precisely how the binding of Fis to \textit{DARS2} depends on the growth rate of the cell remains to be determined. Since $\alpha_{\rm d2}^+$ contributes to the initiation volume only in the high-growth rate regime of overlapping replication forks, while $\alpha_{\rm d2}^-$ only (weakly) contributes to the initiation volume at low growth rates, see also Eqs. \ref{eq:v_initi_non_overlapping_time_varying} and \ref{eq:v_initi_overlapping_time_varying} below, we assume, for simplicity, that the values of $\alpha_{\rm d2}^+$ and  $\alpha_{\rm d2}^-$ are independent of the growth rate. The site \textit{DARS1} was
found to be neither cell-cycle regulated nor growth rate-dependent and
is approximately ten times weaker than \textit{DARS2} \textit{in vitro} \cite{Katayama2017, Kasho2014}. We use a constant activation rate of \textit{DARS1} of $\alpha_{\rm d1}=1.67$~min$^{-1}$ in the LDDR model.\\
\\
As the dissociation constant of the DnaA boxes on the DNA is in the range of $K_{\rm D}^{\rm s} = 1-50$~nM and \textit{datA}, \textit{DARS1} and \textit{DARS2} are chromosomal binding sites for DnaA, we use a dissociation constant of $K_{\rm D}^{\rm datA}= K_{\rm D}^{\rm d1}= K_{\rm D}^{\rm d2}= K_{\rm	D}=50 \, \mu$m$^{-3}$. Less is known about the dissociation constant of RIDA and the acidic phospholipids and in our model we simply use the same dissociation constant of $K_{\rm D}^{\rm RIDA}= K_{\rm D}^{\rm l}= K_{\rm D}=50 \, \mu$m$^{-3}$.

\subsection{The LD model}
\label{sec:LD_model}
In order to disentangle the effect of the different activators and deactivators, we first focus on the low growth rate regime in the LD model. As we argued in the main text, the acidic phospholipids together with \textit{datA} should be the main players at low
growth rates. In this section, we discuss how the LD model gives rise to a constant initiation volume per origin, thus acting as an origin-density sensor (section \ref{sec:LD_initiation_volume}). Then we investigate the effect of
protein synthesis in this model (section \ref{sec:LD_synthesis}). We
show that the results presented in the main text also hold for the
ultra-sensitivity regime (section \ref{sec:LD_ultra_sensitivity}). Details on how we simulate the LD model and propagate equation 4 of the main text are given in section \ref{sec:LDsimdetails}. All parameters used in the LD model
in the main text and in the SI can be found in Table
\ref{tab:LD_LDDR_parameters}.

\begin{table}
	\centering
	\caption{Parameters used in the LD/LDDR model}
	{ \footnotesize
		\begin{tabular}{lrrrr}
			Parameter & name & value (LD) & value (LDDR) & Motivation \\
			\midrule
			$\alpha_{\rm l} \, [l]$ [$\mu $m$^{-3}$ \, h$^{-1}$] & activation rate lipids & 2755 & 750 & combined with $\beta_{\rm datA}, \beta_{\rm datA}^-, \alpha_{\rm d1}, \alpha_{\rm d2}^- $ \\
			& & & & to match $v^\ast$ from \cite{Si2017} \\
			& & Fig. 3B: 27550 & & \\\
			& & Fig. \ref{fig:S14_Optimisation}B: 2142 & & \\\
			$\beta_{\rm datA}$ [h$^{-1}$] & deactivation rate \textit{datA} & 600 & - & \cite{Kasho2013} \\
			& & Fig. 3B: 6000 & & \\
			$K_{\rm D}^{\rm l}$ [$\mu $m$^{-3}$] & dissociation constant lipids & 50 & - & taken to be equal to $K_{\rm D}^{\rm datA}$ \\
			$K_{\rm D}^{\rm datA}$ [$\mu $m$^{-3}$] & dissociation constant \textit{datA} & 50 & - & \cite{Schaper1995} \\
			$K_{\rm D}$ [$\mu $m$^{-3}$] & dissociation constant in LDDR model & - & 50 & \cite{Schaper1995} \\
			$\tau_{\rm datA}^+$ [h] after $t_{\rm i}$ & start high deactivation rate \textit{datA} & - & 0 & \cite{Kasho2013} \\
			$\tau_{\rm datA}^-$ [h] after $t_{\rm i}$ & end high deactivation rate \textit{datA} & - & 0.2 & \cite{Kasho2013} \\
			$\beta_{\rm datA}^+$ [h$^{-1}$] & high deactivation rate \textit{datA} & - & 600 & \cite{Kasho2013} \\
			$\beta_{\rm datA}^-$ [h$^{-1}$] & low deactivation rate \textit{datA} & - & 300 & \cite{Kasho2013} \\
			$[D]_{\rm T}$ [$\mu $m$^{-3}$] & total DnaA concentration & 400 & 400 & \cite{Hansen1991, Speck2001} \\
			$f^\ast$ & critical initiator fraction & 0.75 & 0.75 & \cite{Kurokawa1999, Katayama2001} \\
			& & Fig. \ref{fig:S14_Optimisation}B: 0.5 & & \\\
			$[D]_{\rm ATP,f}^\ast$ [$\mu $m$^{-3}$] & critical free ATP-DnaA concentration & - & switch-titration model: 200 & \cite{Speck2001, Kurokawa1999, Katayama2001} \\
			$\alpha_{\rm d1}$ [h$^{-1}$] & activation rate \textit{DARS1} & - & 100 & \cite{Katayama2017, Kasho2014} \\
			& & & Fig. \ref{fig:SI_dars1_datA_only}: 1200 & \\
			$\tau_{\rm d1}$ [h] after $t_{\rm i}$ & replication time \textit{DARS1} & - & 0.1 & \cite{Katayama2017} \\
			& & & Fig. \ref{fig:SI_dars1_datA_only}: 0.35 & \\
			$\alpha_{\rm d2}^+$ [h$^{-1}$] & high activation rate \textit{DARS2} & - & 643 & combined with $\beta_{\rm rida}$ \\
			& & & Fig. \ref{fig:S10_LDDR_RIDA_adder}: 1930 & to match $v^\ast$ from \cite{Si2017} \\
			& & & Fig. \ref{fig:SI_LDDR_time_traces} (dashed red line): 50 & \\
			$\alpha_{\rm d2}^-$ [h$^{-1}$] & low activation rate \textit{DARS2} & - & 50 & set to arbitrary low value \\
			$\tau_{\rm d2}$ [h] after $t_{\rm i}$ & replication time \textit{DARS2} & - & 0.2 & \cite{Kasho2014} \\
			$\tau_{\rm d2}^+$ [h] after $t_{\rm i}$ & start high activation rate \textit{DARS2} & - & 0.2 & \cite{Kasho2014} \\
			$\tau_{\rm d2}^-$ [h] after $t_{\rm i}$ & end high activation rate \textit{DARS2} & - & 2/3 & \cite{Kasho2014} \\
			$\beta_{\rm rida}$ [h$^{-1}$] & deactivation rate RIDA & - & 500 & \cite{Nakamura2010, Kasho2013, Moolman2014} \\
			&  & & Fig. \ref{fig:S10_LDDR_RIDA_adder}: 1500 & \\
			$\tau_{\rm b}$ [h] after $t_{\rm i}$ & refractory period & 0.17 & 0.17 & \cite{Campbell:1990it,Lu:1994ee, Waldminghaus:2009em} \\
			$T_{\rm C}$ [h] & C-period & 2/3 & 2/3 & \cite{Cooper1968} \\
			$T_{\rm D}$ [h] & D-period & 1/3 & 1/3 & \cite{Cooper1968} \\
			$\alpha$ [h$^{-1}$] & production rate lipids & $955$ & $260$ & combined with $\beta_{\rm datA}, \beta_{\rm datA}^-, \alpha_{\rm d1}, \alpha_{\rm d2}^- $ \\
			& & & & to match $v^\ast$ from \cite{Si2017} \\
			$D_{\rm l}$ & noise strength lipids & $5000$ & $5000$ & set to match CV from \cite{Wallden2016} \\
			&  & & switch-titration model: 1000 & \\
			&  & & Fig. \ref{fig:S14_Optimisation}: $10^6$ & \\
			$\rho$ [$\mu \rm{m}^{-3}$] & number density & $10^6$ & $10^6$ & \cite{Milo2013} \\
			$\phi_0$ & gene allocation fraction of DnaA & $10^{-3}$ & $10^{-3}$ & to match DnaA concentration \\
			& & & Fig. \ref{fig:S18_mutations}: $1.5 \times 10^{-3}$ & reported in \cite{Hansen1991} \\
			$K_{\rm D}^{\rm p}$ [$\mu $m$^{-3}$] & dissociation constant DnaA promoter & 300 & 300 & \cite{Speck1999, Hansen1991} \\
			&  & & switch-titration model: 400 & \\
			$n$ & Hill coefficient DnaA promoter & 5 & 5 & \cite{Speck1999} \\
			$D_{\rm D}$ & noise strength DnaA & 100 & 100 & set to match CV from \cite{Wallden2016} \\
			$D_\eta$ & noise strength RIDA & $100$ & $100$ & set to match CV from \cite{Wallden2016} \\
			$\lambda$ [h$^{-1}$] & growth rate & 0.35-1.73 & 0.35-1.73 & \cite{Si2017, Wallden2016} \\
			& & & Fig. 4: 0.35 & \\
			\bottomrule
		\end{tabular}
	}
	\\
	{Section \ref{sec:switch_parameters} provides further motivation for chosen parameter values.}
	\label{tab:LD_LDDR_parameters}
\end{table}

\subsubsection{The LD model is an origin-density sensor that ensures a constant initiation volume per origin}
\label{sec:LD_initiation_volume}
As the DnaA protein is not actively degraded and its concentration is approximately constant, the production rate of new DnaA proteins
equals the growth rate. If the activation and inactivation rates are much higher than the growth rate and (hence) the production rate of new proteins, then the effect of protein
synthesis and dilution on the concentration of active DnaA can be neglected. In this scenario, the change in the concentration of active DnaA is given by an activation term due to the lipids and a deactivation term due to DDAH:
\begin{equation}
\frac{d[D]_{\rm ATP}}{dt}
= \alpha_{\rm l} \,[l] \, \frac{[D]_{\rm ADP}}{K_{\rm D}^{\rm l} + [D]_{\rm ADP}} - \beta_{\rm datA} \, [n_{\rm ori}] \, \frac{[D]_{\rm ATP}}{K_{\rm D}^{\rm datA}+[D]_{\rm ATP}}
\label{eq:switch_no_synth_conc}
\end{equation}
with the activation and deactivation rates $\alpha_{\rm l}$ and $\beta_{\rm datA}$ and the Michaelis-Menten constants $K_{\rm D}^{\rm l}$ and $K_{\rm D}^{\rm datA}$. As the deactivation site \textit{datA} is located close to the origin, we have used here that their concentrations are equal. The concentrations of constitutively expressed and negatively autoregulated proteins are nearly constant in time (see Fig. \ref{fig:S2_growing_cell_model}~A/B). In addition, experiments have shown that the total DnaA concentration varies by no more than 50\% over a tenfold change in the growth rate \cite{Zheng2020}. We therefore assume that the total concentration $[D]_{\rm T}$ is not only constant in time but also independent of the growth rate. Dividing equation \ref{eq:switch_no_synth_conc} by this total concentration $[D]_{\rm T}$ and using $[D]_{\rm ADP}= [D]_{\rm T}-[D]_{\rm ATP}$ we obtain
\begin{equation}
\frac{df}{dt}
= \tilde{\alpha}_{\rm l} \,[l] \, \frac{1-f}{\tilde{K}_{\rm D}^{\rm l} + 1 - f} - \tilde{\beta}_{\rm datA} \, [n_{\rm ori}] \, \frac{f}{\tilde{K}_{\rm D}^{\rm datA}+f}
\label{eq:switch_simple_fraction_no_synth}
\end{equation}
with the re-normalized activation and deactivation rates $\tilde{\alpha}_{\rm l}= \alpha_{\rm l}/ [D]_{\rm T}$ and $\tilde{\beta}_{\rm datA}= \beta_{\rm datA}/ [D]_{\rm T}$ and the Michaelis-Menten constants $\tilde{K}_{\rm D}^{\rm l}= K_{\rm D}^{\rm l}/ [D]_{\rm T}$ and $\tilde{K}_{\rm D}^{\rm datA}= K_{\rm D}^{\rm datA}/ [D]_{\rm T}$. If the activation and deactivation rates are high compared to the growth rate, the system is well characterized by the steady state ($\frac{df}{dt}=0$). The theoretical prediction of the initiation volume per origin $v^\ast_{\rm th}= 1/[n_{\rm ori}]$ is obtained by setting equation \ref{eq:switch_simple_fraction_no_synth} to zero:
\begin{align}
v_{\rm th}^\ast = \frac{\beta_{\rm datA}}{\alpha_{\rm l} \, [l]} \, \frac{f^\ast}{\tilde{K}_{\rm D}^{\rm datA}+f^\ast} \, \frac{\tilde{K}_{\rm D}^{\rm l} +1 - f^\ast}{1-f^\ast}
\label{eq:v_init_simple_switch}
\end{align}
with the critical initiator fraction $f^\ast= [D]_{\rm ATP}^\ast/[D]_{\rm t}$. 

While it has been believed that the initiation volume is independent of the growth rate \cite{Wallden2016, Si2017}, recent experiments indicate that the initiation volume varies non-monotonically within a 50\% bandwidth over a tenfold change in the growth rate \cite{Zheng2020}. All parameters on the right side of equation \ref{eq:v_init_simple_switch} could in principle vary with the growth rate $\lambda$ of the cell. The initiation volume per origin $v^\ast_{\rm th}$ is constant at all growth rates if all terms on the right side are constant or if the growth rate dependencies of the parameters cancel each other out. While the deactivation rate of DDAH is known to be temporally regulated over the course of the cell cycle, the explicit growth rate dependence has not been studied so far; such a dependence could well be possible, as several proteins like Dia and IHF (whose concentrations could vary with $\lambda$) are involved. A growth-rate dependence of the critical initiation fraction $f^\ast$ has not been reported but could be possible, as two other proteins, Dia and IHF, are involved in the initiation process \cite{Katayama2017, Keyamura2009}. 
The lipid concentration in equation \ref{eq:v_init_simple_switch} stems from a combination of the two types of lipids CL and PG. The cell membrane composition is complex and could depend on the growth rate. Our model predicts however that if all other parameters of equation \ref{eq:v_init_simple_switch} are growth-rate independent, the for replication initiation relevant phospholipid concentration should be approximately constant in order to obtain a constant initiation volume $v^\ast$.

\subsubsection{The effect of protein synthesis}
\label{sec:LD_synthesis}
In this section, we investigate the role of protein synthesis in the LD model and analyse its effect on the amplitude of the oscillations in the active DnaA fraction. As DnaA binds strongly to both ATP and ADP and the concentration of ATP is approximately ten times higher than the concentration of APD in \textit{E. coli} \cite{Katayama2010}, we assume that every newly produced protein binds to ATP right after synthesis. Thus, the change in the total number of DnaA proteins due to protein synthesis equals the change in the ATP-DnaA concentration:
\begin{equation}
\frac{dN_{\rm D}^{\rm T}}{dt} = k_{\rm prod} = \frac{dN_{\rm ATP}^{\rm synth}}{dt}
\label{eq:LD_synthesis_total_number_change}
\end{equation}
where $k_{\rm prod}$ is the effective production rate of ATP-DnaA, which takes into account gene regulation. 
The change in the total concentration of DnaA is given by 
\begin{equation}
\frac{d [D]_{\rm T}}{dt}= \frac{dN_{\rm D}^{\rm T}}{dt} \frac{1}{V} + \frac{d}{dt} \, \left( \frac{1}{V}\right) \, N_{\rm D}^{\rm T} = \frac{k_{\rm prod}}{V} - \lambda [D]_{\rm T}
\label{eq:LD_synthesis_change_in_total_dnaa_conc}
\end{equation}
As in the previous section, we assume that $[D]_{\rm T}$ is constant in time, such that $k_{\rm prod}/V = \lambda  [D]_{\rm T}$.
Using equation \ref{eq:LD_synthesis_total_number_change} and \ref{eq:LD_synthesis_change_in_total_dnaa_conc}, and exploiting that $k_{\rm prod}/V = \lambda  [D]_{\rm T}$, we obtain the following expression for the change in the ATP-concentration due to protein synthesis:
\begin{equation}
\frac{d [D]_{\rm ATP}^{\rm synth}}{dt}= \frac{dN_{\rm ATP}^{\rm synth}}{dt} \frac{1}{V} - \lambda [D]_{\rm ATP} =\frac{k_{\rm prod}}{V} - \lambda [D]_{\rm ATP} = \lambda ([D]_{\rm T} - [D]_{\rm ATP})
\label{eq:switch_active_conc_change}
\end{equation}
By dividing equation \ref{eq:switch_active_conc_change} by the total concentration $[D]_{\rm T}$ and combining it with equation \ref{eq:switch_simple_fraction_no_synth} we obtain the change in the active fraction of the main text:
\begin{equation}
\frac{df}{dt}
= \tilde{\alpha}_{\rm l} \,[l] \, \frac{1-f}{\tilde{K}_{\rm D}^{\rm l} + 1 - f} - \tilde{\beta}_{\rm datA} \, [n_{\rm ori}] \, \frac{f}{\tilde{K}_{\rm D}^{\rm datA}+f} + \lambda (1 - f)
\label{eq:switch_simple_fraction_SI}
\end{equation}
The third term on the right-hand side is the additional activation term that comes from protein synthesis, and the fact that newly synthesized proteins rapidly bind ATP; this term is indeed proportional to the growth rate and decreases linearly with the ATP-DnaA fraction. 

\begin{figure}
	\centering
	\includegraphics[width =0.6\textwidth]{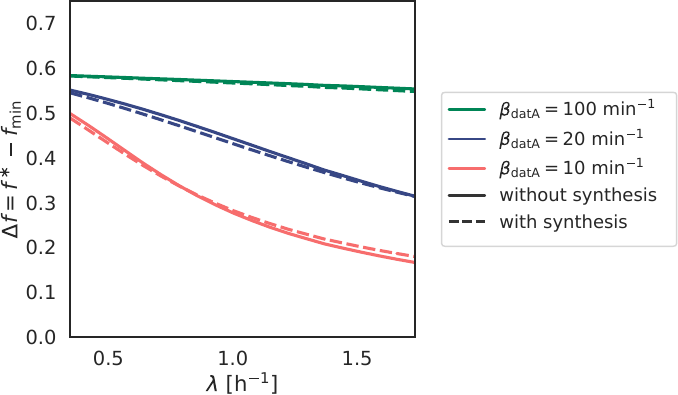}
	\caption{\textbf{Comparison of the LD model with and without synthesis} The oscillation amplitude $\Delta f$ as a function of the growth rate $\lambda$ for different magnitudes of the activation and deactivation rates ($\alpha_{\rm l} = 4.6 \times \beta_{\rm datA}$). The solid lines show the predictions of the LD model that includes the effect of protein synthesis (equation \ref{eq:switch_simple_fraction_SI}) while the dashed lines shows the results of the model that does not (equation \ref{eq:switch_simple_fraction_no_synth}). 
		The amplitude of the oscillations  decreases with the growth rate, but that this dependence is weaker for higher (de)activation rates. }
	\label{fig:SI_LD_synthesis}
\end{figure}
The effect of protein synthesis on the DnaA oscillations depends strongly on the relative magnitude of the activation and deactivation rates and the growth rate. Figure \ref{fig:SI_LD_synthesis} shows however that the effect of protein synthesis (third term right-hand side Equation \ref{eq:switch_simple_fraction_SI}) on the amplitude of the oscillations is small.

\subsubsection{The LD model in the ultra-sensitivity regime}
\label{sec:LD_ultra_sensitivity}
\begin{figure}
	\centering
	\includegraphics[width =0.6\textwidth]{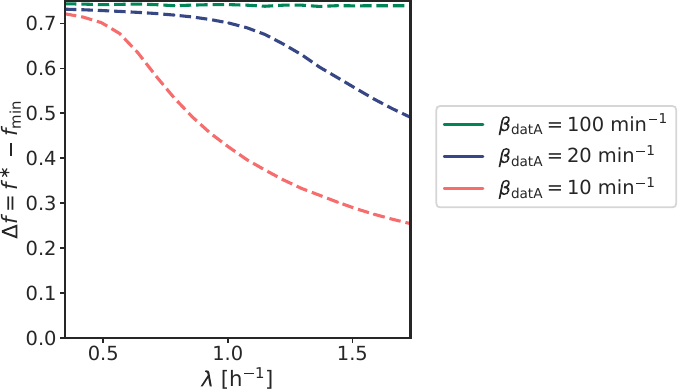}
	\caption{\textbf{The LD model in the ultra-sensitive regime} The amplitude $\Delta f$ of the oscillations in the active fraction $f$ as a function of the growth rate at different magnitudes of the activation and deactivation rates ($\alpha_{\rm l} = 4.6 \times \beta_{\rm datA}$). The amplitude of the oscillations $\Delta f$ becomes small for biologically realistic values of the (de)activation rates, even deep in the ultra-sensitive regime. Here, the dissociation constants $K_{\rm D}^{\rm l}=K_{\rm D}^{\rm datA}= 5 \, \mu$m$^{-3}$ for lipid-mediated activation of DnaA and {\it datA} mediated deactivation, respectively (see Eq. \ref{eq:switch_simple_fraction_SI}), is 10 times lower than that used for Figure 3 of the main text.}
	\label{fig:SI_ultra_sensitive_LD}
\end{figure}

Figure 3 in the main text shows that for biological (de)activation rates the amplitude of the oscillations in the active fraction becomes very small. Here we ask whether this effect could be alleviated by bringing the system deeper into the ultra-sensitivity regime. The ultra-sensitivity can be increased by increasing the difference in the dissociation constants $K_{\rm D}^{\rm l}$ and $K_{\rm D}^{\rm datA}$ with respect to the total DnaA concentration $[D]_{\rm T}$. In the main Figure 3, the dissociation constants of the activator $K_{\rm D}^{\rm l}$ and deactivator $K_{\rm D}^{\rm datA}$ are approximately ten times smaller than the total DnaA concentration $[D]_{\rm T} = 400 \, \mu$m$^3$ (see Table \ref{tab:LD_LDDR_parameters}). The system is thus already in the ultra-sensitive regime. Here we push the system even deeper in the ultra-sensitivity regime by setting the dissociation constants of both activator and deactivator to $K_{\rm D}^{\rm l}=K_{\rm D}^{\rm datA}= 5 \, \mu$m$^{-3}$, almost two orders of magnitude smaller than the total concentration. Figure \ref{fig:SI_ultra_sensitive_LD} shows the amplitude of the oscillations in the active fraction $f$ in this (highly) ultra-sensitive regime. 
The amplitude of the oscillations at high and intermediate (de)activation rates is slightly higher in this deep ultra-sensitive regime. Importantly, however at low rates ($\beta_{\rm datA}= 10$~min$^{-1}$) the amplitude of the oscillations drops for high growth rates like in the less ultra-sensitive regime presented in the main section. Therefore, regardless of the degree of ultra-sensitivity, the experimentally reported activation and deactivation rates are too low to explain the experimentally observed high amplitude oscillations in the active initiator fraction \cite{Kasho2013}. Our modelling predicts that at high growth rates, RIDA and \textit{DARS2} become essential to sustain large amplitude oscillations, as we describe in more detail in section \ref{sec:LDDR_model} on the LDDR model.

\subsubsection{Simulation details of the LD model}
\label{sec:LDsimdetails}
To simulate the LD model, we propagate the fraction of active,
ATP-bound DnaA according to equation \ref{eq:switch_simple_fraction_SI},
which is identical to equation 4 of the main text. The volume grows
exponentially according to $dV/dt = \lambda V$ and the origin
density is given by $[n_{\rm ori}]=n_{\rm ori} / V$. When the
fraction $f$ equals the critical fraction $f^*$, replication is initiated, and the number of origins
$n_{\rm ori}$ doubles. The cell then divides a constant time
$\tau_{\rm cc}$ later. During cell division, the volume and the
number of origins are halved. 

\subsection{The LDDR model}
\label{sec:LDDR_model}
In the LD model, we argued that at low growth rates replication initiation is mainly controlled by the activating lipids and the deactivating site \textit{datA}. At high growth rates ($\lambda > \ln(2)/T_{\rm C}$), the replication forks are overlapping and RIDA is still active at the moment of replication initiation.  Simultaneously, \textit{DARS2} is activated via the protein Fis \cite{Mallik.2006,Flatten2013, Kasho2014}. Here, we present the Lipid-\textit{DatA}-\textit{DARS1/2}-RIDA (LDDR) model where we include all activators and deactivators with their characteristic temporal regulation (section \ref{sec:LDDR_temporally_varying_rates}). Then we show that the LDDR model gives rise to stable cell cycles with large amplitude oscillations in the active fraction at all growth rates (section \ref{sec:LDDR_time_traces}). In order to obtain a constant initiation volume at all growth rates, we however need to make a specific parameter choice (section \ref{sec:LDDR_v_init_const}). The finding that RIDA and \textit{DARS2} are essential for obtaining high amplitude oscillations at high growth rates raises the question whether a system consisting only of \textit{DARS2} and RIDA is not sufficient. We show however, that a model where all activators and
deactivators are located on the chromosome (like \textit{DARS2}-RIDA or \textit{datA}- \textit{DARS1}) does not give rise to stable cell cycles (section \ref{sec:LDDR_datA_dars1_unstable}). 
In section \ref{sec:LDDRsimdetails} we present the simulation details of this model. The parameters of the LDDR model are described in section \ref{sec:switch_parameters} above and their values are listed in Table \ref{tab:LD_LDDR_parameters}. 

\subsubsection{Temporal variations in the (de)activation rates in the LDDR model ensures large amplitude oscillations in the active fraction}
\label{sec:LDDR_temporally_varying_rates}
Experiments indicate that several activators and deactivators are temporally regulated over the course of the cell cycle (section \ref{sec:switch_parameters}). In this section we present how we model this temporal regulation in the LDDR model.

\begin{figure}
	\centering
	\includegraphics[width =0.6\textwidth]{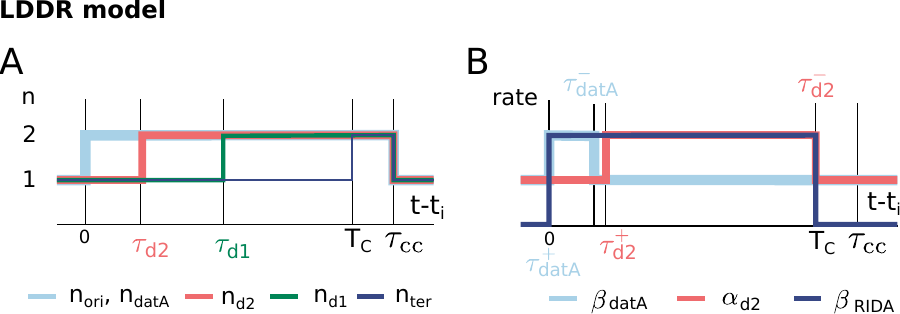}
	\caption{\textbf{The (de)activation rates in the LDDR model are time-dependent.} (A) The number of origins $n_{\rm ori}(t-t_{\rm i})$, \textit{datA} sites $n_{\rm datA}(t-t_{\rm i})= n_{\rm ori}(t-t_{\rm i})$, \textit{DARS2} sites $n_{\rm d2}(t-t_{\rm i})= n_{\rm ori}(t-t_{\rm i}-\tau_{\rm d2})$, \textit{DARS1} sites $n_{\rm d1}(t-t_{\rm i})= n_{\rm ori}(t-t_{\rm i}-\tau_{\rm d1})$ and termini $n_{\rm ter}(t-t_{\rm i})= n_{\rm ori}(t-t_{\rm i}-T_{\rm C})$ per cell as a function of the time after initiation of replication at $t_{\rm i}$. The time to replicate the entire chromosome is $T_{\rm C}$ and the time from the beginning of replication to cell division is $\tau_{\rm cc}$. Shown is the scenario for the low growth-rate regime of non-overlapping replication forks. (B) The cell cycle time dependent rates of \textit{datA}, \textit{DARS2} and RIDA as a function of the same cell cycle as in (A).}
	\label{fig:S7_LDDR_time_dependent_rates}
\end{figure}

In the LDDR model, the number of catalytic RIDA complexes is proportional to the number of origins with a rate $\beta_{\rm RIDA}$ that is only non-zero during the period of active replication $T_{\rm C}$ (Fig. \ref{fig:S7_LDDR_time_dependent_rates} B). The chromosomal sites \textit{DARS1} and \textit{DARS2} are located in the middle of the chromosome and are replicated at constant times $\tau_{\rm d1}$ and $\tau_{\rm d2}$, respectively, after the origin (Fig. \ref{fig:S7_LDDR_time_dependent_rates}~A). The activities of DDAH and \textit{DARS2} are temporally regulated during the cell cycle via binding of the integrating host factor (IHF) \cite{Katayama2017, Kasho2013, Kasho2014, Riber2016}. IHF binding to \textit{datA} increases within about 5-10 minutes, peaks at about 15 minutes, and decreases again about 20-30 minutes after initiation of replication \cite{Katayama2017, Kasho2013}. The binding of IHF to \textit{DARS2} increases after 10 minutes, peaks after 20 minutes and decreases again 30-40 minutes after initiation of replication \cite{Kasho2014, Riber2016}. We model these observations via step functions $\alpha_{\rm d2}(t-t_{\rm i})$ and $\beta_{\rm datA}(t-t_{\rm i})$ with a high and a low rate for \textit{DARS2} and DDAH, respectively, that vary as a function of the time since initiation of replication $t-t_{\rm i}$ (Fig. \ref{fig:S7_LDDR_time_dependent_rates} B). As \textit{DARS2} remains highly active until replication termination, it can counteract the strong deactivator RIDA in the overlapping replication fork regime. \textit{DARS2} is additionally regulated in a growth-rate dependent manner via the protein Fis \cite{Flatten2013, Kasho2014}. As there is no evidence for temporal regulation via Fis, we do not model Fis explicitly. \textit{DARS1} activation is modeled via a constant activation rate $\alpha_{\rm d1}$. 
Summing up, we thus propose the following expression for the change in the ATP-DnaA fraction (see also Fig. \ref{fig:S7_LDDR_time_dependent_rates}~A and B):
\begin{align}
\frac{df}{dt}= & \, \big( \tilde{\alpha}_{\rm l} \,[l]+ \tilde{\alpha}_{\rm d1} \, [n_{\rm ori}(t - \tau_{\rm d1})]
+ \tilde{\alpha}_{\rm d2}(t) \, [n_{\rm ori}(t - \tau_{\rm d2})] \big) \, \frac{1-f}{\tilde{K}_{\rm D} + 1 - f}
-\big( \tilde{\beta}_{\rm datA}(t) + \tilde{\beta}_{\rm rida}(t) \big)  \, [n_{\rm ori}]\, \frac{f}{\tilde{K}_{\rm D}+f}
+ \lambda \, (1 - f)
\label{eq:switch_complex_fraction_SI}
\end{align}
with the re-normalized activation and deactivation rates $\tilde{\alpha}_{\rm l}= \alpha_{\rm l}/ [D]_{\rm T}$, $\tilde{\alpha}_{\rm d1}= \alpha_{\rm d1}/ [D]_{\rm T}$, $\tilde{\alpha}_{\rm d2}= \alpha_{\rm d2}/ [D]_{\rm T}$, $\tilde{\beta}_{\rm datA}= \beta_{\rm datA}/ [D]_{\rm T}$ and $\tilde{\beta}_{\rm rida}= \beta_{\rm rida}/ [D]_{\rm T}$ and the Michaelis-Menten constant $\tilde{K}_{\rm D}= K_{\rm D}/ [D]_{\rm T}$. The parameters are described in section \ref{sec:switch_parameters} above and its values are listed in Table \ref{tab:LD_LDDR_parameters}.

\subsubsection{Large amplitude oscillations in the active fraction over the course of the cell cycle}
\label{sec:LDDR_time_traces}

\begin{figure}
	\centering
	\includegraphics[width =0.6\textwidth]{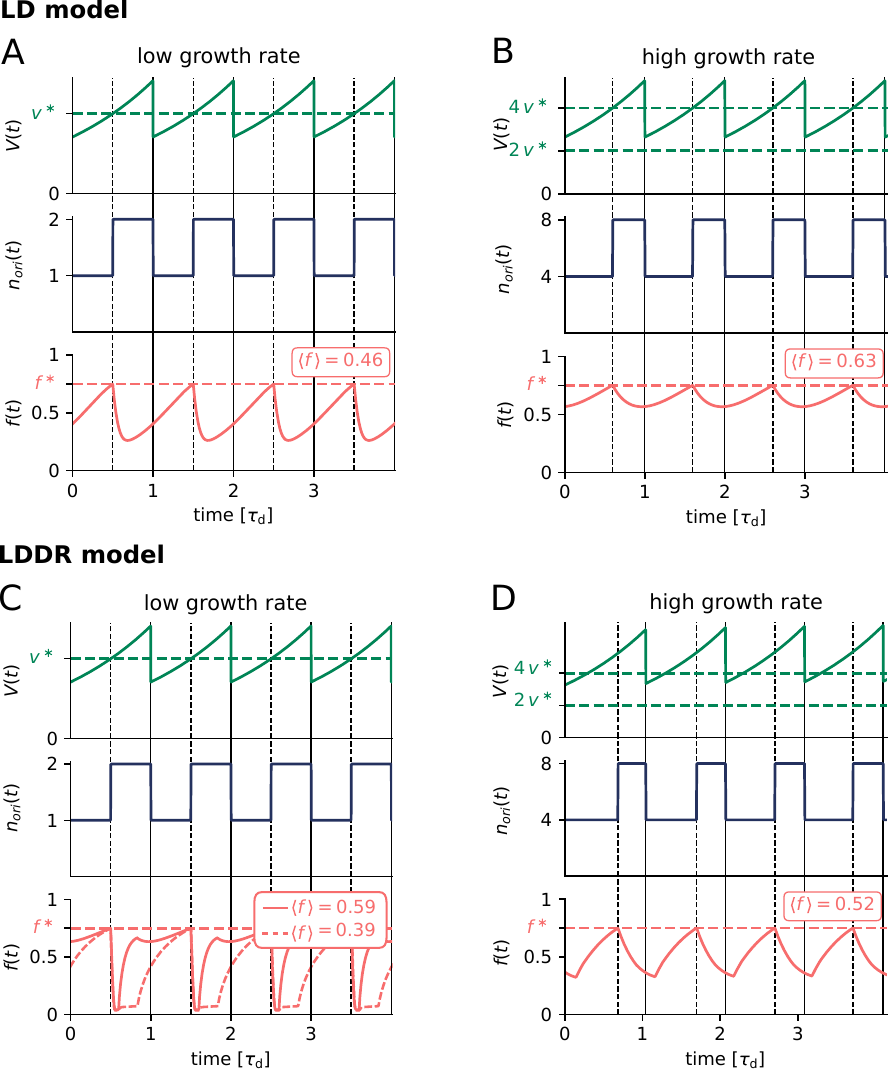}
	\caption{\textbf{The LDDR model ensures high amplitude oscillations in the active fraction even at high growth rates and realistic (de)activation rates} (A, B, C, D) The volume of the cell $V(t)$, the number of origins $n_{\rm ori}(t)$ and the fraction of ATP-DnaA $f(t)$ as a function of time (in units of the doubling time $\tau_{\rm d}$) at a low doubling time of $\tau_{\rm d}=2$~h ($\lambda = 1.35$~h$^{-1}$) (A, C) and a at high doubling time of $\tau_{\rm d}= 25$~min ($\lambda = 1.66$~h$^{-1}$) (B, D). The dashed red line is the critical initiator fraction $f^\ast$ at which replication is initiated. The average active fraction over one cell cycle $\langle f \rangle$ is indicated in red in the third panel. Replication is initiated at a constant volume per origin $v^\ast$ over time (green dashed line). (A, B) While in the LD model at realistic activation and deactivation rates the activating lipids and the deactivating site \textit{datA} generate high-amplitude oscillations at low growth rates, the amplitude of the oscillations becomes very small at high growth rates. (C, D) In the LDDR model, due to the additional temporal modulation of the activation and deactivation rates of \textit{datA}, RIDA and \textit{DARS2}, the amplitude of the oscillations is high both at low and at high growth rates. At low growth rates however, as the activity of the site DARS2 increases a short time after replication initiation ($\tau_{\rm d2}=0.2$~h) compared to the cell doubling time of $\tau_{\rm d}=2$~h, the active fraction is high during a large fraction of the cell cycle, reducing the robustness of the LDDR model at low growth rates. As experiments indicate that the activity of \textit{DARS2} decreases with decreasing growth rate \cite{Mallik.2006, Flatten2013, Kasho2014}, we show in (C) also the time trace of the active fraction when at low growth rates the high activation rate $\alpha_{\rm d2}^+$ of \textit{DARS2} (Fig. \ref{fig:S7_LDDR_time_dependent_rates}) is lower than that at high growth rate; here, we have taken $\alpha_{\rm d2}^+$ to be equal to the low activation rate $\alpha_{\rm d2}^-$ of \textit{DARS2}, such that $\alpha_{\rm d2}^+= \alpha_{\rm d2}^-=50$ (dashed red line). This improves the shape of the oscillations as it leads to a weaker increase in the active fraction upon doubling of \textit{DARS2}.}
	\label{fig:SI_LDDR_time_traces}
\end{figure}

The fact that the (de)activation of DnaA by \textit{DARS2} and RIDA respectively counteract each other in setting the initiation volume (see next section \ref{sec:LDDR_v_init_const}) raises the question what their role is. Our model suggests that the temporal dependence of their activity is essential (Fig. \ref{fig:S7_LDDR_time_dependent_rates} B): right after a new round of replication, the deactivation rate via RIDA is raised but the activation rate via \textit{DARS2} is not. This acts to enhance the amplitude of the oscillations. Indeed, in contrast to the LD model based on lipid-mediated activation and \textit{datA} mediated deactivation only (Fig. \ref{fig:SI_LDDR_time_traces} A, B), the LDDR model gives rise to large amplitude oscillations at all growth rates, even for realistic parameter values (Fig. 3 C in the main text and Fig. \ref{fig:SI_LDDR_time_traces} C, D for time traces). Time-varying activation and deactivation rates in combination with specific positions on the chromosome can thus explain how the cell obtains high amplitude oscillations in the active fraction (Fig. 3 C in the main text).

While the temporal separation of the activities of RIDA and \textit{DARS2} can drastically enhance the amplitude of the oscillations, their fixed delay, together with the fact that they are both coupled to the origin density, can also impede robustness, as shown in Fig. 5C of the main text. More specifically, at low growth rates, the fixed time between replication initiation and replication of the site \textit{DARS2}  is very short ($\tau_{\rm d2}=0.2$~h) compared to the long doubling time of the cell ($\tau_{\rm d}=2$~h). Right after replication initiation, the deactivators \textit{datA} and RIDA lower the active fraction of DnaA and prevent reinitiation. A short time $\tau_{\rm d2}$ later however, the activity of \textit{DARS2} increases and the active fraction rises. In this low growth rate regime, the active fraction is therefore high for a large fraction of the cell cycle, reducing the robustness of the LDDR model at low growth rates (see Fig. 5 C in main text and \ref{fig:SI_full_model_time_traces} J and K). 
Interestingly, there is experimental evidence that the activity of \textit{DARS2} decreases with decreasing growth rate \cite{Mallik.2006, Flatten2013, Kasho2014}. Taking this into account does indeed positively affect the shape of the oscillations, yielding a slower rise in the active fraction when \textit{DARS2} is duplicated (Fig. \ref{fig:SI_LDDR_time_traces} C, dashed red line). Importantly, however, titration naturally enhances the robustness of the switch in the low growth rate regime, by sharpening the oscillations in the concentration of {\em free, active} DnaA (see section \ref{sec:switch_titration_combined}). Moreover, we find that even with a constant, high activity, {\it DARS2} does not significantly affect the initiation volume at low growth rates (see Fig. \ref{fig:S18_mutations}I). Its dominant effect is at high growth rates: in this regime, \textit{DARS2} is essential to vigorously counteract the strong deactivator RIDA, enabling a new round of replication while the old round has not finished yet. For a more detailed comparison against experimental data, we refer to our model validation section \ref{sec:model_validation} and Fig. \ref{fig:S18_mutations}.

\subsubsection{A switch model consisting only of activators and deactivators located on the chromosome does not ensure stable cell cycles}
\label{sec:LDDR_datA_dars1_unstable}
\begin{figure}
	\centering
	\includegraphics[width =0.7\textwidth]{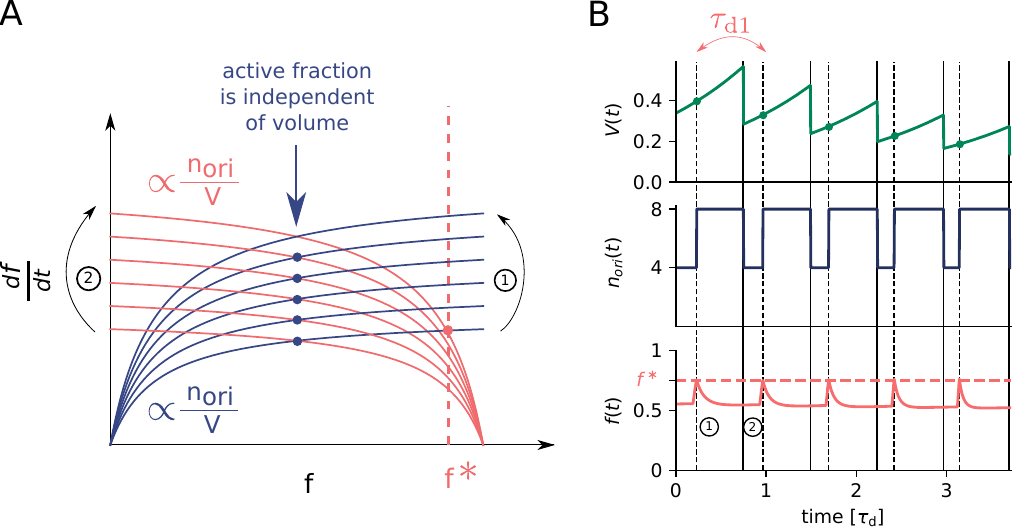}
	\caption{\textbf{A switch model where all activators and deactivators are located on the chromosome does not give rise to stable cell cycles} (A) The activation (red curves) and deactivation rates (blue curve) as a function of the active fraction of the initiator protein $f$ at different moments of the cell cycle. The steady-state active fractions are given by the intersection of the activation and deactivation rates (colored dots). As both the activation and the deactivation rate depend on the origin density, the active fraction becomes volume independent. When the active fraction $f$ equals or is larger than the critical initiation fraction $f^\ast$ (vertical dashed red line), replication is initiated. When replication is initiated (step 1), the number of origin doubles. Due to the vicinity of \textit{datA} to the origin, the deactivation rate doubles right after replication is initiated and the active fraction is reduced to a constant value below the activation threshold. A fixed time $\tau_{\rm d1}= 0.35$~h $=21$~min after replication was initiated, the site \textit{DARS1} is doubled which causes again an increase of the active fraction beyond the critical active fraction $f^\ast$ and a new round of replication is initiated again. (B) The volume of the cell $V(t)$, the number of origins $n_{\rm ori}(t)$ and the fraction of ATP-DnaA $f(t)$ as a function of time (in units of the doubling time $\tau_{\rm d}$) at a doubling time of $\tau_{\rm d}= 0.5$~h $=30$~min. The dashed red line is the critical initiator fraction $f^\ast$ at which replication is initiated. Replication is initiated at a constant time interval $\tau_{\rm ii}= \tau_{\rm d1}$ which, in this example, is smaller than the doubling time of the cell. As a consequence, the volume per origin $v^\ast$ decreases over time and so does the birth volume of the cell. }
	\label{fig:SI_dars1_datA_only}
\end{figure}

The finding that including RIDA and \textit{DARS2} in the model is necessary to ensure the experimentally observed large amplitude oscillations at high growth rates raises the question whether it is also sufficient, meaning the LD model, and more in particular lipid synthesis, is not essential. In this section, we show that eliminating activation and deactivation mechanisms from the full LDDR model only maintains a stable system as long as the lipids with constant activity remain part of the model (Fig. \ref{fig:SI_dars1_datA_only}). When instead all activation and deactivation mechanisms are connected to the chromosome, as in a system combining \textit{DARS1/2} activation with \textit{datA}/RIDA deactivation, both the activation and deactivation rates have the same functional dependence on the volume such that the system cannot sense the origin density anymore.

For a switch consisting only of \textit{datA} and \textit{DARS1}, we obtain the following expression for the change in the active fraction of DnaA: 
\begin{equation}
\frac{df}{dt}= \tilde{\alpha}_{\rm d1} \, [n_{\rm ori}(t-\tau_{\rm d1})] \, \frac{1-f}{\tilde{K}_{\rm D} + 1 - f}
- \tilde{\beta}_{\rm datA} \, [n_{\rm ori}] \, \frac{f}{\tilde{K}_{\rm D}+f} 
\label{eq:simple_switch_fraction_no_lipids}
\end{equation} 
where the site \textit{DARS1} is replicated a time $\tau_{\rm d1}$
after the origin. The concentration of \textit{DARS1} is therefore
proportional to the origin density at an earlier time
$t-\tau_{\rm d1}$. A model where both the activator and deactivator
are proportional to the (time shifted) origin density does not give
rise to stable cell cycles (Fig. \ref{fig:SI_dars1_datA_only} B).
We can understand this observation by plotting the activation and deactivation rates as a function of the active fraction at different moments of the cell cycle (Fig. \ref{fig:SI_dars1_datA_only} A). 
At quasi-steady-state, the active fraction is constant (setting equation \ref{eq:simple_switch_fraction_no_lipids} to zero) and the system will therefore settle to a constant fraction $f$ independent of the volume of the cell. If this fraction $f$ lies above the critical initiation fraction $f^\ast$, replication can be initiated (Fig. \ref{fig:SI_dars1_datA_only} A, red dot). Because of its vicinity to the origin, the site \textit{datA} is replicated right after initiation and reduces the active fraction below the initiation threshold (Fig. \ref{fig:SI_dars1_datA_only} A and B, step 1). A constant time $\tau_{\rm d1}$ after initiation of replication, the site \textit{DARS1} is replicated as well (Fig. \ref{fig:SI_dars1_datA_only} A and B, step 2). The active DnaA fraction rises again rapidly and when it attains the critical initiation fraction $f^\ast$, a new round of replication is initiated. 
The active DnaA fraction thus oscillates between a high and a low ATP-DnaA state at a period given by the time difference in replicating the sites on the chromosome \textit{datA} and \textit{DARS1}, $\tau_{\rm d1}$. This gives rise to a constant initiation period $\tau_{\rm ii}=\tau_{\rm d1}$. 
A system with a constant replication initiation period $\tau_{\rm ii}$, which is thus not coupled to the growth or the volume of the cell, cannot give rise to stable cell cycles; even the smallest deviation of $\tau_{\rm ii}$ from $\tau_{\rm d}$ will inevitably grow and make the system unstable (Fig. \ref{fig:SI_dars1_datA_only} B, green dots in upper panel). If division is coupled to replication initiation via a constant time $\tau_{\rm cc}$, the cell volume does therefore not remain stable after a few generations. 

In summary, a system in which all activators and deactivators are located on the chromosome does not ensure stable cell cycles. This is because the volume dependence of the activation and deactivation rates is then the same, which means that the system cannot sense the origin density. Indeed, to sense the origin density, it is vital that the volume dependence of the activation and deactivation rates is different. 

\subsubsection{Constant initiation volume per origin in the LDDR model}
\label{sec:LDDR_v_init_const}
The LDDR model yields two different predictions for the initiation
volume per origin $v^\ast$ in the quasi-equilibrium regime where
(de)activation is faster than growth, depending on whether RIDA and
\textit{DARS2} are active at the moment of initiation or not. From Eq 5 of the main text (Eq.~\ref{eq:switch_complex_fraction_SI} below), it follows that in the
low growth-rate regime ($\lambda <\ln{2} / T_{\rm C}$) the
replications forks are non-overlapping and the initiation volume
$v^\ast_{\rm no}$ is given by
\begin{align}
v^\ast_{\rm no} = \frac{\beta_{\rm datA}^-}{\alpha_{\rm l} \, [l]} \, \frac{f^\ast}{\tilde{K}_{\beta}+f^\ast} \, \frac{\tilde{K}_\alpha +1 - f^\ast}{1-f^\ast}-\frac{\alpha_{\rm d1}+ \alpha_{\rm d2}^-}{\alpha_{\rm l} [l]}
\label{eq:v_initi_non_overlapping_time_varying}
\end{align}
In the high growth regime ($\lambda >\ln{2} / T_{\rm C}$), the initiation volume per origin is given by 
\begin{align}
v^\ast_{\rm o} = \frac{\beta_{\rm datA}^- + \beta_{\rm rida}}{\alpha_{\rm l} \, [l]} \, \frac{f^\ast}{\tilde{K}_{\beta}+f^\ast} \, \frac{\tilde{K}_\alpha +1 - f^\ast}{1-f^\ast}-\frac{\alpha_{\rm d1}+ \alpha_{\rm d2}^+}{\alpha_{\rm l}\, [l]}
\label{eq:v_initi_overlapping_time_varying}
\end{align}
To obtain the same constant initiation volume at all growth rates, the rate of RIDA and the high activity rate of \textit{DARS2} must be chosen such that they exactly cancel out. By setting $v^\ast_{\rm o} = v^\ast_{\rm no}$ we obtain the following constraint on the high activity rate of \textit{DARS2} as a function of the rate of RIDA: 
\begin{equation}
\alpha_{\rm d2}^+ =\beta_{\rm rida} \, \frac{f^\ast}{\tilde{K}_{\beta}+f^\ast} \, \frac{\tilde{K}_\alpha +1 - f^\ast}{1-f^\ast} + \alpha_{\rm d2}^-
\label{eq:condition_rida_dars2}
\end{equation}
As emphasized above, experiments indicate that the initiation volume varies non-monotonically within a 50\% range over a tenfold change in the growth rate \cite{Zheng2020}. Clearly, a constant initiation volume independent of the growth rate is not a robust property of the LDDR model, but only emerges if the rates are chosen carefully, matching Equation \ref{eq:condition_rida_dars2}. 
Experiments support, however, this prediction: Specifically, our model agrees with the experimental finding that deleting \textit{datA} or disabling RIDA or \textit{DARS2} not only affects the initiation volume, but, importantly, also makes it dependent on the growth rate \cite{Flatten2015, Ogawa2002, Camara2005, Riber2006, Flatten2013, Riber2016}; moreover, while the effect of deleting \textit{datA} is most pronounced at lower growth rates \cite{Flatten2015, Ogawa2002}, disabling RIDA or \textit{DARS2} is more severe at higher growth rates \cite{Camara2005, Riber2006, Flatten2013, Riber2016}.
\subsubsection{Simulation details of the LDDR model}
\label{sec:LDDRsimdetails}
Equation \ref{eq:switch_complex_fraction_SI} describes the dynamics of the active fraction of ATP-bound
DnaA in the LDDR model. The parameters are described in section S2\ref{sec:switch_parameters} above and its values are listed in Table \ref{tab:LD_LDDR_parameters}. Figure \ref{fig:S7_LDDR_time_dependent_rates} illustrates the time-dependence of the rates. The LDDR model is simulated analogously to the LD model as described in section S3\ref{sec:LDsimdetails}, but with $f$ propagated according to equation \ref{eq:switch_complex_fraction_SI}.

\subsection{The switch model can give rise to adder correlations in the initiation volume}
\label{sec:switch_noise}
Recent single-cell experiments have shown that the initiation volume per origin exhibits adder correlations \cite{Si2019, Witz2019}.
In the main text, we have shown that fluctuations in the lipid concentration can give rise to such adder correlations in the initiation volume. In this section, we further analyse the effect of fluctuations in the different components of the switch on the initiation volume. Specifically, we show that fluctuations in the total DnaA concentration could give rise to sizer correlations if the initiator DnaA is negatively autoregulated. The important difference between these two scenarios is the relaxation time of the fluctuations: In the case of the lipids, fluctuations decay with the doubling time of the cell, while negative autoregulation in the total DnaA concentration reduces fluctuations on much faster time scales. As we show below, when the correlation time of the fluctuations in the switch components is shorter than the cell-doubling time, consecutive cell cycles are almost not correlated with each other and we obtain sizer correlations. 

To investigate how fluctuations in the switch components propagate to fluctuations in the initiation volume, it is illuminating to analyse the simpler LD model---indeed, this is the motivation for using the LD rather than the LDDR model for Fig. 4 of the main text. The reason why the LD model is more instructive is that in this model the mapping $f (V )$ between the instantaneous fraction $f (t)$ of active DnaA and the current volume $V (t)$ can be obtained and understood straightforwardly. In the regime where (de)activation is faster than growth, this mapping $f (V )$ is obtained by solving equation \ref{eq:switch_simple_fraction_no_synth} in steady state. As we will show, this mapping $f (V )$ depends in an intuitive manner on the concentrations and activities of the switch components (Fig. \ref{fig:S_switch_adder_total_conc}), such that it becomes clear how fluctuations in these components propagate to fluctuations in the initiation volume. In contrast, in the LDDR model the mapping between the instantaneous fraction $f(t)$ and the instantaneous volume $V(t)$ is non-trivial because the rates of RIDA and \textit{DARS2} are temporally regulated over the course of the cell cycle. Importantly, however, while the LDDR model is less illuminating, the principle remains: fluctuations in the switch components will propagate to fluctuations in the mapping $f(V)$, and these will propagate to fluctuations in the threshold for replication initiation. Indeed, below, in section \ref{sec:LDDRcorr}, we will show that the full LDDR model exhibits the same adder and
sizer correlations as the simpler LD model in response to lipid and DnaA fluctuations, respectively.

Below we first expand on the effect of fluctuations in the lipid concentration as discussed in the main text (section \ref{sec:switch_noise_lipids}). Next, we discuss how fluctuations in the total DnaA concentration propagate to fluctuations in the initiation volume. This transmission depends on whether replication initiation is triggered by a critical fraction or concentration of active DnaA. We thus first address the latter question (section \ref{sec:switch_mapping_total_conc}) before we describe the effect of fluctuations in the total DnaA concentration on correlations in the initiation volume (section \ref{sec:switch_noise_total}). Importantly, in section \ref{sec:LDDRcorr} we show that also in the full LDDR model fluctuations in the switch components can give rise to sizer and adder correlations. Finally, in section \ref{sec:switch_noise_other_components} we demonstrate that not only lipid fluctuations, but also fluctuations in proteins that modulate the activities of \textit{datA}, \textit{DARS1/2} and RIDA can generate adder correlations, supporting the idea that the experimentally observed adder correlations~\cite{Si2019, Witz2019} stem from fluctuations in the components of the DnaA switch.

\subsubsection{Lipid fluctuations give rise to adder correlations in the LD model}
\label{sec:switch_noise_lipids}
In this section, we discuss in more detail the scenario discussed in
the main text of a fluctuating acidic phospholipid concentration
$[l]$. For simplicity, in this section we again keep the total initiator concentration $[D]_{\rm T}$ constant in time and at different growth rates such that initiating replication at a constant ATP-DnaA fraction or concentration is equivalent. In the regime where (de)activation is fast compared to the growth rate, the mapping between the instantaneous fraction $f(t)$ of active DnaA and the current volume $V(t)$ is obtained by solving equation \ref{eq:switch_simple_fraction_no_synth} in steady state. This mapping $f(V)$ is shown for different lipid concentrations in panels A and B of Figure \ref{fig:S_switch_adder_total_conc}, for different degrees of ultra-sensitivity, respectively. It shows that the mapping between the active fraction and the volume depends on the lipid concentration. Since replication is initiated when the active fraction $f$ reaches the critical fraction $f^\ast$ for replication initiation, marked by the horizontal dashed line, fluctuations in the lipid concentration lead to fluctuations in the initiation volume $v^\ast$.

Figure 5A of the main text shows that fluctuations in the lipid concentration lead to adder correlations in the initiation volume $v^\ast$. Here, the lipid concentration is modelled via the following Langevin equation:
\begin{equation}
\frac{d[l]}{dt} = \alpha - \lambda \, [l] + \xi(t).
\label{eq:lipid_concentration_SI}
\end{equation} 
The noise is modelled as Gaussian white noise, $\langle \xi(t) \xi(t^\prime) \rangle = 2 D_{\rm l} \delta (t-t^\prime)$, with the noise strength $D_{\rm l}$ chosen to match the measured variance in the initiation volume of $CV=0.1$ \cite{Wallden2016} (see Table \ref{tab:LD_LDDR_parameters}). The dynamics of the active fraction $f$ is given by equation 4 of the main text (Eq. \ref{eq:switch_simple_fraction_SI} above) and the volume and number of origins are simulated as described in section \ref{sec:LDsimdetails}. To prevent premature reinitiation due to stochastic fluctuations in $f$ immediately after replication initiation, we implement a refractory period of $\tau_{\rm b}=$10 minutes after replication initiation during which replication cannot be reinitiated, mimicking the effect of SeqA \cite{Campbell:1990it,Lu:1994ee,Waldminghaus:2009em}. 

Panel A of Fig. 4 of the main text, reproduced herein
Fig.~\ref{fig:SI_LD_adder_sizer_correlations} A to
facilitate the comparison with other models and sources of
fluctuations, shows that lipid fluctuations give rise to adder
correlations in the added initiation volume between successive initiation events. 

Panels B-D of Fig. 4 of the main text elucidate how fluctuations
in the lipid concentration generate adder correlations in the
initiation volume. Panel B shows that lipid concentration
fluctuations $l(t)\equiv [l](t)$ regress to the mean on a timescale given by the
cell-doubling time $\tau_{\rm d} = \ln(2)/\lambda$. Here, the thin grey lines are time traces
from the simulations, while the solid line is the analytical prediction obtained by
solving equation \ref{eq:lipid_concentration_SI} subject to an initial concentration fluctuation $\delta l_0$:
\begin{align}
\langle \delta l(t) | l_0\rangle &\equiv \langle l(t) |l_0 \rangle - \langle l \rangle,\\
&=\delta l_0 e^{-\lambda t},\\
&=\delta l_0 2^{-t / \tau_{\rm d}}.
\end{align}
where $\langle l(t) |l_0 \rangle $ is the average lipid concentration at time $t$ given an initial concentration $l_0$ at time zero and $\langle l \rangle$ is the average lipid concentration; $\langle\delta l(t) | l_0 \rangle$ is thus the average deviation of the lipid concentration from its mean at time $t$, given an initial concentration fluctuation $l_0$. Panel C of Fig. 4 shows the mapping $v^\ast([l])$ between the initiation volume $v^\ast$ and the lipid
concentration $[l]$, obtained by solving
equation \ref{eq:switch_simple_fraction_no_synth} in steady state; this panel corresponds to panel C of
Fig.~\ref{fig:S_switch_adder_total_conc}. Panel
D of Fig. 4 of the main text demonstrates how the decay of lipid fluctuations shown in panel B (of Fig. 4)
with the mapping $v^\ast([l])$ shown in panel C (of Fig. 4) causes the initiation volume to regress to the mean on the timescale of the doubling time $\tau_{\rm d}$:
\begin{align}
\langle \delta v^\ast_n|v_0^\ast\rangle &\equiv \langle
v_n^\ast|v_0^\ast\rangle - \langle v^\ast\rangle,\\
&= \delta v_0^\ast 2^{-n},
\end{align}
where $\langle v^\ast\rangle$ is the average initiation
volume, $v_0^\ast$ is the initial initiation volume arising from a spontaneous fluctuation, and
$\langle v_n^\ast|v_0^\ast\rangle$ is the average initiation
volume $n$ cell cycles later given that initial initiation volume
$v_0^\ast$. Clearly, fluctuations in the initiation volume relax to
the mean via a geometric series, akin to that observed for the volume
at birth \cite{Taheri-Araghi2015}. Combining
$\langle \delta v^\ast_n | v_0^\ast \rangle= \delta v_0^\ast
2^{-n}$ with the definition of the added initiation volume
$\Delta v^\ast\equiv 2 v^\ast_{n+1} - v^\ast_n$ (see Fig. 4 of main text) shows that the average
added initiation volume $\langle \Delta v^\ast\rangle$,
\begin{align}
\langle \Delta v^\ast\rangle &= 2 \left(\langle v^* \rangle + \langle \delta v^\ast_{n+1} | v_0^\ast \rangle\right) - \left(\langle v^* \rangle + \langle \delta v^\ast_n | v_0^\ast \rangle\right) \\
&=2 \left(\langle v^\ast\rangle +
\delta v_0^\ast 2^{-(n+1)}\right) - \left(\langle v^\ast\rangle + \delta v_0^\ast 2^{-n} \right),\\
&= \langle v^\ast\rangle,
\end{align}
equals the average initiation volume $\langle v^\ast\rangle $,
independent of the initial initiation volume
$v_0^\ast = \langle v^\ast \rangle + \delta v_0^\ast$. Hence, the
initiation volume added between successive cell cycles is independent
of the initiation volume, and equal to the average initiation volume.

\subsubsection{Effect of variations in the total initiator concentration on the initiation volume in the LD model}
\label{sec:switch_mapping_total_conc}
\begin{figure}
	\centering
	\includegraphics[width =0.85\textwidth]{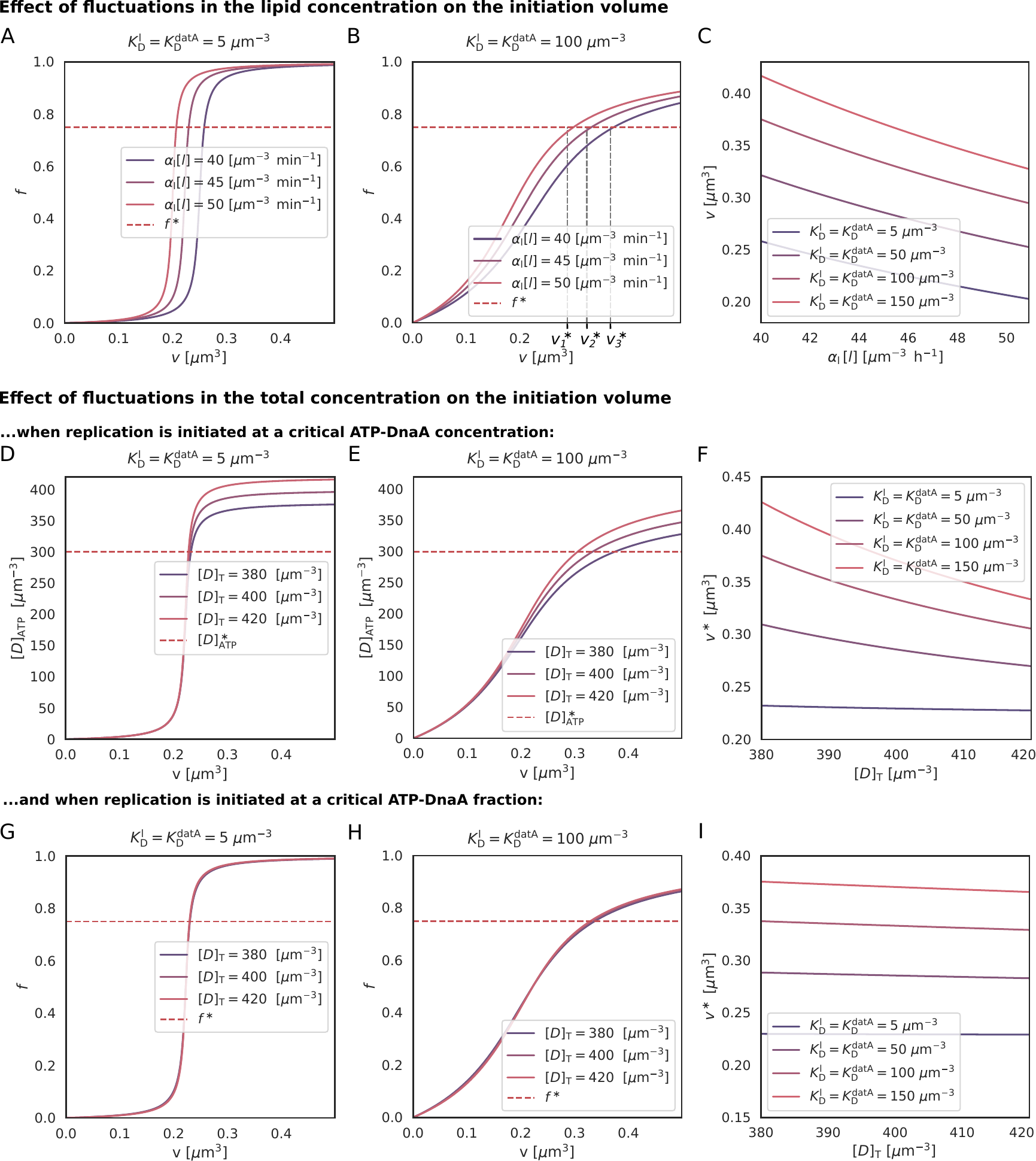}
	\caption{\textbf{Effect of varying lipid and total DnaA	concentration on initiation volume in LD model}
		(A, B) The fraction of active ATP-bound DnaA $f$ as a function of the cell volume per origin $v$ for different lipid concentrations $[l]$, and for two different values of the dissociation constants of the activator and the deactivator of $K_{\rm D}^\alpha=K_{\rm D}^\beta$, respectively. The horizontal dashed line shows the critical fraction $f^\ast$ for replication initiation. Clearly, fluctuations in the lipid concentration lead to fluctuations in the initiation volume $v^\ast$, which is the volume per origin $v$ at which the fraction $f$ equals the critical fraction $f^\ast$. The initiation volume $v^\ast$ as a function of the lipid concentration is shown in panel C, scaling as $1/[l]$ both in the regime of strong ultra-sensitivity (A) and weak ultra-sensitivity (B). (D, E) The concentration of active  ATP-bound DnaA, $[D]_{\rm ATP}$, as a function of the volume per origin $v$ for different values of the total DnaA concentration, both in the regime of strong ultra-sensitivity (D) and weak ultra-sensitivity (E). The horizontal dashed line denotes the critical ATP-DnaA concentration for replication initiation. Fluctuations in the total DnaA concentration generate stronger fluctuations in the initiation volume when the degree of ultra-sensitivity is weaker (panel E). This is highlighted in panel F, which shows the initiation volume as a function of the total DnaA concentration for different values of the dissociation constants. (G, H) The fraction $f$ of active ATP-bound DnaA as a function of the volume per origin $v$ for different total concentrations of DnaA, in the regime of strong (G) and weak (H) ultra-sensitivity. The horizontal dashed line denotes critical fraction $f^\ast$ for replication initiation. The active fraction $f$ depends only weakly on the total DnaA concentration, almost irrespective of the degree of ultra-sensitivity. As a result, the initiation volume is essentially independent of the total DnaA concentration for nearly all values of the dissociation constant (panel I). Replication initiation is thus well protected against noise in the concentration of DnaA. All curves are obtained by solving equation \ref{eq:switch_simple_fraction_no_synth} in steady state. This gives the mapping between the instantaneous concentration $[D]_{\rm ATP}(t)$ or fraction $f(t)$ of active ATP-bound DnaA and the instantaneous volume per origin $v(t)$ when the (de)activation rates are higher than the growth rate.}
	\label{fig:S_switch_adder_total_conc}
\end{figure}

So far, we have assumed that the total DnaA concentration is maintained strictly constant in time and at different growth rates via negative autoregulation. There was thus no difference in initiating replication at a critical concentration $[D]_{\rm ATP}^\ast$ or a critical fraction $f^\ast$. When the total concentration of DnaA is however fluctuating, the concentration of active proteins $[D]_{\rm ATP}$ and the fraction of active proteins $f=[D]_{\rm ATP}/[D]_{\rm T}$ are not directly proportional anymore. This poses a new question: Is replication initiated at a critical ATP-DnaA concentration $[D]_{\rm ATP}^\ast$ or at a critical fraction $f^\ast$? Both scenarios could be possible and have been discussed in literature \cite{Riber2016, Donachie2003}. In the following, we will discuss the effect of fluctuations in the total concentration on the initiation volume in both of these cases in the LD model and in section \ref{sec:validation_DnaA_total_variation} we discuss the effect of variations in the total DnaA concentration in the full switch-titration-SeqA model.
\\
\\
\textbf{Fluctuations in the total concentration can affect the initiation volume if replication is initiated at a critical ATP-DnaA concentration} 
We first consider the case where replication is
initiated at a critical ATP-DnaA concentration $[D]_{\rm ATP}^\ast$. In the LD model for very high (de)activation rates (see equation \ref{eq:switch_no_synth_conc}), the active DnaA concentration can be plotted as a function of the volume of the cell for different total concentrations (Fig. \ref{fig:S_switch_adder_total_conc} D and E). If the dissociation constants of the activator and deactivator are much smaller than the total concentration, the switch is in the ultra-sensitive regime and becomes very steep (Fig. \ref{fig:S_switch_adder_total_conc} D). In this case, the
critical initiation concentration is attained at approximately the same volume per origin independent of the total concentration, as shown in Figure \ref{fig:S_switch_adder_total_conc} F. If the dissociation constants of the activator and deactivator are however in
the same order of magnitude as the total concentration, the ATP-DnaA concentration rises more gradually and attains the critical initiation concentration at different volumes for different total DnaA concentrations (Fig. \ref{fig:S_switch_adder_total_conc} E). Consequently, the initiation volume depends now more strongly on the total DnaA concentration (Fig. \ref{fig:S_switch_adder_total_conc} F).
To summarize, if replication is triggered at a critical ATP-DnaA
concentration $[D]_{\rm ATP}^\ast$ and if the switch is not extremely sharp, we predict a dependence of the initiation volume on the total concentration. We note here that this implies that care should be taken in inferring molecular mechanisms from experiments in which the expression of DnaA is modulated \cite{Hill2012}. In particular, like the initiator accumulation model, also the switch model would predict that the initiation volume decreases as the total DnaA concentration increases.
\\
\\
\textbf{Initiation of replication at a critical ATP-DnaA fraction}
Now, we investigate the scenario where replication is initiated at a critical fraction of ATP-DnaA in the cell, again in the limit where the (de)activation rates are higher than the growth rate (see Eq. \ref{eq:switch_simple_fraction_no_synth}). Interestingly, the fraction as a function of the cell volume in the LD model is essentially independent of the total DnaA concentration (Fig. \ref{fig:S_switch_adder_total_conc} G and H). This finding does also not depend on the steepness of the switch (Fig. \ref{fig:S_switch_adder_total_conc} G and H). The critical fraction $f^\ast$ is attained at an almost
perfectly constant volume per origin (Fig. \ref{fig:S_switch_adder_total_conc} I) at all dissociation constants. The switch mechanism is thus extremely well protected
against variations in the total concentration, if replication is initiated at a critical active fraction of DnaA.

\subsubsection{Negatively autoregulated initiator protein gives rise to sizer correlations in the LD model} 
\label{sec:switch_noise_total}
In this section, we explicitly model the total concentration $[D]_{\rm T}$ and investigate the resulting correlations in the initiation volume. As we have seen in the previous section, the effect of fluctuations in the total concentration is especially high, when replication is initiated at a critical ATP-DnaA concentration and when the system is not too far in the ultra-sensitivity regime. We use a relatively large dissociation constant of $K_{\rm D}^{\rm datA}= K_{\rm D}^{\rm l}=50 \, \mu$m$^{-3}$ in order to obtain a strong dependence of the initiation volume on the total concentration (Fig. \ref{fig:S_switch_adder_total_conc} F). Since the affinities of the two nucleotide binding forms of DnaA to the promoters differ only by a factor of two \cite{Speck1999}, we here make the simplifying assumption that both forms of DnaA have equal affinity for the promoter. As there are five binding sites for DnaA in the promoter region \cite{Speck1999}, we choose a Hill coefficient of $n=5$ in the simulations. The dynamics of the {\em total} number of DnaA is then given by (the same expression as in the AIT model, see equation \ref{eq:AIT_noise_total_conc}):
\begin{equation}
\frac{dN_{\rm D}^{\rm T}}{dt}= \frac{\phi^0\,\lambda \, \rho }{1+\left( \frac{[D]_{\rm T}}{K_{\rm D}^{\rm p}}\right)^n} \, V + \xi_{\rm D} (t).
\label{eq:switch_synthesis_explicit}
\end{equation}
where we have employed the growing cell model with the basal gene allocation fraction $\phi^0$ (see section \ref{sec:ribo_limiting}) and combined it with a noise term $\xi_{\rm D}(t)$ accounting for noise in gene expression (see section \ref{sec:AIT_adder}). The noise is modelled as Gaussian white noise, $\langle \xi_{\rm D}(t) \xi_{\rm D}(t^\prime) \rangle = 2 D_{\rm D} \delta (t-t^\prime)$, where the noise strength $D_{\rm D}$ is tuned to match the measured variance in the initiation volume (see Table \ref{tab:LD_LDDR_parameters}).

The total DnaA concentration is obtained by dividing the total number of DnaA proteins $N_{\rm D}^{\rm T}(t)$ by the explicitly evolved volume $V(t)= V_{\rm b} \, \exp{(t \, \lambda)}$. As newly produced DnaA proteins are more likely to bind ATP than ADP we add the DnaA production term in equation \ref{eq:switch_synthesis_explicit} to the change in the number of active DnaA proteins. The change in the number of ATP-DnaA proteins is computed using
\begin{align}
\frac{dN_{\rm D}^{\rm ATP}}{dt}&=\frac{dN_{\rm D}^{\rm T}}{dt} + \alpha_{\rm l} \,N_{\rm l} \, \frac{[D]_{\rm ADP}}{K_{\rm D}^{\rm l} + [D]_{\rm ADP}} - \beta_{\rm datA} \, n_{\rm ori} \, \frac{[D]_{\rm ATP}}{K_{\rm D}^{\rm datA}+[D]_{\rm ATP}}\\
&= \frac{\phi^0\,\lambda \, \rho}{1+ \left( \frac{[D]_{\rm T}}{K_{\rm D}^{\rm p}}\right)^{n}} V + \xi_{\rm D} (t)+ 
\alpha_{\rm l} \,N_{\rm l} \, \frac{[D]_{\rm ADP}}{K_{\rm D}^{\rm l} + [D]_{\rm ADP}} - \beta_{\rm datA} \, n_{\rm ori} \, \frac{[D]_{\rm ATP}}{K_{\rm D}^{\rm datA}+[D]_{\rm ATP}}
\label{eq:change_atp_dnaa_number}
\end{align}
and the active initiator concentration $[D]_{\rm ATP}(t)$ is obtained by dividing the number of ATP-DnaA proteins $N_{\rm D}^{\rm ATP}(t)$ by the volume $V(t)$. A new round of replication is initiated when the ATP-DnaA concentration reaches the critical concentration for replication initiation $[D]_{\rm ATP}^\ast$; the cell then divides a constant time $\tau_{\rm cc}$ later. During cell division, the volume and the number of DnaA proteins and the number of origins are halved. The rate constants are the same as in the original LD model (see Table \ref{tab:LD_LDDR_parameters}). 

\begin{figure}
	\centering
	\includegraphics[width =0.6\textwidth]{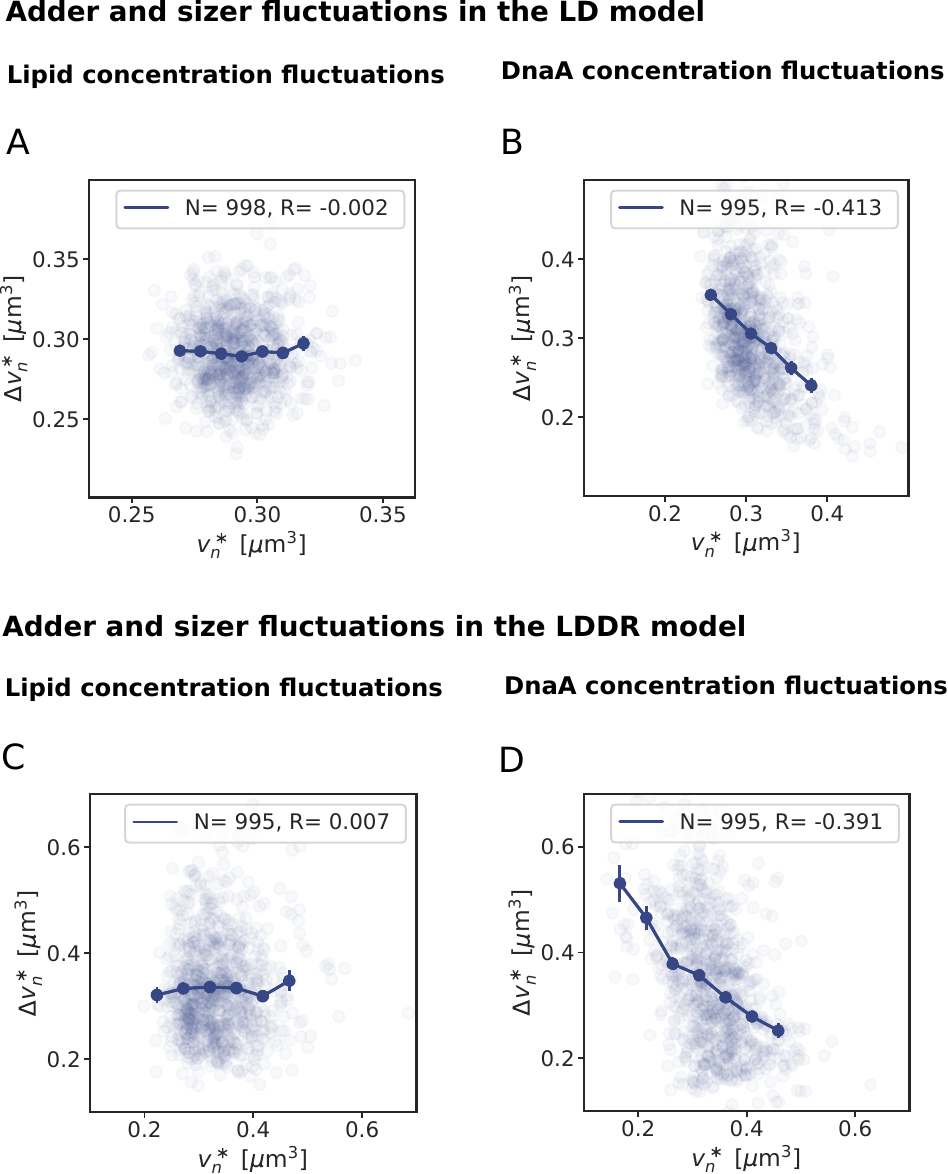}
	\caption{\textbf{Lipid and DnaA concentration fluctuations generate adder and sizer correlations, respectively, in both the LD and LDDR model.} (A, B) Scatter plot of the added initiation volume between successive initiation events, $\Delta v_n^\ast \equiv 2 v_{n+1}^\ast - v_n^\ast$,  and the initiation volume $v_n^\ast$ in the presence of lipid concentration fluctuations (A) and DnaA concentration fluctuations (B) in the LD model. It is seen that in the presence of lipid fluctuations,  $\Delta v_n^\ast$ is independent of $v_n^\ast$ (A), as is characteristic for an adder. In contrast, in the presence of DnaA concentration fluctuations $\Delta v_n^\ast$ is anti-correlated with $v_n^\ast$; a feature characteristic for a sizer. (C, D) Scatter plot of the same data, but for the LDDR model. Also in the LDDR model, lipid fluctuations generate adder correlations in the initiation volume (C), while DnaA concentration fluctuations yield sizer correlations (D). The doubling time in both models is $\tau_{\rm d} = 2$~h, corresponding to non-overlapping replication forks.}
	\label{fig:SI_LD_adder_sizer_correlations}
\end{figure}

To prevent premature reinitiation by stochastic DnaA fluctuations immediately after replication initiation, we also implement a refractory or `eclipse' period of $\tau_{\rm b}=$~10 minutes following replication initiation during which replication cannot be reinitiated, mimicking the effect of SeqA \cite{Campbell:1990it,Lu:1994ee,Waldminghaus:2009em}. The rate constants are the same as in the original LD model (see Table \ref{tab:LD_LDDR_parameters}).
Figure~\ref{fig:SI_LD_adder_sizer_correlations} B shows the result. As fluctuations in the total number are reduced via negative autoregulation within less than one generation, we obtain sizer-like correlations in the initiation volume. We emphasise however that it remains to be verified experimentally how strong the effect of negative autoregulation is.

\subsubsection{Sizer and adder correlations in the full LDDR model} 
\label{sec:LDDRcorr}
The adder or sizer correlations in the initiation volume emerge
from the following four ideas: (i) replication is initiated at a
critical concentration or critical fraction of active, ATP-bound
DnaA; (ii) the mapping between the fraction or concentration of
active ATP-bound DnaA and the volume depends on the concentrations and
activities of the switch components
(Fig. \ref{fig:S_switch_adder_total_conc}) ; (i) and (ii) together
imply that fluctuations in the activities and concentrations of the
switch components will lead to fluctuations in the initiation volume
(Fig. \ref{fig:S_switch_adder_total_conc} C,F,I); (iii) fluctuations in
the initiation volume regress on roughly the same timescale as those
of the switch components, because the mapping between the initiation
volume and the activities or concentrations of the switch components
is fairly linear, certainly when the fluctuations are small enough, and
the rates of activation and deactivation are faster than the growth rate, which they must be generically in order to generate
large-amplitude oscillations in the concentration or fraction of
active DnaA; (iv) adder correlations emerge when this timescale is
set by the growth rate while sizer correlations emerge when this
timescale is significantly faster. These ideas are generic and should apply
not only to the LD model of Fig. 4 in the main text, but
also to the full LDDR model. Here, we show that this is indeed the case.
\\
\\
{\bf Lipid fluctuations generate adder correlations in the LDDR model} Figure \ref{fig:SI_LD_adder_sizer_correlations} C shows the effect of lipid fluctuations in the LDDR model. The model is described by equation~\ref{eq:switch_complex_fraction_SI}, but with the lipid fluctuations modelled in the same way as in the stochastic LD model, see equation \ref{eq:lipid_concentration_SI} or equation 6 of the main text. Like the stochastic LD model, the stochastic LDDR model features an eclipse period of $\tau_{\rm b}=$~10 minutes following replication initiation during which replication cannot be reinitiated \cite{Campbell:1990it,Lu:1994ee,Waldminghaus:2009em}.

Figure~\ref{fig:SI_LD_adder_sizer_correlations} C
demonstrates that also in the full LDDR model, with parameter values
estimated from experimental data (see Table \ref{tab:LD_LDDR_parameters} and section \ref{sec:switch_parameters}), adder correlations in the initiation volume emerge from fluctuations in the lipid concentration. In section \ref{sec:switch_noise_other_components} below, we argue that this effect is much more generic: any switch component that fluctuates on a timescale set by the growth rate, be it lipids, {\it datA}, RIDA, or {\it DARS1/2}, will generate adder correlations in the initiation volume.
\\
\\
{\bf Negatively autoregulated initiator protein gives rise to sizer correlations in the LDDR model} Figure~\ref{fig:SI_LD_adder_sizer_correlations} D  shows the effect of fluctuations in the total concentration of DnaA. The stochastic production of DnaA is modelled in exactly the same way as in the LD model, see equation~\ref{eq:AIT_noise_total_conc}. Combining this with equation~\ref{eq:switch_complex_fraction_SI} yields the following equation for the dynamics of the number of ATP-bound DnaA molecules:
\begin{align}
\frac{dN_{\rm D}^{\rm ATP}}{dt}&= \frac{\phi^0\,\lambda \, \rho }{1+\left( \frac{[D]_{\rm T}}{K_{\rm D}^{\rm p}}\right)^n}  \, V  + \xi_{\rm D} (t)+ 
\left( \alpha_{\rm l} \,N_{\rm l} + \alpha_{\rm d1} \, n_{\rm ori}(t - \tau_{\rm d1})
+ \alpha_{\rm d2}(t) \, n_{\rm ori}(t - \tau_{\rm d2}) \right)\, \frac{[D]_{\rm ADP}}{K_{\rm D}^{\rm l} + [D]_{\rm ADP}} \nonumber\\
&- \left(\beta_{\rm datA} (t)+ \beta_{\rm rida}(t) \right)\, n_{\rm ori} \, \frac{[D]_{\rm ATP}}{K_{\rm D}^{\rm datA}+[D]_{\rm ATP}}.
\label{eq:switch_complex_fraction_noise}
\end{align}
As in the stochastic LD model of section \ref{sec:switch_noise_total}, a new round of replication is initiated when the ATP-DnaA concentration reaches the critical concentration for replication initiation. The cell then divides a constant time $\tau_{\rm cc}$ later. The volume grows exponentially with growth rate $\lambda$ and upon cell division the volume and copy numbers of DnaA and the number of origins are halved. And as for the other stochastic switch models, this model features an eclipse period of $\tau_{\rm b}=$~10 minutes following replication initiation during which replication cannot be reinitiated \cite{Campbell:1990it,Lu:1994ee,Waldminghaus:2009em}.

Figure ~\ref{fig:SI_LD_adder_sizer_correlations} D shows that in the full LDDR model, like in the LD model, DnaA copy number fluctuations give rise to sizer correlations. Negative autoregulation speeds up the regression of the initiation threshold to its mean, turning the system (back) into a sizer. Importantly, it remains to be experimentally verified how strong the effect of negative autoregulation of the protein DnaA is.
\begin{figure}
	\centering
	\includegraphics[width =0.3\textwidth]{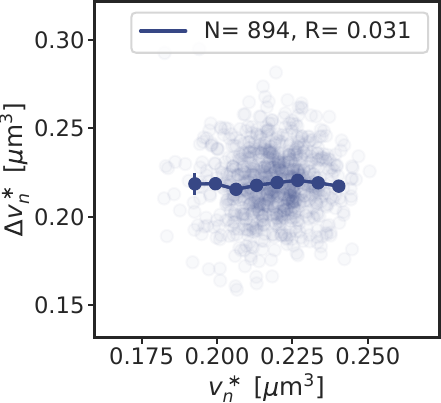}
	\caption{\textbf{RIDA concentration fluctuations
			generate adder correlations in the LDDR model.} Scatter plot of the added initiation volume between successive initiation events, $\Delta v_n^\ast \equiv 2 v_{n+1}^\ast - v_n^\ast$, and the initiation volume $v_n^\ast$. While the lipids and {\it datA} control the initiation volume in the low growth-rate regime, {\it DARS2} and RIDA control the initiation volume in the high growth-rate regime of overlapping replication forks. The Figure shows that in this regime RIDA fluctuations generate adder correlations in the initiation volume, as observed experimentally ~\cite{Si2019, Witz2019}. The cell-doubling time is $\tau_{\rm d}=0.55~{\rm h} \approx 33$~min, corresponding to a growth rate of $\lambda=1.25$~h$^{-1}$. The correlations in the initiation volume in the low growth regime are shown in Fig.~\ref{fig:SI_LD_adder_sizer_correlations}. The dark blue line shows the mean of the binned data and the error bars represent the standard error of the mean (SEM) per bin. The number of data points $N$ and the Pearson correlation coefficient $R$ are indicated.}
	\label{fig:S10_LDDR_RIDA_adder}
\end{figure}

\subsubsection{Fluctuations in other switch components}
\label{sec:switch_noise_other_components}
The activities of \textit{datA}, \textit{DARS1/2} and RIDA are all
influenced by other proteins. IHF affects the activity of \textit{datA}
\cite{Kasho2013} and \textit{DARS2} \cite{Kasho2014}, while Fis
modulates the activity of \textit{DARS2} \cite{Kasho2014}. In
addition, the activity of RIDA is influenced by Hda
\cite{Nakamura2010}. Fluctuations in these proteins will lead to
fluctuations in the respective activation and deactivation rates, just like lipid fluctuations affect the activation rate; in fact, since these proteins are present in (much) lower concentrations than the acidic phospholipids (even though the most potent lipid, cardiolipin, constitutes only a small fraction, 5\%, of the total lipid concentration \cite{Saxena2013}), their fluctuations are likely to be stronger. The fluctuations in the (de)activation rates caused by these proteins will, in turn, generate fluctuations in the concentration or active fraction of ATP-DnaA as a function of the volume, thus causing fluctuations in the initiation size. Because the activation and deactivation rates are typically higher than the growth rate (see section \ref{sec:switch_parameters} and Fig. \ref{fig:SI_LDDR_time_traces}), fluctuations in the initiation
volume regress on the same timescale as that of the fluctuations in the switch components. If the switch components decay on a timescale set by the growth rate, because the proteins are neither degraded actively nor produced via strong feedback control, then their fluctuations will give rise to fluctuations in the initiation volume that relax on the timescale set by the growth rate. These fluctuations will therefore also generate adder correlations in the initiation volume, just like the lipids do. We thus argue that the idea that fluctuations in the switch components can generate adder correlations is general.

To provide support for this idea, we study how fluctuations in the activity of RIDA in the full LDDR model, as induced by e.g. Hda \cite{Nakamura2010}, propagate to fluctuations in the initiation volume. The system is modeled in exactly the same way as in the previous section (S3\ref{sec:LDDRcorr}), except that now the lipid concentration is constant while the activity of RIDA fluctuates:
\begin{align}
\tilde{\beta}_{\rm rida}(t)&= \overline{\tilde{\beta}}_{\rm rida}+\eta(t),\\
\frac{d\eta}{dt}&= -\lambda \eta + \xi_\eta (t).
\end{align}
Here, $\overline{\tilde{\beta}}_{\rm rida}$ is the mean RIDA activity, while $\xi_\eta(t)$ models Gaussian white noise with strength $\langle \xi_\eta(t) \xi_\eta(t^\prime\rangle = 2 D_\eta \delta (t-t^\prime)$, such that the colored noise $\eta$ describes fluctuations with zero mean that decay on a timescale set by the growth rate $\lambda$.

Figure~\ref{fig:S10_LDDR_RIDA_adder} shows that, as anticipated, fluctuations in the activity of RIDA generate adder correlations in the initiation volume. Adder correlations will emerge from fluctuations in switch components that relax on a timescale given by the growth rate. This is one of the central findings of our study. 

\section{Full switch-titration-SeqA model: model validation and predictions}
\label{sec:model_validation_and_prediction}
In this section, we validate our theoretical model by comparing key predictions to experimental observations and then make several novel experimentally testable predictions. We first present a version of the model that contains all elements that we previously discussed separately (section \ref{sec:switch_titration_combined}): We include all so far known activators and deactivators of DnaA as discussed in the LDDR model (see section \ref{sec:LDDR_model}); we model the expression of DnaA explicitly (see section \ref{sec:switch_noise_total}); and we include a fixed number of homogeneously distributed titration sites (section \ref{sec:titration_sites}). Additionally, we now also include a blocked period of DnaA synthesis after replication initiation due to SeqA. We then discuss several experimental results that our full model can qualitatively reproduce (section \ref{sec:model_validation}). Finally, we present experimentally testable predictions that follow from our model (section \ref{sec:model_predictions}).

\subsection{Switch-titration-SeqA model: combining titration with activation switch, as well as blocked synthesis by SeqA}
\label{sec:switch_titration_combined}
In the LD and the LDDR model we argue that replication initiation in \textit{E. coli} is regulated via a switch of the initiator protein DnaA. In the AIT model, we have shown that also the titration mechanism can ensure stable cell cycles at low growth rates. Furthermore, experiments indicate that the protein SeqA plays
an important role in the regulation of replication initiation by not only blocking replication initiation, but also DnaA synthesis during an `eclipse period' of about 10 minutes after replication initiation \cite{Campbell:1990it, Lu:1994ee,Waldminghaus:2009em, Katayama2017}. An open question therefore remains what role the activation switch, the titration sites for DnaA on the chromosome and the protein SeqA play in the regulation of replication initiation. To dissect the effects of these different mechanisms, we first include the homogeneously
distributed titration sites in the LDDR model and show that at low growth rates they transiently lower the concentration of ATP-DnaA that is available for initiating replication at the origin (section \ref{sec:implementation_switch_titration_combined}). The titration sites therefore shape the oscillations in the {\em free concentration} of ATP-bound DnaA. In section \ref{sec:switch_titration_robustness} we show that a concentration cycle, as induced by titration, can generically enhance an  activation cycle, as driven by the switch, by increasing the steepness of the oscillations (the ``gain''), which dampens the propagation of fluctuations in the free concentration of active DnaA to the initiation volume. Titration can thus protect a switch from fluctuations in its components.  We then combine the switch-titration model with SeqA and show that in this full model the oscillations in the free ATP-DnaA concentration are large in all
growth regimes (section \ref{sec:full_model_large_sharp_osc}). Finally, we discuss in which parameter regime either the switch, titration or SeqA is the principal driver in setting the initiation volume (section
\ref{sec:switch_titration_combined_parameter_regimes}).

\subsubsection{Switch-titration model: combining titration with activation switch leads to sharper oscillations in the free active DnaA concentration at low growth rates}
\label{sec:implementation_switch_titration_combined}
In this section, we first explain how the titration sites can be included in the switch model. Then we show that titration helps the DnaA activation switch by shaping the oscillations in the free ATP-DnaA concentration at low growth rates. At high growth rates, in the overlapping-replication fork regime, the homogeneously distributed titration sites cannot generate large oscillations in the {\em total} free concentration $[D]_{\rm T,f}$ of DnaA ($[D]_{\rm ADP,f}+[D]_{\rm ATP,f}$) and the oscillations in the active free DnaA concentration $[D]_{\rm ATP,f}$ are mainly generated by the switch.

In order to combine the activation switch with titration, we model the change in the total number of DnaA proteins explicitly. Using the growing cell model (see section \ref{sec:ribo_limiting}), 
the change in the total number of DnaA proteins is then given by equation \ref{eq:switch_synthesis_explicit} (with the noise set to zero in the mean-field model) as explained in section \ref{sec:switch_noise_total}. 
As before in section \ref{sec:switch_noise_total} we make the simplifying assumption that the two nucleotide forms of DnaA have the same affinity for the promoter.
The total DnaA concentration is obtained by dividing the total number of DnaA proteins $N_{\rm D}^{\rm T}(t)$ by the explicitly evolved volume $V(t)= V_{\rm b} \, \exp{(t \, \lambda)}$. As newly produced DnaA proteins are more likely to bind ATP than ADP we add the DnaA production term to the change in the number of active DnaA proteins. The change in the number of ATP-DnaA proteins is then given by equation \ref{eq:switch_complex_fraction_noise} and the active initiator concentration $[D]_{\rm ATP}(t)$ is obtained by dividing the number of ATP-DnaA proteins $N_{\rm D}^{\rm ATP}(t)$ by the volume $V(t)$ (see section \ref{sec:switch_noise_total} for more details).

While equation \ref{eq:switch_complex_fraction_noise} describes how the {\em total} concentration of ATP-DnaA is modelled, it does not describe the dynamics of the {\em free cytosolic} ATP-DnaA concentration. Both ATP-DnaA and ADP-DnaA have relatively high affinity for the approximately 300 DnaA boxes per chromosome \cite{Roth1998}. We therefore assume here equal affinity of ATP-DnaA and ADP-DnaA to these titration sites.
Exploiting this, we can then use equation \ref{eq:free_proteins} to calculate the free concentration of DnaA, $[D]_{\rm T, f}$, given the total concentration $[D]_{\rm T}$ of DnaA and the total concentration of titration sites $[s]_{\rm T}$, as described in section \ref{sec:titration_sites}.
Like in the AIT model (section \ref{sec:AIT_ribo_limiting}), only the DnaA proteins that are not bound to the titration sites $[D]_{\rm T, f}$ can repress the production of new initiator proteins.  
In the presence of high-affinity titration sites, we assume that replication is initiated when the free concentration of ATP-DnaA in the cell reaches a critical threshold $[D]_{\rm ATP, f}^\ast$. Exploiting the equal affinities of the two nucleotide binding states of DnaA for the titration sites, and fast binding and unbinding dynamics (see section \ref{sec:titration_sites}), the fraction $g= [D]_{\rm ATP, f}/ [D]_{\rm T, f}$ of the concentration of free ATP-DnaA $[D]_{\rm ATP, f}$ over the concentration of free total DnaA $[D]_{\rm T, f}$ is equal to the fraction of the total ATP-DnaA concentration over the total DnaA concentration per cell $f= [D]_{\rm ATP}/[D]_{\rm T}$. The free ATP-DnaA concentration is therefore given by the concentration of free DnaA $[D]_{\rm T, f}$ times the active fraction of DnaA $f$:
\begin{equation}
[D]_{\rm ATP, f}(t) = [D]_{\rm T, f}(t) \times f(t)
\label{eq:critical_free_active_conc}
\end{equation}
It remains an open question whether all DnaA proteins or only the freely diffusing DnaA can be activated and deactivated via the switch of the LDDR model. Both scenarios could be envisioned: While it might seem more natural to assume that only free DnaA can be activated or deactivated, also DnaA that is bound to titration sites might be in contact with the acidic phospholipids or with the site \textit{datA} via supercoiled DNA. Additionally, as RIDA is moving along the entire chromosome during DNA replication, every titration site will be in the proximity of RIDA once and bound DnaA could be inactivated at that moment. Importantly, however, when the affinities of the two nucleotide bound forms of the DnaA to the titration sites are equal and the binding dynamics are fast, the active fraction in the cytoplasm $g$ equals the total active fraction $f$, irrespective of whether the activation and deactivation reactions happen only in the cytoplasm or also on the DNA. This question only affects the magnitude of the activation and deactivation rates: If only the free DnaA can be (de)activated by the components of the switch, activation and deactivation rates become lower because fewer DnaA proteins are available. If the dissociation constants of the activators and deactivators are however lower than the free DnaA concentration, the system remains in the ultra-sensitivity regime and the titration sites affect the magnitude of the (de)activation rates of the switch only weakly. We therefore here assume out of simplicity that all DnaA, no matter whether bound or unbound to titration sites, can be (de)activated by the switch components. 

\begin{figure}
	\centering
	\includegraphics[width =0.8\textwidth]{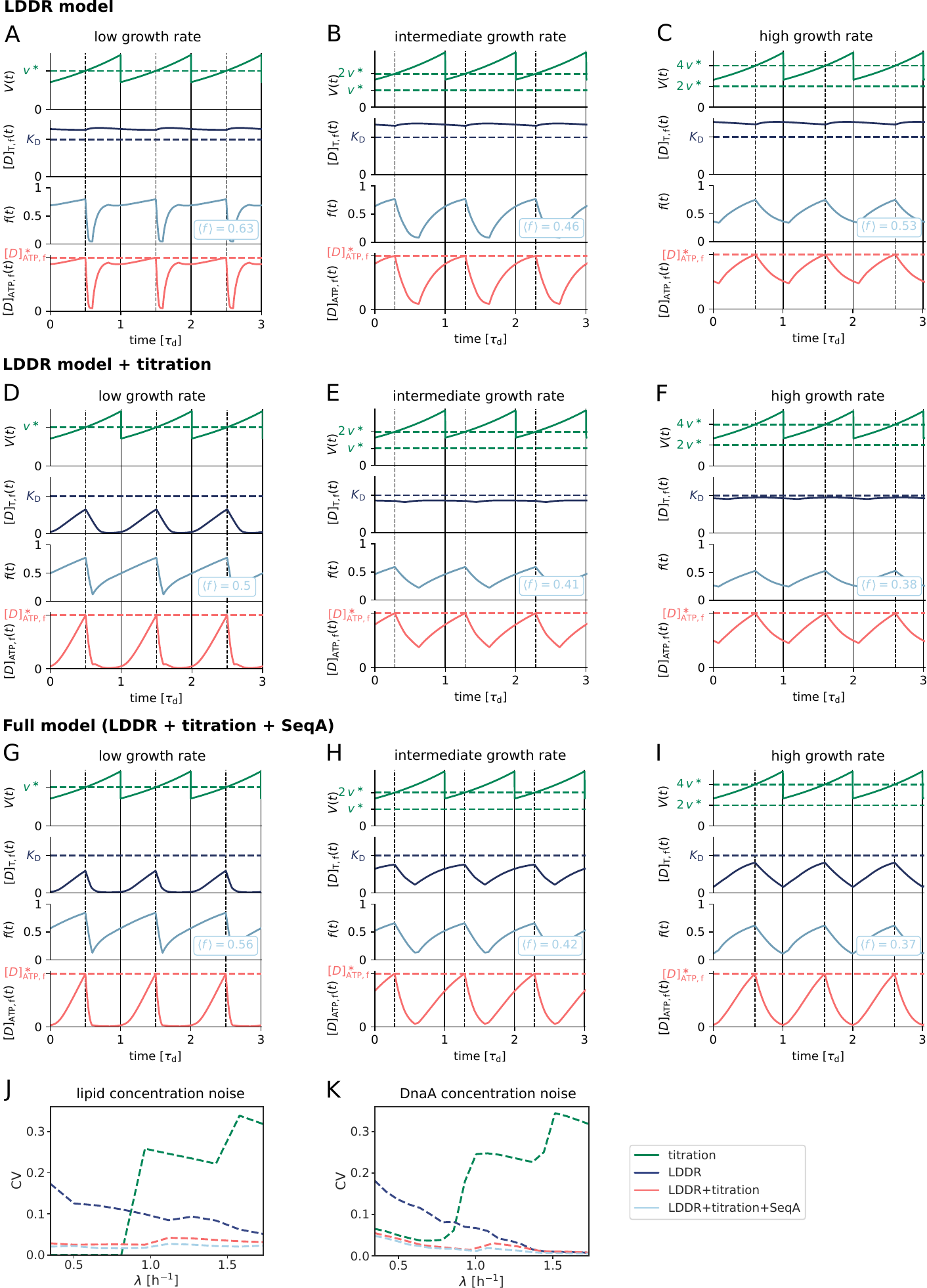}
	\caption{\textbf{Combining the activation switch with tiration sites and SeqA gives rise to large amplitude oscillations in the active free DnaA concentration at all growth rates} (A-I) The volume of the cell $V(t)$, the free DnaA concentration $[D]_{\rm T, f}$, the fraction of DnaA $f(t)$ that is bound to ATP (irrespective of whether the DnaA is in the cytoplasm or on the titration sites) and the concentration of free ATP-DnaA $[D]_{\rm ATP, f}(t)$ as a function of time (in units of the doubling time $\tau_{\rm d}$) for $\tau_{\rm d}= 2$~h (A, D, G),  $\tau_{\rm d}= 35$~min (B, E, H) and $\tau_{\rm d}= 25$~min (C, F, I)). The average active fraction over one cell cycle $\langle f \rangle$ is indicated in light blue in the third panel. While combining the LDDR model with titration helps to shape the oscillations at low growth rates, it does not significantly affect the free DnaA concentration at high growth rates. Adding the effect of SeqA has no strong impact at low growth rates, but strongly increases the oscillations in the free total DnaA concentration $[D]_{\rm T, f}$ at intermediate and high growth rates. (J, K) The coefficient of variation ${\rm CV}= \sigma /\mu$ with the standard deviation $\sigma$ and the average initiation volume $\mu=\langle v^\ast \rangle$ as a function of the growth rate for different models in the presence of noise in the lipid concentration (J) or in the DnaA concentration (K). The large coefficient of variation in the LDDR model at low growth rates is reduced significantly by the titration sites. Conversely, the LDDR model prevents the reinitiation events that inevitably occur at high growth rates in the AIT model. Adding the effect of SeqA to transiently block DnaA synthesis increases the robustness of the system even further. Clearly, the model that combines all mechanisms initiates replication most precisely.}
	\label{fig:SI_full_model_time_traces}
\end{figure}

Comparing Figure \ref{fig:SI_full_model_time_traces} A and D shows that including the titration sites in the LDDR model leads to sharper oscillations in the \textit{free concentration} of ATP-DnaA at low growth rates. In the LDDR model, the ATP-DnaA concentration first decreases strongly after replication initiation due to the combined action of the site \textit{datA} and RIDA and then rises again when the activation sites DARS1/2 are being doubled. At low growth rates, the fixed doubling time of e.g. DARS2 $\tau_{\rm d2}$ is much shorter than the cell-doubling time $\tau_{\rm d}$ ($\tau_{\rm d2}=0.2$~h $\ll \tau_{\rm d}=2$~h), leading to a relatively high ATP-DnaA concentration during most of the cell cycle. In Figure 5 C of the main text we have shown that the resulting slow rise in the ATP-DnaA concentration towards the initiation threshold leads to large variations in the initiation volume in the presence of noise in the lipid concentration. Figure \ref{fig:SI_full_model_time_traces} D (and Fig. 5B) shows that including titration sites can significantly sharpen the oscillations in the free ATP-DnaA concentration at low growth rates. As in this growth regime, titration sites are synthesized faster than DnaA proteins, the free concentration drops rapidly after replication initiation and remains low during most of the cell cycle. Only when all titration sites have been filled begins the free concentration to rise and replication can be initiated (Fig. \ref{fig:SI_full_model_time_traces} D). At intermediate and high growth rates however, the rate at which new titration sites are being synthesized after initiation is comparable or even lower than the synthesis rate of new DnaA proteins. In this regime, the oscillations in the free DnaA concentration become much weaker and the shape of the oscillations in the free ATP-DnaA concentration is dominated by the switch (Fig. \ref{fig:SI_full_model_time_traces} E/F). In this regime, the DnaA activation switch is the dominant pacemaker.

\subsubsection{On the necessity of combining titration with activation: How titration can enhance the switch}
\label{sec:switch_titration_robustness}
{\bf Switch is necessary} In the main text and in the previous section, we argued that the combined switch-titration system is more robust to fluctuations than either mechanism alone. How general is this prediction? Our model shows that a titration-based model with a homogeneous titration-site distribution is not sufficient for controlling replication initiation at all growth rates (see section \ref{sec:AIT_homogeneous_titration}). The failure of the titration-based mechanism arises from the different scaling of the protein synthesis and the titration-site formation rate with the growth rate. The synthesis rate of DnaA scales with the growth rate,  see Eqs. \ref{eq:alphap} and \ref{eq:AIT_production_rate_growing_cell}. In contrast, the titration-site formation rate is set by the DNA replication rate, which varies only very weakly with the growth rate \cite{Si2017}. The different scaling of these two timescales with the growth rate means that a titration-based mechanism must fail inevitably at sufficiently high growth rates --- because this prediction is based on a comparison of two timescales, it is insensitive to the details of the model (see also section \ref{sec:AIT_homogeneous_titration}). A robust prediction from our analysis is, therefore, that a titration-based mechanism is not sufficient for generating robust replication-initiation cycles at all growth rates; another mechanism is essential. As we show in the main text (Fig. 2~C/D), transient suppression of protein synthesis by SeqA after replication initiation \cite{Campbell:1990it} can prevent reinitiation events at high but not at intermediate growth rates. We thus argue that the switch is essential to generate robust replication cycles at all growth rates.

\begin{figure}
	\centering
	\includegraphics[width =0.7\textwidth]{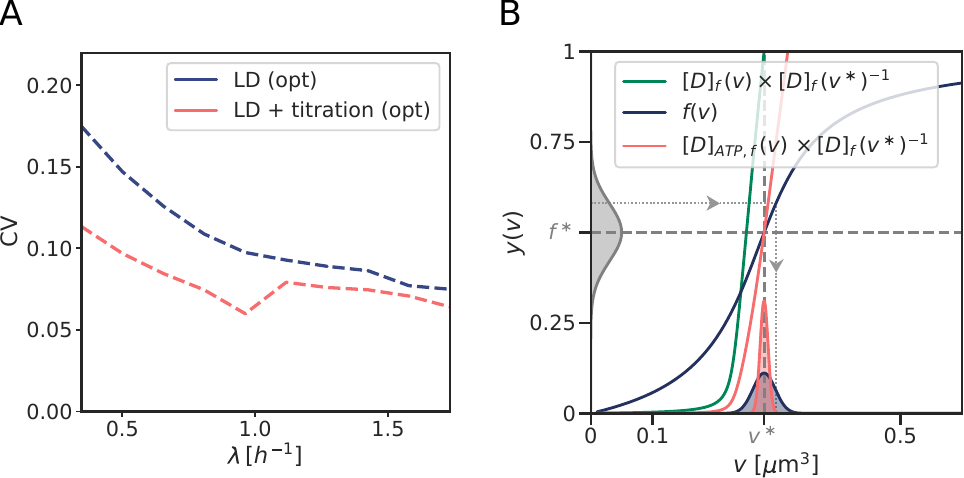}
	\caption{\textbf{Combining the activation switch with titration enhances robustness of replication initiation in the presence of noise in the lipid concentration at all growth rates} (A) The system that combines a switch (LD) with titration has a lower Coefficient of Variation (CV) in the initiation volume than the system based on the switch only. This is because a concentration cycle, as induced by titration, can generically enhance an activation cycle, as driven by the switch, as illustrated in the panel on the right. (B) The total concentration of cytoplasmic DnaA, $[D]_{\rm T, f}$, normalized by its value at the initiation volume $v^\ast$, as a function of the cell volume per origin $v$ (green line); the active fraction of DnaA, $f$, as a function of $v$ (blue line); the cytoplasmic concentration of active DnaA, $[D]_{\rm ATP, f}=[D]_{\rm T,f}\times f$, normalized by its value at $v^\ast$, as a function of $v$ (red line). It is seen that the gain in the combined system, given by the slope of $[D]_{\rm ATP, f}$ (red line) at $v=v^\ast$ is higher than that of the switch-only system, given by the slope of $f$ (blue line) at $v=v^\ast$: multiplying an activation cycle $f(v)$ with a concentration {\em cycle} $[D]_{\rm T,f}(v)$, instead of a constant concentration, leads to sharper oscillations in $[D]_{\rm ATP,f}(v)$. As a result, fluctuations in $f$ and hence  $[D]_{\rm ATP,f}$ propagate less to fluctuations in the initiation volume: the distribution of $v^\ast$ in the combined system (red distribution x-axis) is narrower than that of the switch-only system (blue distribution x-axis). The mapping $f(v)$ is obtained by solving Eq. \ref{eq:switch_simple_fraction_no_synth} in steady state and $[D]_{\rm T,f}(v)$ is obtained by solving Eq. \ref{eq:free_proteins}. See Table \ref{tab:LD_LDDR_parameters} for parameters.}
	\label{fig:S14_Optimisation}
\end{figure}

{\bf Switch may be sufficient, but titration can enhance the switch} But could a switch be sufficient? Clearly, our work shows it could: while a system based on titration alone must fail at high growth rates, one based on a switch alone could generate stable replication cycles at all growth rates. Nonetheless, the experiments indicate that the system combines a switch with titration \cite{Roth1998, Schaper1995, Hansen2018}. What could be the benefit of adding titration to the switch? One possibility is that it makes the switch more robust to fluctuations in the switch components. Fig. 5 of the main text shows that adding titration to the LDDR model can indeed enhance the precision of replication initiation. The LDDR model contains however many parameters. Here, we argue that a concentration cycle, as created by titration, can {\em generically} enhance an activation cycle, as induced by the switch. To show this, we consider a minimal model of the switch, namely the LD model. We will optimize this system by minimizing the coefficient of variation in the initiation volume, subject to plausible experimental constraints. We then add to this optimal switch the titration mechanism, {\em keeping all the parameters of the switch (and also the titration system) the same.}  As Fig. \ref{fig:S14_Optimisation}~A shows, the combined system is more robust than the optimal system based on only the switch. Clearly, adding titration makes it possible to beat the precision limit of the switch. We then show mathematically how a concentration cycle can, generically, enhance an activation cycle: it can increase the sharpness of the oscillations, the ``gain'', which means that fluctuations in the cytoplasmic concentration of active DnaA propagate less to fluctuations in the initiation volume (see Fig. \ref{fig:S14_Optimisation}~B). This underscores the principal finding of our study: while the switch helps titration by preventing premature reinitiation at high growth rates, titration can help the switch by sharpening the oscillations, increasing the precision of replication initiation.

Concretely, a major source of noise in the switch are the fluctuations in the activation and deactivation components; the switch is inherently fairly robust to fluctuations in the total concentration of DnaA (see Fig. \ref{fig:S_switch_adder_total_conc}). We therefore study the coefficient of variation (CV) in the initiation volume arising from lipid fluctuations. In our minimal model of the switch, the LD model, the experimental constraints are: the initiation volume \cite{Si2019, Zheng2020, Elf2007} and the maximum (de)activation rates \cite{Kasho2013} (see Table S2). The optimization parameters are the critical fraction $f^\ast$ for replication initiation, and the dissociation constants $\tilde{K}_{\rm D}^{\rm l}= K_{\rm D}^{\rm l}/ [D]_{\rm T}$ and $\tilde{K}_{\rm D}^{\rm datA}= K_{\rm D}^{\rm datA}/ [D]_{\rm T}$. The noise strength $D_{\rm l}$ only sets the scale for the CV of the initiation volume, and does not affect the outcome of the optimization procedure; it is set such that the CV is comparable to that measured experimentally at high growth rates (see Fig. \ref{fig:S14_Optimisation}~A). The parameters of the titration system, the number of titration sites $n_{\rm s}$ and the titration-site affinity $K_{\rm D}^{\rm s}$, are taken to be the same as in the main text and in the rest of the SI (see Table \ref{tab:RIT_AIT_parameters}). We then add this titration system to the optimal switch, keeping the parameters of the optimal switch and the titration system the same; also the total DnaA concentration is the same in all three systems. There is only one parameter that remains to be specified in the combined system, which is the threshold for replication initiation, $[D]_{\rm ATP}^\ast$; this is set such that the initiation volume matches that observed experimentally.

{\bf How titration can help the switch} Fig. \ref{fig:S14_Optimisation}~A shows that adding titration to the best switch reduces the CV in the initiation volume. Like any cellular system \cite{Elowitz.2002, Govern:2014ef}, the robustness of the DnaA activation switch is fundamentally limited by constraints on protein copy numbers and reaction rates and externally induced (extrinsic) fluctuations in these quantities. Our work shows that if these limit the precision of the switch in controlling replication initiation, then adding titration to the switch is a useful strategy to lift these limitations. Titration can help the switch, because a concentration cycle, as generated by titration, can generically enhance the precision of an activity cycle, as driven by the switch. To see this, we consider the mapping between the  concentration of cytoplasmic, active DnaA, $[D]_{\rm ATP, f}$, and the volume per origin $v$ of the cell,  see Fig. \ref{fig:S14_Optimisation}~B.   The former is given by $[D]_{\rm ATP, f} = [D]_{\rm T, f} \times f$, where $[D]_{\rm T,f}$ is the total concentration of cytoplasmic DnaA (i.e, active and inactive) and $f$ is the fraction of active DnaA, see Eq. \ref{eq:critical_free_active_conc}. The general idea is then that oscillations in $[D]_{\rm T, f}$ (green line of Fig. \ref{fig:S14_Optimisation}~B), as induced by titration, can conspire with the oscillations in the active fraction $f$ (blue line of Fig. \ref{fig:S14_Optimisation}~B), as driven by the switch, to generate sharper oscillations in $[D]_{\rm ATP, f}$ (red line of Fig. \ref{fig:S14_Optimisation}~B); and these sharper oscillations mean that fluctuations in $[D]_{\rm ATP, f}$ lead to smaller fluctuations in the initiation volume, as illustrated in Fig. \ref{fig:S14_Optimisation}~B.

To make this mathematically concrete, we will exploit that the fluctuations in the switch components cause fluctuations in the active fraction $f(v)$ and not the total cytoplasmic DnaA concentration $[D]_{\rm T,f}(v)$, which is controlled by titration. The variance of the fluctuations in $[D]_{\rm ATP, f}$ that arise from fluctuations in the switch components is then given by
	\begin{equation}
	\sigma^2_{[D]_{\rm ATP, f}} = [D]_{\rm T, f}^2\sigma^2_f ,
	\label{eq:sigDf}
	\end{equation}
	where $\sigma^2_f$ is the variance of the fluctuations in the active fraction resulting from the switch. Linearizing the input-output relation $[D]_{\rm ATP, f}(v)$, and using the rules of error propagation \cite{Govern:2014ef}, the variance in the  initiation volume is given by
	\begin{align}
	\sigma^2_{v^\ast} &= \frac{\sigma^2_{[D]_{\rm ATP, f}}}{g_{D\to v}^2},
	\label{eq:varv*}
	\end{align}  
	where $g_{D\to v} = d [D]_{\rm ATP,f} / dv$ is the ``gain'', which determines how fluctuations in $[D]_{\rm ATP,f}$ propagate to variations in the volume per origin $v$, see Fig. \ref{fig:S14_Optimisation}~B;  both the numerator and denominator of Eq. \ref{eq:varv*} are  evaluated at $v=v^\ast$. Noting that  $[D]_{\rm ATP, f}(v) = [D]_{\rm T, f} (v)\times f(v)$ and using Eq. \ref{eq:sigDf} we find that
	\begin{align}
	\sigma^2_{v^\ast} &= \frac{[D]_{\rm T, f}^2\sigma^2_f }{[D]_{\rm T, f}^2 (df/dv)^2 + (d [D]_{\rm T, f}/dv)^2 f^2 + 2 [D]_{\rm T, f} f (d [D]_{\rm T, f}/dv) (df / dv)}\\
	&= \frac{\sigma^2_f }{ (df/dV)^2 + (d [D]_{\rm T, f}/dv)^2 f^2/[D]_{\rm T, f}^2 + 2 [D]_{\rm T, f} f (d [D]_{\rm T, f}/dV) (df / dv)/[D]_{\rm T, f}^2}\label{eq:varv*main}\\
	&\leq \frac{\sigma^2_f}{(df/dv)^2} = \sigma^{2,f}_{v^\ast}.
	\end{align}
	Importantly, $df/dv$ and $\sigma^2_f$ are properties of the switch, and, in comparing the combined to the switch-only system, are evaluated at the same  initiation volume $v=v^\ast$ in the two systems. As a result,  $\sigma^{2,f}_{v^\ast}$ is the variance in the initiation volume of the switch-only system, in which  the oscillations in the active, cytoplasmic DnaA concentration, $[D]_{\rm ATP,f}(v)$, are only driven by the activation cycle $f(v)$ and $d [D]_{\rm T, f}/dv$ is zero.  Eq. \ref{eq:varv*main}, therefore, shows that by matching the concentration cycle to the activation cycle, such that the total cytoplasmic concentration $[D]_{\rm T, f}$ rises when the active fraction $f(v)$ rises and both $d [D]_{\rm T, f}/dv$ and $df/dv$ are non-zero,  the concentration cycle of titration can help the activation cycle of the switch by reducing the variance in the initiation volume. 

\subsubsection{Full switch-titration-SeqA model: combining switch, titration and SeqA in the full model leads to sharp, high amplitude oscillations in the free active DnaA concentration at all growth rates}
\label{sec:full_model_large_sharp_osc} 
In the full model, we combine the LDDR model with titration sites and the effect of SeqA to transiently block DnaA production. In section \ref{sec:implementation_switch_titration_combined} we have demonstrated that adding titration to the switch leads to sharper oscillations in the free ATP-DnaA concentration at low, but not at high growth rates, as compared to oscillations driven by the switch alone. Including the effect of SeqA to transiently block DnaA synthesis allows for sharp oscillations in the total free DnaA concentration also at intermediate and high growth rates (Fig. \ref{fig:SI_full_model_time_traces} H and I): By transiently blocking DnaA production after replication initiation, the newly produced titration sites can lower the free DnaA concentration. When the blocked period is over, the free DnaA concentration increases again as new proteins are synthesized faster than titration sites in this regime. These oscillations in the free concentration and the oscillations in the active fraction together lead to large and sharp oscillations in the free ATP-DnaA concentration at high and intermediate growth rates (compare red curve in Fig. \ref{fig:SI_full_model_time_traces} E/F to H/I). Indeed, adding SeqA to the titration-switch model lowers the variance in the initiation volume even more, especially at high growth rates in the presence of noise in the lipid concentration (Fig. \ref{fig:SI_full_model_time_traces} J) or in the DnaA concentration (Fig. \ref{fig:SI_full_model_time_traces} K). The latter result on the containment of DnaA expression noise is particularly interesting: the full model that combines all mechanisms is more precise than all other models that combine only a subset of mechanisms, for all growth rates. This really shows that these three mechanisms - protein activation, titration, periodic suppression of protein synthesis - act synergistically. We thus conclude that the combination of titration, DnaA activation switch and SeqA yields robust oscillations in the concentration of active DnaA over the full range of growth rates. Recent experiments support the idea that \textit{E. coli} combines titration with an activation switch since removing either mechanism alone still yields stable cycles, but with increased variation in the initiation size \cite{Knoppel2021}. 

\subsubsection{Whether titration, the activation switch or SeqA sets the initiation volume depends on the parameter regime}
\label{sec:switch_titration_combined_parameter_regimes} 

In the following, we present how in different parameter regimes either titration, the switch or the effect of blocking DnaA synthesis via SeqA can determine the initiation volume. In the full switch-titration-SeqA model, replication is initiated at a critical free, active concentration $[D]_{\rm ATP, f}^\ast$ (equation \ref{eq:critical_free_active_conc}). Therefore, both the oscillations in the free concentration and the active fraction contribute to reaching the critical initiation threshold. This observation makes it possible to steer the system from a switch-dominated to a titration-dominated regime, by controlling the thresholds of the respective mechanisms. Here, we discuss the effect of modulating the titration threshold, by varying either the basal synthesis rate or the number of titration sites. Specifically, in the full model, the dissociation constant of the promoter $K_{\rm D}^{\rm p}$ must again, as in the AIT model, be higher than that of the origin $K_{\rm D}^{\rm ori}$, such that the free cytosolic concentration can reach the critical concentration $[D]_{\rm ATP,f}^\ast$ necessary for replication initiation.  Contrary to the AIT model, it is now however possible to attain the (total) free concentration set by the promoter $K_{\rm D}^{\rm p}$ without reaching the critical initiation threshold $[D]_{\rm ATP, f}^\ast$, if the active fraction remains sufficiently low to prevent initiation. Conversely, the titration sites could prevent replication initiation even when the active fraction $f$ is high, by lowering the free concentration for a significant fraction of the cell cycle. Therefore, it depends on the number of titration sites and on the rate at which these sites are filled up, whether the oscillations in the active fraction or the oscillations in the free concentration set the initiation volume.

\begin{figure}
	\centering
	\includegraphics[width =0.95\textwidth]{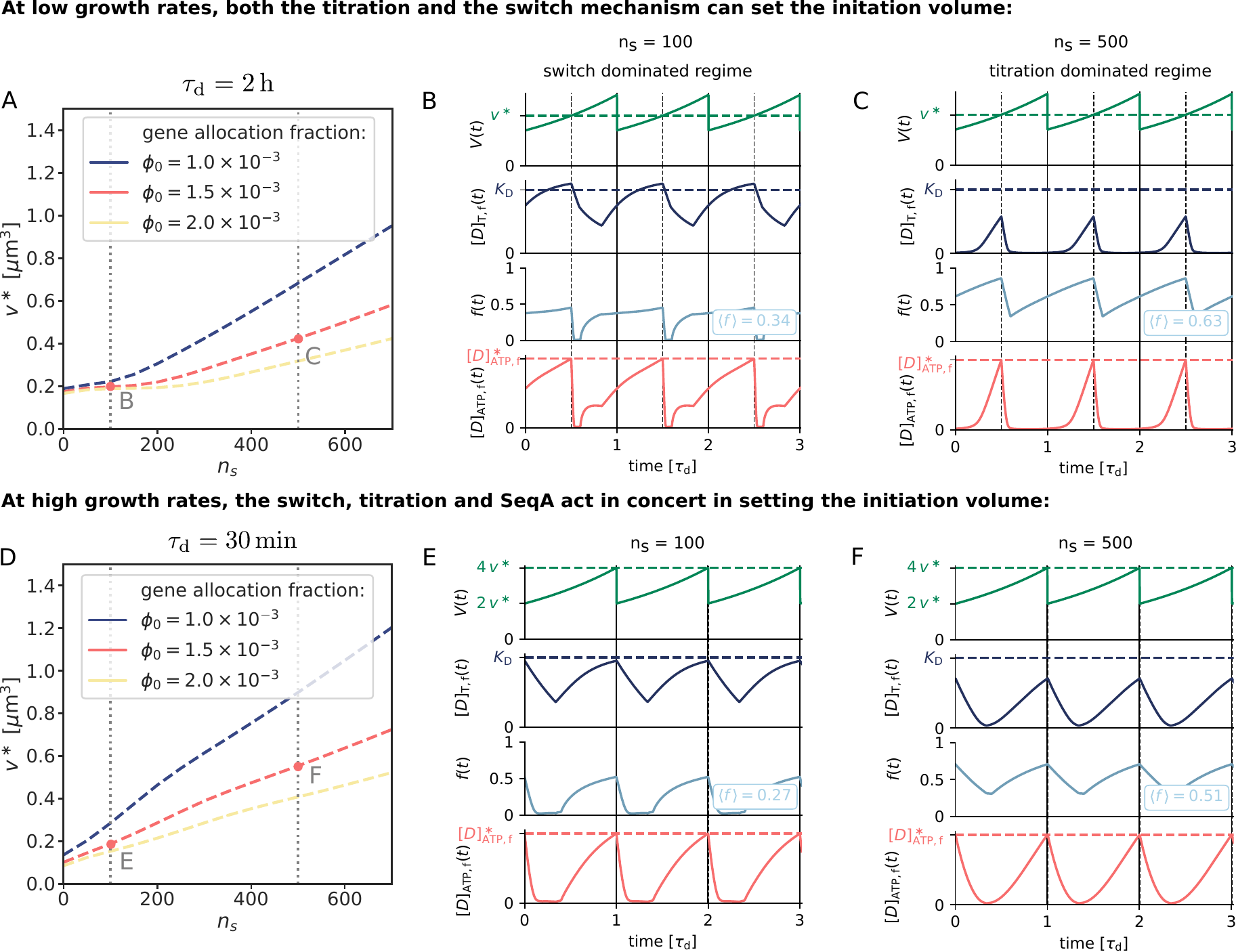}
	\caption{{\footnotesize \textbf{While at low growth rates both titration and DnaA activation can set the initiation volume, at high growth rates the switch, titration and SeqA act in concert in regulating replication initiation} (B, C, E, F) The volume of the cell $V(t)$, the free DnaA concentration $[D]_{\rm T, f}$, the fraction of DnaA $f(t)$ that is bound to ATP (irrespective of whether the DnaA is in the cytoplasm or on the titration sites) and the concentration of free ATP-DnaA $[D]_{\rm ATP, f}(t)$ as a function of time (in units of the doubling time $\tau_{\rm d}= 2$~h (B, C) and $\tau_{\rm d}= 30$~min (E, F)). The blue dashed line is the dissociation constant $K_{\rm D}$ of the DnaA to its promoter. The average active fraction over one cell cycle $\langle f \rangle$ is indicated in light blue. The dashed red line is the critical free ATP-DnaA concentration $[D]_{\rm ATP, f}^\ast$ at which replication is initiated. Replication is initiated at a constant volume per origin $v^\ast$ over time (green dashed line). The details of the simulations are described in section \ref{sec:switch_titration_combined}.
			(A, D) The initiation volume $v^\ast$ as a function of the number of titration sites per chromosome $n_{\rm s}$ for different gene allocation fractions $\phi_0$. (A) At low growth rates, the initiation volume exhibits two regimes: When the number of titration sites is small compared to the total concentration the initiation volume is constant and equal to the initiation volume set by the LDDR model in the absence of titration sites. At higher numbers of titration sites the initiation volume increases linearly with the number of titration sites as predicted by the AIT model (equation \ref{eq:v_init_titration_ait_model}). The onset of the titration-dominated regime is shifted to higher numbers of titration sites per chromosome with increasing gene allocation fraction. As the blocked period is short compared to the doubling time of the cell ($\tau_{\rm b} = 10$~min $\ll \tau_{\rm d}=2$~h), SeqA only plays a minor role at low growth rates. (B, C) Two time traces of the parameter regimes indicated by red dots in figure (A) are shown. At low growth rates, the time $T_{\rm C}$ to replicate the entire DNA is relatively short compared to the doubling time of the cell ($T_{\rm C}=40$~min $< \tau_{\rm d}=2$~h). Right after replication initiation, new titration sites are synthesized faster than new DnaA proteins and the free concentration of DnaA drops. At a small number of titration sites per origin $n_{\rm s}$ ($n_{\rm s}=100$) the free concentration however only drops weakly and quickly recovers a constant total concentration set by the dissociation constant $K_{\rm D}^{\rm p}$ of the DnaA promoter (second panel in B). In this regime, the switch dominates the oscillations in the free, active DnaA concentration (lowest panel in B). When the number of titration sites is however very high ($n_{\rm s}=500$), the free DnaA concentration is low during most of the cell cycle (second panel in C). When all titration sites are filled does the free concentration begin to rise. As the active fraction is already high at this large volume (third panel in C), the rise in the free concentration dominates the timing of replication initiation in this regime. (D) At a higher growth rate ($\tau_{\rm d}=30$~min), the initiation volume increases strongly with the number of titration sites for all basal rates and the two regimes (switch vs. titration-dominated) that we found at low growth rates have disappeared. Contrary to the low growth rate regime, the blocked-synthesis period takes up a significant fraction of the cell cycle and the blocked period, the switch and the titration sites act together in setting the initiation volume. (E, F) Two time traces of the parameter regimes indicated by red dots in figure (D) are shown. At high growth rates, blocking DnaA synthesis for 10 minutes in combination with synthesizing new titration sites causes a drop in the free concentration after replication initiation. For a small number of titration sites ($n_{\rm s}=100$) the drop in the free concentration is relatively small and the oscillation in the free, active concentration of DnaA (lowest panel in E) are still dominated by the switch. For large numbers of titration sites, 
			the oscillations in the free, active fraction are shaped both by titration in combination with blocked production and the switch (lowest panel in F). Importantly, as titration together with SeqA gives rise to reinitiation events at this growth rate (Fig. 5C, region of non-zero coefficient of variation), the switch is essential for ensuring stable cycles in this regime.
	}}
	\label{fig:SI_switch_titration_combined}
\end{figure}

We first consider the low growth rate regime. In the absence of titration sites, the free DnaA concentration is set by the dissociation constant of the DnaA promoter $K_{\rm D}^{\rm p}$ due to negative autoregulation. In this switch-only scenario, replication is initiated at a nearly constant, critical ATP-DnaA fraction
$f^\ast = [D]_{\rm ATP, f}^\ast/[D]_{\rm T, f}$ because the total free concentration is nearly constant. When we include a finite but small number of titration sites, the production of new titration sites after replication initiation transiently lowers the free DnaA concentration at low growth rates (Fig. \ref{fig:SI_switch_titration_combined} B, panel two). 
As the number of titration sites per chromosome is however small, they are filled up quickly (especially when the basal production rate as set by $\phi_0$ is large) and the free concentration rises quickly until it is again constant (Fig. \ref{fig:SI_switch_titration_combined} B, panel two). As in this regime the free concentration of DnaA is essentially constant in time before the active fraction rises at a volume as set by the balance between the activation and deactivation rates, the change in the free ATP-DnaA is dominated by the change in the ATP-DnaA fraction and the critical volume per origin $v^\ast$ is mainly set by the switch (Fig. \ref{fig:SI_switch_titration_combined} B, panel four). In this scenario, the titration sites play a supporting role in preventing premature reinitiation by lowering the free concentration after replication initiation, but they do not set the initiation volume. Indeed, at low numbers of titration sites, the initiation volume remains constant as a function of the number of titration sites and equal to the initiation volume in the absence of titration sites (Fig. \ref{fig:SI_switch_titration_combined} A). However, by increasing the number of titration sites while keeping the basal synthesis rate as set by $\phi_0$ constant, the time to fill up the larger number of titration sites per origin increases. Now, the free DnaA concentration remains low for a longer time (Fig. \ref{fig:SI_switch_titration_combined} C). When the cell reaches the critical volume at which the active fraction as set by the switch rises, the free concentration is still very low and prevents replication initiation. Only when the titration sites fill up at a larger volume does the free DnaA concentration rise and is replication initiated. As expected for the AIT model, in this regime the initiation volume increases linearly with the number of titration sites (compare prediction of equation \ref{eq:v_init_titration_ait_model} to Fig. \ref{fig:SI_switch_titration_combined} A): the initiation volume is now dominated by titration. For a fixed number of titration sites, the system can be brought into the titration-dominated regime by decreasing the basal synthesis rate. Conversely, by increasing the basal synthesis rate, the fixed number of titration sites can be filled up more rapidly and the system is driven into the switch-dominated regime, where the initiation volume is  dominated by the rise in the active fraction rather than the rise in the free concentration.

At high grow rates, the replication forks overlap and the time to replicate all titration sites is longer than the inter-initiation time set by the doubling time of the cell. 
Upon varying the total number of titration sites per origin, we therefore do not observe the same transition as at low growth rates from a switch to a titration-dominated regime (Fig. \ref{fig:SI_switch_titration_combined} D). Instead, the initiation volume rises with the number of titration sites per origin at all basal synthesis rates. As the blocked period by SeqA makes up 1/3 of the cell cycle in this regime, it causes a significant drop in the free DnaA concentration that increases with increasing number of titration sites. This leads to an increasing initiation volume for higher numbers of titration sites. The initiation volume is thus in this regime set by a combined effect of the switch, the titration sites and the relatively long blocked period. 
It is however important to note that the switch is essential for the stability, as in this regime removing the switch results in reinitiation events as shown in Fig. 2C of the main text.

\textbf{To summarize,} an activation switch is able and hence sufficient to generate stable DnaA oscillations at all growth rates, but titration helps to raise the amplitude of these oscillations. It can do so at all growth rates but most predominantly at low growth rates. Indeed, at low growth rates a titration-based mechanism is sufficient but in the regime of overlapping forks rates it needs another mechanism. At sufficiently high growth rates 
($\lambda > 1.4$~h$^{-1}$), SeqA based repression of DnaA synthesis can play this role, but in the crossover regime ($1.0 < \lambda < 1.4$~h$^{-1}$) this no longer suffices. In this regime, the switch, according to our model, becomes essential.

\textbf{What is the biologically relevant regime?} The number of titration sites has been estimated to be around 300, and Figure \ref{fig:SI_full_model_time_traces} G/H/I shows that in this regime, the switch and titration act in concert to generate DnaA oscillations. This is perhaps not too surprising because in this regime these mechanisms act synergistically to raise the amplitude of the oscillations in the concentration of free, ATP-bound DnaA. However, we emphasize that at this stage we regard this as a model prediction, which can be tested experimentally by varying the number of titration sites: as Figure \ref{fig:SI_switch_titration_combined} shows, this makes it possible to drive the system from a titration-based regime to a switch-based one. Interestingly, very recent experiments do support the prediction that the system combines titration with activation because removing either mechanism only still yields stable cell cycles \cite{Knoppel2021}. Moreover, as we show in the next section, our full model agrees, without any additional fitting, quantitatively with experiments in which the effective number of titration sites (\ref{sec:validation_plasmids_oriC}) or the total DnaA concentration (\ref{sec:validation_DnaA_total_variation}) was increased, further supporting the idea that the system combines titration with activation to drive robust replication cycles.

\subsection{Model validation}
\label{sec:model_validation}
In this section, we compare several key predictions from our  full switch-titration-SeqA model against experimental data. First, we discuss several mutants in which activators and deactivators of the DnaA activation switch have been deleted or modified (section \ref{sec:switch_validation_mutations}). In section \ref{sec:validation_plasmids_oriC}, we then show that our model is in quantitative agreement with experiments in which the number of titration sites in the cell was varied via oriC-plasmids. We next demonstrate that our full model can reproduce experiments in which the total DnaA concentration was varied (section \ref{sec:validation_DnaA_total_variation}). Then we show (in section \ref{sec:validation_oscillations}) that we can reproduce the observation of Si et al. \cite{Si2019} that externally driven oscillations in the DnaA concentration can transform an initiation adder into an initiation sizer. Finally, we show that the results of our model on the correlations in the inter-initiation volume are robust to different types of coupling between the replication and the division cycle (section \ref{sec:division_separate}).

\begin{figure}
	\centering
	\includegraphics[width =0.77\textwidth]{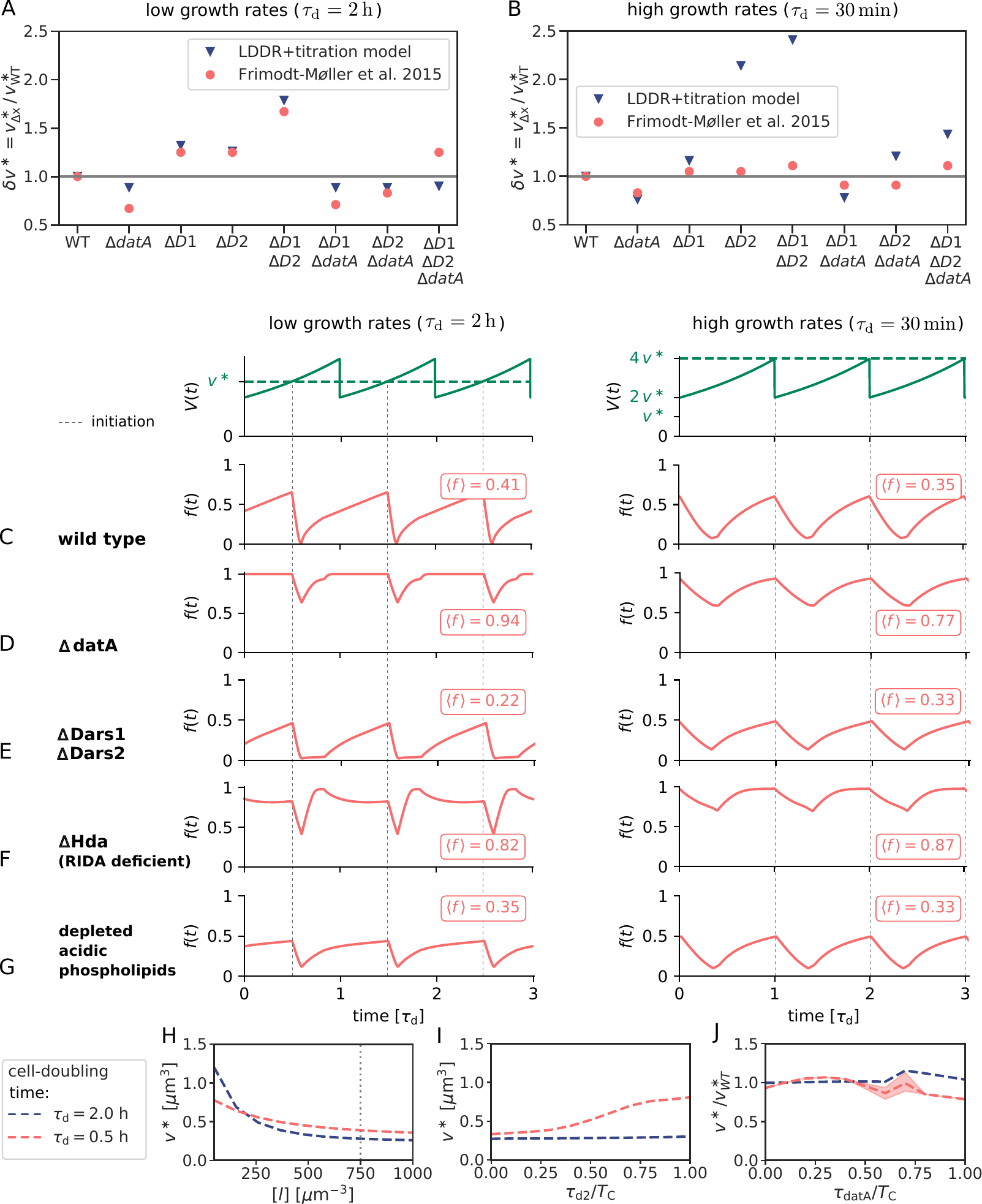}
	\caption{{\footnotesize \textbf{The effects of deleting different combinations of the switch (de)activators in the simulations on the initiation volume and the time traces of the active fraction are very similar to experiments, both at high and at low growth rates.} All results in this figure are obtained with the full model containing titration, LDDR switch and SeqA as explained in section \ref{sec:switch_titration_combined}. (A, B) For various mutants $x$ with initiation volume $v^\ast_{\rm \Delta x}$ we show the relative change of the initiation volume $\delta v^\ast= v^\ast_{\rm \Delta x}/v^\ast_{\rm WT}$ with respect to the initiation volume of the wild type $v^\ast_{\rm WT}$ obtained from the simulations of the switch-titration model (as explained in section \ref{sec:switch_titration_combined}) with a blocked production for $\tau_{\rm b}=10$~min (blue triangles) and compare it to the experimentally determined relative change in the average volume per number of origin obtained experimentally by Frimodt-M{\o}ller et al. \cite{Frimodt-Moller2015} (red circles); as shown in the SI of Si et al. \cite{Si2017}, there is a direct mapping between the average initiation volume per origin and the average volume per average number of origin. (A) At low growth rates, the effects of deleting different combinations of activators and deactivators on the change in the initiation volume agree very well with the experimentally observed relative change in the average volume per average number of origin. (B) Also at high growth rates, the simulations agree well with experimental observations. (C, D, E, F, G) The volume of the cell $V(t)$ and the active fraction $f(t)$ as a function of time (in units of the doubling time $\tau_{\rm d}$) for low and high growth rates. Replication is initiated at a critical free ATP-DnaA concentration $[D]_{\rm ATP, f}^\ast$ (see section \ref{sec:switch_titration_combined}) and the system gives rise to a constant initiation volume per origin $v^\ast$ over time (green dashed line). For brevity and as it is a quantity that is typically measured in experiments, we here only show the fraction of active, ATP-bound DnaA, and not the concentration. The average active fraction over one cell cycle $\langle f \rangle$ is indicated in red. Deleting DnaA activators (\textit{DARS1/2} and lipids) tends to increase the average active fraction, while deleting DnaA deactivators (\textit{datA} and Hda) tends to lower the active fraction, as observed experimentally. (H) Depleting the lipid concentration $[l]$ leads to an increase in the initiation volume $v^\ast$ with a stronger effect at low growth rates than at high growth rates. (I) When varying the relative position of the site \textit{DARS2} on the chromosome (where 0 is at the origin and 1 at the terminus), the initiation volume increases when \textit{DARS2} is moved towards the terminus at high growth rates but not at low growth rates. (J) Moving the site \textit{datA} on the chromosome towards the terminus leads to a weak relative increase of the initiation volume per origin $v^\ast$ compared to the wild type initiation volume $v^\ast_{\rm WT}$ at low growth rates and to a relative decrease at high growth rates. The red shaded area indicates the maximal and minimal initiation volume per origin occuring at a given \textit{datA} position at high growth rates. }}
	\label{fig:S18_mutations}
\end{figure}

\subsubsection{Effect of mutations in the activators and deactivators of DnaA}
\label{sec:switch_validation_mutations}
In this section, we discuss the experimentally reported effect of mutations in the different (de)activators of DnaA and we compare them to the predictions of our model. We use the full model that combines the LDDR model, with titration sites, and a blocked period via SeqA during which both DnaA synthesis and replication initiation are blocked (see section \ref{sec:switch_titration_combined}).
\\
\\
\textbf{Effect of mutations in the chromosomal deactivation site \textit{datA}}
The important role of the chromosomal site \textit{datA} in the regulation of replication initiation has been described in many studies \cite{Kitagawa1996, Kitagawa1998, Ogawa2002, Nozaki2009, Kasho2013, Kasho2014, Flatten2015, Frimodt-Moller2016}.
Deletion of \textit{datA} increases the cellular levels of ATP-DnaA as compared to wild-type cells by 5-10\% \cite{Kitagawa1998, Katayama2001, Kasho2013}. While the size and morphology of these mutant cells were indistinguishable from wild-type cells \cite{Kitagawa1998, Flatten2015}, replication is initiated untimely, resulting in a broad distribution of number of chromosomes per cell \cite{Kitagawa1998}. The average number of chromosomes per cell mass increases in a growth rate-dependent manner \cite{Kitagawa1998}. When \textit{datA} is removed, the initiation volume per origin becomes smaller at low growth rates, while at high growth rates there is almost no effect of deleting \textit{datA} \cite{Flatten2015}. These findings indicate that \textit{datA} is important to prevent over-initiation and asynchronous initiation \cite{Kitagawa1998}. Originally, it was believed that \textit{datA} influences replication initiation negatively by titrating large amounts of DnaA to the chromosome \cite{Kitagawa1996, Kitagawa1998}. More recently it was however demonstrated that \textit{datA} can hydrolyze ATP-DnaA and the strong effect of deleting this chromosomal region was attributed to the resulting higher concentration of ATP-DnaA in the cell \cite{Kasho2013}. Our switch-titration-SeqA model can qualitatively reproduce these results: deleting \textit{datA} results in a lower initiation volume both at high and at low growth rates, although at low growth rates not as pronounced as in the experiments (Fig. \ref{fig:S18_mutations} A and B). The time traces of the mutant show that the effect of deleting \textit{datA} on the active fraction of DnaA is especially severe at low growth rates, as RIDA is only active for a small fraction of the cell cycle in this regime (Fig. \ref{fig:S18_mutations} D, left panel). At high growth rates, the active fraction still exhibits regular oscillations, however at a higher average active fraction (Fig. \ref{fig:S18_mutations} D, right panel). This finding agrees well with the experimental observation that deleting \textit{datA} raises the active DnaA fraction and that at high growth rates cells without \textit{datA} were shown to still exhibit regular temporal oscillations in the active fraction \cite{Katayama2001}.

It has been found experimentally that the chromosomal loci \textit{DARS1}, \textit{DARS2} and \textit{datA} are conserved among several \textit{E. coli} strains \cite{Frimodt-Moller2015}. This finding suggests that their positions on the chromosome play an important role in the regulation of replication initiation. Focusing on {\it datA}, it has been observed that moving {\it datA} towards the terminus can have two effects: 1) change the initiation volume and 2) cause premature reinitiation. Concerning the first observation, it has been observed that placing {\it datA} near the terminus at a cell doubling time of $\tau_{\rm d}\approx 40$~min decreases the initiation volume per origin $v^\ast$ \cite{Frimodt-Moller2016}. Also our model predicts that at high growth rates the initiation volume decreases when {\it datA} is moved towards the terminus, see Fig. \ref{fig:S18_mutations}~J. This finding can be understood via the change in the effective gene dosage \cite{Frimodt-Moller2016}: translocating \textit{datA} to the terminus lowers the effective gene dosage of \textit{datA} and thereby reduces the effective deactivation rate, leading to a lower initiation volume per origin \cite{Frimodt-Moller2016}. This effect is especially strong at high growth rates, as in this regime the doubling time is shorter than the time to replicate the entire chromosome: changing the position of {\it datA} has then a large effect on the average {\it datA} copy number. In contrast, at lower growth rates, it takes longer before replication is initiated during the cell cycle, and the doubling of the copy number thus happens later; moving the position will thus have a smaller effect on the effective copy number. Indeed, our model predicts that at low growth rates the initiation volume does not decrease; in fact, it increases, albeit weakly (Fig. \ref{fig:S18_mutations}J). This is because of a second, {\em spatio-temporal} effect (rather than a change in the {\em average} copy number), as revealed by our model: moving {\it datA} towards the terminus means that {\it datA} will be doubled {\it later} in the cell cycle and therefore also the activity of \textit{datA} will increase later in the cell cycle. This means that, if a new round of replication has not yet been initiated, the active DnaA concentration will reach the initiation threshold later in the cell cycle, {\em increasing} the initiation volume. There are thus two competing effects, and which one dominates depends on the growth rate. To our knowledge, it has not been measured how the initiation volume changes when {\it datA} is translocated towards the terminus at low growth rates; we thus regard this as a novel prediction from our model. Concerning the second observation, the premature reinitiation events: experiments have shown that moving {\it datA} towards the terminus leads to premature reinitiation at high growth rates (i.e., for $\tau_{\rm d}\approx 28$~min \cite{Kitagawa1998}), but not at low growth rates (i.e., not for $\tau_{\rm d}\approx 40$~min\cite{Frimodt-Moller2016} and $\tau_{\rm d}\approx 50$~min \cite{Kitagawa1998}). Our model can qualitatively reproduce these observations (see Fig. \ref{fig:S18_mutations}~J). According to our model, these premature reinitiations are due to the second, spatio-temporal, effect associated with moving {\it datA} towards the terminus: this perturbs the shape of the temporal oscillations as generated by \textit{datA}, \textit{DARS2} and RIDA, such that the active fraction first decreases after initiation due to a high activity via RIDA, then rises due to an increase in \textit{DARS2} activity and then decreases {\em again} due to a high \textit{datA} activity before the active free concentration rises and reaches the initiation threshold. At higher growth rates, these ``double'' oscillations can lead to premature reinitiation events (Fig. \ref{fig:S18_mutations}~J, shaded area). In contrast, at low growth rates, the effect of titration on shaping the oscillations of the active free concentration is much stronger, and this can protect the system from premature reinitiation (Fig. \ref{fig:S18_mutations} J, blue line). Our model thus predicts that the effects of moving the position of {\it datA} are highly non-trivial and growth-rate dependent; to test our predictions, measurements of time traces of the active fraction, albeit experimentally difficult, would be ideal.
\\
\\
\textbf{Effect of mutations in the chromosomal activation sites \textit{DARS1} and \textit{DARS2}}
The two chromosomal sequences \textit{DARS1} and \textit{DARS2} can regenerate ATP-DnaA from ADP-DnaA via nucleotide exchange, resulting in replication initiation both \textit{in vitro} and \textit{in vivo} \cite{Fujimitsu2009, Kasho2014}. Introducing extra copies of \textit{DARS1} and \textit{DARS2} increases the overall ATP-DnaA level and leads to a decrease of the average volume per origin \cite{Fujimitsu2009}. Deletion of either \textit{DARS1} or \textit{DARS2} reduces the ATP-DnaA concentration in the cell by 14\% and 30\% \cite{Fujimitsu2009}, respectively, and leads to delayed and in some cases asynchronous initiation as compared to wild type cells \cite{Fujimitsu2009, Kasho2014, Frimodt-Moller2015, Riber2016}. These findings show the important role that \textit{DARS1} and \textit{DARS2} play in replication initiation.  While \textit{DARS1} is not known to require an additional protein to activate DnaA, the more potent site \textit{DARS2} requires the proteins IHF and Fis for its functioning \cite{Kasho2014}. Experiments indicate that Fis is more abundant in cells at high growth rates \cite{Flatten2013, Kasho2014} and the effect of deleting Fis leads to severe over-initiation at high growth rates but had almost no effect at low growth rates \cite{Flatten2013}. Combining this observation with the finding that \textit{DARS2} requires Fis for re-activating DnaA, \textit{DARS2} seems to play a more important role at high than at low growth rates. Our model qualitatively reproduces these experimental observations: Both at high and at low growth rates deletion of \textit{DARS1}, \textit{DARS2} or both lead to an increase in the initiation volume as observed experimentally (Fig. \ref{fig:S18_mutations} A and B). We also note that the simulations overestimate this effect at higher growth rates, which may indicate that the activation rate of DARS1/2 depends on the growth rate, because, e.g., of the growth-rate dependence of the additional regulators like IHF and Fis. Nonetheless, deleting {\it DARS1/2} reduces the amplitude of the oscillations, especially at high growth rates, in line with the observation that {\it DARS2} is particularly important at higher growth rates (Fig. \ref{fig:S18_mutations} E, right panel).

Intriguingly, similarly to \textit{datA}, the relative position of \textit{DARS1/2} with respect to the origin and the terminus is conserved in various genomes of different sizes and strains \cite{Frimodt-Moller2015}. This suggests that also the position of the site \textit{DARS2} on the chromosome plays a role for the regulation of replication initiation \cite{Frimodt-Moller2015, Frimodt-Moller2016, Kitagawa1998}. Indeed, experiments show that relocating the sites \textit{DARS1} and \textit{DARS2} to different positions on the chromosome affects the replication cycle \cite{Frimodt-Moller2016, Inoue2016}. Translocating \textit{DARS2} to the terminus had the strongest effect and resulted in asynchronous initiation and a lower origin concentration as compared to the wild type. Interestingly, the degree to which the replication cycle was affected by translocating \textit{DARS2} was growth-rate dependent: At high growth rates e.g. for cells growing in LB/glucose medium at 42 $^\circ$C ($\tau_{\rm d}=18-10$~min), the origin/mass ratio was 87\% of the wild type, almost as low as for cells lacking \textit{DARS2} entirely (with origin/mass ratio of 82\% of the wild type) \cite{Inoue2016}. At low growth rates however, e.g. for cells grown in M9/glucose/caa medium at 30 $^\circ$C ($\tau_{\rm d}=71-77$~min), translocating \textit{DARS2} to the terminus only led to a slight decrease of the origin to mass ratio (with origin/mass ratio of 95\% of the wild type) \cite{Inoue2016}. Our model provides a novel explanation for these experimental observations: Within our LDDR model, \textit{DARS2} plays the important role of compensating the strong deactivator RIDA. As described in the main text, putting {\it DARS2} near the origin would immediately counteract the effect of RIDA after a new round of replication has been initiated; while this would keep the initiation volume constant as a function of the growth rate, it would also nullify the effect of RIDA on raising the amplitude of the oscillations. On the other hand, to keep the initiation volume nearly independent of the growth rate \cite{Zheng2020}, the activity of {\it DARS2} must be high to counteract RIDA before a new round is started; putting {\it DARS2} too close to the terminus would make this impossible at sufficiently high growth rates, such that then the initiation volume goes up. This is indeed precisely what our simulations show: shifting the position of \textit{DARS2} towards the terminus increases the initiation volume (Fig. \ref{fig:S18_mutations} I) at high growth rates, but not at low growth rates. Our model, therefore, can not only reproduce the observation that moving {\it DARS2} towards the terminus increases the initiation volume, but also that this effect is stronger for higher growth rates.
\\
\\
\textbf{Effect of mutations in the deactivation mechanism RIDA}
Regulatory Inactivation of DnaA (RIDA) is a mechanism promoting ATP hydrolysis in a replication coupled manner by the formation of a complex with the protein Hda and the DNA-loaded clamp \cite{Katayama2010, Kurokawa1999, Kato2001}.
It is the predominant mechanism by which reinitiation events in \textit{E. coli} are prevented \cite{Camara2005, Katayama2010}. RIDA inactivation via the deletion of the Hda gene or inactivation of the clamp increases the cellular ATP-DnaA level to 70–80\% of the total number of DnaA molecules \cite{Kurokawa1999, Kato2001}. RIDA deficient cells initiate replication asynchronously and earlier than wild type cells at smaller initiation volume \cite{Kitagawa1998, Fujimitsu2008, Kasho2013, Riber2016}. In the simulations we cannot address the effect of asynchronous initiation of replication, because in our mean-field model, by construction, all origins are fired simultaneously when the critical free ATP-DnaA concentration is attained. In combination with the titration mechanism, we obtain stable cell cycles even though the time traces of the active fraction are strongly disturbed. At low growth rates, the active fraction first decreases after replication initiation due to the duplication of \textit{datA} and the blocked protein synthesis, but then rises quickly to its maximal value due to the strong activation via the sites \textit{DARS1} and \textit{DARS2} (Fig. \ref{fig:S18_mutations} F, left panel). Also at high growth rates, the active fraction is very high throughout most of the cell cycle (Fig. \ref{fig:S18_mutations} E, right panel). Both observations agree well with the experimentally reported strong increase of the ATP-DnaA level in RIDA deficient cells.
\\
\\
\textbf{Effect of mutations in the acidic phospholipids}
In section \ref{sec:role_of_lipids} we discuss the experimental evidence for the role of the lipids in DnaA reactivation. Here, we discuss the predictions from our full model.
It predicts that depleting acidic lipids weakens DnaA activation, which increases the initiation volume per origin, see Fig. \ref{fig:S_switch_adder_total_conc} and Fig. \ref{fig:S18_mutations} H. Our model also predicts that this effect is stronger at lower growth rates (Fig. \ref{fig:S18_mutations} H). From the time-traces of the full model in Figure \ref{fig:S18_mutations} G we furthermore predict that in cells with depleted acidic phospholipids the average fraction of ATP-DnaA is reduced and the amplitude of the oscillations in the active fraction decreases.  These are clear predictions that could be tested experimentally, using mutants in which the {\it pgsA} gene is brought under the control of an inducer \cite{Xia1995, Fingland2012}. The specific predictions from our model in combination with experiments should make it possible to elucidate the contested role of the acidic lipids in replication initiation, and may clarify the apparent contradictions between the studies of Refs. \cite{Shiba2004,Shiba2012,Camsund2020} and those of Refs. \cite{Saxena2013, Zheng2001, Fingland2012} (see also section \ref{sec:role_lipids_experiments}.
\subsubsection{Effect of varying the number of titration sites per origin on the initiation volume per origin}
\label{sec:validation_plasmids_oriC}
Our full model predicts that varying the number of titration sites per origin affects the initiation volume and that the degree by which the initiation volume changes depends on the relative change in the number of titration sites per origin, the growth-rate regime, and whether the system is dominated by the switch, titration or SeqA (Fig. \ref{fig:SI_switch_titration_combined} and section \ref{sec:switch_titration_combined_parameter_regimes}). In the following, we compare this prediction to experiments. Christensen et al. \cite{Christensen1999} showed that introducing different multicopy plasmids that contain different variants of the oriC region changes the initiation volume. While oriC does not affect the switch, because it is neither an activator nor a deactivator of DnaA, it does contain several high and low-affinity binding sites for DnaA and as such could affect the titration mechanism. Christensen et al. observed that for all growth rates studied, plasmids with a higher number of DnaA boxes cause a larger decrease in the chromosomal origin concentration and hence a larger increase in the initiation volume per origin \cite{Christensen1999}. This finding is consistent with the predictions from our full model: the initiation volume increases  with increasing number of titration sites per origin {(Fig. \ref{fig:SI_switch_titration_combined})}. Remarkably, the agreement is not only qualitative, but even nearly quantitative, without any additional fitting. Specifically, the  degree to which the initiation volume changes depends on the relative change in the number of titration sites per origin and the growth rate. The strain carrying plasmid pFHC946, which contained all of the DnaA boxes from the {\it oriC} region, showed about a 10\% increase in the initiation volume per origin at a high growth rate, and a nearly 20\% change in the initiation volume per origin at a low growth rate \cite{Christensen1999}. To compare this observation to the predictions from our model, we note that at the high growth rate, the number of pFHC946 plasmids per origin is about 13 while at the low growth rate it is about 57 (Table 1 of Ref. \cite{Christensen1999})). To estimate the total number of DnaA titration sites that these plasmid copies carry, we consider that plasmid pFHC946, like the chromosomal oriC region, contains two high-affinity DnaA binding sites (R1 and R4) and one intermediate affinity site R2, resulting in 2-3 medium-high affinity titration sites per plasmid copy. The total number of titration sites on the plasmid copies combined is thus $13 \times (2-3) \approx 25 - 40$ at high growth rates and $57 \times (2-3) \approx 100 - 150$ at low growth rates. Our model predicts that at high growth rates (Fig. \ref{fig:SI_switch_titration_combined}D, red line), a change in the number of titration sites from $n_{\rm s} = 300$, corresponding to wild-type cells with no extra plasmids, to $n_{\rm s} \approx 350$, corresponding to the pFHC946 strain, causes a relative change in the initiation volume of about $10\%$, in quantitative agreement with the experiments of Christensen et al. At low growth rates, our model predicts (Fig. \ref{fig:SI_switch_titration_combined} A, red line) that a change in the number of titration sites, from $300$ in wild-type to $400 - 450$ in the pFHC946 cells, cause a change in the initiation volume of about $22 - 35\%$, in near quantitative agreement with the reported change of about $20\%$.

\subsubsection{Effect of varying the total DnaA concentration on the initiation volume per origin}
\label{sec:validation_DnaA_total_variation}
In this section, we show that the effect of varying the total DnaA concentration on the initiation volume per origin in our full model is in good agreement with experimental observations. The total DnaA concentration in \textit{E. coli} cells can be varied by introducing plasmids containing inducible {\it dnaA} promoters into cells \cite{Atlung1993, Hill2012}, by replacing the native {\it dnaA} promoter with an inducible promoter \cite{Knoppel2021} or by repressing the native {\it dnaA} promoter further using tunable CRISPR interference \cite{Si2017}. All of these experiments reported a negative dependence of the initiation volume on the total DnaA concentration \cite{Atlung1993, Hill2012, Knoppel2021, Si2017}. Furthermore, Zheng et al. measured the average DnaA concentration and the initiation mass per origin at different growth rates and also reported a negative correlation between these two variables \cite{Zheng2020}. In a recent paper by Flatten et al. \cite{Flatten2015} it was however reported that upon increasing the total DnaA concentration two-fold, the average volume per number of origins decreased only very weakly \cite{Flatten2015}. Upon increasing the total DnaA concentration even further (up to 35 times the wild type concentration), also Flatten et al. observed a decrease in the volume per number of origins in combination with excessive and asynchronous initiations \cite{Flatten2015}. Finally, Si et al. \cite{Si2019} found that oscillatory perturbation of the total DnaA concentration affected the initiation volume per origin and that the initiation volume was anti-correlated with the total DnaA concentration. These results clearly show that DnaA synthesis plays a non-negligible role in regulating the initiation volume and that in all experiments, the initiation volume per origin is negatively correlated with the total DnaA concentration.

\begin{figure}
	\centering
	\includegraphics[width =0.4\textwidth]{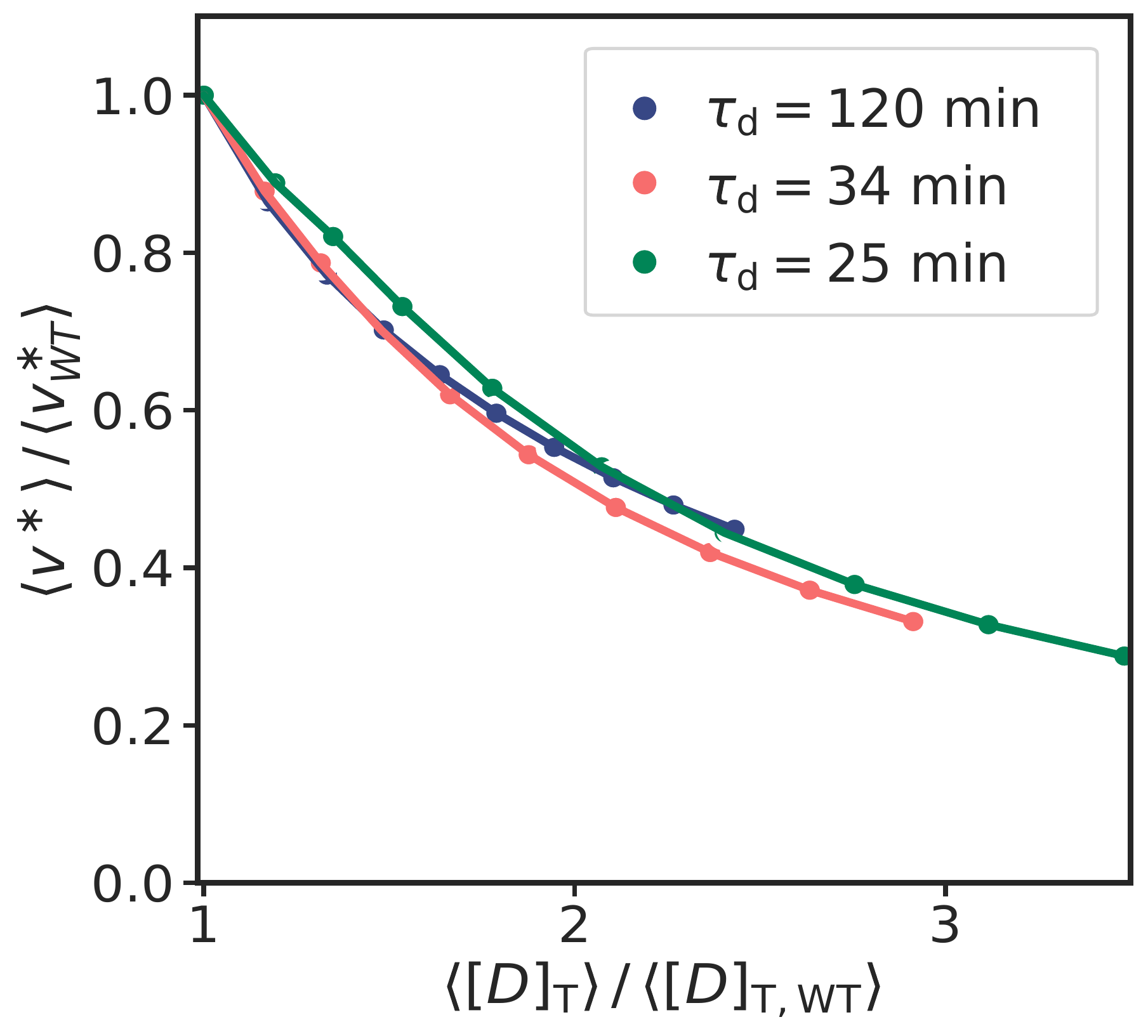}
	\caption{\textbf{The full model can reproduce the experimental observation that the initiation volume per origin is negatively correlated with the total DnaA concentration.} The average initiation volume per origin $\langle v^\ast \rangle$ as a function of the average total concentration $\langle [D]_{\rm T} \rangle$ for three different cell doubling times $\tau_{\rm d}$. The total DnaA concentration is varied by adding a constant DnaA production term according to equation \ref{eq:DnaA_over_expression_constant} to the full model (as described in sections \ref{sec:implementation_switch_titration_combined} and \ref{sec:full_model_large_sharp_osc}) and varying the external gene allocation fraction $\phi_0^{\rm ext} = 0-2.25 \times 10^{-3}$. Both the average initiation volume per origin and the average total concentration have been normalized by their value in the absence of external DnaA expression, $\langle v^\ast_{\rm WT} \rangle$ and $\langle [D]_{\rm T, WT} \rangle$, respectively. In line with various experiments \cite{Atlung1993, Hill2012, Si2017, Hansen2018, Zheng2020, Knoppel2021}, the average initiation volume per number of origins decreases with increasing average total DnaA concentration. See Table \ref{tab:LD_LDDR_parameters} for parameters. }
	\label{fig:S17_vary_total_conc}
\end{figure}

To test the effect of a varying the total DnaA concentration in our full model, we mimick the plasmid experiments of Hill et al. \cite{Hill2012} and of Atlung et al. \cite{Atlung1993} by adding an external production term of DnaA to our full Switch-titration-SeqA model (as described in sections \ref{sec:implementation_switch_titration_combined} and \ref{sec:full_model_large_sharp_osc})
	\begin{equation}
	\frac{dN^{\rm ext}_{\rm D,ATP}}{dt} = \phi_0^{\rm ext} \, \lambda \, \rho \, V
	\label{eq:DnaA_over_expression_constant}
	\end{equation}
	with the external gene allocation fraction $\phi_0^{\rm ext}$, the number density $\rho$ and the volume of the cell $V$. Increasing the external gene allocation fraction $\phi_0^{\rm ext}$ leads to an increase in the average DnaA concentration in the cell. We find that in the full model the initiation volume decreases with increasing total concentration for a wide range of growth rates, in excellent agreement with experiments (see Fig. \ref{fig:S17_vary_total_conc}). The initiation volume per origin is approximately inversely proportional to the total DnaA concentration, such that for a doubling of the DnaA concentration the initiation volume is approximately halved. This is consistent with the finding of Zheng et al. who observe that for a decrease of the total concentration of about 20\%, the average initiation mass per number of origins increases by about 20\% (see Extended Data Fig. 5 of \cite{Zheng2020}). Also Atlung and Hansen \cite{Atlung1993} find an almost linear increase in the number of origins per mass (the inverse of $v^\ast$) with increasing total DnaA concentration up to at least two times of the wild type DnaA total concentration; at higher total concentrations, the average number of origins per mass reached a plateau, but this could be explained by a reduced replication speed due to too high replication initiation rates \cite{Atlung1993, Hansen2018}. Therefore, our simulations are in good quantitative agreement with several experimental findings.
\subsubsection{Externally driven oscillation in the DnaA concentration can turn an initiation-adder into an initiation-sizer}
\label{sec:validation_oscillations}
In this section, we test whether our full switch-titration-SeqA model can explain the experimental result by Si et al. \cite{Si2017} that dynamically perturbing the DnaA concentration can turn an initiation-adder into an initiation-sizer. In section \ref{sec:switch_noise} we found that in the LDDR model noise in a negatively autoregulated DnaA production rate gives rise to sizer correlations in the initiation volume, while noise in the (de)activators of DnaA gives rise to adder correlations. In the initiator titration model, noise in the initiator protein gives however rise to adder correlations (Fig. \ref{fig:S4_AIT_adder} B). These findings open two questions: First, what correlations in the initiation volume arise from DnaA gene expression noise when both the switch and the titration sites contribute to setting the initiation volume? We demonstrate that while in the titration-dominated regime (e.g. for high numbers of titration sites or low DnaA synthesis rate, see Fig. \ref{fig:SI_switch_titration_combined}) noise in DnaA gene expression gives rise to adder correlations, in the switch-dominated regime it can give rise to sizer-correlations provided that negative autoregulatinon is strong enough to reduce the relaxation time of DnaA concentration fluctuations. The second question is what correlations arise when we combine a source of noise that gives rise to sizer correlations with one that gives rise to adder correlations? We show that when combining DnaA noise in the switch-dominated regime, thus giving sizer noise, with adder-noise in the lipid concentration, it depends on the strength of the respective sources of noise whether the system exhibits adder or sizer correlations. Finally, we show that our switch-titration model can reproduce the recent experimental results by Si et al. \cite{Si2019}: By starting from a system in which noise in both the lipids and the DnaA concentration contribute to adder correlations in the initiation volume, dynamically perturbing the DnaA concentration gives rise to sizer noise in the initiation volume.

\begin{figure}
	\centering
	\includegraphics[width =0.6\textwidth]{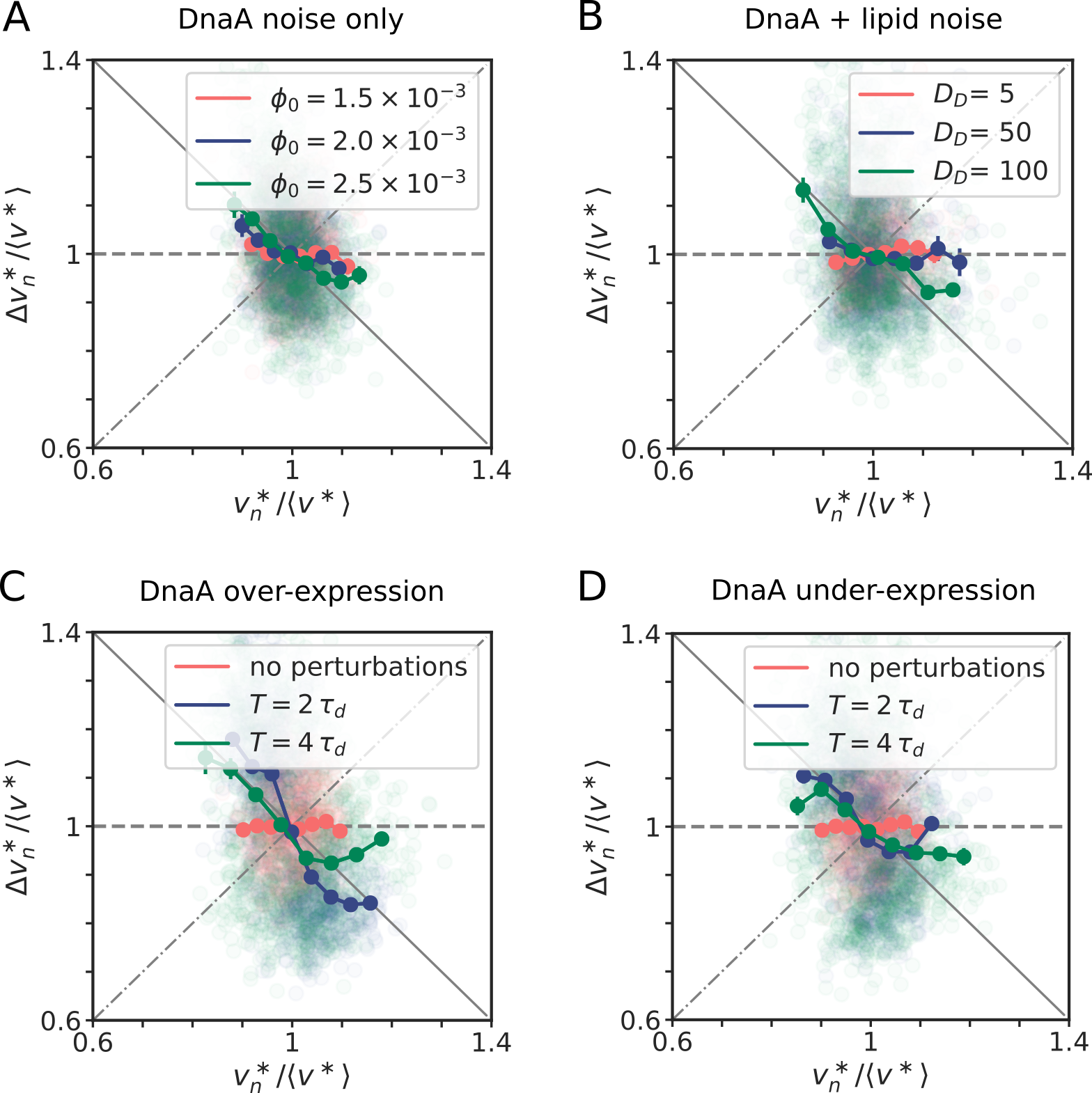}
	\caption{\textbf{Externally driven oscillations in the DnaA concentration can turn an initiation-adder into an initiation-sizer} (A, B, C, D) Scatter plot of the added initiation volume between successive initiation events, $\Delta v_n^\ast \equiv 2 v_{n+1}^\ast - v_n^\ast$, and the initiation volume $v_n^\ast$. Both the x and the y axis are normalized by the average initiation volume $\langle v^\ast \rangle$ at the respective parameters. For comparison, the dashed line is a perfect adder, the solid line a perfect sizer and the dash-dotted line a perfect timer. The doubling time in all plots is $\tau_{\rm d}=2$~h. (A) In the switch-titration model it depends on the basal rate of DnaA gene expression whether noise in the DnaA concentration (with noise strength $D_{\rm D}=100$) generates an initiation adder or sizer. At a low gene allocation fraction $\phi_0$ and thus a low DnaA basal rate, the DnaA promoter is weakly repressed for most of the cell cycle because of the long time it takes to fill all titration sites; consequently, DnaA noise gives rise to adder correlations (red data points). By increasing the gene allocation fraction $\phi_0$ of DnaA and thus the basal DnaA expression rate, the system moves from the titration-dominated regime to the switch-dominated regime. In the switch-dominated regime, the titration sites are filled up more quickly and the DnaA promoter is repressed during most of the cell cycle. As explained in the LDDR model, negative autoregulation then generates sizer-correlations in the initiation volume (blue and green data points). (B) Starting from a system that is in the switch-dominated regime ($\phi_0=2.5 \times 10^{-3}$) and in which DnaA noise gives rise to sizer-correlations in the intiation volume, we now add noise in the lipid concentration (see section \ref{sec:switch_noise_lipids}). Whether the system is then an initiation-adder or an initiation-sizer depends on the respective strengths of the noise in the DnaA and lipid concentration. For a noise strength of the lipids of $D _{\rm l} = 1000$ and a noise strength of DnaA of $D_{\rm D} = 5$, the adder noise from the lipids is dominant and the resulting correlations in the initiation volume are an adder (red data points). With increasing DnaA noise strength, the correlations in the initiation volume become more and more sizer-like (blue and green data points).
		(C) Now we consider a system where both noise in the DnaA and in the lipid concentration give rise to adder correlations ($\phi_0=1.5 \times 10^{-3}$, $D_{\rm D} = 100$ and $D _{\rm l} = 1000$, red data points). Similar to the over-expression experiments performed by Si et al. \cite{Si2019}, we add external oscillations in DnaA in order to turn the adder into a sizer. We express DnaA both via its endogeneous promoter according to equation \ref{eq:AIT_noise_total_conc} and via externally driven oscillations according to equation \ref{eq:oscillating_dnaA}. For a period of $T= 2 \, \tau_{\rm d}$ (blue data points), which is the optimal period for inducing negative auto-correlations in the initiation volume, the adder correlations are strongly sizer-like. When DnaA is induced with a period of $T= 4 \, \tau_{\rm d}$ like in the experiments by Si et al. \cite{Si2019} (green data points), the adder (still) becomes sizer-like as observed experimentally \cite{Si2019}. (D) By removing expression from the endogeneous promoter of DnaA and combining a basal constitutive DnaA expression rate with oscillations according to equation \ref{eq:oscillating_dnaA_under_exression} with $\phi_0^{\rm ext} = 1.5 \times 10^{-3}$, the switch-titration model can reproduce the under-expression experiments of Si et al. \cite{Si2019}. Again, we can turn an unperturbed system that exhibits adder correlations (red data points, same as in C) into a system that is more sizer-like (green and blue data points).}
	\label{fig:S9_2_combine_adder_and_sizer_noise}
\end{figure}

First, we study the correlations in the initiation volume in the switch-titration model when noise in the DnaA concentration is the only source of noise. We model noise in DnaA gene expression according to equation \ref{eq:AIT_noise_total_conc} as explained in section \ref{sec:switch_noise_total}. Whether noise in the DnaA concentration then gives rise to adder or sizer correlations in the initiation volume depends on whether the system is in the switch or in the titration-dominated regime (Fig. \ref{fig:S9_2_combine_adder_and_sizer_noise} A). In the switch-dominated regime, the basal rate is sufficiently high to fill up the titration sites before the next round of replication is initiated. In this regime, DnaA is negatively autoregulated during most of the cell cycle and fluctuations in the DnaA concentration are reduced rapidly, when, as assumed here, autoregulation is sufficiently strong. Then, the correlations in the initiation volume are sizer-like, as explained in the main text (\ref{fig:S9_2_combine_adder_and_sizer_noise} A, green data points). We can move from the switch to the titration-dominated regime by either increasing the number of titration sites or by decreasing the basal DnaA synthesis rate (see Fig. \ref{fig:SI_switch_titration_combined}). In the titration-dominated regime, the DnaA promoter is almost not repressed during most of the cell cycle. In this regime, fluctuations in the total DnaA concentration are reduced via dilution on a time scale set by the doubling time of the cell, giving rise to adder correlations in the initiation volume (\ref{fig:S9_2_combine_adder_and_sizer_noise} A, red data points).

Now we want to study a system that combines fluctuations that tend to generate sizer-like correlations with fluctuations that tend to generate adder-like correlations. To this end, we will study a system that is in the switch-dominated regime, where DnaA generates sizer-like fluctuations and the lipids generate adder-like correlations. Figure \ref{fig:S9_2_combine_adder_and_sizer_noise} B demonstrates that a combination of sizer-noise in the DnaA and adder-noise in the lipid concentration can give rise to both sizer or adder noise, depending on the relative strengths of the respective sources of noise. At a low magnitude of the noise in DnaA as compared to the lipids, the fluctuations in the initiation volume are dominated by the fluctuations in the lipid concentration and we find an initiation adder (Fig. \ref{fig:S9_2_combine_adder_and_sizer_noise} B, red data points). By increasing the noise strength of the DnaA production rate while keeping the lipid noise strength constant, the observed adder correlations in the initiation volume become sizer-like. From this finding we predict that if replication initiation in \textit{E. coli} is dominated by the switch rather than titration, the dominant source of noise generating adder correlations in the initiation volume is set by fluctuations in the concentration of (de)activators of DnaA such as the lipids, RIDA (via Hda) and \textit{datA} (via IHF). Conversely, if the system is in the titration-dominated regime, the adder correlations can arise from noise {\em both} the switch (de)activators and DnaA.

Now, we want to test whether in the switch-titration model we can turn an initiation adder into an initiation sizer like demonstrated in recent experiments by Si et al. \cite{Si2019}. Starting from the observation that the initiation volume increases with decreasing DnaA concentration, Si et al. dynamically perturbed the DnaA concentration of cells and measured the effect on the initiation volume. Inducing DnaA with a period of $T= 2 \, \tau_{\rm d}$ causes cells that have initiated at a larger than average volume in generation $n$ to initiate at a smaller than average volume in generation $n+1$. Si et al. therefore proposed that periodically inducing the production of DnaA with a period given by $T= 2 \, \tau_{\rm d}$ would lead to negative auto-correlations of the initiation volume, thus breaking the for an adder typical mother-daughter auto-correlation of 1/2. In their experiments Si et al. had to use a period of $T= 4 \, \tau_{\rm d}$ due to the high induction and dilution time of DnaA. They find that the periodic induction of DnaA with a period of $T= 4 \, \tau_{\rm d}$ can indeed break the initiation adder and the initiation volume showed weak sizer correlations.

Starting from an unperturbed system where both the lipid and the DnaA noise give rise to adder correlations in the initiation volume (Fig. \ref{fig:S9_2_combine_adder_and_sizer_noise} C, red data points) we now include the effect of strong externally driven oscillations in the DnaA concentration. 
In the experiments by Si et al. \cite{Si2019}, a strain carrying extra \textit{dnaA} under the $P_{\rm lac}$ promoter on plasmids was used to induce DnaA. In the over-expression experiments, DnaA was expressed both by its endogenous promoter on the chromosome and by the inducer controlled extra \textit{dnaA} on the plasmid. In our simulations, we mimic these DnaA over-expression experiments by expressing DnaA as described in section \ref{sec:switch_noise_total} from the chromosome and adding the following deterministic oscillation in the expression rate of ATP-DnaA proteins to the system:
\begin{equation}
\frac{dN^{\rm ext}_{\rm D,ATP}}{dt} = a \, (\cos{(2 \, \pi / T \, t)} + 1)
\label{eq:oscillating_dnaA}
\end{equation}
where $a$ is the amplitude and $T$ is the period of the oscillations. We find that externally driven oscillations according to equation \ref{eq:oscillating_dnaA} with a period of $T= 4 \, \tau_{\rm d}$ lead to strong sizer correlations (Fig. \ref{fig:S9_2_combine_adder_and_sizer_noise} C, green data points). The amplitude of the oscillations is sufficiently high and thus, the dominant source of noise are not the lipid or the unperturbed DnaA concentration fluctuations anymore, but the driven oscillations in the DnaA concentration. At a period of $T= 2 \, \tau_{\rm d}$, the effect becomes even stronger and we see that the system oscillates as predicted between a high and a low initiation volume, thus creating strong sizer correlations (Fig. \ref{fig:S9_2_combine_adder_and_sizer_noise} C, blue data points).

Si et al. additionally performed DnaA under-expression experiments by removing the endogeneous expression of DnaA and using the DnaA expressed from the plasmid as the only source of DnaA expression. We model this by removing expression of DnaA via the endogeneous promoter according to equation \ref{eq:AIT_noise_total_conc}. Instead, we add a basal production rate following again the growing cell model of gene expression and combine it with the oscillations in the DnaA number as in equation \ref{eq:oscillating_dnaA}:
\begin{equation}
\frac{dN^{\rm ext}_{\rm D,ATP}}{dt} = \phi_0^{\rm ext} \, \lambda \, \rho \, V + a \, (\cos{(2 \, \pi / T \, t)} + 1)
\label{eq:oscillating_dnaA_under_exression}
\end{equation}
Again we find that using this modified DnaA gene expression rate, we obtain sizer-correlations in the initiation volume  when the system is driven at the optimal period of $T= 2 \, \tau_{\rm d}$ (Fig. \ref{fig:S9_2_combine_adder_and_sizer_noise} D, blue data points). Also using the same period $T= 4 \, \tau_{\rm d}$ as in the under-expression experiments Si et al. \cite{Si2019} lead to weak sizer correlations (Fig. \ref{fig:S9_2_combine_adder_and_sizer_noise} D, green data points). We have shown that our switch-titration model can explain how an unperturbed system that has adder correlations in the initiation volume can become sizer-like by strong dynamic perturbations of DnaA.

\subsubsection{Loosening the coupling between replication initiation and division}
\label{sec:division_separate}
According to experiments at the population level, the time from the initiation of replication until cell division, the cycling time $\tau_{\rm cc}$, is approximately constant \cite{Si2017}.  In the main text, we therefore assumed that $\tau_{\rm cc}$ is constant. This allowed us to study the cell cycle entirely from the perspective of the replication cycle. Experiments show, however, that this is an
oversimplification \cite{Michelsen:2003ku, Adiciptaningrum2015, Wallden2016, Micali2018, Micali2018_2, Si2019, Witz2019} and that cell division  is more loosely coupled to the replication cycle \cite{Micali2018, Micali2018_2, Si2019, Witz2019, LeTreut2021, Witz:2020fb}. Of particular interest are two recent single-cell studies, by Si et al. \cite{Si2019} and Witz et al. \cite{Witz2019},
respectively. Both studies indicate that the cell cycle consists of two adders, a DNA replication adder and a cell-division adder. Both studies also agree on the nature of the replication adder: the data of both studies unequivocally show that the added volume between successive initiation events is independent of the initiation volume, as our model also predicts (Fig. 4, Figs \ref{fig:SI_LD_adder_sizer_correlations} and
\ref{fig:S10_LDDR_RIDA_adder}).  However, the authors of these two studies come to different conclusions concerning the nature of the division adder \cite{ Si2019, Witz2019, LeTreut2021, Witz:2020fb}.  By employing a statistical framework with stochastic simulations, Witz et al. conclude that the second adder concerns the added volume between replication initiation and cell division \cite{Witz2019}.  Si et al. showed that by inducing oscillatory perturbations in the concentration of DnaA, the adder correlations in the replication initiation volume can be destroyed, while the adder on the level of cell division remains intact; they
conclude that the division adder concerns the added volume from birth to division and suggest that cell division is controlled by a separate molecular mechanism \cite{Si2019}.

We emphasise that the central question of our manuscript is how replication initiation is regulated---not how cell division is controlled, nor how this is coupled to replication initiation. Naturally, our assumption that the time $\tau_{\rm cc}$ between cell division and replication initiation is constant will affect the correlations between the initiation volume and cell division, since this directly couples division to replication initiation. The pertinent question is, however, whether the adder correlations in the initiation volume remain robust to this  assumption. 

\begin{figure}
	\centering
	\includegraphics[width =0.83\textwidth]{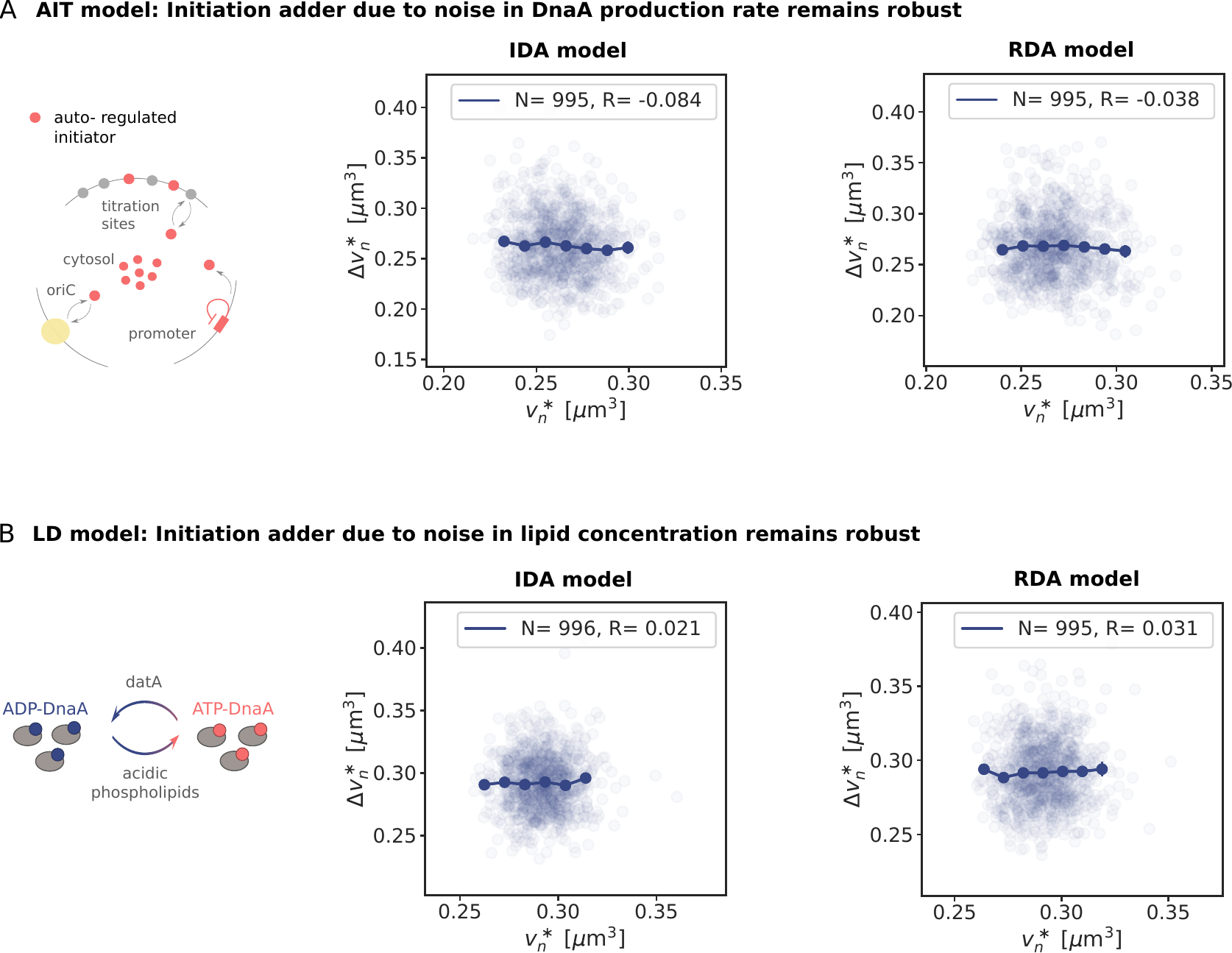}
	\caption{\textbf{The adder correlations in the initiation volume in the AIT and the LD model are robust to a more loose coupling between the division and the replication cycle.} (A, B) The added volume per origin between consecutive replication initiation events, $\Delta v^\ast_{\rm n}= 2 \, v^\ast_{\rm n+1}-v^\ast_{\rm n}$, as a function of the initiation volume $v^\ast_{\rm n}$. The dark blue lines show the mean of the binned data and the error bars represent the standard error of the mean (SEM) per bin. The number of data points $N$ and the Pearson correlation coefficient $R$ are indicated. The doubling time in all plots is $\tau_{\rm d}=2$~h. (A) In the AIT model, gene expression noise in the initiator protein gives rise to adder correlations in the initiation volume even if the division cycle is coupled more loosely to replication initiation: While in the IDA model, cell division is triggered completely independently via a separate division adder, in the RDA model division is triggered when an on average constant volume has been added from replication initiation to cell division (compare to Fig. \ref{fig:S4_AIT_adder}). (B) In the LD model in the presence of noise in the lipid concentration (according to equation 6 in the main text), the added volume per origin between successive initiation events remains independent of the initiation volume, both when the division cycle is controlled via the IDA and the RDA model (compare to Fig. 4 B in the main text or to Figure \ref{fig:SI_LD_adder_sizer_correlations} A).}
	\label{fig:SI_separate_division}
\end{figure}

To address these questions, we compare the results of our models of the main text in which replication is coupled to cell division via a constant time $\tau_{\rm cc}$ between these two events, to the predictions of two other models in which replication is coupled to cell division either via the model of Si et al. or that of Witz et al. These two alternative models contain the same molecular, mechanistic description of replication initiation as our models presented in the main text. And like our models, they describe cell division and its coupling to replication initiation phenomenologically.  The models differ, however, in the nature and strength of this coupling between cell division and replication. While in our model cell division is tightly coupled to replication initiation, with a constant $\tau_{\rm cc}$ between these two events, in the other two models the coupling is more loose. The first of these two models is based on that of Si {\it et al.} \cite{Si2019,LeTreut2021}, which, following \cite{Witz:2020fb}, we call the Independent Double Adder (IDA) model. In this model, the cell division
cycle is completely independent of the replication cycle. Cells divide when a Gaussian distributed volume $\Delta_{\rm IDA}$ with mean $\mu_{\rm IDA}= \langle V_{\rm b} \rangle$ and a standard deviation $\sigma_{\rm LDA}$ ( with coefficient of variation ${\rm CV}_{\rm IDA}= \sigma_{\rm IDA} / \mu_{\rm IDA}=0.1$) has been added to the birth volume, independent of the cell size at birth. As the replication and the division cycle are not coupled in this model, it could happen that a cell attempts to divide before replication has finished. To prevent this biologically unrealistic scenario from happening, we impose in the simulations that replication must be finished before a cell can divide. This scenario however only happens extremely rarely. In the second model, based on that of Witz et al. and called the Replication Double Adder (RDA) model \cite{Witz2019, Witz:2020fb}, cells divide when a Gaussian distributed volume $\Delta_{\rm RDA}$ with mean $\mu_{\rm RDA}= \langle V_{\rm d} \rangle - \langle v^\ast \rangle = \langle v^\ast \rangle \, \left( \exp(\lambda \, \tau_{\rm cc}) - 1\right)$ and a standard deviation $\sigma$ (with coefficient of variation
${\rm CV}_{\rm RDA} = \sigma_{\rm RDA}/\mu_{\rm RDA}=0.1$) has been added since replication initiation, independent of the initiation volume. In this model, the coupling between replication and division is thus of intermediate strength.

At the mean-field level, all results on the initiation
volume should be independent of the type of division control, as the initiation volume in the accumulation and in the switch model is determined by concentrations of proteins which do not change upon cell division. 
We show below that, in addition to these mean-field observations,
the adder correlations of the initiation volume obtained in the AIT and the LD model remain unchanged when cell division is controlled by either an independent adder running from cell birth to division, as in the IDA model, or by an adder running from replication initiation to cell division, as in the RDA model.

We re-evaluate all obtained correlations between consecutive cell cycles both in the AIT and the LD model in the case where cell division is controlled according to the IDA or RDA model, as described above. We find that while correlations between the initiation volume and the birth volume are different, as expected, the previously obtained correlations between consecutive initiation volumes per origin remain unchanged, for both models. Specifically, in the AIT
model, gene expression noise in the initiator protein production rate still gives rise to adder correlations in the initiation volume per origin, both in the IDA and RDA model (Fig. \ref{fig:SI_separate_division} A). In the LD model, fluctuations in the lipid concentration result again in adder correlations for the initiation volume, both in the IDA and RDA model (Fig. \ref{fig:SI_separate_division} C). 
We conclude that our principal finding that fluctuations in
switch components can generate adder correlations in the initiation volume is robust: these correlations depend on the correlation time of the fluctuations in the switch components, but do not depend on the specific type of coupling of the replication cycle to the division cycle.

\subsection{Novel predictions from the full switch-titration model}
\label{sec:model_predictions}
We here summarize the novel predictions from our full switch-titration model. Some of these have already been discussed in the previous subsection, when they naturally followed from the model validation.

\begin{enumerate}
	\item We predict that at low growth rates the activation switch and the titration mechanism act synergistically to raise the amplitude of the oscillations in the concentration of free, ATP-bound DnaA  (see section \ref{sec:switch_titration_combined_parameter_regimes}). Whether the system is in the titration or a switch-dominated regime could be tested by varying the number of titration sites and the basal initiator production rate, as Fig. \ref{fig:SI_switch_titration_combined} shows. While in the switch-dominated regime the initiation volume should be relatively independent of the number of titration sites, in the titration-dominated regime the initiation volume should increase linearly with the number of titration sites (Fig. \ref{fig:SI_switch_titration_combined} A). This is a strong, robust prediction from our model that could be tested experimentally. Since the titration sites have a characteristic sequence \cite{Schaper1995, Roth1998} it should be possible to vary the number of titration sites on the chromosome, making it possible to steer the system between a switch-dominated regime to a titration-dominated one.
	\item We furthermore predict that due to the random distribution of titration sites on the chromosome, at intermediate and high growth rates the titration mechanism alone cannot ensure stable cell cycles. At high growth rates, the help of SeqA is sufficient to generate robust cell cycles.  Yet, at intermediate growth rates the switch becomes essential for preventing premature reinitiation events (Figs. 2 and 5C of main text). The predicted dependence of the importance of the activation switch on the growth rate could be tested using mutants in which the switch is effectively turned off, for example by deleting \textit{datA} \cite{Kasho2013} and deactivating RIDA \cite{Kurokawa1999, Katayama1998, Kato2001}. Our model predicts that stable replication cycles can be generated solely by titration and SeqA at low and high growth rates, but not at intermediate growth rates. To further study the interplay between titration, activation, and blocked synthesis by SeqA, it would be of interest to remove the latter mechanism; this may be achieved by removing the GATC site in the promoter region of DnaA, which is necessary for binding SeqA \cite{Lu:1994ee}. Our model predicts that removing this mechanism has a much larger effect at high than at low growth rates. 
	\item We predict that if replication initiation in \textit{E. coli} is dominated by the switch rather than titration, the dominant source of noise generating adder correlations in the initiation volume is set by fluctuations in the concentration of (de)activators of DnaA such as the lipids, RIDA (via Hda) and \textit{datA} (via IHF). Conversely, if the system is in the titration-dominated regime, the adder correlations can arise both from the switch components and DnaA.
	
	\item Our model predicts that the initiation volume is inversely proportional to the total DnaA concentration,  see Fig. \ref{fig:S17_vary_total_conc}. More specifically, it predicts that this relation holds over a wide range of growth rates.
	
	\item According to our switch model, the initiation volume scales inversely proportional to the acidic phospholipid concentration. We predict that in the switch-dominated regime the initiation volume should decrease with increasing lipid concentration and that the effect of depleting the lipids on the initiation volume is stronger at low than at high growth rates (Fig. \ref{fig:S18_mutations} H).
\end{enumerate}

\section{The role of the lipids}
\label{sec:role_of_lipids}

The capacity of the switch to act as an origin-density sensor
	hinges on the idea that the activation and deactivation rates have different functional dependencies on the origin density. In the switch models, all deactivation rates are proportional to the origin density, but not all activation reactions are: while the rate of DARS1/2 activation is proportional to the origin density, the rates of DnaA activation via protein synthesis and the acidic phospholipids are not. However, the importance of the lipids for reactivating DnaA \textit{in vivo} remains unclear \cite{Shiba2004, Shiba2012,Camsund2020}. Moreover, there is evidence that the effect of the lipids depends on oriC
	\cite{Crooke1992}. Here, in section \ref{sec:no_lipids}, we analyze the importance of the lipids for DnaA reactivation, both for the switch-only models (LD and LDDR) and for the full model. Then, in section \ref{sec:lipids_oriC}, we discuss the role of oriC in lipid-mediated DnaA reactivation. The principal findings of these sections are: lipid-mediated DnaA activation is essential to the switch, both in the LD and in the LDDR model: taking the lipids out entirely, dramatically lowers the amplitude of the oscillations, such that in the presence of biochemical noise they will likely not persist.  Interestingly, however, the full model, which includes titration and SeqA, is robust to removing DnaA activation via the lipids. While a model based on titration alone is only stable at low growth rates and a system based on only a lipid-deficient switch produces merely very weak oscillations at any growth rate, the combined system yields robust oscillations at all growth rates.
\\

\subsection{The importance of lipid-mediated DnaA rejuvenation}
\label{sec:no_lipids}
While the \textit{in vitro} evidence that acidic phospholipids enhance the release of nucleotides is compelling \cite{Sekimizu1988,Crooke1992}, the \textit{in vivo} experiments paint a mixed picture \cite{Xia1995, Fingland2012, Shiba2004, Shiba2012,Camsund2020}. In the next section, \ref{sec:role_lipids_experiments}, we will first discuss these experiments. We will then study the importance of lipid-mediated DnaA activation by taking out the lipids, first in the LD model (\ref{sec:no_lipids_LD}), then in the LDDR model (\ref{sec:no_lipids_LDDR}), and then in the full model (\ref{sec:no_lipids_full}). We find that while the lipids are essential for generating high-amplitude oscillations in the switch-only models, the full model is remarkably robust: removing the lipids from the full model does reduce the precision of replication initiation, but not dramatically.

\subsubsection{Experiments}
\label{sec:role_lipids_experiments}
Acidic phospholipids from the cell membrane promote dissociation of both ADP and ATP from DnaA very effectively \cite{Sekimizu1988}, and DnaA can be reactivated by exchange of the bound nucleotide \textit{in vitro} in the presence of ATP and oriC \cite{Crooke1992, Castuma.1993}. Depleting acidic phospholipids \textit{in vivo} can lead to growth arrest \cite{Xia1995, Heacock1989} and inhibit initiation at oriC \cite{Fingland2012}, which supports the idea that lipids can reactivate DnaA by promoting the exchange of bound ADP for ATP. More specifically, Fingland et al. observed that by bringing \textit{pgsA} under an inducible promoter acid-lipid depleted cells slowed down their growth and that as they began to slow down in their growth, the number of origins per cell, as determined via run-out experiments using rifampicin and cephalexin, decreased relative to cells that continue to synthesize acidic phospholipids \cite{Fingland2012}. Also the DNA content and cell mass decreased. These observations could be explained by the decrease in the growth rate, which tends to reduce the average number of origins per cell, DNA content, and cell mass \cite{Si2017}. Several studies however showed that the reported growth arrest of acidic phospholipid-deficient cells can be suppressed by overexpression of a mutant form of DnaA (L366K) which contains mutations in the membrane-binding domain \cite{Saxena2013, Zheng2001, Fingland2012}. Moreover, the growth arrest can be suppressed by mutating rnhA, which allows the cell to bypass normal oriC-dependent initiation via a process called recA-dependent constitutive stable DNA replication \cite{Xia1995}.  Both of these findings strongly indicate that the growth arrest is caused by the inability to initiate replication \cite{Xia1995, Saxena2013, Zheng2001, Fingland2012}.

In contrast, Shiba et al. found that the lethal effect of a {\it pgsA} null mutation, which causes a complete lack of the major acidic phospohlipids, phosphatidylglycerol and cardiolipin, is alleviated by mutations that change the membrane structure \cite{Shiba2004, Shiba2012}. Moreover, a recent study by Camsund et al. \cite{Camsund2020} reported that while downregulating \textit{pgsA} reduced the growth rate of the cell (rather than growth arrest), the initiation volume per origin was not significantly affected. From this finding, the authors concluded that, in contrast to the earlier reports \cite{Saxena2013, Zheng2001, Fingland2012}, the lipids are not vital for replication initiation. Motivated by these observations, we therefore analyze in the next sections the behavior of the LD, LDDR, and the full model, when lipid-mediated DnaA activation is taken out completely.

\subsubsection{LD model}
\label{sec:no_lipids_LD}
The change of the active fraction as a function of time for a system where DnaA is activated only via DnaA synthesis and deactivated via the chromosomal site \textit{datA} is given by:
	\begin{equation}
	\frac{df}{dt}
	= \lambda (1 - f) - \tilde{\beta}_{\rm datA} \, [n_{\rm ori}] \, \frac{f}{\tilde{K}_{\rm D}^{\rm datA}+f}
	\label{eq:switch_simple_fraction_no_lipids_SI}
	\end{equation}
	As we again combine an origin-density dependent deactivation rate with a constant activation rate (for a given growth rate), initiating replication at a critical active fraction $f^\ast$ results in stable replication cycles with the initiation volume per origin $v^\ast$, see Fig. \ref{fig:SI_LD_only_synthesis} A and B. However, the amplitude of the oscillations is very low (Fig. \ref{fig:SI_LD_only_synthesis} B). 
	Protein synthesis yields an activation rate that is linear in the active DnaA fraction $f$ (see Eq. S40 and Fig. \ref{fig:SI_LD_only_synthesis} A). In contrast, lipid-mediated activation is non-linear in $f$. This is important, because together with the non-linear deactivation rate, it gives rise to an ultra-sensitive activation mechanism (see Fig. 3A of the main text). Removing lipid-mediated activation from the model thus eliminates the ultra-sensitive activation mechanism, which reduces the amplitude of the oscillations in $f$ (compare Fig. \ref{fig:SI_LD_only_synthesis} A with Fig. 3A).

Yet, there is a second effect, which is best illustrated in the quasi-equilibrium limit, where the critical initiation volume per origin $v^\ast$, obtained by setting $df/dt = 0$ in equation \ref{eq:switch_simple_fraction_no_lipids_SI}, is given by 
	\begin{equation}
	v^\ast =  \frac{\tilde{\beta}_{\rm datA}}{\lambda} \, \frac{ f^\ast}{(\tilde{K}_{\rm D}^{\rm datA}+f^\ast) \, (1 - f^\ast) }
	\label{eq:v_init_LD_no_lipids}
	\end{equation}
	The key point to note is that the activation rate is now set by the growth rate $\lambda$, which is much lower than the experimentally measured \textit{datA}-mediated deactivation rate $\tilde{\beta}_{\rm datA}$. Hence, to obtain an initiation volume that is consistent with what is observed experimentally the initiation threshold $f^\ast$ would have to be lowered drastically. However, this would strongly reduce the amplitude of the oscillations. Alternatively, the deactivation rate $\tilde{\beta}_{\rm datA}$ itself would have to be much lower. Yet, this would also strongly reduce the amplitude of the oscillations  (Fig. \ref{fig:SI_LD_only_synthesis} B).  In the presence of biochemical noise, these small oscillations would therefore not be sufficient to ensure stable cell cycles. Indeed, to generate large amplitude rhythms, not only the deactivation but also the activation rate needs to be larger than the growth rate. This is what lipid-mediated activation can accomplish.\\

\subsubsection{LDDR model}
\label{sec:no_lipids_LDDR}
We have also analyzed the LDDR model without lipid-mediated DnaA activation. This system is described by Eq. \ref{eq:switch_complex_fraction_SI} but without the first lipid term on the right-hand side. Crucially, in order for the switch to be stable, the activation and deactivation rates must have different functional dependencies on the origin density, see section \ref{sec:LDDR_datA_dars1_unstable}. Moreover, the \textit{DARS1/2} activation rate scales with the origin density like the rates of {\em both} deactivation mechanisms, \textit{datA} and RIDA. Taken together, these two observations mean that, in the absence of origin-independent lipid-mediated DnaA activation, the \textit{DARS1/2} activation rate cannot be too large: it must be smaller than the protein synthesis rate, which is then the only mechanism of DnaA activation or deactivation that is independent of the origin density; otherwise, the net, overall rates of both activation and deactivation would scale with the origin density, making the system unstable (see Fig. \ref{fig:SI_dars1_datA_only} B). Since the protein synthesis rate is set by the growth rate, this means that the activation and deactivation rates must be smaller than or comparable to the growth rate, at all growth rates. The net result is that a LDDR model without the lipids exhibits oscillations of only weak amplitude, for all growth rates (see Fig. \ref{fig:SI_LD_only_synthesis} C/D).

\subsubsection{Full model}
\label{sec:no_lipids_full}
\begin{figure}
	\centering
	\includegraphics[width =0.75\textwidth]{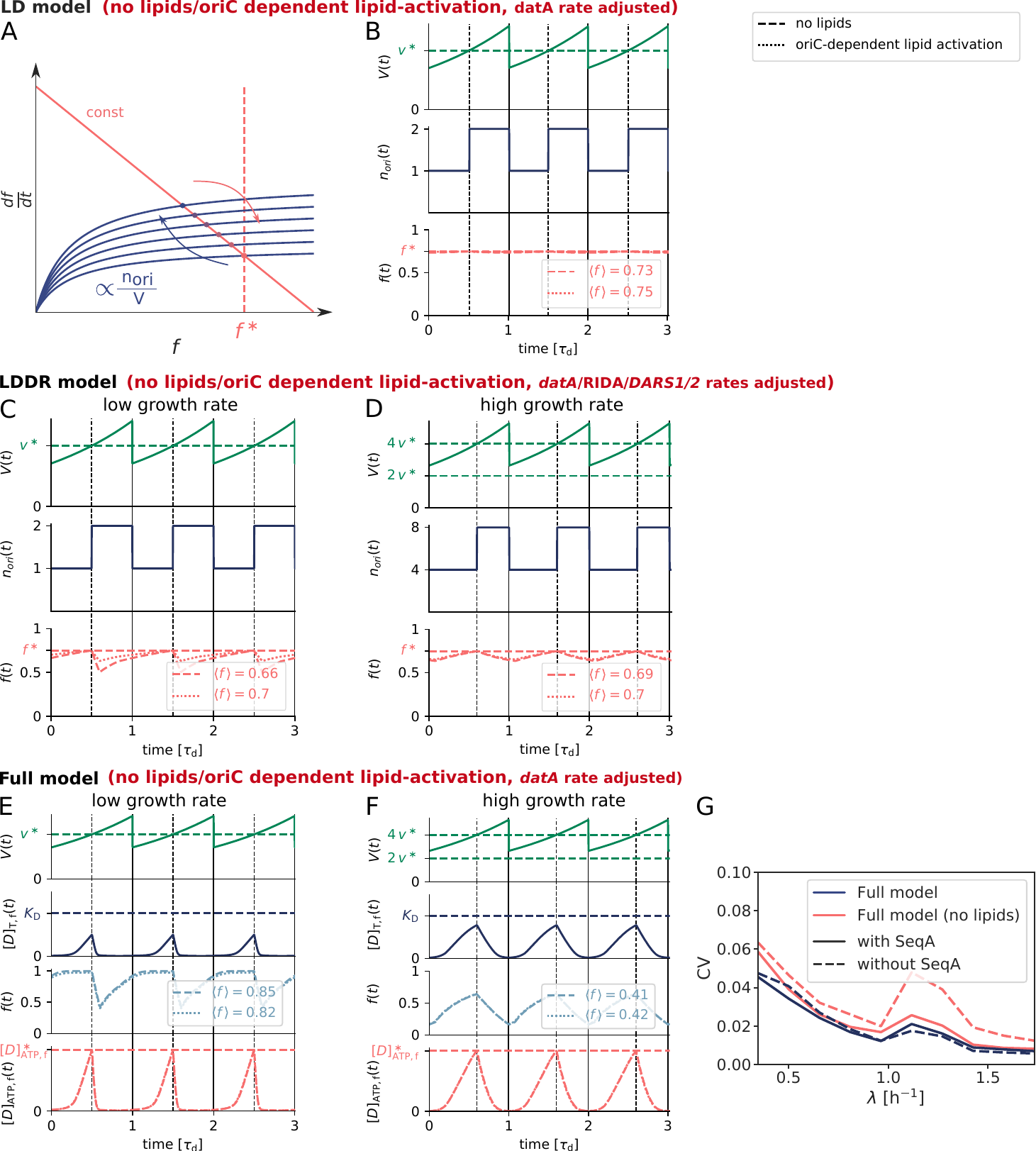}
	\caption{{\footnotesize \textbf{Effect of removing the lipids and of oriC-dependent lipid-activation in the LD, the LDDR and the full model} (A) Combining activation via DnaA synthesis with deactivation via \textit{datA}: The constant activation rate (red line) and the origin density-dependent deactivation rate (blue curve) as a function of the active fraction of the initiator protein $f$ at different moments of the cell cycle (see equation \ref{eq:switch_simple_fraction_no_lipids_SI}). 
				Contrary to the LD model, where both activation and deactivation are non-linear (see Fig. 3A), here the activation rate is linear in $f$, eliminating the ultra-sensitivity and thus giving rise to a smaller amplitude in the active fraction $f$. (B, C, D, E, F) The volume of the cell $V(t)$, the number of origins $n_{\rm ori}(t)$, the fraction of ATP-DnaA $f(t)$, the free DnaA concentration $[D]_{\rm T,f}(t)$ and the concentration of free ATP-DnaA $[D]_{\rm ATP,f}(t)$ as a function of time (in units of the doubling time $\tau_{\rm d}$) with $\tau_{\rm d}=2$~h (B, C, E) and $\tau_{\rm d}=25$~min (D, F). Replication is initiated at a critical initiator fraction $f^\ast$ in the LD and LDDR model (red dashed line in B, C, D) and at a critical free, ATP-DnaA concentration $[D]_{\rm ATP,f}^\ast$ (red dashed line in E, F) in the full model. The average active fraction over one cell cycle $\langle f \rangle$ is indicated and replication is initiated at a constant volume per origin $v^\ast$ over time (green dashed line). (B) LD model (no lipids): With DnaA synthesis (set by the growth rate $\lambda = 0.35$~h$^{-1}$) being the only activator, the rate of \textit{datA} ($\beta_{\rm datA}=20$~h$^{-1}$) must be small in order to ensure that replication is initiated at the experimentally observed initiation volume per origin $v^\ast$. Due to low (de)activation rates and the lack of ultra-sensitivity, the amplitude of the oscillations becomes extremely small and would not be sufficient to generate stable cell cycles in the presence of noise. LD model (oriC-lipids): Similarly, when lipid-activation is oriC-dependent (see equation \ref{eq:switch_lipid_ori_SI}), the amplitude of the oscillations in the active fraction becomes small due to a small effective deactivation rate from \textit{datA} and the lipids (dotted line, $\alpha_{\rm l}=750$~h$^{-1}$ and $\beta_{\rm datA}=600$~h$^{-1}$). (C, D) LDDR model (no lipids): In the LDDR model, not only the rate of \textit{datA} ($\beta_{\rm datA}^{-}=20$~h$^{-1}$, $\beta_{\rm datA}^{+}=120$~h$^{-1}$), but also the rates of \textit{DARS1/2} and RIDA must be low ($\beta_{\rm RIDA}=100$~h$^{-1}$, $\alpha_{\rm d2}^{-}=10$~h$^{-1}$, $\alpha_{\rm d2}^{+}=128$~h$^{-1}$, $\alpha_{\rm d1}^{+}=10$~h$^{-1}$), because otherwise the replication cycles become unstable as described in Figure \ref{fig:SI_dars1_datA_only} and section \ref{sec:LDDR_datA_dars1_unstable}. Low (de) activation rates thus result in weak oscillations in the active DnaA fraction $f$ both at low (C) and at high growth rates (D). LDDR model (oriC-lipids): When lipid-activation is oriC-dependent, the amplitude of the oscillations in the active fraction is again small due to the small effective deactivation rate via \textit{datA} and the lipids (dotted line, $\alpha_{\rm l}=100$~h$^{-1}$, $\beta_{\rm datA}^{-}=120$~h$^{-1}$ and $\beta_{\rm datA}^{+}=220$~h$^{-1}$). (E, F) Full model (see section \ref{sec:switch_titration_combined}): While the rate of \textit{datA} must again be small (same \textit{datA} rates as in C and D) to ensure the experimentally observed initiation volume, homeostasis (stability) is ensured via titration, allowing for high rates of RIDA and \textit{DARS1/2} (set to same values as in original full model). The full model without lipids (dashed lines) or with oriC-dependent lipid activation (dotted lines) thus gives rise to stable cell cycles with large amplitude oscillations both at low (E) and at high growth rates (F). (G) The coefficient of variation ${\rm CV}= \sigma /\mu$ with the standard deviation $\sigma$ and the average initiation volume $\mu=\langle v^\ast \rangle$ as a function of the growth rate for different models in the presence of noise in the DnaA concentration. Removing the lipids from the full model raises the coefficient of variation, especially at intermediate growth rates in the absence of SeqA (red dashed line).}}
	\label{fig:SI_LD_only_synthesis}
\end{figure}

The full model, including titration, SeqA, and the switch without lipid-mediated activation, is given by Eq. \ref{eq:switch_complex_fraction_noise}, but with the lipid term in \ref{eq:switch_complex_fraction_noise} removed. Crucially, in the presence of titration and SeqA, the rates of \textit{DARS1/2} and RIDA do not have to be dialed down to make the system stable, as is the case for the LDDR model without the lipids; only the rate of \textit{datA}, which is no longer balanced by the lipids, needs to be reduced, to keep the initiation volume consistent with experiments (indeed, merely taking out the lipids raises the initiation volume, see Fig. \ref{fig:S18_mutations}G/H). This system is surprisingly robust, for the full range of growth rates, see Fig. \ref{fig:SI_LD_only_synthesis} E/F. This is particularly interesting, because a system with {\em only} titration and SeqA or with {\em only} a (lipid-deficient) switch, cannot generate robust oscillations at all growth rates, while the full system, which combines all three mechanisms, can. The antagonism between \textit{DARS1/2} and RIDA generates large oscillations in the fraction, titration and SeqA generate large oscillations in the concentration, and the interplay between \textit{DARS1/2}-RIDA and titration-SeqA ensures stability. In particular, homeostasis is ensured by titration,  making the unstable lipid-devoid switch stable, while, conversely, the lipid-devoid switch prevents the reinitiation events that inevitably happen in the titration-SeqA-only system at intermediate growth rates (Figs. 2 and 5C). We do find that a full model with lipid-mediated DnaA activation is more robust than a full model without it, but the difference is surprisingly small (Fig. \ref{fig:SI_LD_only_synthesis} G, compare solid blue to solid red line). Yet, in the full model without SeqA (switch + titration), a switch without lipids can no longer fully prevent reinitiations in the intermediate growth rate regime (Fig. \ref{fig:SI_LD_only_synthesis} G, compare dashed blue to dashed red line).

\subsection{The role of oriC in lipid-mediated DnaA reactivation}
\label{sec:lipids_oriC}

In all the switch models discussed above and in the main text, we have assumed that lipid-mediated DnaA rejuvenation is independent of the origin density. However, there is evidence that the effect of the lipids depends on oriC \cite{Crooke1992}. In section \ref{sec:role_oriC_experiments} below, we first discuss these experiments in more detail. We then discuss the picture that emerges from these experiments, and the different DnaA rejuvenation scenarios that could be envisioned. We then discuss a scenario in which the rate of DnaA reactivation is proportional to the origin density. We show that an activation rate that is proportional to the origin density effectively lowers the \textit{datA} deactivation rate, thereby reducing the amplitude of the oscillations. Hence, our modelling predicts that if lipids are to play an activating role, the activation rate should not depend on the origin density, or at least not linearly.

\subsubsection{Experiments}
\label{sec:role_oriC_experiments}
{\bf Experimental observations} While acidic phospholipids like cardiolipin (CL) and phosphatidylglycerol (PG) enhance the release of ADP and ATP from DnaA \cite{Sekimizu1988,Yung1988,Crooke1992}, it has also been reported that CL blocks the binding of ATP to DnaA \cite{Sekimizu1988}. Moreover, while phospholipids can restore replication activity of DnaA bound to ADP \cite{Sekimizu1988}, it has also been observed that CL can inactivate nucleotide-free DnaA for replication initiation \cite{Sekimizu1988}. Furthermore, incubating DnaA-ADP or DnaA-ATP in the presence of phospholipids inhibits the binding of DnaA-ADP/ATP to oriC \cite{Crooke1992}, and adding phospholipids speeds up the dissociation of DnaA-ADP/ATP from oriC, albeit weakly \cite{Crooke1992}. Of particular interest is the observation that while incubating DnaA-ADP with the lipids and oriC restores replication activity (irrespective of whether the other replication components were added initially or later) \cite{Crooke1992}, first incubating DnaA-ADP with the lipids and then adding oriC later does not \cite{Crooke1992}; the temporal order in which components are added thus appears to matter. In addition, lipids lower the apparent affinity between DnaA and ATP, yet the presence of oriC during the incubation of DnaA with lipids restores it \cite{Crooke1992}. Consistent with this observation, the stimulating effect of the lipids on the release of nucleotides is weakened by the binding of DnaA-ADP/ATP to oriC \cite{Crooke1992}.

{\bf Emerging picture} The picture that emerges from these
	studies is that acidic phospholipids and nucleotides mutually
	exclude each other in (and thus compete for) binding DnaA. This could explain the
	observation that lipids both promote the release and inhibit the
	binding of nucleotides \cite{Sekimizu1988}; in turn, the
	release-promoting effect could explain why lipids can restore
	replication-initiation activity \cite{Sekimizu1988,Crooke1992}. In
	addition, the lipids also compete with oriC for binding DnaA. This
	may explain why adding lipids impedes the binding of DnaA to oriC \cite{Crooke1992} and
	hence replication initiation \cite{Sekimizu1988,Crooke1992}, and why it
	stimulates the dissociation of DnaA-ADP/ATP from oriC \cite{Crooke1992} (the
	fact that this effect is weak indicates that the complex of oriC and
	DnaA is very stable). At the same time, DnaA retains its high
	affinity for nucleotides when bound to oriC \cite{Crooke1992}. These
	observations are consistent with structural studies \cite{Saxena2013}, which show that DnaA binds the nucleotides and DNA
	(oriC) via different domains, while it binds the lipids via
	the domain (III) that also binds the nucleotides (hence their mutually
	exclusive binding), yet via a site that is on the border with the domain (IV) that binds
	oriC; the latter could explain why oriC and lipid binding are
	mutually
	exclusive.

{\bf Open questions} The biggest open question is how to
	reconcile the observation that lipids in a mixture of DnaA-ADP,
	oriC, ATP and other replication components can initiate DNA
	synthesis (Table I of \cite{Crooke1992}), while ADP is much less
	likely to be released from DnaA by the lipids when the DnaA-ADP is
	bound to oriC (Figs. 3 and 4 of \cite{Crooke1992}). The observation
	of Crooke et al. that the temporal order in which oriC and lipids
	are added to DnaA-ADP \cite{Crooke1992} is particularly confusing,
	because the reactions involved are association-dissocation reactions
	(i.e. DnaA-lipid binding, DnaA-oriC binding, and
	DnaA-(oriC)-nucleotide binding), which can reach thermodynamic
	equilibrium; and in the equilibrium state, the temporal order in
	which the components have been added should be irrelevant. It is
	conceivable that the reactions do not reach equilibrium, which would
	make it hard to interpret the results. It is also possible that
	the results depend in a non-trivial manner on the (relative)
	concentrations of the components, especially because the reactions
	involve competitive binding. This is particularly pressing, because
	the concentrations in the \textit{in vitro} experiments are likely to be very different from the effective concentrations \textit{in vivo}; the latter
	depend not only on the concentrations of the lipids in the membrane,
	but also on the membrane area and the cytoplasmic volume of the
	cell.

{\bf Alternative activation scenarios}. To make progress, we
	have analyzed alternative lipid-mediated DnaA activation
	scenarios. Besides the model of the main text (scenario 1), which
	assumes that DnaA is reactivated in the cytoplasm after it has
	dissociated from the membrane (see section
	\ref{sec:switch_parameters}), we considered two other scenarios: 2)
	DnaA-ADP binds to oriC, and then DnaA-ADP-oriC interacts with the
	lipids, leading to the exchange of ADP for ATP; 3) DnaA-ADP binds
	the lipids, ADP is released, oriC moves to the DnaA bound to the
	lipids, causing ATP to bind. 
	Scenario 1 of the main text is
	consistent with the observation that lipids can stimulate the
	release of ADP \cite{Sekimizu1988}, that DnaA can bind ATP
	\cite{Sekimizu1988,Crooke1992}, and that DnaA-ATP can bind to oriC
	\cite{Katayama2010, Nishida:2002dp, Speck2001}; the only open question is on what timescale DnaA dissociates from the membrane \cite{Aranovich2006, Garner1998}. The second scenario is consistent with the observation that incubating DnaA-ADP with the lipids and oriC restores replication activity \cite{Crooke1992}, but, as mentioned above, it seems at odds with the observation that lipids are less likely to induce the release of ADP when DnaA-ADP is bound to oriC \cite{Crooke1992}. Moreover, there is no evidence that oriC is associated with the membrane \cite{Knoppel2021}. The same criticism applies to scenario 3. 
	We therefore believe that the scenario of the main text, scenario 1, is the most likely scenario.\\

Yet, while scenarios 2 and 3 appear less likely, they also cannot be ruled out. Importantly, in scenarios 2 and 3 the rejuvenation of DnaA is contingent on oriC, making the lipid-mediated activation reaction dependent on the origin density. Below we therefore discuss a switch in which the activation rate scales with the origin density. Given the ambiguous experimental results discussed above, and the lack of quantitative, time-series data like that obtained for the Kai system \cite{Nishiwaki-Ohkawa2014,Paijmans2017}, we have not developed a detailed mathematical model that includes the competitive DnaA binding between the lipids, nucleotides, and oriC, but rather a coarse-grained model similar to that of the main text.

\subsubsection{Switch with lipid-mediated activation rate that depends on the origin density}
The dynamics of the active fraction in the LD model of the switch, yet with a lipid-mediated activation rate that depends on the origin density, is given by
	\begin{align}
	\frac{df}{dt}
	&= \lambda (1 - f) + \tilde{\alpha}_{\rm l} \,[l] \,  [n_{\rm ori}] \,\frac{1-f}{\tilde{K}_{\rm D}^{\rm l} + 1 - f}  - \tilde{\beta}_{\rm datA} \, [n_{\rm ori}] \, \frac{f}{\tilde{K}_{\rm D}^{\rm datA}+f}\\
	&= \lambda (1 - f) - \, [n_{\rm ori}] \left(\tilde{\beta}_{\rm datA}\frac{f}{\tilde{K}_{\rm D}^{\rm datA}+f} - \tilde{\alpha}_{\rm l} \,[l] \, \frac{1-f}{\tilde{K}_{\rm D}^{\rm l} + 1 - f} \right) 
	\label{eq:switch_lipid_ori_SI}
	\end{align}
Clearly, a lipid-mediated activation rate that scales with the origin density tends to renormalize the \textit{datA} de-activation rate: it indeed tends to lower the effective deactivation rate. Importantly, at the mathematical level, the system therefore becomes similar to the LD model in which the lipids have been taken out completely, with one activation rate that is independent of the origin density and one effective deactivation rate that scales with the origin density. This effective deactivation rate, from \textit{datA} and the lipids, needs to balance the activation rate from protein synthesis. And since even at high growth rates, the protein synthesis rate is relatively low as compared to the measured \textit{datA} deactivation rate (see \cite{Kasho2013} and Table \ref{tab:LD_LDDR_parameters}), the effective deactivation rate must be low as well in order to yield an initiation volume that is consistent with experiments. This yields oscillations of low amplitude, similar to the oscillations of the lipid-independent model (Fig. \ref{fig:SI_LD_only_synthesis} B).

We have also considered an LDDR model in which the lipid-mediated activation rate is proportional to the origin density. However, precisely as in the LD model, this merely renormalizes, i.e. lowers, the \textit{datA} deactivation rate. The system therefore becomes mathematically similar to the LDDR model in which lipid-mediated DnaA activation is taken out entirely, with only one term that is independent of the origin density, namely DnaA activation via protein synthesis; and, again, because the protein synthesis rate is low, all other activation and deactivation rates must be low. The system therefore only exhibits very weak oscillations in the fraction of active DnaA, very similar indeed to those of the lipid-independent model  (Fig. \ref{fig:SI_LD_only_synthesis} C/D, dotted lines in third panel).

For completeness, we have also considered a full model with a lipid-mediated DnaA activation rate that is proportional to the origin-density, but as found for the LD and LDDR model, the results are similar to the full model with the lipids taken out completely (Fig. \ref{fig:SI_LD_only_synthesis} E/F, dotted lines in third panel). Indeed, the full model is surprisingly robust to taking out the lipids fully or making its effect scale with the origin density, although the precision of replication initiation is highest when the effect of the lipids is independent of the origin density. Lipids can thus enhance replication initiation by promoting the exchange of DnaA-bound ADP for ATP, but only if the associated activation rate is independent of the origin density.

\end{document}